\documentclass[twocolumn]{aastex631}

\usepackage{amsmath}
\usepackage[caption=false]{subfig}
\usepackage{graphicx}
\usepackage{placeins}
\usepackage{multirow}

\newcommand{\MgII}{\ion{Mg}{2}}
\newcommand{\HeII}{\ion{He}{2}}
\newcommand{\CIII}{\ion{C}{3}]}
\newcommand{\CIV}{\ion{C}{4}}
\newcommand{\OIII}{\ion{O}{3}]}

\begin{document}

\title{EPOCHS III: Unbiased UV continuum slopes at $6.5<z<13$ from combined PEARLS GTO and public JWST NIRCam imaging}

\author[0000-0003-0519-9445]{Duncan Austin}
\affiliation{Jodrell Bank Centre for Astrophysics, Alan Turing Building, University of Manchester, Oxford Road, Manchester M13 9PL, UK}

\author[0000-0003-1949-7638]{Christopher J. Conselice}
\affiliation{Jodrell Bank Centre for Astrophysics, Alan Turing Building, University of Manchester, Oxford Road, Manchester M13 9PL, UK}
    
\author[0000-0003-4875-6272]{Nathan J. Adams}
\affiliation{Jodrell Bank Centre for Astrophysics, Alan Turing Building, University of Manchester, Oxford Road, Manchester M13 9PL, UK}

\author[0000-0002-4130-636X]{Thomas Harvey}
\affiliation{Jodrell Bank Centre for Astrophysics, Alan Turing Building, University of Manchester, Oxford Road, Manchester M13 9PL, UK}

\author[0009-0009-8105-4564]{Qiao Duan}
\affiliation{Jodrell Bank Centre for Astrophysics, Alan Turing Building, University of Manchester, Oxford Road, Manchester M13 9PL, UK}

\author[0000-0002-9081-2111]{James Trussler}
\affiliation{Jodrell Bank Centre for Astrophysics, Alan Turing Building, University of Manchester, Oxford Road, Manchester M13 9PL, UK}

\author[0000-0002-3119-9003]{Qiong Li}
\affiliation{Jodrell Bank Centre for Astrophysics, Alan Turing Building, University of Manchester, Oxford Road, Manchester M13 9PL, UK}

\author[0009-0003-7423-8660]{Ignas Juodžbalis}
\affiliation{Jodrell Bank Centre for Astrophysics, Alan Turing Building, University of Manchester, Oxford Road, Manchester M13 9PL, UK}
\affiliation{Kavli Institute for Cosmology, University of Cambridge, Cambridge, Madingley Road, Cambridge, CB3 0HA}

\author[0000-0003-2000-3420]{Katherine Ormerod}
\affiliation{Jodrell Bank Centre for Astrophysics, Alan Turing Building, University of Manchester, Oxford Road, Manchester M13 9PL, UK}
\affiliation{Astrophysics Research Institute, Liverpool John Moores University, 146 Brownlow Hill, Liverpool, L3 5RF}

\author[0000-0002-8919-079X]{Leonardo Ferreira}
\affiliation{Department of Physics \& Astronomy, University of Victoria, Finnerty Road, Victoria, British Columbia, V8P 1A1, Canada}

\author[0009-0008-8642-5275]{Lewi Westcott}
\affiliation{Jodrell Bank Centre for Astrophysics, Alan Turing Building, University of Manchester, Oxford Road, Manchester M13 9PL, UK}

\author[0009-0005-0817-6419]{Honor Harris}
\affiliation{Jodrell Bank Centre for Astrophysics, Alan Turing Building, University of Manchester, Oxford Road, Manchester M13 9PL, UK}

\author[0000-0003-3903-6935]{Stephen M. Wilkins}
\affiliation{Astronomy Centre, Department of Physics and Astronomy, University of Sussex, Brighton, BN1 9QH, UK}

\author[0000-0003-0883-2226]{Rachana Bhatawdekar}
\affiliation{European Space Agency (ESA), European Space Astronomy Centre (ESAC), Camino Bajo del Castillo s/n, 28692 Villanueva de la Cañada, Madrid, Spain}

\author[0000-0002-6089-0768]{Joseph Caruana}
\affiliation{Department of Physics, University of Malta, Msida MSD 2080, Malta}
\affiliation{Institute of Space Sciences \& Astronomy, University of Malta, Msida MSD 2080, Malta}

\author[0000-0001-7410-7669]{Dan Coe} 
\affiliation{Space Telescope Science Institute, 3700 San Martin Drive, Baltimore, MD 21218, USA}
\affiliation{Association of Universities for Research in Astronomy (AURA) for the European Space Agency (ESA), STScI, Baltimore, MD 21218, USA}
\affiliation{Center for Astrophysical Sciences, Department of Physics and Astronomy, The Johns Hopkins University, 3400 N Charles St. Baltimore, MD 21218, USA}

\author[0000-0003-3329-1337]{Seth H. Cohen} 
\affiliation{School of Earth and Space Exploration, Arizona State University,
Tempe, AZ 85287-1404, USA}

\author[0000-0001-9491-7327]{Simon P. Driver} 
\affiliation{International Centre for Radio Astronomy Research (ICRAR) and the
International Space Centre (ISC), The University of Western Australia, M468,
35 Stirling Highway, Crawley, WA 6009, Australia}

\author[0000-0002-9816-1931]{Jordan C. J. D'Silva} 
\affiliation{International Centre for Radio Astronomy Research (ICRAR) and the
International Space Centre (ISC), The University of Western Australia, M468,
35 Stirling Highway, Crawley, WA 6009, Australia}
\affiliation{ARC Centre of Excellence for All Sky Astrophysics in 3 Dimensions
(ASTRO 3D), Australia}

\author[0000-0003-1625-8009]{Brenda Frye} 
\affiliation{Department of Astronomy/Steward Observatory, University of Arizona, 933 N Cherry Ave,
Tucson, AZ, 85721-0009, USA}

\author[0000-0001-6278-032X]{Lukas J. Furtak}
\affiliation{Physics Department, Ben-Gurion University of the Negev, P.O. Box 653, Be'er-Sheva 84105, Israel}

\author[0000-0001-9440-8872]{Norman A. Grogin} 
\affiliation{Space Telescope Science Institute,
3700 San Martin Drive, Baltimore, MD 21218, USA}

\author[0000-0001-6145-5090]{Nimish P. Hathi}
\affiliation{Space Telescope Science Institute, 3700 San Martin Drive, Baltimore, MD 21218, USA}

\author[0000-0002-4884-6756]{Benne W. Holwerda}
\affiliation{Department of Physics and Astronomy, University of Louisville, Natural Science Building 102, Louisville KY 40292, USA}

\author[0000-0003-1268-5230]{Rolf A. Jansen} 
\affiliation{School of Earth and Space Exploration, Arizona State University, Tempe, AZ 85287-1404, USA}

\author[0000-0002-6610-2048]{Anton M. Koekemoer} 
\affiliation{Space Telescope Science Institute,
3700 San Martin Drive, Baltimore, MD 21218, USA}

\author[0000-0001-6434-7845]{Madeline A. Marshall} 
\affiliation{National Research Council of Canada, Herzberg Astronomy \&
Astrophysics Research Centre, 5071 West Saanich Road, Victoria, BC V9E 2E7,
Canada}
\affiliation{ARC Centre of Excellence for All Sky Astrophysics in 3 Dimensions
(ASTRO 3D), Australia}

\author[0000-0001-6342-9662]{Mario Nonino} 
\affiliation{INAF-Osservatorio Astronomico di Trieste, Via Bazzoni 2, 34124
Trieste, Italy} 

\author[0000-0002-6150-833X]{Rafael {Ortiz~III}} 
\affiliation{School of Earth and Space Exploration, Arizona State University,
Tempe, AZ 85287-1404, USA}

\author[0000-0003-3382-5941]{Nor Pirzkal} 
\affiliation{Space Telescope Science Institute,
3700 San Martin Drive, Baltimore, MD 21218, USA}

\author[0000-0003-0429-3579]{Aaron Robotham} 
\affiliation{International Centre for Radio Astronomy Research (ICRAR) and the
International Space Centre (ISC), The University of Western Australia, M468,
35 Stirling Highway, Crawley, WA 6009, Australia}

\author[0000-0003-0894-1588]{Russell E. Ryan, Jr.} 
\affiliation{Space Telescope Science Institute,
3700 San Martin Drive, Baltimore, MD 21218, USA}

\author[0000-0002-7265-7920]{Jake Summers} 
\affiliation{School of Earth and Space Exploration, Arizona State University, Tempe, AZ 85287-1404, USA}

\author[0000-0001-9262-9997]{Christopher N. A. Willmer} 
\affiliation{Steward Observatory, University of Arizona,
933 N Cherry Ave, Tucson, AZ, 85721-0009, USA}

\author[0000-0001-8156-6281]{Rogier A. Windhorst}
\affiliation{School of Earth and Space Exploration, Arizona State University,
Tempe, AZ 85287-1404, USA}

\author[0000-0001-7592-7714]{Haojing Yan} 
\affiliation{Department of Physics and Astronomy, University of Missouri,
Columbia, MO 65211, USA}

\author[0000-0003-1096-2636]{Erik Zackrisson}
\affiliation{Observational Astrophysics, Department of Physics and Astronomy, Uppsala University, Box 516, SE-751 20 Uppsala, Sweden}
\affiliation{Swedish Collegium for Advanced Study, Linneanum, Thunbergsv\"a{}gen 2, SE-752 38 Uppsala, Sweden}



\begin{abstract}

We present an analysis of rest-frame UV continuum slopes, $\beta$, using a sample of 1011 galaxies at $6.5<z<13$ from the EPOCHS photometric sample collated from the GTO PEARLS and public ERS/GTO/GO (JADES, CEERS, NGDEEP, GLASS) \emph{JWST} NIRCam imaging across $178.9~\mathrm{arcmin}^2$ of unmasked blank sky. We correct our UV slopes for the photometric error coupling bias using $200,000$ power law SEDs for each $\beta=\{-1,-1.5,-2,-2.5,-3\}$ in each field, finding biases as large as $\Delta\beta\simeq-0.55$ for the lowest SNR galaxies in our sample. Additionally, we simulate the impact of rest-UV line emission (including Ly$\alpha$) and damped Ly$\alpha$ systems on our measured $\beta$, finding biases as large as $0.5-0.6$ for the most extreme systems. We find a decreasing trend with redshift of $\beta=-1.51\pm0.08-(0.097\pm0.010)\times z$, with potential evidence for Pop.~III stars or top-heavy initial mass functions (IMFs) in a subsample of 68 $\beta+\sigma_{\beta}<-2.8$ galaxies. At $z\simeq11.5$, we measure an extremely blue $\beta(M_{\mathrm{UV}}=-19)=-2.73\pm0.06$, deviating from simulations, indicative of low-metallicity galaxies with non-zero Lyman continuum escape fractions $f_{\mathrm{esc, LyC}}\gtrsim0$ and minimal dust content.
The observed steepening of $\mathrm{d}\beta/\mathrm{d}\log_{10}(M_{\star} / \mathrm{M}_{\odot})$ from $0.22\pm0.02$ at $z\simeq7$ to $0.81\pm0.13$ at $z\simeq11.5$ implies that dust produced in core-collapse supernovae (SNe) at early times may be ejected via outflows from low mass galaxies.
We also observe a flatter $\mathrm{d}\beta/\mathrm{d}M_{\mathrm{UV}}=0.03\pm0.02$ at $z\simeq7$ and a shallower $\mathrm{d}\beta/\mathrm{d}\log_{10}(M_{\star} / \mathrm{M}_{\odot})$ at $z<11$ than seen by \emph{HST}, unveiling a new population of low mass, faint, galaxies reddened by dust produced in the stellar winds of asymptotic giant branch (AGB) stars or carbon-rich Wolf-Rayet binaries.

\end{abstract}



\section{Introduction} \label{sec:intro}

The first year of study with the James Webb Space Telescope \citep[\emph{JWST},][]{Gardner2023} has led to the discovery of a multitude of high-redshift galaxy candidates in the first billion years of cosmic evolution \citep{Castellano2022, Finkelstein2022, Harikane2022, Naidu2022, Yan2022, Adams2023, Atek2023-SMACS, Austin2023, Bouwens2023, Donnan2023, Leung2023} discovered in deep near-infrared (NIR) imaging using the Near-InfraRed Camera \citep[NIRCam,][]{Rieke2005, Rieke2023} instrument. In addition, several studies including \citet{Tang2023, Fujimoto2023, Nakajima2023}; and \citet{Bunker2023-NIRSpec} have spectroscopically confirmed galaxies using JWST's Near InfraRed Spectrograph (NIRSpec) Micro Shutter Assembly \citep[MSA,][]{Ferruit2022, Jakobsen2022, Rawle2022, Boker2023}, with the most distant sources found at $z\sim12-13$ \citep[e.g.][]{Curtis-Lake2023, Wang2023, Castellano2024}.

In this early JWST data, measurements of the ultraviolet luminosity function (UV LF) have revealed an overabundance in intrinsically UV bright sources at $z>10$ \citep{Adams2023-EPOCHS-II, Donnan2023, Finkelstein2022-CEERSKP1, Bouwens2022c, PerezGonzalez2023, Castellano2023-GLASS-XIX, McLeod2023, Leung2023, Finkelstein2023,Donnan2024}, which may be explained by either the mapping of UV luminosity to host halo mass \citep[e.g.][]{Mason2023}, an increased star formation efficiency \citep[SFE, e.g.][]{Inayoshi2022, Mirocha2023}, or minimal dust obscuration at high redshift \citep{Ferrara2023, Ziparo2023}. Further investigation into the rest-frame UV properties of these galaxies is therefore of the utmost importance to help distinguish between these plausible scenarios.

The rest-frame UV continuum slope, $\beta$, is commonly calculated via the power law $f_{\lambda}\propto\lambda^{\beta}$ \citep{Calzetti1994} (henceforth C94), and is a key diagnostic for understanding the properties of the stellar continuum, providing a tracer of massive O/B-type main sequence stars and their surrounding HII nebular regions. It is primarily dependent on the galactic dust content, meaning the UV dust attenuation ($A_{\mathrm{UV}}$) can be estimated from $\beta$ via the empirical \citet{Meurer1999} relation, $A_{\mathrm{UV}}=4.43 + 1.99\beta$. The $100~\mathrm{Myr}$ star formation rates (SFRs) of star-forming galaxies (SFGs) can therefore be estimated from dust-corrected intrinsic UV magnitudes, $M_{\mathrm{UV}}$, under an assumed luminosity to SFR conversion \citep[$\kappa_{\mathrm{UV}}$, see][]{Madau2014}. Additionally, $\beta$ has a more minor dependence on stellar ages and metallicities \citep[see e.g.][]{Bouwens2012, Tacchella2022}, especially in bursty star formation history (SFH) models, evidence for which has been observed in a spectroscopically confirmed $z=7.3$ galaxy by \citet{Looser2023}.

Since $\beta$ is closely related to the dust content in galaxies, we can use it to study the global build up of galactic dust. It is expected that the dominant dust production mechanism at early times is nucleation in the ejecta of core-collapse supernovae \citep[SNe, e.g.][]{Todini2001, Bianchi2009, Marassi2019}, although dust destruction processes such as sputtering, sublimation, and grain-grain collisions in the resulting reverse shock are not well constrained even in the local Universe \citep[e.g.][]{Bocchio2016, Micelotta2018, Kirchschlager2019, Kirchschlager2024}. As well as core-collapse supernovae (SNe), dust production also occurs in the stellar winds of $0.8-8~\mathrm{M}_{\odot}$ asymptotic giant branch (AGB) stars \citep[e.g.][]{Zhukovska2008, Gail2009, Hofner2018}, $12-30~\mathrm{M}_{\odot}$ red super giant (RSG) stars \citep[][]{Levesque2006}, and $\gtrsim30~\mathrm{M}_{\odot}$ Wolf-Rayet (WR) stars that are carbon-rich (WC stars) with an OB companion \citep[e.g.][]{Lau2020, Lau2022}, although the accompanying SN is expected to very quickly destroy dust produced in both RSGs and WCs. Dust reprocessing is also expected in the interstellar medium (ISM) via astration and depletion \citep[e.g.][]{Draine1979, Draine2009, Draine2011}, although these processes are not yet fully understood (see \citet{Schneider2023} for a  thorough review). The dust chemistry and grain size distribution impact the extinction curve shape \citep{Salim2020}, which is often assumed to follow the \citet{Calzetti2000}, LMC \citep{Gordon2003}, or SMC \citep{Pei1992} laws. Recent results regarding the relation between IR excess ($\mathrm{IRX}=\log_{10}(L_{\mathrm{IR}}/L_{\mathrm{UV}})$) and $\beta$ (the IRX-$\beta$ relation) from the Atacama Large Millimeter/submillimeter Array (ALMA) Reionization Era Bright Emission Line Survey \citep[REBELS,][]{Bouwens2022} data have, however, demonstrated the suitability of a ``Calzetti-like'' attenuation curve at $z\simeq7$ \citep{Bowler2024}.

Prior to the onset of JWST, much effort had been made to measure $\beta$ in large, unbiased, photometric galaxy samples at $z \sim 4-10$ collated from deep imaging by Hubble Space Telescope's (HST) Advanced Camera for Surveys (ACS), Near Infrared Camera and Multi-Object Spectrometer \citep[NICMOS, e.g.][]{Hathi2008}, and Wide Field Camera 3 (WFC3) Infra-Red (IR) instruments. The most notable work at $z=7-10$ has been conducted with WFC3-IR, where $\beta$ is measured either directly from F105W (Y-band), F125W (J-band), F140W (JH-band) and F160W (H-band) colors \citep{Bouwens2009, Bouwens2010, Finkelstein2010, McLure2011, Dunlop2012, Dunlop2013, Bouwens2014a, Wilkins2016, Bhatawdekar2021}, or from the best-fitting SED template \citep[e.g.][]{Finkelstein2012} in the 10 C94 filters designed to omit UV absorption/emission features present in galaxy spectra. These studies find relatively red $\beta\sim-1.9$ at $z\simeq4$ \citep[e.g.][]{Hathi2013} and moderately blue $\beta\sim -2.2 / -2.4$ at $z\simeq7$ in galaxies with $M_{\mathrm{UV}}\simeq-19$, symbolic of low-metallicity, relatively dust-free systems, corroborating predictions from simulations \citep{Wilkins2012, Wilkins2013, GonzalezPerez2013}.

More recently, the redder JWST/NIRCam filters have given access to the rest-frame UV at $z\gtrsim10$, allowing for both the first $\beta$ measurements at these redshifts and improved $\beta$ measurements from vastly deeper rest-frame NUV/optical coverage for galaxies at $z\gtrsim5$. Photometric $\beta$ measurements have been made in abundance with JWST \citep{Topping2022, Cullen2023a, Nanayakkara2023, Morales2023}, with \citet{Topping2023} and \citet{Cullen2023b} observing, on average, extremely blue UV slopes at $z\gtrsim10$, implying little-to-no dust presence at these epochs. In addition, JWST studies have found tentative evidence for $\beta<-3$ in \textit{individual} galaxies at $z\gtrsim10$ \citep[e.g.][]{Atek2023-SMACS, Austin2023, Cullen2023a, Furtak2023-SMACS}, although these may be a result of either photometric scatter or the known $\beta$ bias from the increased selectability of faint blue galaxies compared to their redder counterparts \citep[see e.g.][]{Rogers2013}. These ultra-blue sources may, however, provide evidence for exotic Pop.~III stellar populations \citep{Zackrisson2011} and/or top-heavy/bottom-light IMFs \citep[e.g.][]{Rasmussen2023, Steinhardt2023-HOT-IMFs}. As well as this, these blue galaxies require high Lyman Continuum (LyC) escape fractions, $f_{\mathrm{esc, LyC}}$, needed to adequately reduce the reddening of the continuum by free-free, free-bound and two-photon nebular emission \citep{Schaerer2002, Schaerer2003, Raiter2010, Byler2017, Topping2022}. These $f_{\mathrm{esc, LyC}}$ are unfortunately impossible to measure directly due to the opacity of the intergalactic medium (IGM) in the Epoch of Reionization (EoR). Several indirect tracers have been suggested including blue UV $\beta$ slopes \citep{Zackrisson2013, Chisholm2022}, low [OIII]+H$\beta$ equivalent widths \citep[EWs,][]{Topping2022, Endsley2023}, small sizes \citep{Mascia2023-GLASS}, strong {\MgII}-$\lambda\lambda2796,2803$ \citet{Chisholm2020, Katz2022}, etc, although it is likely that a combination of these are required to accurately measure $f_{\mathrm{esc, LyC}}$ \citep[e.g.][]{Choustikov2023}. This is necessary to estimate the relative importance of low mass galaxies, which are expected to contribute a significant fraction of the total ionizing photon budget of galaxies needed to reionize the neutral IGM by $z\simeq6$ \citep[see e.g.][]{Bouwens2015-EoR, Planck2016, Planck2020}.

In this work we use the EPOCHS v1 sample photometric galaxy sample, presented in EPOCHS-I (Conselice et al. in prep.), comprised from deep public and PEARLS GTO \citep[PI: R. Windhorst \& H. Hammel, PIDs: 1176 \& 2738,][]{Windhorst2023} JWST/NIRCam imaging. The content of this paper is as follows. We outline our data sources and cataloguing procedure in \autoref{sec:data} before calculating UV properties in \autoref{sec:calculating_UV_properties}. $\beta$ biases are discussed in \autoref{sec:PL_beta_bias} and $\beta$ scaling relations are presented in \autoref{sec:Results}. We discuss the dust implications and possibility of exotic Pop.~III and top-heavy IMF scenarios in \autoref{sec:Discussion} before concluding in \autoref{sec:conclusions}. 

$\Lambda$CDM cosmological parameters of $\Omega_m=0.3$, $\Omega_{\Lambda}=0.7$, $H_0=70\text{ kms}^{-1}\text{Mpc}^{-1}$, and AB magnitudes \citep{Oke74, Oke83} are assumed throughout this analysis.

\section{Data and cataloguing} \label{sec:data}

\subsection{PEARLS imaging}
\label{sec:pearls_imaging}

In this work, we utilize blank field JWST/NIRCam photometric imaging from the Prime Extragalactic Areas for Reionization Science \citep[PEARLS; PI: R. Windhorst \& H. Hammel, PIDs: 1176 \& 2738,][]{Windhorst2023} GTO survey. This includes 3 NIRCam parallel fields surrounding the Hubble Frontier Field \citep[HFF,][]{Lotz2017} MACS J0416.1-2403 (hereafter MACS-0416) lensing cluster at $z=0.396$ and the El Gordo \citep[$z=0.87$,][]{Menanteau2012} cluster parallel. In addition, we also include data from the 4 spokes of the North Ecliptic Pole Time Domain Field \citep[NEP-TDF,][]{Jansen2018}. As part of the PEARLS survey design, these fields contain imaging in 4 short wavelength (SW; F090W, F115W, F150W, F200W) and 4 long-wavelength (LW; F277W, F356W, F410M, F444W) NIRCam filters. In addition, we also include bluer F606W imaging from the HST ACS Wide Field Channel (WFC) in the NEP-TDF from the GO-15278 (PI: R.~Jansen) and GO-16252/16793 \citep[PIs: R.~Jansen \&
N.~Grogin, see][]{OBrien2024} HST programs. At the time of writing, the first 2 NEP-TDF spokes, the first epoch of MACS-0416, and the El Gordo cluster are publicly available from the Minkulski Archive for Space Telescopes (MAST)\footnote{An overview of PEARLS data and papers, as well as access to both independently reduced NIRCam imaging and cluster lens models are available at \url{https://sites.google.com/view/jwstpearls}.}.

\subsection{Public ERS and GO imaging}
\label{sec:public_imaging}

As part of the EPOCHS sample, we include public ERS and GO imaging from the Grism Lens-Amplified Survey from Space \citep[GLASS; PI: T. Treu, PID: 1324,][]{Treu2022-GLASS}, data release 1 of deep JWST Advanced Deep Extragalactic Survey imaging in GOODS-South \citep[JADES-Deep-GS; PI: D. Eisenstein, PID: 1180,][]{Eisenstein2023,Rieke2023}, Cosmic Evolution Early Release Science \citep[CEERS; PI: S. Finkelstein, PID: 1345,][]{Bagley2023-CEERS} and Next Generation Deep Extragalactic Exploratory Public \citep[NGDEEP; PIs: S. Finkelstein, C. Papovich \& N. Pirzkal, PID: 2079,][]{Bagley2023-NGDEEP} surveys. Photometric NIRCam imaging is available in F115W, F150W, F200W, F277W, F356W, and F444W for every survey. Where available we include the F090W (GLASS and JADES) wideband filter and the F410M (CEERS and JADES) and F335M (JADES) medium band NIRCam filters. 

In addition, we also include bluer data from HST ACS/WFC for CEERS and NGDEEP to somewhat compensate for the lack of blue F090W NIRCam photometry. We use ACS/WFC data in the F606W and F814W filters from the Cosmic Assembly Near-IR Deep Extragalactic Legacy Survey \citep[CANDELS; PIs: S. Faber \& H. Ferguson, DOI: 10.17909/T94S3X, ][]{Grogin2011-CANDELS,Koekemoer2011-CANDELS} covering the Extended Groth Strip \citep[EGS,][]{Groth1994} for CEERS, and deep F435W, F606W, and F814W ACS/WFC data from v2.5 of the Hubble Legacy Fields \citep[HLF,][]{Whitaker2019-HLF} covering the HUDF-Par2 for NGDEEP.

\subsection{NIRCam data reduction pipeline}
\label{sec:nircam_data_reduction}

Our data reduction pipeline is similar to that of \citet{Ferreira2022, Adams2023}; and \citet{Austin2023}, and is identical to that used in the rest of the EPOCHS series. A detailed account of this pipeline is presented in \citet{Adams2023-EPOCHS-II} (henceforth EPOCHS-II), which is summarized below.

We use version 1.8.2 of the official JWST data reduction pipeline \citep{Bushouse2022}\footnote{\url{https://github.com/spacetelescope/jwst}} and Calibration Reference Data System (CRDS) v1084, which includes improved in-flight LW flat fielding which dramatically deepens these images compared to CRDS v0995. Between stage 1 and 2 we subtract templates of `wisps' in F150W and F200W (SW artefacts produced by reflection off the top secondary mirror support strut, which are most prominent in the B4 detector) using the official STScI templates.\footnote{\url{https://stsci.app.box.com/s/1bymvf1lkrqbdn9rnkluzqk30e8o2bne}} We do not remove `claws' (artefacts from nearby bright foreground stars within the susceptibility region) in the NIRCam imaging, but instead mask these post reduction. After stage 2, we apply Chris Willott's $1/f$ noise correction.\footnote{See \url{https://github.com/chriswillott/jwst}} Before stage 3 we perform background subtraction on individual `cal.fits' frames, which consists of an initial flat background subtraction followed by a further 2D background subtraction using {\tt photutils} \citep{Bradley2022}. This skips the sky subtraction step in stage 3 and allows for quicker background subtraction assessment and fine-tuning. Post stage 3, we use the {\tt tweakreg} part of the DrizzlePac\footnote{\url{https://github.com/spacetelescope/drizzlepac}} python package to first align the F444W image onto the Gaia-DR3 \citep[Data Release 3,][]{GAIA-DR2,GAIA-DR3} derived World Co-ordinate System (WCS) before matching all remaining filters to this WCS. We finally pixel-match to the F444W image with the use of {\tt astropy reproject} \citep{Hoffmann2021}\footnote{\url{https://reproject.readthedocs.io/en/stable/}} to make drizzled images with a pixel scale of $0\farcs03$/pixel.

\subsection{Catalogue creation and photo-\texorpdfstring{$z$}{z}'s}
\label{sec:catalogues_and_photo_zs}

To detect sources in each field, we perform forced photometry in apertures detected in the inverse-variance weighted stacked F277W, F356W, and F444W reduced images using {\tt SExtractor} \citep{Bertin1996-SExtractor} with the setup given in EPOCHS-II to produce an initial photometric catalogue. This includes band fluxes and magnitudes in both Kron \citep[``FLUX\_AUTO'' and ``MAG\_AUTO'',][]{Kron1980} and 0\farcs32 diameter circular (``FLUX\_APER'' and ``MAG\_APER'') apertures as well as an estimate of the half light radius (``FLUX\_RADIUS''), among other quantities. Once photometric fluxes have been measured for our JWST/NIRCam images, we correct for the missing flux that is spread beyond our aperture by the point spread function (PSF) using the simulated {\tt WebbPSF} \citep{Perrin2012-WebbPSF, Perrin2014-WebbPSF} PSFs. Our $0\farcs32$ apertures contain approximately $70-80\%$ of the total point source flux to get the best balance of depth, number of pixels used and minimize the aperture correction obtained from the potentially uncertain WebbPSF models.

For the Hubble images with differing dimensions to our reduced JWST imaging, we use the {\tt photutils} \citep{Bradley2022} python module to perform forced photometry in the same $0\farcs32$ diameter circular apertures from the F277W, F356W, and F444W JWST/NIRCam stack. We also confirm that the fluxes we obtain are consistent with those measured using {\tt sep} \citep[{\tt SExtractor} for python;][]{Barbary2016-sep}. We PSF correct our Hubble ACS/WFC photometry using the PSF models given in \citet{Bohlin2016}.

To calculate image depths in each band, we randomly place $0\farcs32$ diameter apertures in empty regions defined by both the {\tt SExtractor} segmentation map and our band dependent image masks. These image masks are manually created to mask out stellar diffraction spikes, a $50-100$ pixel border near the shallower image edges, as well as any remaining image artefacts, such as stray `snowballs', `wisps', or `claws'. Each aperture used in these depth calculation is forced to not overlap any other aperture as well as being at least $1\arcsec$ away from any source. The flux measurements of the nearest 200 apertures to each detected source are used to estimate a ``local depth'' and hence provide a photometric error defined as the Normalized Median Absolute Deviation (NMAD) of measured fluxes. We choose to adopt this flux error as opposed to the underestimated {\tt SExtractor} output as this more appropriately deals with the correlated noise in the photometric images. To account for potential JWST/NIRCam zero point (ZP) uncertainties, we also impose a $10\%$ minimum flux error in each band. \autoref{tab:5sigma_depths} shows the average $5\sigma$ depths obtained in each NIRCam band for the JWST surveys used in this work. These are calculated as the NMAD of all apertures in empty image regions as opposed to the average local depth to avoid aperture double counting.

We next calculate photometric redshifts (photo-$z$'s) for our EPOCHS sample using the {\tt{EAZY-py}} \citep{Brammer2008-EAZY} spectral energy distribution (SED) fitting code using the standard ``tweak\_fsps'' templates produced using the flexible stellar population synthesis (FSPS) package \citep{Conroy2010} combined with set 1 and 4 from the bluer \citet{Larson2023} template set (referred to henceforth as ``fsps\_larson'') with stronger emission line treatment. The 3 pure stellar templates from \citet{Larson2023} (set 1) use v2.2.1 of the Binary Populations and Spectral Synthesis \citep[BPASS;][]{Eldridge2017-BPASS, Stanway2018-BPASS, Byrne2022-BPASS} stellar population synthesis (SPS) model for ages $\{10^6, 10^{6.5}, 10^7\}$~Gyr at a fixed metallicity $Z_{\star}=0.05~\mathrm{Z}_{\odot}$ while adopting a standard \citet{Chabrier2003} IMF. In addition, template set 4 includes nebular continuum and line emission calculated from {\tt CLOUDY} v17 \citep{Ferland2017-CLOUDY} with Ly$\alpha$ removed, ionization parameter $\log U=-2$, $Z_{\mathrm{gas}}=Z_{\star}$, Lyman continuum escape fraction $f_{\mathrm{esc, LyC}}=0$, $n_\mathrm{H}=300~\mathrm{cm}^{-2}$ gas cloud hydrogen density, and spherical geometry. We run the {\tt EAZY-py} photo-z fitting twice with redshift both free ($0<z<25$) as well as a `low-redshift' run, where the redshift is limited to be $z<6$, in order to determine the $\chi^2$ of the best-fitting low-z solution. This is required for our selection criteria outlined in \autoref{sec:selection} and has previously been performed for high-$z$ galaxy identification in the past \citep[e.g.][]{Hainline2023b}.

\subsection{High-\texorpdfstring{$z$}{z} sample selection}
\label{sec:selection}

One potential source of contamination in our EPOCHS sample, other than low-z galaxy interlopers, comes from Milky Way Y- and T-type brown dwarfs. To account for these, we fit the {\tt Sonora Bobcat} brown dwarf templates \citep{Marley2021-SonoraBobcat}, to our candidate galaxies via least squares regression find the best-fitting solution. In addition, we note that our sample may well contain ``little red dots'' (LRDs), as found by \citep[e.g.][]{Matthee2023, Kokorev2024} which are hypothesized to be active galactic nuclei \citep[AGN; see][]{Labbe2023}. The impact of contamination by these LRDs on the results of this work is presented in \autoref{sec:little_red_dots} and an analysis of the impact on the global stellar mass function (GSMF) is given in EPOCHS-IV \citep{harvey2024epochs}.

To select a robust sample of high redshift galaxies from our photometric catalogues, we use similar selection criteria as used in EPOCHS-I/II/IV, outlined below:
\begin{enumerate}
    \item There must be at least 1 photometric band entirely bluewards of the Ly$\alpha$ break at $\lambda_{\mathrm{rest}}=1216~\mathrm{\AA}$. The limits used here are defined by the upper and lower 50\% transmission limits of the filters taken from the Spanish Virtual Observatory \citep[SVO;][]{Rodrigo2020-SVO} filter profile service. This constrains us to $z\gtrsim6.5$ when using JWST/NIRCam data alone.
    \item Any bands entirely bluewards of the Ly$\alpha$ break must be non-detected at the $3\sigma$ confidence level, with bands straddling the break having no SNR requirements.
    \item The two bands directly redwards (but not straddling) the Ly$\alpha$ break must be detected at $>5\sigma$, and all other redwards NIRCam widebands must be detected at $>2\sigma$. 
    \item The integral of the output {\tt EAZY-py} probability density function (PDF) must have greater than $60\%$ contained within 10\% of the maximum likelihood of the {\tt EAZY-py} redshift posterior.
    \item The best-fit {\tt EAZY-py} (redshift free) solution must satisfy $\chi^2_{\mathrm{red.}} < 3$, and the difference between the `redshift free' and `low redshift' best-fitting solutions must satisfy $\Delta \chi^2>4$.
    \item The {\tt SExtractor} half-light radius must be greater than or equal to $1.5$~pixels in the LW F277W, F356W, and F444W JWST/NIRCam images to ensure the removal of F200W drop-out (dithered) hot-pixels.
    \item If the {\tt SExtractor} half-light radius is smaller than the PSF full width at half maximum (FWHM) in F444W, then the difference between the best-fitting brown dwarf solution and the `redshift free' {\tt EAZY-py} run must satisfy $\Delta \chi^2_{\mathrm{red.}}>4$.
\end{enumerate}

In addition to the above selection criteria, we also remove rogue candidates from our sample by eye. This check was done independently by authors DA, TH, QL, and NA. This splits our sample into ``certain'', ``uncertain'' and ``rejected'' candidates. In total we have 1214 unmasked galaxies and 59 brown dwarf candidates in the full EPOCHS catalogue, which also includes the blank NIRCam parallel fields of the SMACS-0723 and CLIO\footnote{The CLIO cluster is identified at $z=0.42$ within the Galaxy and Mass Assembly \citep[GAMA,][]{Driver2011, Liske2015} survey and imaged with NIRCam as part of the GTO PEARLS JWST program.} strong gravitational lensing clusters. Our ``rejected'' subsample contains 49 sources removed completely by eye ($4\%$ of the full sample) which are mainly sources containing obvious LW hot pixels which just pass our criterion 6 above. This leaves 111 ``uncertain'' galaxy candidates ($9\%$ of the full sample) and 1054 ``certain'' galaxy candidates. These ``uncertain'' candidates mainly comprise sources that are contaminated by stray wisps, claws, or diffuse foreground light. When limiting this to the fields and redshift range used in this work ($6.5<z<13$), and removing all galaxies at $6.5<z<7.5$ in fields without bluer HST ACS/WFC data (El Gordo, MACS-0416, and GLASS) due to the extreme blue $\beta$ biases present (see \autoref{sec:power_law_beta}), we are left with 1011 galaxies in our EPOCHS-III sample covering $178.9~\mathrm{arcmin}^2$ of unmasked blank sky area (excluding CLIO and SMACS-0723 included in the full EPOCHS sample). Full catalogues and an associated README file will be distributed as part of EPOCHS-I.

\begin{table*}[]
    \centering
    \caption{$5\sigma$ depths (in $0\farcs32$ diameter apertures) and areas of the blank field surveys used in the EPOCHS-III sample. These differ from those presented in EPOCHS-I as we limit our redshift range to $z<13$ and also do not use the GTO PEARLS CLIO and ERS SMACS-0723 parallel fields in this analysis. This sample also differs slightly to EPOCHS-II/EPOCHS-IV due to our exclusion of galaxies at $6.5<z<7.5$ in El Gordo, MACS-0416, and GLASS due to the large $\beta$ biases associated with this redshift range. We note that NGDEEP is split into three regions; one has deep HST ACS/WFC coverage (4.03~arcmin$^2$) and another has shallower ACS/WFC coverage (1.28~arcmin$^2$), leaving 0.98~arcmin$^2$ with JWST NIRCam data alone. These HST areas and depths for NGDEEP are collated from the work of \citet{Austin2023}.}
    \setlength{\tabcolsep}{2pt}
    \begin{tabular}{c|cc|ccccccccc|c}
    \toprule
    \multirow{2}{*}{Survey} & \multicolumn{2}{c}{HST ACS/WFC} & \multicolumn{9}{c|}{JWST NIRCam} & Area / \\
    & F606W & F814W & F090W & F115W & F150W & F200W & F277W & F335M & F356W & F410M & F444W & arcmin$^2$ \\
    \hline
    NEP-TDF & 28.74 & - & 28.50 & 28.50 & 28.50 & 28.65 & 29.15 & - & 29.30 & 28.55 & 28.95 & 57.32 \\
    MACS-0416 & - & - & 28.67 & 28.62 & 28.49 & 28.64 & 29.16 & - & 29.33 & 28.74 & 29.07 & 12.30 \\ 
    El Gordo & - & - & 28.23 & 28.25 & 28.18 & 28.43 & 28.96 & - & 29.02 & 28.45 & 28.83 & 3.90 \\
    CEERS P1-8,10 & 28.60 & 28.30 & - & 28.70 & 28.60 & 28.89 & 29.20 & - & 29.30 & 28.50 & 28.85 & 60.31 \\
    CEERS P9 & 28.31 & 28.32 & - & 29.02 & 28.55 & 28.78 & 29.20 & - & 29.22 & 28.50 & 29.12 & 6.08 \\
    NGDEEP & 29.20/30.30 & 28.80/30.95 & - & 29.78 & 29.52 & 29.48 & 30.28 & - & 30.22 & - & 30.22 & 6.29 \\
    JADES-Deep-GS & 29.07 & - & 29.58 & 29.78 & 29.68 & 29.72 & 30.21 & 29.58 & 30.17 & 29.64 & 29.99 & 22.98 \\
    GLASS & - & - & 29.14 & 29.11 & 28.86 & 29.03 & 29.55  & - & 29.61 & - & 29.84 & 9.76 \\
    \botrule
    \end{tabular}
    \label{tab:5sigma_depths}
\end{table*}

\subsection{Completeness and contamination}
\label{sec:Completeness_contamination}

To estimate the completeness and contamination in our sample, we utilize 5 realizations of the JAdes extraGalactic Ultradeep Artificial Realizations \citep[{\tt JAGUAR};][]{Williams2018} mock catalogue of star-forming galaxies (SFGs). The mock photometry in the catalogue is generated from a family of {\tt{BEAGLE}} \citep{Chevallard2016-BEAGLE} SEDs incorporating the \citet{BC03} SPS and \citet{Gutkin2016} nebular emission models. The distribution of redshifts and masses of these SEDs are selected to match the mass functions of \citet{Tomczak2014}, extrapolated to match the $z>4$ HST UV luminosity functions (UV LFs) of \citet{Bouwens2015,Bouwens2016,Calvi2016,Stefanon2017,Oesch2013,Oesch2018}, with $M_{\mathrm{UV}}$ calculated from the $M_{\mathrm{UV}}-M_{\star}$ relations from 3D-HST \citep{Skelton2014,Momcheva2016}. The UV $\beta$ slopes in the catalogue are calculated from the $M_{\mathrm{UV}}-\beta$ relationship given in \citet{Bouwens2009,Bouwens2015}.

For each survey used in this work, we scatter the mock {\tt JAGUAR} photometry within errors set by the $1\sigma$ measured depth with a minimum 10\% flux error in each filter before running through our selection procedure outlined in \autoref{sec:selection}. The absolute magnitude, $M_{\mathrm{UV}}$, and UV continuum slope, $\beta$, were calculated for the selected sample at the fixed {\tt{EAZY}} redshifts (as explained in \autoref{sec:calculating_UV_properties}), with stellar masses calculated with {\tt{Bagpipes}} for the {\tt JAGUAR} catalogue using the setup in \autoref{tab:pipes_priors} taking the same bandpass filters and average depths from the CEERS survey. 

To account for the SED modelling differences between {\tt{BEAGLE}} and {\tt{Bagpipes}}, we calculate a median mass difference, $\langle\Delta M_{\star}\rangle=\log_{10}(M_{\star,\mathrm{obs}}/M_{\star,\mathrm{int}})$, between observed and intrinsic stellar masses in the CEERS {\tt{JAGUAR}} catalogue and apply these scaling factors to estimate observed stellar masses from the {\tt{JAGUAR}} catalogues simulating other EPOCHS-III fields. For our assumed {\tt{Bagpipes}} lognormal SFH, we find the differences between correctly identified high-$z$ SFGs, $\langle\Delta M_{\star,\mathrm{high-}z}\rangle=0.244$, and contaminant Balmer break interlopers, $\langle\Delta M_{\star,\mathrm{interlopers}}\rangle=0.527$. A closer look at the mass differences for this sample using both the {\tt{Bagpipes}} \citep{Carnall2018} and {\tt{Prospector}} \citep{Johnson2021} Bayesian SED fitting tools is presented in EPOCHS-IV, although we note the large dependence of SFH assumption on the stellar masses we measure.

\begin{figure}
    \centering
    \includegraphics[width=0.45\textwidth]{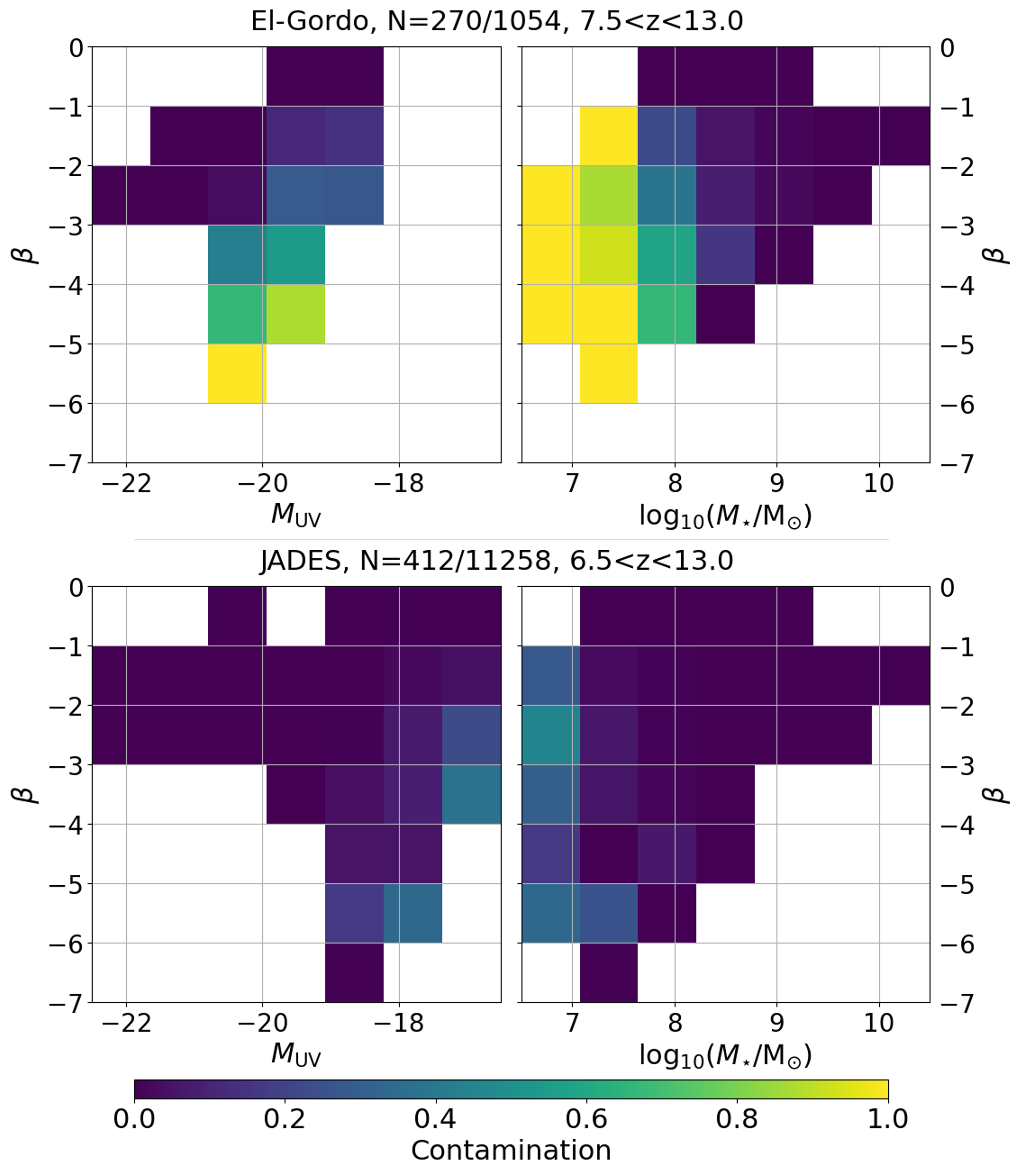}
    \caption{Contamination from lower redshift SFGs in the El Gordo (upper panels) and JADES-Deep-GS (lower panels) fields for the entire redshift range used in this work ($7.5<z<13.0$ for NIRCam only fields and $6.5<z<13.0$ otherwise) as a function of $\Theta=(M_{\mathrm{UV}}, \beta)$ (left panels) and $\Theta=(\log_{10}M_{\star}, \beta)$ (right panels) calculated from the {\tt JAGUAR} mock galaxy catalogue \citep{Williams2018}. The observed $\beta$ plotted here is calculated using the photometric power law method. The total number of selected low-z Balmer break interlopers and selected galaxies is shown in the plot titles. As expected the most contamination occurs in the faintest, lowest mass bins which becomes larger in the bluer bins due to the favouring selection of redder galaxies at these intrinsic magnitudes/masses. JADES-Deep-GS has less contamination in total which occurs in lower mass/fainter bins due to its greater depth ($m_{\mathrm{F277W}}\simeq30.2$) and increased number of filters (HST/ACS\_WFC/F606W and JWST/NIRCam/F335M) compared to El Gordo ($m_{\mathrm{F277W}}\simeq29.0$).}
    \label{fig:contamination}
\end{figure}

For each survey, we calculate the contamination from the respective {\tt JAGUAR} catalogues following
\begin{equation}
    \mathrm{Cont}(\Theta_{\mathrm{obs}})=\frac{N_{\mathrm{selected,interlopers}}(\Theta_{\mathrm{obs}})}{N_{\mathrm{selected}}(\Theta_{\mathrm{obs}})} \mathrm{ ,}
    \label{eq:contamination}
\end{equation}
in bins of $\Theta=\{(M_{\mathrm{UV}}, \beta), (M_{\star}, \beta)\}$ in intrinsic and observed galaxy property frames for completeness and contamination respectively. In theory, the derived quantities in \autoref{eq:contamination} are also redshift dependent, however we decide not to bin the contamination by redshift due to the relatively small size of our {\tt JAGUAR} catalogues. The contamination for our shallowest and deepest surveys (El Gordo and JADES-Deep-GS, where a summary of survey depths is given in EPOCHS-I/II/IV and \autoref{tab:5sigma_depths}), are shown in \autoref{fig:contamination}.

We also use our {\tt JAGUAR} catalogues to provide an estimate of the sample completeness, calculated as
\begin{equation}
    \mathrm{Comp}(\Theta_{\mathrm{int}})=\frac{N_{\mathrm{selected}}(\Theta_{\mathrm{int}})}{N_{\mathrm{total}}(\Theta_{\mathrm{int}})}
    \label{eq:completeness}
\end{equation}
in the parameter spaces $\Theta=\{(z, M_{\mathrm{UV}}, \beta), (z, M_{\star}, \beta)\}$, where we bin the completeness in the redshift bins used in this work ($6.5<z<8.5$, $8.5<z<11$, and $11<z<13$). We plot both $20\%$ completeness contours (red) and our {\tt JAGUAR} simulation limits (black) in \autoref{fig:UV_slope_vs_M_UV_z_split} and \autoref{fig:UV_slope_vs_mass_z_split}. It is noteworthy that although the effective sky area across our 5 {\tt JAGUAR} realizations ($\sim605~\mathrm{arcmin}^2$) far exceeds the sky coverage of our EPOCHS-III galaxy sample, there still exists very few high mass galaxies in our highest redshift bin. This is most-likely due to the rapid fall-off at the high-mass end of the {\tt JAGUAR} GSMF, which could be somewhat resolved by running more {\tt JAGUAR} realizations or using a simulation which more accurately match the high mass end of updated JWST GSMFs. As well as this, the mock catalogues have a limited range of $\beta$ covered almost entirety by our SED fitting template set, making 20\% completeness contours difficult to extend past mere faint, low mass detection limits in the parameter spaces probed for this work.

\section{Calculating UV properties}
\label{sec:calculating_UV_properties}

In this section we discuss the two methods of calculating $\beta$ from the photometric data. The first method calculates $\beta$ via a power law fit ($f_{\lambda}\propto\lambda^{\beta}$) to the wideband photometric fluxes for rest-frame wavelengths $1250<\lambda_{\mathrm{rest}}/\mathrm{\AA}<3000$. The second fits the 10 C94 filters using the same power law to the Bayesian SED template posterior for each source. These are discussed in \autoref{sec:power_law_beta} and \autoref{sec:pipes_beta_calc} respectively. We compare these methods in \autoref{sec:photometric_beta_comparison} and to spectroscopic $\beta$ in \autoref{sec:spectroscopic_beta_comparison}. Since the Bayesian SED fitting method is systematically biased against measuring the bluest $\beta$ due to the implicit prior on $\beta$, we favour the photometric power law method in this work.

\subsection{Calculating \texorpdfstring{$\beta$}{Beta slopes} from photometric fluxes}
\label{sec:power_law_beta}

Since the launch of JWST, we now have deep, high resolution imaging in the Near InfraRed (NIR) redwards of $1.6~\mu\mathrm{m}$, making the direct calculation of $\beta$ using multi-band photometry (as opposed to SED fitting) the preferred method. Where previously we could not calculate $\beta$ in this way at $z\gtrsim8.5$ using only the reddest F160W HST/WFC3IR filter, the vastly improved rest-frame UV coverage from the JWST/NIRCam F200W, F277W, and F356W widebands now makes this possible at these high redshifts.

\begin{figure}
    \centering
    \includegraphics[width=0.45\textwidth]{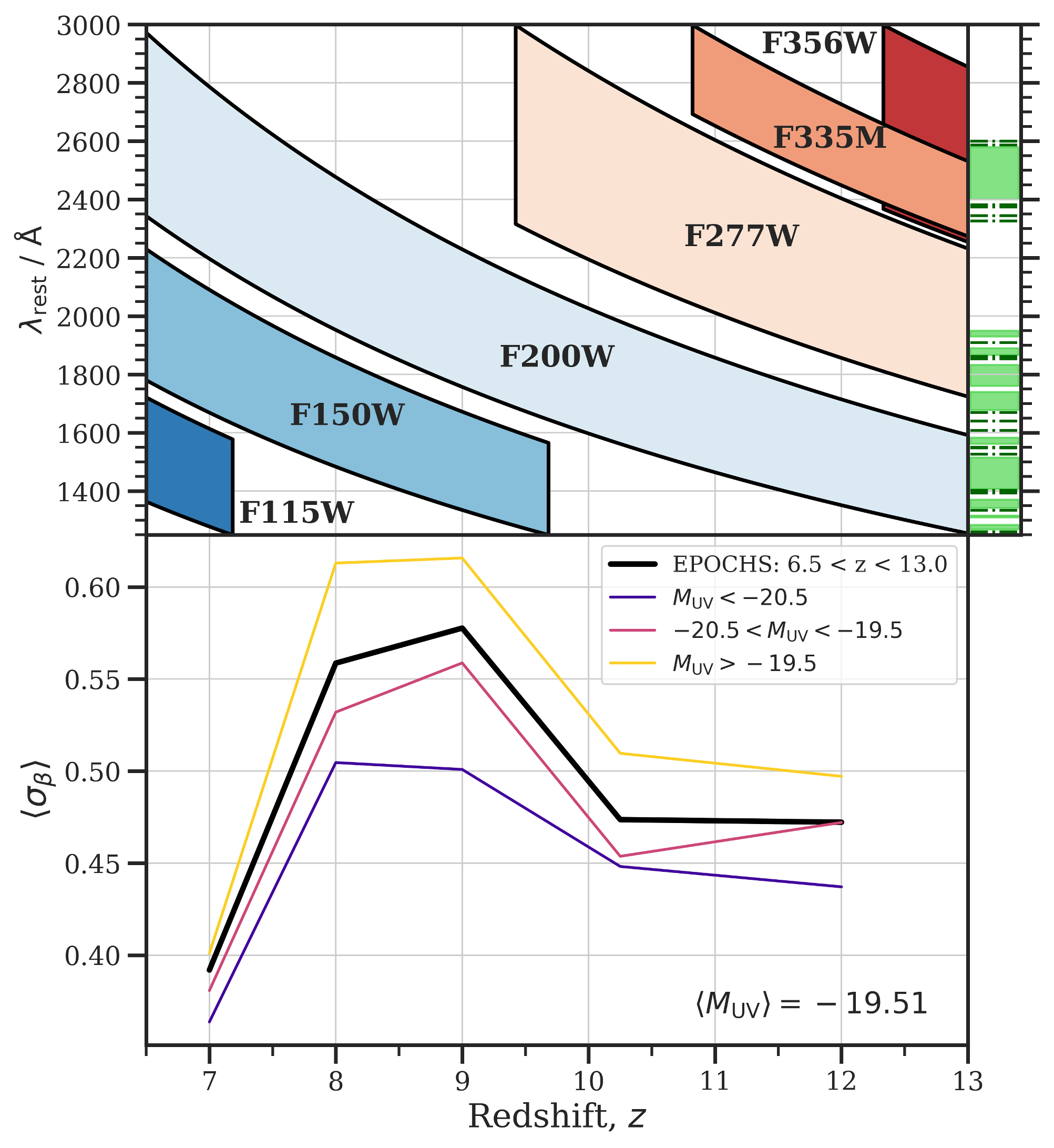}
    \caption{\textbf{Top:} Rest-frame UV coverage of the JWST/NIRCam filters used in this work as a function of both redshift and $\lambda_{\mathrm{rest}}$. The right-hand side shows the rest-wavelength coverage of the 10 C94 filters in shaded lime green which are used to avoid prominent rest frame UV nebular emission lines shown as dash-dotted dark green lines in this same side-plot. \textbf{Bottom:} Average power law measured $\beta$ errors, $\langle \sigma_{\beta} \rangle$ as a function of both redshift and $M_{\mathrm{UV}}$. We plot our EPOCHS-III sample, with $\langle M_{\mathrm{UV}} \rangle = -19.51$, in black as well as in 3 $M_{\mathrm{UV}}$ bins ($M_{\mathrm{UV}}<-20.5$ (purple), $-20.5<M_{\mathrm{UV}}<-19.5$ (pink), and $M_{\mathrm{UV}}>-19.5$ (yellow)). As expected, we observe a trend of decreasing $\sigma_{\beta}$ with increasing intrinsic UV magnitude due to the increased apparent magnitude (and hence decreased photometric errors) of the brighter sources in our sample.}
    \label{fig:Rest_frame_UV_filters}
\end{figure}

In this work, we fit a power law of the form $f_{\lambda}\propto\lambda^{\beta}$ to the rest-frame UV photometry, as done by e.g.\, \citet{Rogers2014, Bouwens2014a} in the HST era and more recently by e.g.\, \citet{Cullen2023a, Topping2022} using early JWST ERS/ERO NIRCam data. We define which filters are used in this fitting procedure as those which fall entirely within $1250<\lambda_{\mathrm{rest}}~/~\mathrm{\AA}<3000$ as given by the same 50\% flux limits as used in our selection criteria in \autoref{sec:selection}, where the rest frame wavelengths are derived from the best-fitting {\tt EAZY-py} redshift. For filter $i$, this is calculated as
\begin{equation}
    f_{\lambda, i} \propto\frac{\int_{0}^{\infty}T_i(\lambda)\lambda^{1 + \beta}\mathrm{d}\lambda}{\int_{0}^{\infty}T_i(\lambda)\lambda\mathrm{d}\lambda}
\end{equation}
in the photon-counting convention, where $T_i$ is the filter transmission profile for filter $i$ taken from the SVO filter profile service. The filters used are plotted as a function of redshift and rest wavelength coverage in the top panel of \autoref{fig:Rest_frame_UV_filters}.

From these power law fits, we calculate $M_{\mathrm{UV}}$ from the bandpass averaged flux in a $100~\mathrm{\AA}$ wide top-hat filter centred on $\lambda_{rest}=1500~\mathrm{\AA}$. We correct these $M_{\mathrm{UV}}$ by a factor $-2.5\log_{10}(\mathrm{``FLUX\_AUTO''/``FLUX\_APER''})$ in the band with effective wavelength closest to $\lambda_{\mathrm{obs}}=1500(1+z)~\mathrm{\AA}$ to account for extended sources when required. Occasionally {\tt{SExtractor}} over-corrects faint extended sources, so in cases where ``FLUX\_AUTO''/``FLUX\_APER''$>10$, we instead switch to using the F444W band to apply the correction which avoids the issue.

In the lower panel of \autoref{fig:Rest_frame_UV_filters} we plot the mean $\beta$ errors, $\langle\sigma_{\beta}\rangle$, as a function of redshift for the entirely of the EPOCHS-III sample (black) and in 3 magnitude bins ($M_{\mathrm{UV}}<-20.5$, $-20.5<M_{\mathrm{UV}}<-19.5$, and $M_{\mathrm{UV}}>-19.5$). Here we outline two main reasons for the trend in $\sigma_{\beta}$ across our sample: the number of JWST/NIRCam bands present in the rest-frame UV and the SNR ratio of each galaxy in these aforementioned filters. From \autoref{fig:Rest_frame_UV_filters} we see that for the majority of our redshift range of interest ($7.2<z<9.4$ and $9.5<z<12.3$) there are only two rest-UV widebands, whereas this increases to three at $6.5<z<7.2$ and $12.3<z<13$. In the JADES-Deep-GS data, the inclusion of the F335M medium band instead gives this dataset three rest-UV filters at $10.8<z<12.3$ and four at $12.3<z<13$. It is clear to see that the increased number of rest-UV bands reduces $\sigma_{\beta}$ for our entire sample from $\sigma_{\beta}\simeq0.4$ at $z<7.2$ to $\sigma_{\beta}\simeq0.55$ at $7.2<z<9.4$ and $\sigma_{\beta}\simeq0.47$ at $z>9.4$. We also note that objects with brighter apparent UV magnitudes have reduced $\sigma_{\beta}$ due to their smaller photometric errors. This is apparent in the lower panel of \autoref{fig:Rest_frame_UV_filters}, where the average for the full sample with $\langle M_{\mathrm{UV}}\rangle=-19.51$ lies between the $-20.5<M_{\mathrm{UV}}<-19.5$ and $M_{\mathrm{UV}}>-19.5$ bins.

\subsection{Calculating \texorpdfstring{$\beta$}{Beta slopes} via Bayesian SED fitting with Bagpipes}
\label{sec:pipes_beta_calc}

In addition to measuring $\beta$ with a power-law fit to the rest-frame UV photometry, we run the {\tt{Bagpipes}} Bayesian SED fitting code \citep{Carnall2018} to measure $\beta$ from the best-fit SED to all of the available photometric fluxes rather than just those in the rest-frame UV. We use the 2016 version of the \citet{BC03} (BC03) SPS models assuming a \citet{Kroupa2001} IMF and \citet{Calzetti2000} dust attenuation law with redshift fixed to that measured by {\tt EAZY-py}. The \citet{Inoue2014} model for InterGalactic Medium (IGM) attenuation is assumed and we do not model the Ly$\alpha$ damping wing due to the uncertain nature of the patchy IGM during reionization. We assume a lognormal SFH parametrized by the timescales for SF onset ($t_{\mathrm{start}}$), SF peak ($t_{\mathrm{max.}}$), and lognormal FWHM. The stellar mass ($M_{\star}$), metallicity ($Z$), V-band dust attenuation ($A_V$), nebular ionization parameter ($U$), and Lyman continuum escape fraction ($f_{\mathrm{esc, LyC}}$), are given wide priors in base 10 logarithmic space (equivalent to uniform priors on the logged parameter). We fix the age of the nebular birth clouds to $t_{\mathrm{BC}}=10~\mathrm{Myr}$ which are also assumed to have no additional dust attenuation (as in the default {\tt Bagpipes} setup). A summary of the fixed and free parameters of our {\tt Bagpipes} fits, as well as the underlying assumptions made about the family of SEDs we are using to perform the analysis, are given in \autoref{appendix:spectroscopy_tab}.

\begin{table}[]
    \centering
    \caption{Summary of fixed and free parameters used in our {\tt Bagpipes} Bayesian SED fitting procedure. Parameter names, descriptions, and prior distributions/limits for the parameters defining the stellar, nebular and dust properties, as well as the lognormal star formation history are outlined. We note that we have recomputed the default {\tt Bagpipes} BC03 SPS grids using {\tt CLOUDY} v17.03 to extend the range of $\log U$ to $-1$.}
    \setlength{\tabcolsep}{2pt}
    \begin{tabular}{cccc}
    \toprule
    Parameter & Prior & Limits/Value & Description \\
    \hline
    \multicolumn{4}{c}{Redshift} \\
    \hline
    $z$ & Fixed & $z_{\mathrm{phot}}$ & {\tt EAZY-py} photo-z \\
    \multicolumn{4}{c}{Star formation history (lognormal)} \\
    \hline
    $t_{\mathrm{start}}$ & $\log_{10}$ & $(1~\mathrm{Myr}, t_U)$ & Time of SF onset \\
    $t_{\mathrm{peak}}$ & $\log_{10}$ & $(10^{-3}, 15)$~Gyr & Time of peak SF \\
    FWHM & $\log_{10}$ & $(10^{-3}, 15)$~Gyr & SFH FWHM \\
    \multicolumn{4}{c}{Stellar properties} \\
    \hline
    $\log_{10}(M_{\star}/\mathrm{M}_{\odot})$ & Uniform & $(5, 12)$ & Stellar mass formed \\
    $\log_{10}(Z_{\star})$ & Uniform & $(-6, 1)$ & Stellar metallicity \\
    $Z_{\mathrm{gas}}$ & Fixed & $Z_{\star}$ & Gas-phase metallicity \\
    \multicolumn{4}{c}{Nebular properties} \\
    \hline
    $\log_{10}(U)$ & Uniform & $(-3, -1)$ & Ionization parameter \\
    \multirow{2}{*}{$f_{\mathrm{esc, LyC}}$} & \multirow{2}{*}{$\log_{10}$} & \multirow{2}{*}{$(0.001, 1)$} & Lyman continuum \\
     & & & escape fraction \\
    $t_{\mathrm{BC}}$ & Fixed & $10$~Myr & Birth cloud lifetime \\
    \multicolumn{4}{c}{Dust properties} \\
    \hline
    $A_V$ & $\log_{10}$ & $(10^{-4}, 10)$ & V-band attenuation \\
    \multirow{2}{*}{$\tau_{\mathrm{BC}}$} & \multirow{2}{*}{Fixed} & \multirow{2}{*}{0} & Birth cloud \\
    & & & optical depth \\
    \botrule
    \end{tabular}
    \label{tab:pipes_priors}
\end{table}

Several physical properties are calculated from these {\tt Bagpipes} runs, including $\beta$, $M_{\mathrm{UV}}$, and $M_{\star}$. $\beta$ is calculated by fitting the same power law function as in \autoref{sec:power_law_beta} to posterior spectrum in the 10 C94 top-hat filters and $M_{\mathrm{UV}}$ is calculated in the same way as in \autoref{sec:power_law_beta}. The posterior on stellar mass, $M_{\star}$, is an output from the base {\tt{Bagpipes}} code which is subsequently corrected for extended sources by ``FLUX\_AUTO''/``FLUX\_APER'' in the F444W band where appropriate which traces the rest-frame optical light.

We use the stellar masses calculated in our {\tt Bagpipes} run when determining the contamination likeliness of each galaxy in \autoref{sec:Completeness_contamination} and when observing trends between $\beta$ and $M_{\star}$ in \autoref{sec:beta_mass}. Whilst we do not consider the majority of the constrained {\tt Bagpipes} parameters in this work, it is noteworthy that the $f_{\mathrm{esc, LyC}}$ values in \autoref{fig:photometric_beta_comparison} are illustrative only and are not well constrained due to their degeneracy with stellar metallicities and galaxy ages derived from the assumed lognormal SFH.

\subsection{Comparison of photometric \texorpdfstring{$\beta$}{Beta slope} methods}
\label{sec:photometric_beta_comparison}

\begin{figure}
    \centering
    \includegraphics[width=0.49\textwidth]{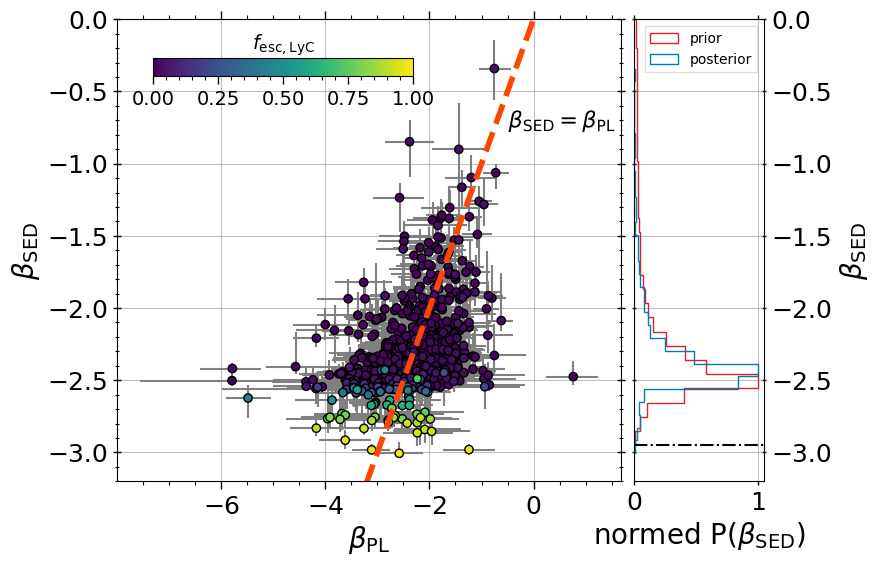}
    \caption{\textbf{Left:} Comparison of photometrically measured $\beta$ slopes for both the SED fitting and bias corrected power law methods for our EPOCHS-III sample colored by {\tt Bagpipes} derived $f_{\mathrm{esc, LyC}}$. $\beta_{\mathrm{PL, corr}}=\beta_{\mathrm{SED}}$ is shown in dashed orange, which does not represent the best-fit to the data. \textbf{Right:} UV slope prior (red) and posterior probability density (blue) for $\beta_{\mathrm{SED}}$. We see that the posterior $\beta_{\mathrm{SED}}$ somewhat follows the prior in that the extreme blue $\beta$ values are largely avoided due to the limited range of parameter space explored by our {\tt Bagpipes} template set. The black dot-dashed line represents the bluest achievable $\beta_{\mathrm{SED}}=-2.95$.}
    \label{fig:photometric_beta_comparison}
\end{figure}

In this section, we compare and contrast our two different photometric methods to measure $\beta$ in order to determine which values to use in the analysis performed in this work. A comparison of $\beta$ measurements (with the power law $\beta$ corrected for the biases explored in \autoref{sec:PL_beta_bias}) is shown in the left-hand panel of \autoref{fig:photometric_beta_comparison}.

The SED fitting method uses all available photometric data and produces more precise values of $\beta$ for individual galaxies due to the mitigation of photometric errors arising when fitting a power law to a small number of flux measurements. In addition, using the SED method means that $\beta$ values are calculated using the same underlying assumptions as the stellar masses, leading to more consistent conclusions. In addition, the power law method is more susceptible to photometric scatter (see \autoref{sec:photo-z_coupling_bias}), rest-frame UV contamination by nebular emission lines (see \autoref{sec:emission_line_bias}), as well as the redshift and filterset-dependent unequal coverage of the rest-frame UV (\autoref{fig:Rest_frame_UV_filters}).

Photometric SED fitting for $\beta$ is the favoured method of \citet{Finkelstein2012, Bhatawdekar2021, Morales2023}, however it has previously been noted that this method is systematically biased by the limited range of $\beta$ allowed by the choice of IMF and SPS model \citep[e.g.][]{Rogers2013, Cullen2023a}. This primarily impacts the range of allowed blue $\beta$ slopes which biases the SED fitted $\beta$ redwards (i.e. less negative values). Whilst our template set reaches as blue as $\beta=-2.95$ for the youngest ($t\simeq1$~Myr), most metal poor ($Z_{\star}=10^{-6}~Z_{\odot}$), dust-free, stellar dominated ($f_{\mathrm{esc, LyC}}=1$) galaxies, the likelihood of measuring blue $\beta$ slopes via SED fitting is impacted by the implicit $\beta$ prior (determined as a complex function of our other input priors outlined in \autoref{tab:pipes_priors}). 

We plot this $\beta$ prior in the right-hand panel of \autoref{fig:photometric_beta_comparison} and compare to the $\beta$ posterior from our SED fitting method. It is clear that while the bluest achievable $\beta$, shown as the black dash-dotted line, is sufficiently blue, the probability of measuring $\beta\lesssim-2.7$ becomes increasingly unlikely using the {\tt{Bagpipes}} setup given in \autoref{tab:pipes_priors}. We therefore choose to adopt the power law photometric $\beta$ method in the rest of the analysis done in this paper as to avoid this curtailing towards bluer $\beta$. We test whether our $\beta$ prior is strongly dependent on SFH by re-running our SED fitting procedure with the ``continuity bursty'' SFH parametrization used in EPOCHS-IV and originally by \citet{Leja2019} and \citet{Tacchella2022}, finding minimal differences in the $\beta$ value at which this prior falls off. Re-running instead using templates derived from BPASS as opposed to BC03 pushes this soft limit bluewards by $\sim0.1-0.2$ and the lowest possible UV slope to $\beta=-3.1$; the value of these $\beta$ limits are therefore impacted most strongly by choice of SPS model.

\subsection{Spectroscopic comparison}
\label{sec:spectroscopic_beta_comparison}

To assess the accuracy of our photometric $\beta$ measurements, we compare to those derived from spectroscopic measurements using the low resolution $R\sim100$ NIRSpec PRISM/CLEAR disperser-filter combination from the DAWN JWST Archive (DJA).\footnote{\url{https://dawn-cph.github.io/dja/}} An outline of the data reduction process using the public {\tt msaexp}\footnote{\url{https://github.com/gbrammer/msaexp}} tool is given in \citet{Heintz2023}.

We cross-match our photometric sample from CEERS and JADES-Deep-GS with those with robust spectroscopic redshifts in DJA finding a total of 55 matches, with 41 having robust NIRSpec redshifts (selected using ``grade == 3''). This sample consists of 15 galaxies from CEERS, 7 from the CEERS DDT \citep[PI: P. Arrabal-Haro, PID: 2750,][]{ArrabalHaro2023a, ArrabalHaro2023b}, 8 from JADES program 1180 \citep[PI: D. Eisenstein, PID: 1180,][]{Eisenstein2023}, and 11 from JADES program 1210 \citep[PI: N. L$\ddot{\mathrm{u}}$tzgendorf, PID: 1210,][]{Bunker2023-NIRSpec}. A summary of the spectroscopically measured $\beta$ and previous literature studies publishing these sources are given in \autoref{tab:spec_beta}.

Additionally, we search the DJA for medium ($R\sim1,000$) and high resolution ($R\sim2,700$) spectra, finding matches for 4 CEERS galaxies in the medium resolution F100LP/G140M, F170LP/G235M, and F290LP/G395M filter-disperser pairs (CEERSP3-559,2668,9946 and CEERSP6-7138). 4 JADES-Deep-GS galaxies have medium resolution F070LP/G140M, F170LP/G235M and F290LP/G395M (IDs 12248, 15023, 15705, and 26579) grating spectra only and 6 more (IDs 14738, 15297, 18531, 18605, 19523, and 21391) have both medium resolution and additional high resolution F270LP/G395H grating spectra.

\begin{figure}
    \centering
\includegraphics[width=0.45\textwidth]{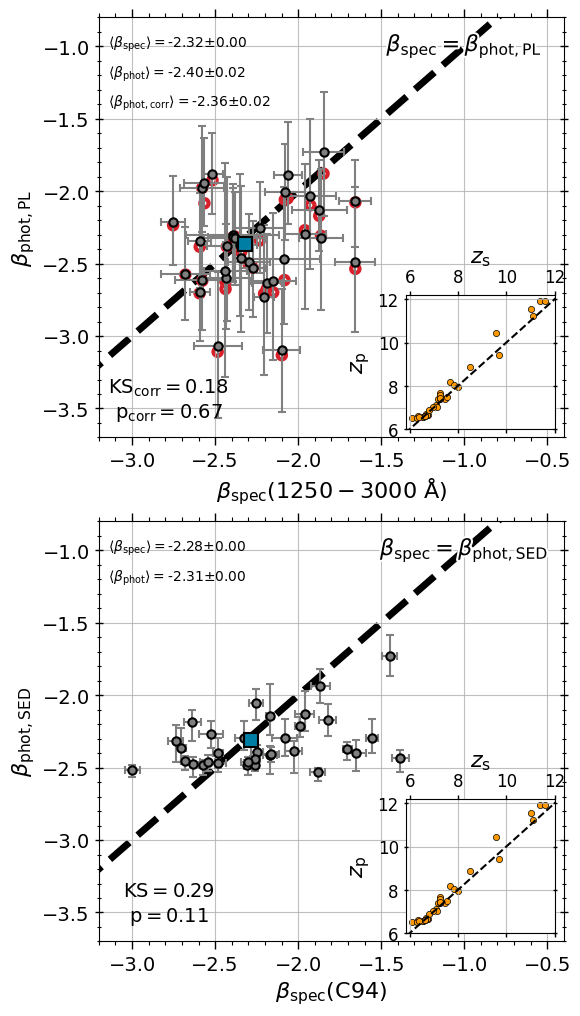}
    \caption{Comparison of redshifts and $\beta$ slopes for our $6.5<z<13$ EPOCHS photometric sample cross-matched with $>3\sigma$ UV continuum detected PRISM/CLEAR NIRSpec spectroscopic results from the DAWN JWST Archive (DJA). \textbf{Top:} Power law measured $\beta$ that are both corrected (grey circles) and uncorrected (hollow red circles) for the $\beta$ bias in \autoref{sec:PL_beta_bias} against spectroscopic $\beta$ measured over $1250<\lambda_{\mathrm{rest}}~/~\mathrm{AA}<3000$. The blue square shows the inverse-variance weighted mean for our bias corrected results. \textbf{Bottom:} SED measured $\beta$ from our {\tt{Bagpipes}} fitting against spectroscopic $\beta$ measured in the 10 C94 filters. Inverse-variance weighted mean values are given in the upper left and a photometric vs spectroscopic redshift comparison is given in the lower right. $\beta_{\mathrm{phot}}=\beta_{\mathrm{spec}}$ is shown as a thick dashed line. Two-sided KS-test results and corresponding p-values are shown in the lower left of both panels, where power law $\beta$ in the upper panel has been bias corrected.}
    \label{fig:beta_spec_comparison}
\end{figure}

We calculate spectroscopic $\beta$ slopes by fitting a power law in $f_{\lambda}$ to the NIRSpec PRISM spectra both in the 10 C94 filters and in the wavelength range $1250<\lambda_{\mathrm{{rest}}}~/~\mathrm{\AA}<3000$. In addition, rest-frame UV signal-to-noise ratios (SNRs) are calculated over this same wavelength range. We plot our spectroscopic $\beta$ measurements against photometric $\beta$ calculated using both the power law (described in \autoref{sec:power_law_beta}) and SED fitting methods (see \autoref{sec:pipes_beta_calc}) in \autoref{fig:beta_spec_comparison}, including only those galaxies with UV continuum SNR$_{\mathrm{UV}}>3$.

\autoref{fig:beta_spec_comparison} also shows the comparison of spectroscopic and photometric redshifts which, if significantly different, may provide a large impact on the measured $\beta$. We categorize the photo-z error by a quantity $\epsilon$, where
\begin{equation}
    1 + \epsilon = \frac{1 + z_p}{1 + z_s} = \frac{\lambda_\mathrm{p}}{\lambda_\mathrm{s}}
    \label{eq:epsilon}
\end{equation}
also indicates a wavelength shift between photometric and spectroscopic observed frame SED features at $\lambda_{\mathrm{p}}$ and $\lambda_\mathrm{s}$ respectively. We find that 40/41 galaxies in this spectroscopic sample fall within $\lvert\epsilon\rvert<0.1$, hence we conclude that this is not a significant factor in the differences between $\beta$ measurements here.

We measure the inverse-variance weighted mean and standard error for our different $\beta$ measurements, finding photometric results of $\langle \beta_{\mathrm{phot, SED}}\rangle = -2.31 \pm 0.00$ and $\langle \beta_{\mathrm{phot, PL}}\rangle = -2.40 \pm 0.02$. In addition, we also correct the power law $\beta$ measurements for the $\beta$ bias explored in \autoref{sec:PL_beta_bias} which reddens the inverse-variance weighted mean by $0.04$ to $\langle \beta_{\mathrm{phot, PL, corr}}\rangle = -2.36 \pm 0.02$. To assess whether our photometric and spectroscopic $\beta$ measurements are statistically likely to arise from the same underlying distribution (which they should since they are the same galaxies), we perform a two-sided Kolmogorov-Smirnoff (KS) test using the ``scipy.stats.ks\_2samp'' \citep{scipy.stats.ks2_samp} python function. For our power law photometric $\beta$ we measure $\mathrm{KS}=0.18$ with a corresponding $\mathrm{p}=0.67$ after bias corrections compared to $\mathrm{(KS, p)}=(0.29, 0.11)$ for the SED fitting photometric $\beta$, suggesting that the SED fitting results likely contain larger systematic errors.

Since the SED fitting method uses the same wavelength coverage as the spectroscopic results, we may naively expect these results to match closely, however they remain systematically red (with $6.8\sigma$ significance) due to the limited parameter space covered by the SED template set. This is portrayed by the {\tt{Bagpipes}} $\beta$ prior in the right-hand panel of \autoref{fig:photometric_beta_comparison}. Even though the photometric power law fits for $\beta$ do not have complete coverage over $1250<\lambda_{\mathrm{rest}}~/~\mathrm{\AA}<3000$ (with redshift and filter-set dependence shown in \autoref{fig:Rest_frame_UV_filters}), they disagree at just the $2.0\sigma$ significance level with the spectroscopic results. In addition, we note that the difference could be compounded by additional $\beta$ biases caused by strong rest-frame UV emission lines, Ly$\alpha$ emission, or damped Ly$\alpha$ (DLA) systems that we do not correct for in this work (see \autoref{sec:PL_beta_bias}).

In the rest of this work, we favour the power law photometric $\beta$ calculation method due to the fewer systematic biases associated with measuring $\beta$ in this way, albeit at the expense of greater photometric $\sigma_{\beta}$ errors.

\section{UV continuum slope biases}
\label{sec:PL_beta_bias}

\subsection{The impact of strong rest-frame UV emission lines}
\label{sec:emission_line_bias}

\begin{figure}
    \centering
    \includegraphics[width=0.45\textwidth]{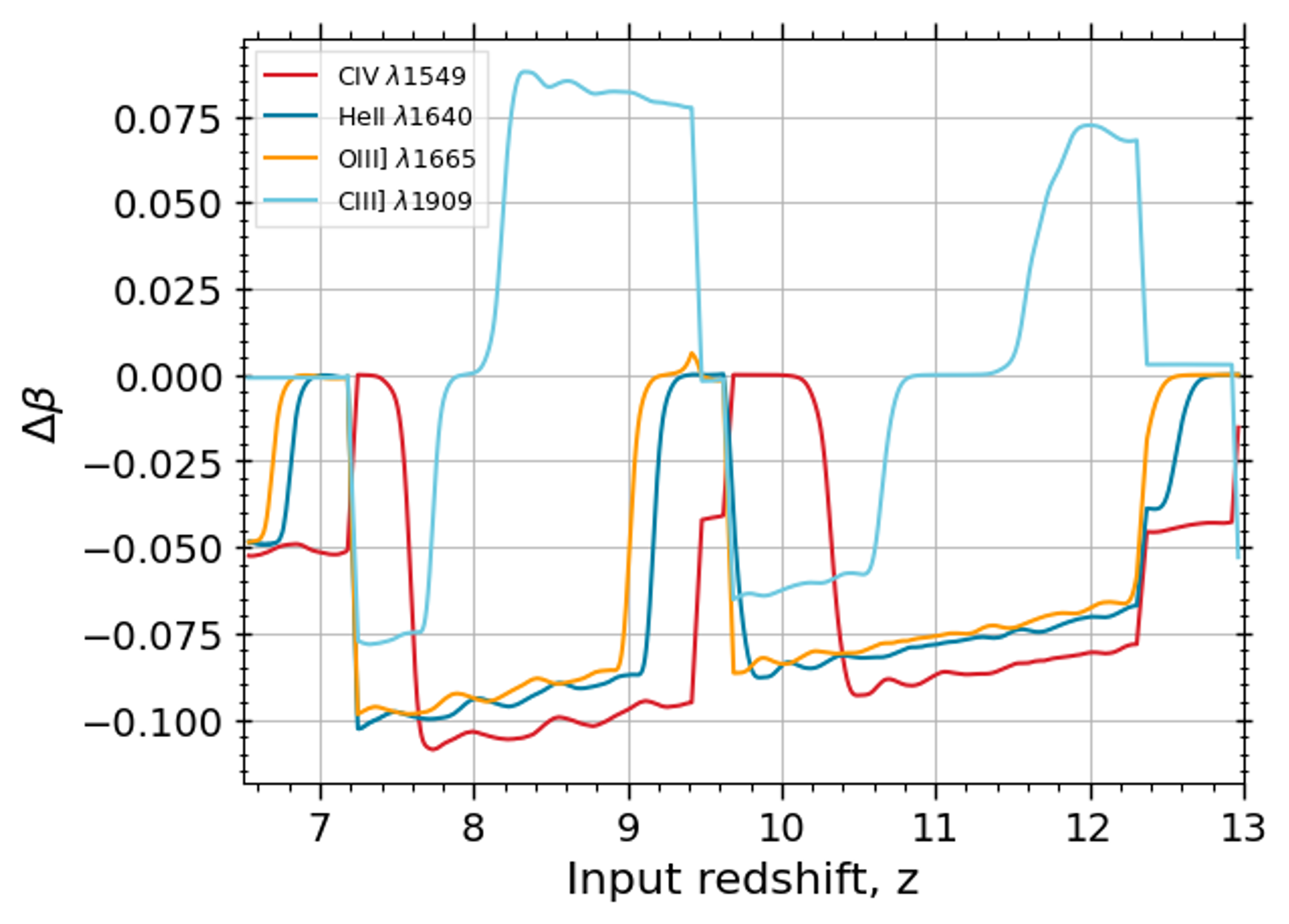}
    \caption{$\beta$ bias from rest-frame UV line emission. Shown are results assuming rest-frame EWs of $10~\mathrm{\AA}$ for {\CIV}~$\lambda1549$, {\HeII}~$\lambda1640$, {\OIII}~$\lambda1665$, and {\CIII}~$\lambda1909$ applied to a set of $10,000$ power law SEDs with $\beta_{\mathrm{int}}=-2.5$ evenly spaced across $6.5<z<13$. These biases scale linearly with emission line EW and are true under the assumption that $\epsilon=0$, or that the emission line does not impact the photometrically derived redshift.}
    \label{fig:UV_EM_line_bias}
\end{figure}

It is well known that modelling the rest-frame UV as a power law is not strictly correct due to the presence of the $2175~\mathrm{\AA}$ dust bump \citep{Stecher1965, Hensley2023} from carbonaceous dust grains (see \citet{Witstok2023} for more details), as well as rest-frame UV emission lines and nebular continuum emission in SFGs. In this section, we simulate the impact of rest-UV emission lines on $\beta$. We produce $10,000$ mock power law SEDs evenly spaced in redshift between $6.5<z<13$ at fixed intrinsic $\beta_{\mathrm{int}}=-2.5$, with photometric errors calculated assuming the galaxy has $m_{\mathrm{UV}}=26$ with each filter having a $5\sigma$ depth of $m_{\mathrm{AB}}=30$. Doppler broadened rest-frame UV emission lines with fixed Doppler parameter, $b=150~\mathrm{kms}^{-1}$, and rest-frame equivalent widths (EWs) $1-25$~\AA\ were individually added to the SEDs before calculating bandpass-averaged fluxes in the standard 8 PEARLS JWST/NIRCam bands. From these photometric fluxes, we then measured $\beta$ using our preferred power law method at the fixed input redshift with the bias in $\beta$ measured as $\Delta\beta_{\mathrm{line}} = \beta_{\mathrm{line}} - \beta_{\mathrm{int}}$. We do this for the rest-UV emission lines {\CIV}~$\lambda1549$, {\HeII}~$\lambda1640$, {\OIII}~$\lambda1665$, and {\CIII}~$\lambda1909$ which contaminate the F115W, F150W, and F200W SW NIRCam wideband filters. The resulting bias as a function of input redshift is shown in \autoref{fig:UV_EM_line_bias} for $\mathrm{EW}_{\mathrm{rest}}=10$~\AA{}. We note that this simulation is independent of $\beta$, $m_{\mathrm{UV}}$, and Doppler $b$, and that the biases scale linearly with rest-frame EW such that
\begin{equation}
\Delta\beta(\mathrm{EW}_{\mathrm{rest}})=\mathrm{EW}_{\mathrm{rest}}\frac{\Delta\beta(\mathrm{EW}_{\mathrm{rest}} = 10~\mathrm{\AA})}{10} \mathrm{ .}
\end{equation}

Previous observations of SFGs from the ground have found rest-UV emission line equivalent widths (EWs) as high as 27~{\AA} for the semi-forbidden {\CIII}~$\lambda1909$ \citep{Rigby2015, Stark2017} and $\sim8-10$~{\AA} for {\OIII}~$\lambda1665$ and {\HeII}~$\lambda1640$ \citep{Nanayakkara2019, Saxena2020} at $z\sim4$. Strong {\HeII}~$\lambda1640$ emitters are often associated with AGN when {\CIV}~$\lambda1549$ emission is also present \citep[e.g.][]{Feltre2016}, although it has been shown that BPASS SPS models can explain the observed ratios through star-formation alone \citep{Xiao2018}. In addition, \citet{Jaskot2016} find that the BPASS models show that {\CIII}~$\lambda1909$ is the strongest line bluewards of $2700$~{\AA}, which becomes more prominent in young, metal-poor systems with high ionization parameters. 

More recently, several authors have reported strong rest-frame UV emission lines at $z>6$ in NIRSpec spectroscopic data. At $z\simeq8-8.5$, \citet{Tang2023} find {\CIII}~$\lambda1909$ EWs as large as $12-16~\mathrm{\AA}$, with \citet{Hsiao2023} and \citet{Fujimoto2023} finding highly ionized bubbles at $z=8.5-13$ and $z=10.17$ in the Ultradeep NIRSpec and NIRCam Observations before the Epoch of Reionization survey \citep[UNCOVER; PIs: I. Labbe \& R. Bezanson; PID: 2561;][]{Bezanson2022} and the MACS-0647 lensing cluster. Recent JWST observations have also detected strong {\CIV}~$\lambda1549$ due to the presence of hard ionizing fields at early times, with \citet{Topping2024} observing rest-frame {\CIV}~$\lambda1549$ EW of $34~\mathrm{\AA}$ in a gravitationally lensed $z=6.11$ object behind the RXCJ2248 lensing cluster \citep[see also][]{Mainali2017, Schmidt2017}. \citet{Castellano2024} also find the strongest EoR {\CIV}~$\lambda1549$ EW of $45.8\pm1.2~\mathrm{\AA}$ in the spectroscopically confirmed GHZ2/GLASS-z12 object at $z=12.34$; this source was initially detected in GLASS NIRCam photometry by \citet{Naidu2022} and \citet{Castellano2022} and is included in our EPOCHS-III sample (see EPOCHS-I, Figs.~2/3) and provides evidence for UV line emission bias in our sample. Additionally, \citet{Charbonnel2023, Cameron2023-GNz11, Senchyna2023} also observe strong Nitrogen abundance in the well-studied GN-z11 \citep{Oesch2016, Bunker2023-NIRSpec}.

While we cannot determine rest-frame UV line EWs in our sample using broadband photometry alone, these studies give us an idea of the expected $\beta$ biases and ionization conditions in reionization era galaxies. The rest-UV line EWs and intrinsic $\beta$ slopes measured in the C94 filters for the blue \citet{Larson2023} templates used in this work are given in \autoref{tab:Larson23_UV_line_EWs}. Due to the strong rest-frame UV emission lines in these templates, there is a redshift and filter-set dependent bias between the power law fit to the photometry and from the SED in the C94 filters. Even when accounting for the emission lines, we find that using the C94 filters biases $\beta$ red by up to 0.1 compared to our power law method. We attribute this to the increased wavelength coverage of the C94 filters towards the nebular continuum two-photon turnover in the youngest \citet{Larson2023} templates quantified by $\Delta\beta_{\mathrm{neb}} = \beta_{\mathrm{C94}}(f_{\mathrm{esc, LyC}} = 0) - \beta_{\mathrm{C94}}(f_{\mathrm{esc, LyC}} = 1)$.

\begin{table}[]
    \centering
    \begin{tabular}{c|ccc}
    \toprule
    $\log_{10}$(Template Age~/~yr) & 6.0 & 6.5 & 7.0 \\
    \hline
    EW$_{\mathrm{rest}}$(Ly$\alpha$) / \AA & 0.0 & 0.0 & 0.0 \\EW$_{\mathrm{rest}}$({\CIV}~$\lambda1549$) / \AA & 5.6 & 2.0 & 0.5 \\
    EW$_{\mathrm{rest}}$({\HeII}~$\lambda1640$) / \AA & 0.3 & 1.0 & 1.1 \\
    EW$_{\mathrm{rest}}$({\OIII}~$\lambda1665$) / \AA & 5.2 & 3.6 & 1.0 \\
    EW$_{\mathrm{rest}}$({\CIII}~$\lambda1909$) / \AA & 28.7 & 21.5 & 7.3 \\
    $\beta_{\mathrm{C94}}(f_{\mathrm{esc, LyC}} = 0)$ & -2.36 & -2.32 & -2.52 \\
    $\beta_{\mathrm{C94}}(f_{\mathrm{esc, LyC}} = 1)$ & -3.11 & -2.82 & -2.77 \\
    $\Delta\beta_{\mathrm{neb}}$ & 0.75 & 0.50 & 0.25 \\
    \botrule
    \end{tabular}
    \caption{Rest-frame EWs of the most prominent UV emission lines from the blue \citet{Larson2023} templates used in our SED fitting procedure (set 4) for ages ranging $1-10$~Myr. There is no Ly$\alpha$ present in these templates and {\CIII}~$\lambda1909$ is consistently the strongest of these lines which appears especially potent at the youngest ages. Intrinsic $\beta$ slopes calculated in the 10 C94 filters for $f_{\mathrm{esc,LyC}}=\{0,1\}$ (i.e. including and excluding CLOUDY nebular emission) are also given. For a decreasing galaxy age, the reddening of the underlying continuum by nebular emission ($\Delta\beta_{\mathrm{neb}}$) increases in addition to the stellar continuum becoming bluer.}
    \label{tab:Larson23_UV_line_EWs}
\end{table}

\subsection{Proximate damped Ly\texorpdfstring{$\alpha$}{a} systems and the Ly\texorpdfstring{$\alpha$}{a} damping wing}
\label{sec:DLA_bias}

When traversing the IGM in the EoR, Ly$\alpha$ photons from the source galaxy are absorbed by neutral hydrogen along the line-of-sight to the observer. This, combined with attenuation from the Ly$\alpha$ forest bluewards of 1216~\AA, produces the characteristic red-skewed Ly$\alpha$ damping wing. The most widely used prescription of the Ly$\alpha$ damping wing is presented in \citet{Miralda-Escude1998}. This model assumes that the IGM has a constant (volume-averaged) neutral hydrogen fraction, $x_{\mathrm{HI}}$, between the source redshift, $z_{\mathrm{gal}}$, and that the end of reionization occurs at $z_{\mathrm{Re, end}}$, and neglects the patchy nature of reionization. The Ly$\alpha$ optical depth through the neutral IGM, $\tau_{\mathrm{IGM, Ly\alpha}}(\lambda_{\mathrm{obs}}, z_{\mathrm{gal}}, x_{\mathrm{HI}}, R_{\mathrm{b}}, \tau_{\mathrm{GP}}, z_{\mathrm{Re, end}})$\footnote{It is noteworthy that $z_{\mathrm{Re, end}}$ has a minor impact on calculated values of $\tau_{\mathrm{IGM, Ly\alpha}}$ in the \citet{Miralda-Escude1998} model.}, is also given as a function of $R_{\mathrm{b}}$, the radius of the surrounding ionized HII bubble, and $\tau_{\mathrm{GP}}$, the cosmology-dependent \citet{Gunn-Peterson1965} optical depth. Using NIRSpec data from the first year of {\emph{JWST}} operation, large ionized bubbles with $R_{\mathrm{b}}\gtrsim1$~cMpc have been inferred from these Ly$\alpha$ damping wings at high-redshift \citep{Whitler2023c, Witstok2023-LAEs, Umeda2023}, with a large implied contribution from faint galaxies with $M_{\mathrm{UV}}>-16$ in large-scale galaxy overdensities \citep{Lu2024}. Recent comparisons to theorized damping wing profiles assuming a more realistic patchy reionization process \citep[e.g.][]{Keating2023a} suggest that the presence of small amounts of residual HI as low as $x_{\mathrm{HI}}\sim10^{-5}$ \citep{McQuinn2016} within these ionized bubbles and strong Ly$\alpha$ emission \citep[seen observationally by][]{Saxena2023a} may instead be responsible for these large $R_{\mathrm{b}}$ values. Due to the uncertainties in modelling this damping wing, we choose not to quantify any associated $\beta$ biases. We do note, however, that this is likely to bias our $\beta$ slightly redwards due to the softening of the Ly$\alpha$ break leading to minor redshift overestimations when SED fitting using templates that do not include this dampening.

In addition, DLA systems with high column densities of neutral hydrogen $N_{\mathrm{HI}}>2\times10^{20}~\mathrm{cm}^{-2}$ \citep{Lanzetta2000}, arising from dense gas clouds in the vicinity of starburst galaxies, have been observed at high-redshift both in quasar \citep{Totani2006} and galaxy spectra \citep{Heintz2023}. These systems act to soften the Lyman-break and increase photometric Lyman-Balmer break degeneracy, decreasing photo-z accuracy. The UV $\beta$ slopes of these systems are often biased red due to the reduction in flux associated with the first band redwards of $1250~\mathrm{\AA}$. In addition, photo-z's are often overestimated when SED fitting codes confuse the reduction in flux in the first band redwards of the Lyman-$\alpha$ break with the break itself. At certain filter-set dependent redshifts, $\Delta\beta_{\mathrm{DLA}}$ becomes smaller with increasing $N_{\mathrm{HI}}$ when this band is no longer included in the rest-frame UV range used to calculate $\beta$. We do not attempt to identify DLAs in our selection procedure, but instead simulate the impact on simple power law SED models. 

We produce $10,000$ mock power law SEDs with $\beta=-2.5$ and $m_{\mathrm{UV}}=26$ are produced with redshifts evenly spaced in the interval $6.5<z<13$ assuming \citet{Inoue2014} IGM attenuation post-reionization. These SEDs are then propagated through the ISM with fixed Ly$\alpha$ velocity offset, $\Delta v_{\mathrm{Ly}\alpha} = 0$, Doppler $b=150~\mathrm{kms}^{-1}$, and a range of neutral hydrogen column densities $\log_{10}(N_{\mathrm{HI}} / \mathrm{cm}^{-2})=\{ 21, 21.5, 22, 22.5, 23, 23.5\}$. The DLA optical depth is calculated as
\begin{equation}
    \tau_{\mathrm{DLA}}(\lambda_{\mathrm{rest}})=CaN_{\mathrm{HI}}H(a, x(\lambda_{\mathrm{rest}} + \Delta v_{\mathrm{Ly\alpha}}, b)) \mathrm{,}
\end{equation}
where $C$ and $a$ are the Ly$\alpha$ photon absorption constant and damping parameters, and $H(a, x(\lambda, b))$ is the Voijt-Hjerting approximation to the Voigt profile \citep{Tepper-Garcia2006} which is dependent on the Doppler parameter, ${b=\sqrt{2k_BT_{\mathrm{HI}}/m_p}}$, via 
\begin{equation}
x(\lambda, b)=\frac{c(\lambda-\lambda_{\mathrm{Ly\alpha}})}{b\lambda_{\mathrm{Ly\alpha}}}\mathrm{.}  
\end{equation}
This Voigt profile is insensitive to the choice of $b$ for DLAs with large $N_{\mathrm{HI}}$ where the Lorentzian wings from the naturally broadened Ly$\alpha$ dominate. 

We then calculate bandpass-averaged fluxes from the standard 8-band PEARLS JWST/NIRCam filterset and re-run through our {\tt EAZY-py} photo-z fitting procedure before re-calculating $\beta$ using the photometric power law method at the derived photo-z to calculate the bias $\Delta\beta_{\mathrm{DLA}}=\beta_{\mathrm{DLA}}-\beta_{\mathrm{int}}$. It is noteworthy that in this process we include only photometric bands at $\lambda_{\mathrm{rest}}<3000$~\AA\ to avoid the non-power law nature of the Balmer jump at $3646$~\AA, Balmer break at $\sim4000$~\AA, and strong rest-optical nebular emission lines including [OIII]/H$\beta$.

\begin{figure}
    \centering
    \includegraphics[width=0.49\textwidth]{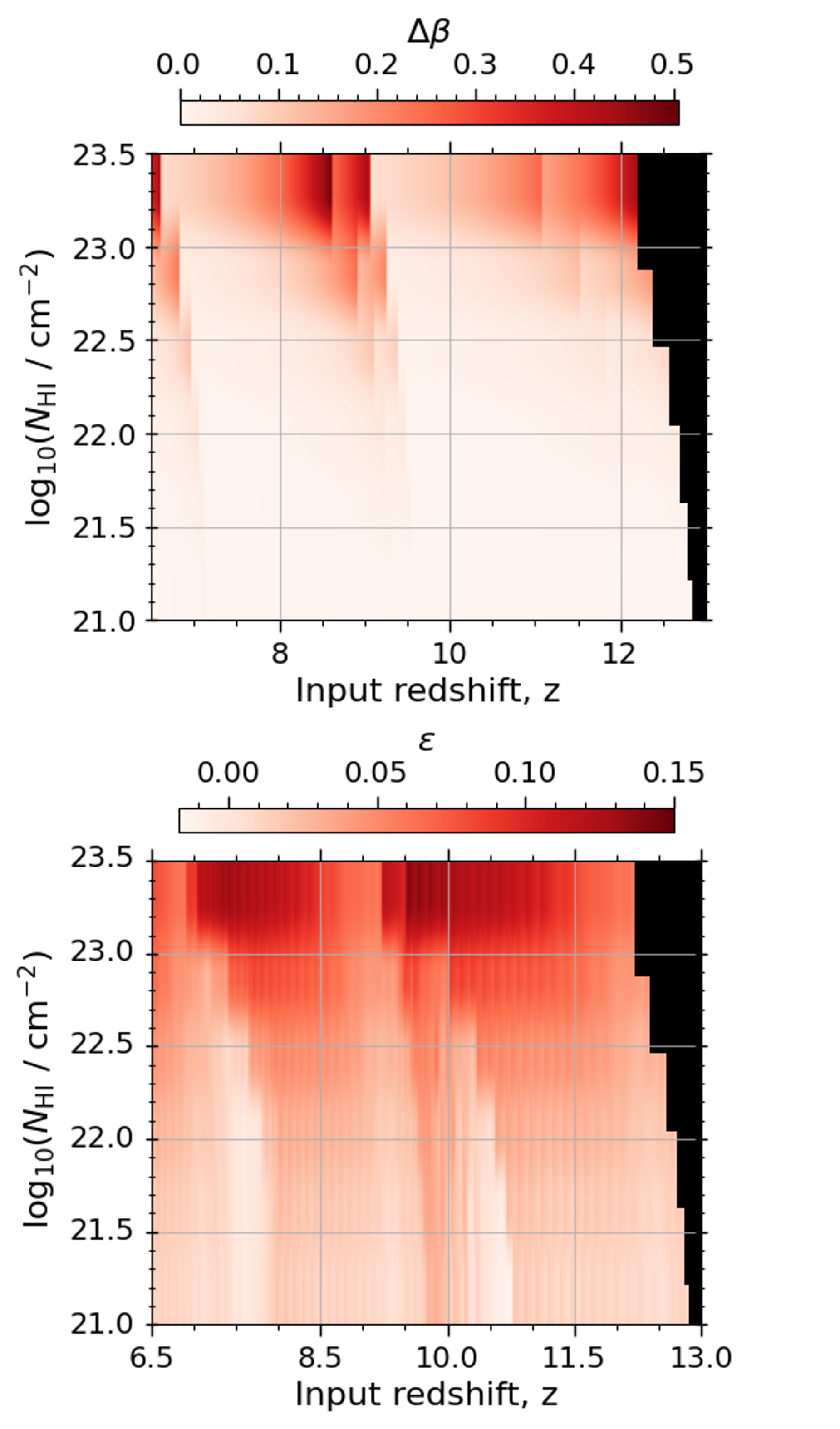}
    \caption{\textbf{Left:} Power law measured $\beta$ bias as a function of input redshift and DLA HI column density, $N_{\mathrm{HI}}$. Biases as large as $0.5$ redwards are observed in DLAs with the largest column densities, $N_{\mathrm{HI}}=10^{23.5}~\mathrm{cm}^{-2}$, at redshifts $z\sim\{6.5, 9, 12\}$. \textbf{Right:} DLA $\beta$ bias as a function of photo-z error, $\epsilon$, as defined in \autoref{eq:epsilon}, with red values showing overestimated photo-z, reaching a maximum of $15\%$. The black region towards the highest redshifts shows the region that would fail the observed $z<13$ cut due to the photo-z overestimation.}
    \label{fig:DLA_beta_bias}
\end{figure}

We plot $\Delta\beta_{\mathrm{DLA}}$ as a function of $N_{\mathrm{HI}}$, in \autoref{fig:DLA_beta_bias}, finding $\Delta \beta_{\mathrm{DLA}}=0.5$ even in the most extreme DLA scenarios with $N_{\mathrm{HI}}=10^{23.5}~\mathrm{cm}^{-2}$ , meaning that our sample may well be biased red by DLAs in some specific cases. This upper $N_{\mathrm{HI}}$ limit is set by the required HI column density should the \citet{Cameron2023-nebular} objects be contrived DLAs as opposed to nebular dominated systems. As of now, the highest DLA column densities observed in the early Universe ($z>10$) are approximately $\log_{10}(N_{\mathrm{HI}}~/~\mathrm{cm}^{-2})=22-22.5$ \citep{Heintz2023, Umeda2023, D'Eugenio2023}, and large spectroscopic studies have recently been performed to assess the abundance of these DLA systems \citep{Heintz2024}. Much is still to be determined about these systems, however, including whether large neutral gas reservoirs can form in the presence of strong winds, producing larger Ly$\alpha$ velocity offsets $\Delta v_{\mathrm{Ly\alpha}} \gtrsim +400~\mathrm{kms}^{-1}$, as well as how common DLAs are at high-redshift where gas masses are expected to be larger. We conclude that it is likely that some of our sources are biased red, however the extent of this in our photometric sample is largely unknown.

\subsection{The impact of Ly\texorpdfstring{$\alpha$}{a} emission}
\label{sec:lyman_alpha_beta_bias}

Although Lyman Alpha Emitters (LAEs) are known to exist at low-redshift, during the EoR the observed Ly$\alpha$ rest-frame EWs are expected to reduce due to absorption by neutral Hydrogen along the line-of-sight (i.e. the Lyman-$\alpha$ damping wing). Although the exact nature of this damping wing, and indeed the size of ionized bubbles in the early Universe, is largely unknown, several studies have observed high EW Ly$\alpha$ systems at $z>6$ in recent spectroscopic JWST data \citep[e.g.][]{Saxena2023a, Saxena2023b, Nakane2023, Witstok2023-LAEs, Chen2024, Napolitano2024}. In this section, we study the impact of these LAEs on measured $\beta$ slopes in JWST/NIRCam wideband studies. Even though both the number density \citep{Haiman2002, Malhotra2006} and EW evolution \citep{Tang2024} of these EoR LAEs has already been studied, we do not quantify the average bias in our sample as the selection efficiency of LAEs by our specific EPOCHS criteria in \autoref{sec:selection} is not well known.

The impact of Ly$\alpha$ emission on photometrically measured $\beta$ slopes has in fact already been studied by \citet{Rogers2013} who find a blue bias that increases with EW. Their work analyses the bias in $\beta$ measured from a power law fit to HST/WFC3IR F105W (Y$_{105}$), F125W (J$_{125}$), F140W (J$_{140}$), and F160W (H$_{160}$) photometry at $6.5<z<7.5$, finding $-0.8\lesssim\Delta\beta_{\mathrm{Ly\alpha}}\lesssim-0.5$ at $z\simeq7$ for rest-frame EWs of $50-100~\mathrm{\AA }$. In this section, we analyse the bias in $\beta$ measured using JWST/NIRCam widebands which may be systematically different from the $\beta$ measured using Y$_{105}$J$_{125}$J$_{140}$H$_{160}$ HST/WFC3IR photometry.

\begin{figure*}
    \centering
    \includegraphics[width=0.95\textwidth]{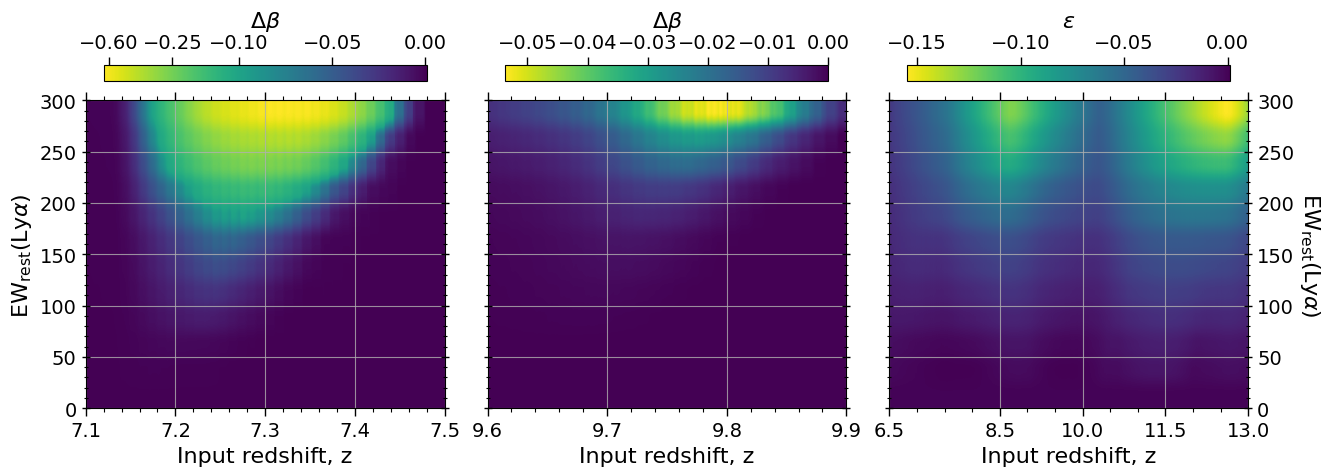}
    \caption{\textbf{Left \& centre:} $\beta$ bias as a function of rest-frame Ly$\alpha$ EW at input redshift $z\simeq7.3$ (left panel) and $z\simeq9.8$ (centre panel). It is worth noting the colorbar scaling in these two panels, where the bias at $z\simeq7.3$ is far greater than at $z\simeq9.8$ even before accounting for the reduction in observed EW due to absorption in the neutral IGM along the line-of-sight. \textbf{Right:} Photo-z error as a function of input redshift and Ly$\alpha$ EW. We see that the photo-z is underestimated by as much as $\epsilon=-0.15$ (i.e. 15\%) in the case of strong Ly$\alpha$ emission at $z\simeq \{8.5, 12.5\}$.}
    \label{fig:Lya_bias}
\end{figure*}

We calculate the impact of Ly$\alpha$ emission on the measured $\beta$ slopes using a similar method to that used to calculate the DLA bias in \autoref{sec:DLA_bias}. $10,000$ mock power law SED templates are created with fixed $\beta=-2.5$, $m_{\mathrm{UV}}=26$ at redshifts $6.5<z<13$ for each rest-frame Ly$\alpha$ $\mathrm{EW}=\{ 5, 10, 20, 30, 40, 50, 75, 100, 150, 200, 300 \}$~\AA\ including Ly$\alpha$ forest IGM attenuation following the \citet{Inoue2014} prescription, with the upper EW limit set by the largest observable values in \citet{Saxena2023a}. Bandpass-averaged fluxes are generated for the standard 8 JWST/NIRCam filters used in PEARLS, with errors produced assuming a $5\sigma$ depth of $m_{\mathrm{AB}}=30$, and the photometry is run through {\tt EAZY-py} to determine photo-z's. $\beta$ is then calculated by the power law method using the rest-frame UV filters determined by the {\tt EAZY-py} photo-z and compared to the intrinsic $\beta=-2.5$ to determine $\Delta\beta_{\mathrm{Ly\alpha}}$, with results shown in \autoref{fig:Lya_bias}.

In order to account for the increased flux in the first redwards band of the Ly$\alpha$ break due to this Ly$\alpha$ emission, SED fitting procedures with underestimated Ly$\alpha$ emission typically underestimate photo-z's (right panel of \autoref{fig:Lya_bias}). This introduces the band containing Ly$\alpha$ into the rest-frame UV and hence biases $\beta$ blue. As can be seen from \autoref{fig:Lya_bias}, there are two redshift ranges of interest (left and central panels) where Ly$\alpha$ emission may produce significant $\Delta\beta_{\mathrm{Ly\alpha}}\simeq-0.6$. At $z=7.3$, the photo-z underestimation induced by the Ly$\alpha$ emission causes the F090W filter to be incorrectly identified as the dropout filter, with the highly-elevated F115W flux density now strongly biasing $\beta$ blue. The scenario is similar at $z=9.8$ instead with the F115W and F150W filters, although the effect is a lot weaker than at $z=7.3$ due to the gap between the F115W and F150W filters not present between F090W and F115W resulting in weaker Ly$\alpha$ throughput and hence little impact on $\Delta\beta$. We note that this estimate of bias is likely an upper limit due to the additional redshift constraints provided by photometric data at $\lambda_{\mathrm{rest}}>3000$~\AA.

The bias methodology mentioned above falls short of a complete implementation of Ly$\alpha$ due to our neglect of Doppler broadening due to HI velocity dispersion, $\sigma_{\mathrm{Ly\alpha}}$, galactic dynamics/outflows shifting the major Ly$\alpha$ peak by $\Delta v_{\mathrm{Ly\alpha}}$ or doubly or triply peaked emission. The profiles of LAEs are explored in more detail using simulations \citep{Blaizot2023}, however this is not expected to have a major impact on our bias simulation results. We also do not consider the impact of the Ly$\alpha$ damping wing, the shape of which is not well known due to the unknown volume averaged neutral fraction of HI as a function of redshift, $\langle x_{\mathrm{HI}}(z)\rangle$, and the patchy reionization process. A further discussion of the Ly$\alpha$ damping wing is presented in \autoref{sec:DLA_bias}.

\subsection{Photometric error coupling bias}
\label{sec:photo-z_coupling_bias}

The most widely reported $\beta$ bias is the so-called ``photometric error coupling bias'' \citep[termed by][]{Bouwens2012} which biases faint objects approaching our detection limit blue. This bias impacts high redshift galaxy samples which are commonly selected based on SNR requirements either side of the Ly$\alpha$ break at $1216~\mathrm{\AA}$ or a specific Ly$\alpha$ break strength, both of which favour the up-scattering of photometric measurements at luminosities just below the survey detection limit. In contrast, galaxies where the first photometric band at $\lambda_{\mathrm{rest}}>1216~\mathrm{\AA}$ scatters down have a reduced Ly$\alpha$ break strength and a more prominent Balmer break photo-z solution and hence are less likely to be selected in our sample. 

This bias has been analysed extensively for $\beta$ measurements using HST/WFC3IR F125W, F140W and F160W fluxes in the 2009 Hubble Ultra Deep Field \citep[HUDF09; PI: G. Illingworth, HST PID: 11563, see][]{Bouwens2011-UVLF}, HUDF12 \citep[PI: R. Ellis, HST PID: 12498,][]{Ellis2013-HUDF12, Koekemoer2013-HUDF12} and CANDELS \citep{Koekemoer2011-CANDELS, Grogin2011-CANDELS} campaigns at $z\sim6-8$ \citep[e.g.][]{Finkelstein2012, Bouwens2012, Dunlop2012, Rogers2013, Dunlop2013, Bouwens2014a}, as well as in the HFF MACS-0416 lensing cluster \citep{Bhatawdekar2021}. More recently, this has been done for large photometrically selected JWST samples using both power law \citep{Cullen2023a, Cullen2023b} and SED fitting \citep{Morales2023} measurements of $\beta$.

\begin{figure}
    \centering
    \includegraphics[width=0.45\textwidth]{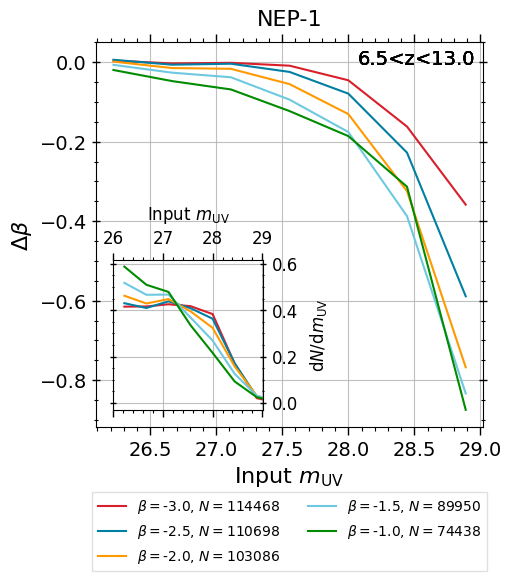}
    \caption{$\beta$ bias as a function of $m_{\mathrm{UV}}$ for the NEP-TDF field calculated from $200,000$ power law SEDs with intrinsic $\beta=\{-1, -1.5, -2, -2.5, -3\}$ shown in green, light blue, yellow, dark blue, and red respectively. The number of selected SEDs is given in the legend for each intrinsic $\beta$, with bluer $\beta$ (at fixed $m_{\mathrm{UV}}$) more efficiently selected. The normalized number density as a function of $m_{\mathrm{UV}}$ is plotted in the lower left inset. Since the redder SEDs are less efficiently selected, they exhibit the greatest $\beta$ bias, which can reach as large as $-0.85$ for $\beta=-1$, $m_{\mathrm{UV}}=29$ in the NEP.}
    \label{fig:beta_bias_NEP_m_UV}
\end{figure}

\begin{figure}
    \centering
    \includegraphics[width=0.49\textwidth]{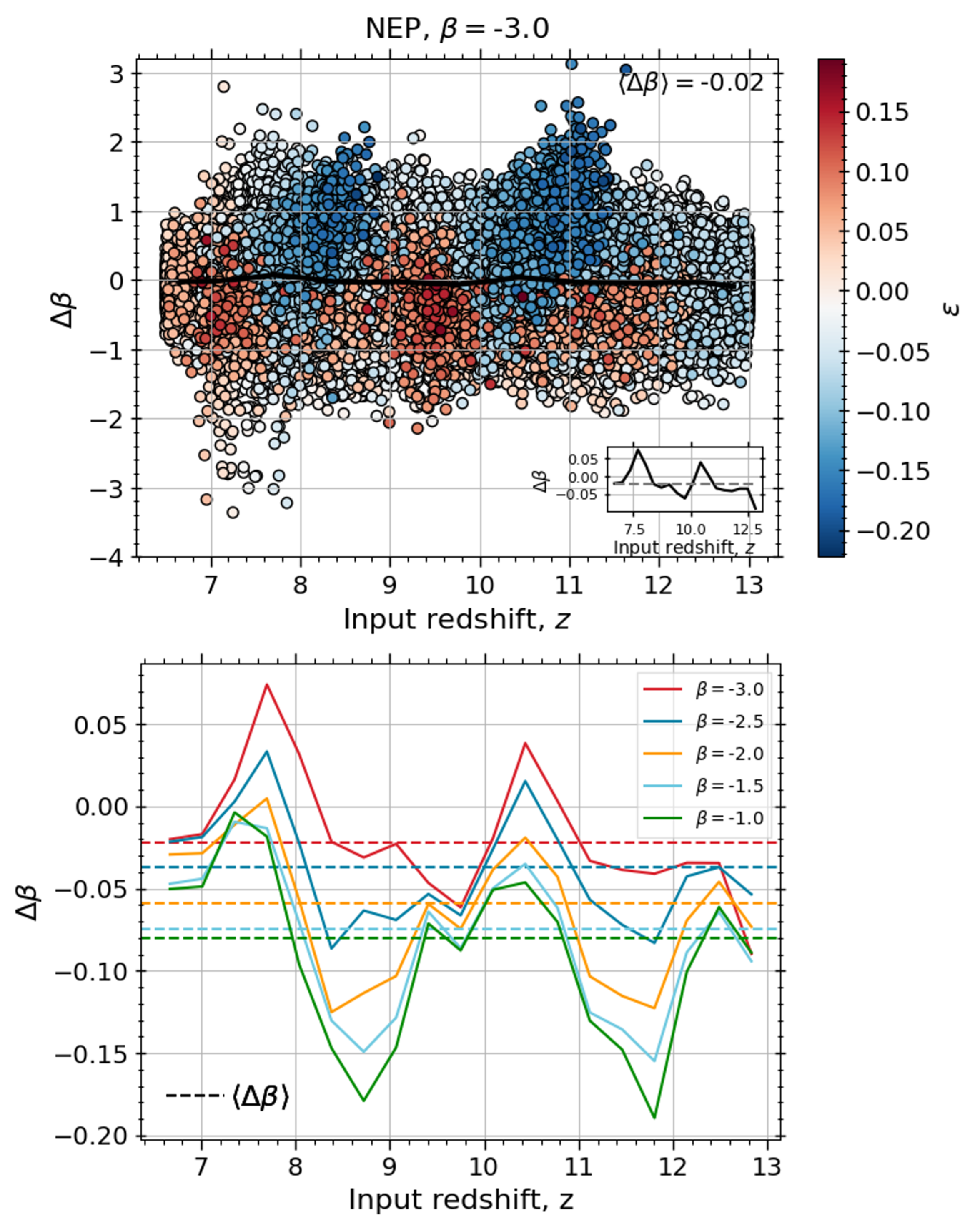}
    \caption{\textbf{Top:} $\Delta\beta$ for galaxies with intrinsic $\beta=-3$ in the NEP-TDF as a function of input redshift. The mock galaxies are colored by the photometric redshift error, where blue(red) points have underestimated(overestimated) photo-z's respectively. The overplotted thick black line shows the average bias in the selected sample, a zoomed in version of which is shown in the lower right. \textbf{Bottom:} $\beta$ bias as a function of $m_{\mathrm{UV}}$ in the NEP-TDF from $200,000$ power law SEDs with results for intrinsic $\beta=\{-1, -1.5, -2, -2.5, -3\}$ shown in green, light blue, yellow, dark blue, and red respectively. Median $\Delta\beta$ values, $\langle\Delta\beta\rangle$, are shown as horizontal dashed lines which become bluer with reddening intrinsic $\beta$.}
    \label{fig:beta_bias_NEP_z}
\end{figure}

To analyse this photometric error coupling bias we produce $200,000$ power law SEDs for each intrinsic $\beta=\{-1, -1.5, -2, -2.5, -3\}$, which appropriately match the UV slopes in our observed data, for each field used in this work. Each SED is next randomly assigned an intrinsic $m_{\mathrm{UV}}$ in the interval $26<m_{\mathrm{UV}}<30$ and bandpass-averaged fluxes are measured with the appropriate filterset for the field. Photometric errors are calculated exactly as for the real data, assuming each galaxy has local depth given by the average $5\sigma$ depths in the field with a 10\% minimum error floor to reflect the same NIRCam ZP tolerance allowed in our real photometry. We next scatter our photometric flux measurements within their respective Gaussian errors and re-run through the same high-$z$ {\tt EAZY-py} photo-z calculation and selection as in \autoref{sec:catalogues_and_photo_zs}. As before, we re-measure $\beta$ using the power law method and calculate $\Delta\beta$ for selected objects only. Results of this simulation for the NEP-TDF field are shown as a function of input $m_{\mathrm{UV}}$ in \autoref{fig:beta_bias_NEP_m_UV} and redshift in \autoref{fig:beta_bias_NEP_z}. 

In \autoref{fig:beta_bias_NEP_m_UV}, we see that bluer objects are inherently more selectable with the EPOCHS selection criteria; approximately 57\% of mock galaxies with intrinsic $\beta=-3$ are selected compared to 37\% with $\beta=-1$, meaning redder faint galaxies are more likely biased blue compared to their intrinsically bluer counterparts in the NEP at a fixed input $m_{\mathrm{UV}}$. The reasoning for this is two-fold, namely that faint red galaxies more regularly fail the $5\sigma$ SNR cut in the first band entirely at $\lambda_{\mathrm{rest}}>1216~\mathrm{\AA}$ (criteria 3 in \autoref{sec:selection}) and are more likely confused with Balmer break galaxies at lower redshift (criteria 4 and 5 in \autoref{sec:selection}) due to their shallower Ly$\alpha$ breaks. This trend is observed across all of the fields in this study.

In \autoref{fig:beta_bias_NEP_z}, we observe both a red bias at $z\simeq7.6,10.5$ and a blue bias at $z\simeq8.6,11.8$ caused mainly by photo-z inaccuracy. At these redshifts, the Ly$\alpha$ break passes between the F090W/F115W and F115W/F150W bands, leading to incorrect photo-z measurements \citep[the second of these is also noted in][]{Cullen2023b}. We note that these redshift-dependent effects can be mitigated considerably by the inclusion of deep HST/WFC3IR and medium band JWST/NIRCam photometry.

\subsection{Summary of bias correction factors}
\label{sec:beta_bias_summary}

In this section, we have outlined potential photometric $\beta$ biases from UV line emission, LAEs, and DLAs using power law SEDs and the 8 wideband PEARLS NIRCam filters. These features, however, are not detectable in our EPOCHS-III photometric data and we therefore do not correct for these biases in this work. The $\beta$ correction factors used in this work are estimated based on the redshift and $m_{\mathrm{UV}}$ dependence of the photometric error coupling bias only. This is calculated for each individual galaxy following
\begin{equation}
    \Delta\beta_{\mathrm{tot}} = \Delta\beta_{z, i} + \Delta\beta_{m_{\mathrm{UV}}, i} - \langle\Delta\beta\rangle_i \mathrm{.}
    \label{eq:beta_bias}
\end{equation}
Here $\Delta\beta_{\mathrm{tot}}$ is the total $\beta$ bias, $\Delta\beta_{z, i}$ and $\Delta\beta_{m_{\mathrm{UV}}, i}$ are the photometric error coupling biases in terms of redshift and $m_{\mathrm{UV}}$ in field $i$, and $\langle\Delta\beta\rangle_i$ is the median bias included to avoid double counting. For our EPOCHS-III sample, we measure an average bias measurement of $\langle \Delta \beta \rangle = -0.06^{+0.06}_{-0.10}$. While this measurement seems like a minor correction, we note that at the extremes we measure galaxies with $\Delta \beta = -0.55$ and $\Delta \beta = 0.24$. It is also clear from \autoref{fig:beta_bias_NEP_m_UV} that the bias is greater for redder galaxies that are fainter and have lower masses due to the preferential up-scattering of red compared to blue galaxies at a given $m_{\mathrm{UV}}$. This means the bias corrections made in this work will mostly act to redden already red galaxies, increasing the slope of our measured $\beta-M_{\mathrm{UV}}$ and decreasing the slope of our $\beta-M_{\star}$ relations in \autoref{sec:beta_MUV} and \autoref{sec:beta_mass} respectively.

The use of power law SEDs, however, does not describe the full picture from the real Universe. Additional constraints in the rest-frame optical (i.e. nebular emission lines, the $\lambda_{\mathrm{rest}}=3646~\mathrm{\AA}$ Balmer jump, or the Balmer break at $\lambda_{\mathrm{rest}}\simeq4000~\mathrm{\AA}$) will increase the accuracy of photo-z measurements. We conclude, therefore, that the bias corrections as a result of photo-z uncertainties are likely overestimated in all fields due to this necessary simplification. In addition, we note that this technique neglects possible degeneracies that may exist between $\Delta\beta_{z, i}$ and $\Delta\beta_{m_{\mathrm{UV}}, i}$, and that an even more computationally expensive procedure is required to adequately measure these using a more statistically significant sample size. 

\section{Results}
\label{sec:Results}

\subsection{Redshift evolution}
\label{sec:beta_z}

\begin{figure}[h!]
    \centering
    \includegraphics[width=0.45\textwidth]{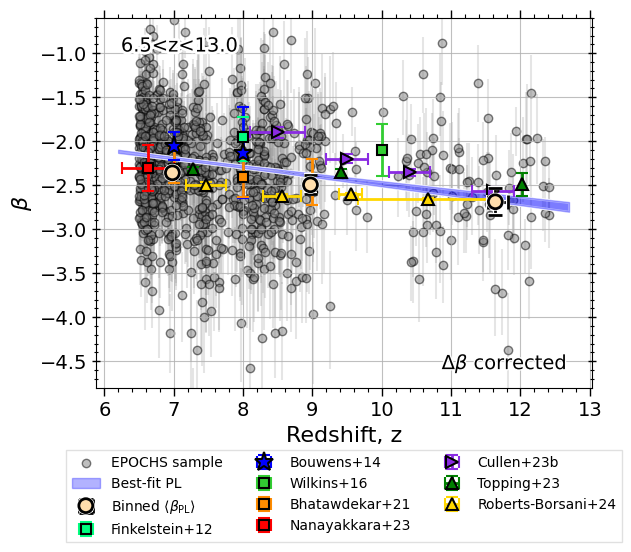}
    \caption{Bias corrected power law $\beta$ evolution as a function of $z$. Black background points show measurements for individual galaxies and beige circular points show our bootstrapped median points. The $16^{\mathrm{th}}-84^{\mathrm{th}}$ percentiles of the posterior power law fit to the data, as given in \autoref{eq:beta_z_fit_params}, is shown in blue. We compare our results to the observational HST studies of \citet{Bouwens2014a} (blue stars), \citet{Wilkins2016} (lime green squares), \citet{Bhatawdekar2021} (orange squares) and more recent JWST work by \citet{Nanayakkara2023} (red squares), \citet{Cullen2023b} (purple triangles), \citet{Topping2023} (dark green triangles), and \citet{Roberts-Borsani2024} (yellow triangles).}
    \label{fig:Beta_vs_z}
\end{figure}

We now determine whether the decreasing trend of $\beta$ with redshift found in HST data \citep[e.g.][]{Finkelstein2012, Bouwens2014a, Bhatawdekar2021} extends to $z>10$ in JWST data. We plot our bias corrected power law $\beta-z$ relation for the EPOCHS-III sample in \autoref{fig:Beta_vs_z}, with circular beige points showing bootstrapped median data points at $z\simeq7$, $z\simeq9$ and $z\simeq11.5$. For each of our $10,000$ bootstraps we 1) randomly scatter each galaxy within their redshift and $\beta$ PDFs, 2) bin the galaxies in redshift and randomly select (with replacement) a number of galaxies chosen from a Gaussian centred on the number of galaxies in the bin with width given by the Poisson error on this, 3) calculate median of each bin, 4) calculate median and $16^{\mathrm{th}}-84^{\mathrm{th}}$ percentiles of the bootstrapped median values in each bin. We also follow this method to calculate bootstrapped median data points in \autoref{sec:beta_MUV} and \autoref{sec:beta_mass}. Our bootstrapped median data points, as well as the corresponding median $\langle M_{\mathrm{UV}}\rangle$ values are given in \autoref{tab:beta_z_beta_corr}.

\begin{table}[]
\centering
\caption{Bootstrapped binned results of median $z$ and bias corrected power law $\beta$ for our EPOCHS-III sample as plotted in \autoref{fig:Beta_vs_z}. We show median $M_{\mathrm{UV}}$ values in each redshift bin with the errors representing the $16^{\mathrm{th}}-84^{\mathrm{th}}$ percentiles of the $M_{\mathrm{UV}}$ distribution. It is clear that at higher redshift our EPOCHS-III selection criteria selects on average intrinsically brighter sources.}
\label{tab:beta_z_beta_corr}
\begin{tabular}{cccc}
\toprule
    $z$ bin & $\langle z \rangle$ & $\langle \beta \rangle$ & $\langle M_{\mathrm{UV}}\rangle$ \\ 
\hline
    $6.5<z<8.5$ & $6.98^{+0.03}_{-0.04}$ & $-2.36^{+0.03}_{-0.03}$ & $-19.27^{+0.79}_{-0.75}$ \\
    $8.5<z<11$ & $8.97^{+0.04}_{-0.03}$ & $-2.50^{+0.11}_{-0.10}$ & $-19.54^{+0.51}_{-0.74}$ \\
    $11<z<13$ & $11.64^{+0.19}_{-0.10}$ & $-2.69^{+0.15}_{-0.15}$ & $-19.79^{+0.53}_{-0.56}$ \\
\botrule
\end{tabular}
\end{table}

We find that the individual data points in \autoref{fig:Beta_vs_z} are best fit by the power law
\begin{equation}
\beta=-1.51\pm0.08-(0.097\pm0.010)\times z \mathrm{,}
\label{eq:beta_z_fit_params}
\end{equation}
implying a decreasing $\beta$ at earlier cosmic times. The negative slope of $\beta-z$ is steeper than the JWST photometric results of \citet{Topping2023} and shallower than those of \citet{Cullen2023b}, who measure $\mathrm{d}\beta/\mathrm{d}z=-0.030^{+0.024}_{-0.029}$ and $\mathrm{d}\beta/\mathrm{d}z=-0.28^{+0.04}_{-0.04}$ respectively at similarly high redshifts, although it remains consistent with the $\mathrm{d}\beta/\mathrm{d}z=-0.10\pm0.06$ derived at $z<6$ by \citet{Bouwens2014a}. These trends are dependent on both the method used to measure $\beta$ and the average $M_{\mathrm{UV}}$ of the galaxy sample. 
The differences may be explained by the dependence of $\mathrm{d}\beta/\mathrm{d}z$ on the average intrinsic UV brightness of the galaxy sample, as noted by \citet{Topping2023}, with brighter $\langle M_{\mathrm{UV}} \rangle$ samples showing a steeper evolution. Our $\langle M_{\mathrm{UV}} \rangle = -19.35$ is brighter than that of \citet{Topping2023}, who measure $-18.61 < \langle M_{\mathrm{UV}} \rangle < -18.16$ in their $0.78-1.15~\mu\mathrm{m}$ dropout samples from deep JADES data in GOODS-South. At $z\simeq\{9.5, 10.5, 11.5\}$, \citet{Cullen2023b} measure $\langle M_{\mathrm{UV}} \rangle = \{-18.9\pm0.8, -19.5\pm0.7, -19.1\pm0.5\} $ in their combined sample, although their $\langle M_{\mathrm{UV}} \rangle = -21.2\pm1.6$ at $7.5<z<9$ is significantly brighter due to their use of much wider ground based COSMOS/UltraVISTA data which drives their measured $\mathrm{d}\beta/\mathrm{d}z$ steeper as a result of their redder $\langle \beta \rangle$ at $z\simeq8-8.5$. The stacked spectroscopic results of \citet{Roberts-Borsani2024} at $z>5$ yield $\mathrm{d}\beta/\mathrm{d}z=-0.06\pm0.01$, deviating from our results by $\simeq2.6\sigma$

We test the $M_{\mathrm{UV}}$ dependence of our derived $\mathrm{d}\beta/\mathrm{d}z$ by splitting our sample into bright ($M_{\mathrm{UV}}<-19.5$) and faint ($M_{\mathrm{UV}}>-19.5$) sub-samples before repeating our $\beta-z$ power law fitting procedure. In our faint ($\langle M_{\mathrm{UV}} \rangle = -18.94$) and bright ($\langle M_{\mathrm{UV}} \rangle = -19.92$) sub-samples we measure $\mathrm{d}\beta/\mathrm{d}z = -0.108\pm0.015$ and $\mathrm{d}\beta/\mathrm{d}z = -0.092\pm0.013$ respectively. It is clear that there exists no clear evolution in $\beta-z$ with $M_{\mathrm{UV}}$ in our sample which we attribute to the flatter $\mathrm{d}\beta/\mathrm{d}M_{\mathrm{UV}}$ found in this work compared to \citet{Cullen2023b}, \citet{Topping2023} and \citet{Bouwens2014a}. This highlights the susceptibility of conclusions related to the evolution of sample averaged $\beta$ to minor differences in SFG selection criteria between different studies. A further discussion regarding trends with $M_{\mathrm{UV}}$ is presented in \autoref{sec:beta_MUV}.

Our derived $\beta-z$ slope implies that, over cosmic time, there is an increasing metal enrichment and dust content within galaxies. We compare our results to the representative UV continuum slope of dwarf galaxies enriched by Pop.~III stars given as $\langle \beta \rangle=-2.51\pm0.07$ by \citet{Jaacks2018}. Our \autoref{eq:beta_z_fit_params} suggests that this transition occurs at a redshift $z=10.3$, although we note that this is later than suggested by the $\beta-z$ fits of \citet{Cullen2023b} and \citet{Topping2023} who measure $z=11.1$ and $z=14.3$ respectively.

\subsection{Correlations with \texorpdfstring{$M_{\text{UV}}$}{UV Magnitude}}
\label{sec:beta_MUV}

We next analyse the trends of observed $\beta$ with $M_{\mathrm{UV}}$, with $\beta$ calculated using both methods explained in \autoref{sec:calculating_UV_properties}. We correct our power law $\beta$ results using the average correction factors outlined in \autoref{sec:photo-z_coupling_bias} based on interpolated $m_{\mathrm{UV}}$, $\beta$, and $z$. Results using bias corrected power law $\beta$ measurements are shown in \autoref{fig:UV_slope_vs_M_UV_z_split}, where we split the galaxies into our three redshift bins ($6.5<z<8.5$, $8.5<z<11$, and $11<z<13$). We calculate bootstrapped median $\langle M_{\mathrm{UV}}\rangle$ and $\langle\beta\rangle$ for our sample (in 2, 3, and 4 magnitude bins respectively for the $6.5<z<8.5$, $8.5<z<11$, and $11<z<13$ redshift bins) following the same procedure introduced in \autoref{sec:beta_z}, with $\langle\beta\rangle$ and $\langle M_{\mathrm{UV}}\rangle$ for each bin given in \autoref{tab:beta_M_UV_beta_corr}. Additionally, we quantify the amplitude and slope of these relations by fitting a power law of the form
\begin{equation}
    \beta = \frac{\mathrm{d}\beta}{\mathrm{d}M_{\mathrm{UV}}}(M_{\mathrm{UV}} + 19) + \beta(M_{\mathrm{UV}} = -19)
    \label{eq:Beta_M_UV_fit_func}
\end{equation}
to each individual data point in our three redshift bins to allow direct comparison with recent JWST studies on $\beta$ \citep[e.g.][]{Topping2023, Cullen2023b}. The fitted amplitude at $M_{\mathrm{UV}}=-19$, $\beta(M_{\mathrm{UV}}=-19)$, and slope, $\mathrm{d}\beta/\mathrm{d}M_{\mathrm{UV}}$, for fits using SED fitting $\beta$ as well as both bias corrected and uncorrected power law $\beta$ are given in \autoref{tab:beta_MUV_beta_mass_fit_params} and plotted in relation to other observation results in \autoref{fig:Beta_vs_M_UV_fit_params}.

We compare our results with binned observational studies from both HST \citep[][]{Finkelstein2012, Dunlop2012, Bouwens2012, Dunlop2013, Bouwens2014a, Bhatawdekar2021} and JWST \citep[][]{Topping2023, Cullen2023b, Roberts-Borsani2024} in the left hand panels of \autoref{fig:UV_slope_vs_M_UV_z_split}. In addition, we also plot individual galaxies from both \citet{Wilkins2016} and \citet{Morales2023} as starred points. In addition to observational comparisons, we also compare to a wide range of simulated results from THESAN \citep[][$z=\{7, 9\}$]{Kannan2022-THESAN}, SC SAM GUREFT \citep[][$z=\{7, 9, 11\}$]{Yung2023-SC-SAM-GUREFT, Yung2024-SC-SAM-GUREFT}, DELPHI \citep[][$z=\{7.1, 9.1, 12.2\}$]{Mauerhofer2023-DELPHI}, FLARES \citep[][$z=\{7, 9, 11\}$]{Lovell2021_FLARES_I, Vijayan2021_FLARES_II, Wilkins2023_FLARES_V}, DREaM \citep[][$z=\{6.5, 8.5, 11\}$]{Drakos2022-DREaM}, and SIMBA-EoR \citep[][$z=9$]{Dave2019-SIMBA-EoR, Wu2020-SIMBA-EoR}. It is challenging to identify the reasons for the differences between these simulations due to the differing initial assumptions made and techniques used. For the most part, however, there is no large discrepancy between the simulations at $6.5<z<8.5$, although they begin to diverge at $11<z<13$. This is most likely due to the fact that these simulations are often calibrated to observational results, more of which are available from HST at $z\simeq7$ than at $z\simeq12$. When comparing to these studies, our results show a far flatter $\mathrm{d}\beta/\mathrm{d}M_{\mathrm{UV}}$ in our lowest redshift bin at $6.5<z<8.5$ as well as much bluer $\beta$ slopes at $11<z<13$ than the simulations (also seen in \citet{Cullen2023b}).

\subsubsection{Flat \texorpdfstring{$\mathrm{d}\beta/\mathrm{d}M_{\mathrm{UV}}$ at $6.5<z<8.5$}{dbeta/dMUV at 6.5 \textless z \ textless 8.5}}
\label{sec:beta_MUV_flat_slope_low_z}

The flatter observed $\mathrm{d}\beta/\mathrm{d}M_{\mathrm{UV}}$ at $6.5<z<8.5$ could be due to a number of factors. In order for this slope to be flattened compared to other results, we could have either discovered a new population of faint red galaxies and/or missed a large population of faint blue galaxies compared to other studies. In our bias simulations in \autoref{sec:PL_beta_bias}, we have found that our SED fitting templates favor the selection of blue galaxies over red ones, making it unlikely that we have missed a large sample of blue galaxies unless they have $\beta<-3.1$ (the bluest \citet{Larson2023} template), which is challenging to reproduce with standard SPS models and IMFs. Our sample covers less sky area than many of these other HST studies and even the JWST work of \citet{Cullen2023b} which contains bright $z>7.5$ objects from COSMOS/UltraVISTA \citep[see][]{Donnan2023} meaning we almost certainly miss the rarer bright red galaxies at $6.5<z<8.5$ in this work compared to others. As well as this, we note that we could have either over-corrected for the $\beta$ bias, which can reduce our $\mathrm{d}\beta/\mathrm{d}M_{\mathrm{UV}}$ by $\sim0.04$ towards the other observational results or we could be impacted by sample contamination (this is explored in more detail in \autoref{sec:contamination_impact}).

\subsubsection{Blue \texorpdfstring{$\beta(M_{\mathrm{UV}}=-19)$ at $11<z<13$}{Beta (MUV = -19) at 11 \textless z \textless 13}}
\label{sec:beta_MUV_blue_beta_highz}

In our highest redshift bin, we observe a bluer $\beta$ than predicted by the SC SAM GUREFT, DELPHI, FLARES, and DREaM simulations. Previous JWST observations by \citet{Topping2023} and \citet{Cullen2023b} find $\beta(M_{\mathrm{UV}}=-19)=\{-2.42\pm0.13, -2.63\pm0.09\}$ at $z\simeq12$ and $11<z<12$ respectively, whereas we observe bluer values of $\beta(M_{\mathrm{UV}}=-19)=-2.73\pm0.06$. We note that this may well be exacerbated by sample contamination, which we assess in \autoref{sec:contamination_impact}, although for the rest of this subsection we explore the physical interpretation of this ultra-blue \textit{average} $\beta$ measurement.

\begin{figure*}[h!]
    \centering
    \includegraphics[width=0.9\textwidth]{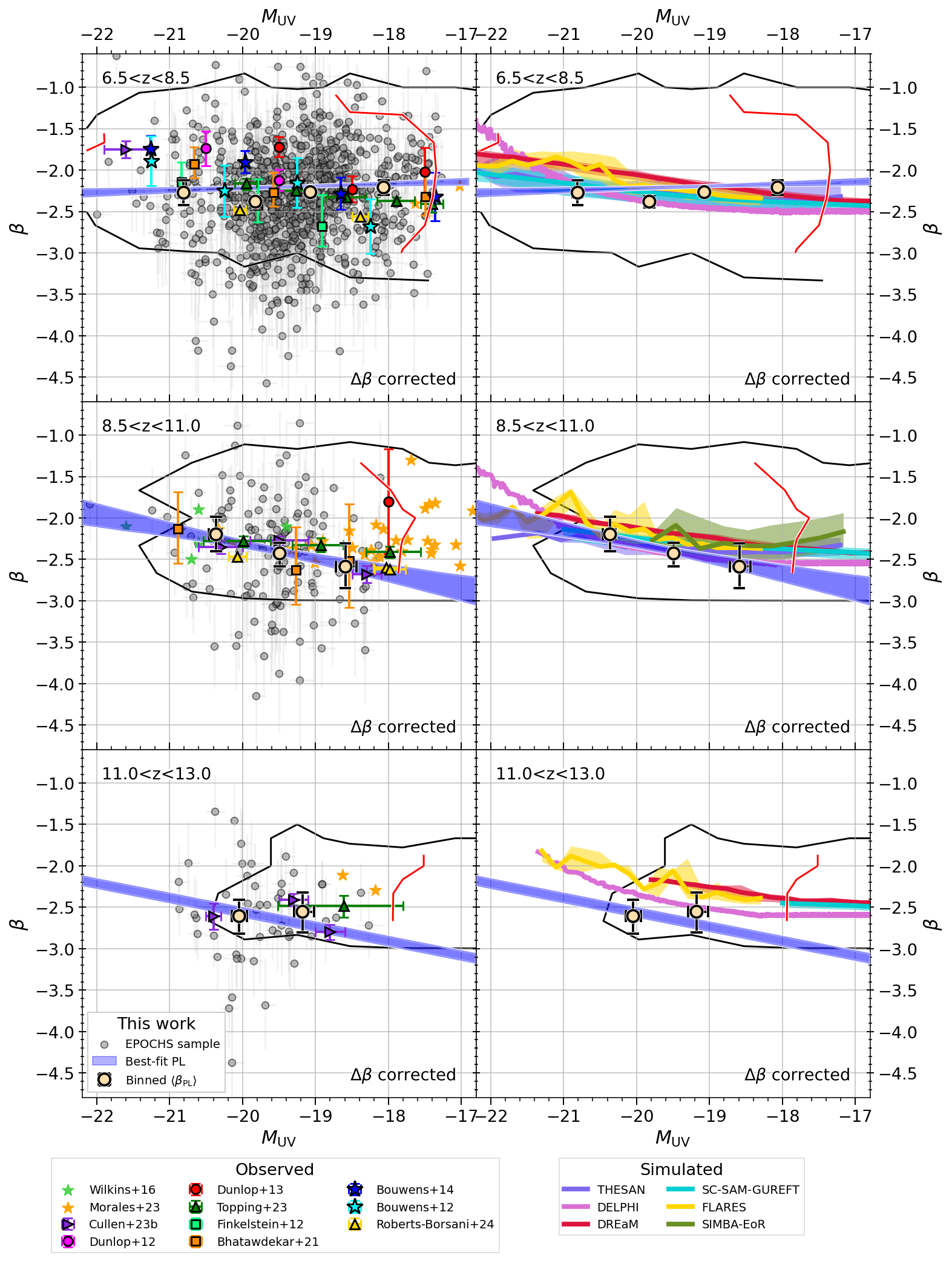}
    \caption{$\beta-M_{\mathrm{UV}}$ split by redshift for $6.5<z<8.5$ (upper), $8.5<z<11.0$ (middle) and $11.0<z<13.0$ (lower). The shaded blue line shows the $16^{\mathrm{th}}-84^{\mathrm{th}}$ percentile of the fit to the power law $\beta$ bias-corrected EPOCHS-III sample (black background points), with the slope at $11.0<z<13.0$ bin fixed to that at $8.5<z<11.0$. Comparisons to observations \citep[][]{Finkelstein2012, Dunlop2012, Bouwens2012, Dunlop2013, Bouwens2014a, Wilkins2016, Bhatawdekar2021, Topping2023, Morales2023, Cullen2023b, Roberts-Borsani2024} and simulations (THESAN \citep[][$z=\{7, 9\}$]{Kannan2022-THESAN}, SC SAM GUREFT \citep[][$z=\{7, 9, 11\}$]{Yung2023-SC-SAM-GUREFT, Yung2024-SC-SAM-GUREFT}, DELPHI \citep[][$z=\{7.1, 9.1, 12.2\}$]{Mauerhofer2023-DELPHI}, FLARES \citep[][$z=\{7, 9, 12\}$]{Lovell2021_FLARES_I, Vijayan2021_FLARES_II, Wilkins2023_FLARES_V}, DREaM \citep[][$z=\{6.5, 8.5, 11\}$]{Drakos2022-DREaM}, and SIMBA-EoR \citep[][$z=9$]{Dave2019-SIMBA-EoR, Wu2020-SIMBA-EoR}) are plotted in the left and right panels respectively. The black lines show limits of the {\tt{JAGUAR}} catalogue which mimics the CEERS observations, and the red lines show the 20\% completeness contours.}
    \label{fig:UV_slope_vs_M_UV_z_split}
\end{figure*}

We use the FLARES simulations to test the impact of changing $f_{\mathrm{esc, LyC}}$ and $A_{\mathrm{UV}}$ of both $\beta(M_{\mathrm{UV}}=-19)$ and $\mathrm{d}\beta/\mathrm{d}M_{\mathrm{UV}}$ in our highest redshift $11<z<13$ bin. The results from FLARES that are plotted in \autoref{fig:UV_slope_vs_M_UV_z_split} assume $f_{\mathrm{esc, LyC}}=0$ with the dust attenuation switched on. We find that turning this dust attenuation off both changes $\beta(M_{\mathrm{UV}}=-19)$ by $\simeq-0.2$ to $\beta(M_{\mathrm{UV}}=-19)\simeq-2.53$ and flattens the slope from $\simeq-0.2$ to $\simeq0.0$. The FLARES $\mathrm{d}\beta/\mathrm{d}M_{\mathrm{UV}}$ predictions are approximate only as their $\beta-M_{\mathrm{UV}}$ does not strictly follow a power law. Since our measured $11<z<13$ $\mathrm{d}\beta/\mathrm{d}M_{\mathrm{UV}}\simeq-0.2$ (which is fixed to the slope in the $8.5<z<11$ bin) is similar to the dusty FLARES simulations, we conclude that turning off the dust law completely does not adequately solve the problem. If we additionally switch off the galactic nebular emission (i.e. by setting $f_{\mathrm{esc, LyC}}=1$), we obtain a significantly bluer $\beta(M_{\mathrm{UV}}=-19)\simeq-2.78$ and shallower $\mathrm{d}\beta/\mathrm{d}M_{\mathrm{UV}}\simeq-0.05$. We therefore conclude that a combination of a reduction in dust attenuation and increase in Lyman continuum escape fractions is expected towards higher redshifts. If the v2.2.1 BPASS SPS models \citep{Stanway2018-BPASS} and \citet{Chabrier2003} IMF assumed in FLARES are indeed correct, then a galaxy at $M_{\mathrm{UV}}=-19$ is expected to be completely dust free with $f_{\mathrm{esc, LyC}}=1$ in order to match our observations.

The FLARES multi-component dust model calculates the UV optical depth by including contributions from both the ISM, $\tau_{\mathrm{ISM}}$ as well as additional attenuation from young stars in birth clouds (BC) $<10$~Myr old, $\tau_{\mathrm{BC}}$ \citep{CharlotFall2000}. These V-band optical depths are calibrated by the dust-to-metal \citep[DTM; see equation 15 from][]{Vijayan2019} and integrated line-of-sight metal column densities (for the ISM optical depth), and spatially resolved stellar metallicities (for $\tau_{\mathrm{BC}}$). In addition, the ISM normalizing factor $\kappa_{\mathrm{ISM}}$ is chosen to match the $z=5$ \citet{Bouwens2015} UV LF, and the BC normalizing factor $\kappa_{\mathrm{BC}}$ is chosen to match the $z=5$ $\beta$ observations from \citet{Bouwens2012, Bouwens2014a} and the $z=8$ [OIII]~$\lambda\lambda4959,5007$+H$\beta$ EW distribution from \citet{DeBarros2019}. Since we expect there to be some contribution from dust in order to retain the $\mathrm{d}\beta/\mathrm{d}M_{\mathrm{UV}}$ slope at these high-redshifts, there may exist either a lag between dust and metal production and/or a change in grain shape, size, and composition (impacting $\kappa_{\mathrm{ISM}}/\kappa_{\mathrm{BC}}$). A more in depth discussion of the dust implications of this work is presented in \autoref{sec:Dust_implications}.

\begin{table}[]
\centering
\caption{Bootstrapped binned results of median $M_{\mathrm{UV}}$ and bias corrected power law $\beta$ for our EPOCHS-III sample as plotted in \autoref{fig:UV_slope_vs_M_UV_z_split}. We show results at $6.5<z<8.5$ in four $M_{\mathrm{UV}}$ bins, as well as in three and two bins at the higher redshifts $8.5<z<11$ and $11<z<13$ respectively.}
\label{tab:beta_M_UV_beta_corr}
\begin{tabular}{ccc}
\toprule
    $M_{\mathrm{UV}}$ bin & $\langle M_{\mathrm{UV}}\rangle$ & $\langle \beta \rangle$ \\ 
\hline 
\multicolumn{3}{c}{$6.5<z<8.5$} \\ 
\hline 
$M_{\mathrm{UV}}<-20.5$ & $-20.81^{+0.06}_{-0.07}$ & $-2.27^{+0.15}_{-0.15}$ \\
$-20.5<M_{\mathrm{UV}}<-19.5$ & $-19.82^{+0.03}_{-0.04}$ & $-2.38^{+0.07}_{-0.07}$ \\
$-19.5<M_{\mathrm{UV}}<-18.5$ & $-19.07^{+0.03}_{-0.03}$ & $-2.27^{+0.05}_{-0.06}$ \\
$M_{\mathrm{UV}}>-18.5$ & $-18.07^{+0.05}_{-0.05}$ & $-2.21^{+0.09}_{-0.09}$ \\
\hline 
\multicolumn{3}{c}{$8.5<z<11.0$} \\ 
\hline 
$M_{\mathrm{UV}}<-20.0$ & $-20.36^{+0.08}_{-0.10}$ & $-2.20^{+0.21}_{-0.20}$ \\
$-20.0<M_{\mathrm{UV}}<-19.0$ & $-19.49^{+0.05}_{-0.05}$ & $-2.43^{+0.13}_{-0.16}$ \\
$M_{\mathrm{UV}}>-19.0$ & $-18.59^{+0.15}_{-0.13}$ & $-2.59^{+0.28}_{-0.26}$ \\
\hline 
\multicolumn{3}{c}{$11.0<z<13.0$} \\ 
\hline 
$M_{\mathrm{UV}}<-19.5$ & $-20.05^{+0.11}_{-0.10}$ & $-2.61^{+0.20}_{-0.21}$ \\
$M_{\mathrm{UV}}>-19.5$ & $-19.18^{+0.16}_{-0.12}$ & $-2.55^{+0.24}_{-0.25}$ \\
\botrule
\end{tabular}
\end{table}

\begin{figure}[h!]
    \centering
    \includegraphics[width=0.45\textwidth]{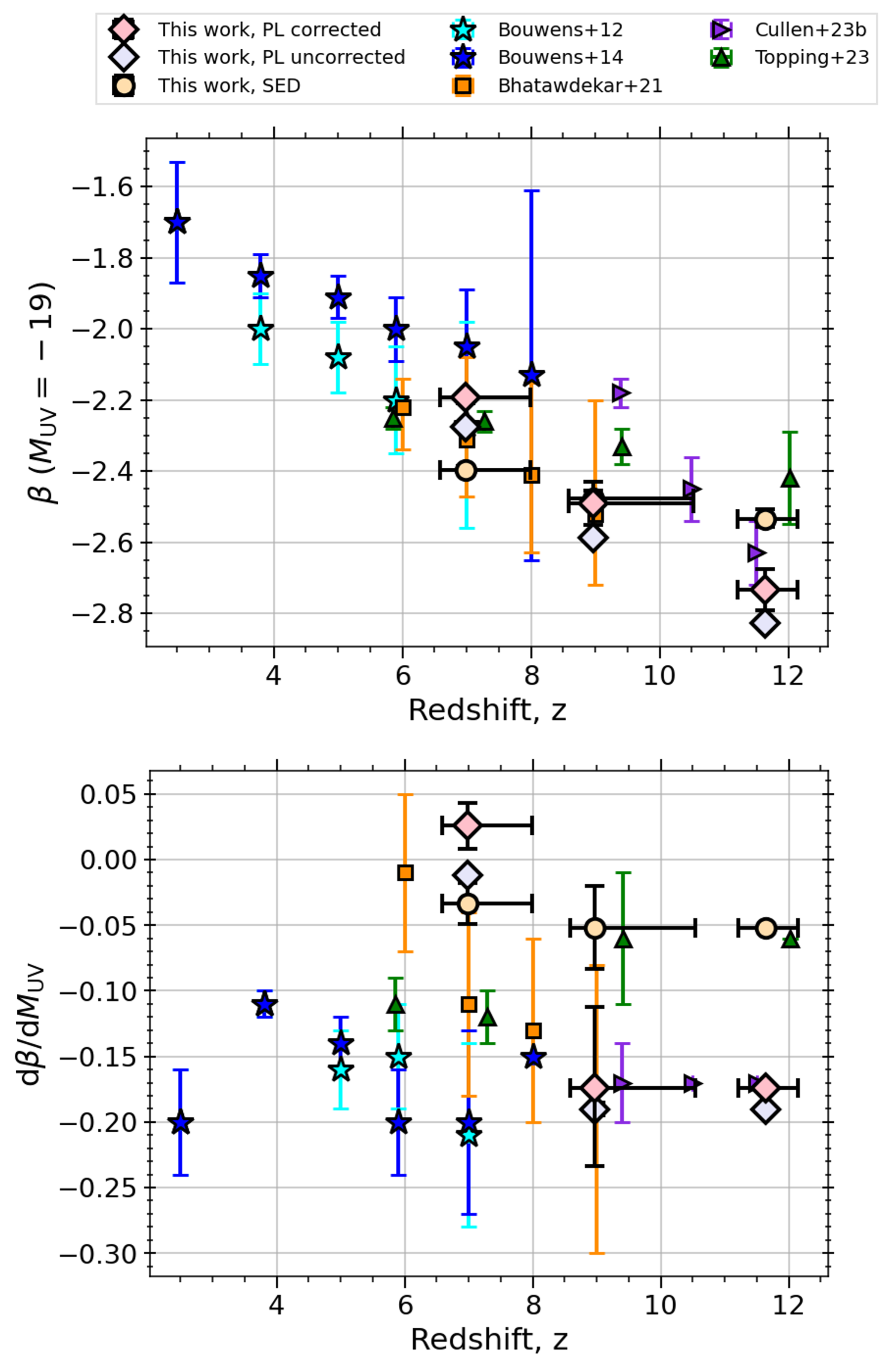}
    \caption{Amplitude (upper panel) and power law slope (lower panel) of the $\beta-M_{\mathrm{UV}}$ relations in \autoref{fig:UV_slope_vs_M_UV_z_split} showing the redshift evolution of $\beta(M_{\mathrm{UV}}=-19)$ and $\mathrm{d}\beta/\mathrm{d}M_{\mathrm{UV}}$ respectively, as defined in \autoref{eq:Beta_M_UV_fit_func}. $\beta$ results for both the bias corrected/uncorrected power law (pink/turquoise diamonds) and SED fitting (beige circles) methods are shown. Observational results collated from the literature from HST \citep{Bouwens2012, Bouwens2014a, Bhatawdekar2021} and JWST \citep{Cullen2023b, Topping2023} are also plotted.}
    \label{fig:Beta_vs_M_UV_fit_params}
\end{figure}

\subsection{Comparisons with stellar mass}
\label{sec:beta_mass}

The relationship between the UV spectral slope and stellar mass of galaxies has been studied in deep blank field surveys since the emergence of HST/WFC3IR data at $1.0<\lambda_{\mathrm{obs}}/\mu\mathrm{m}<1.6$ in the early 2010s. However, with the launch of JWST we now have access to rest-frame optical data up to $z\sim7-8$ allowing for better constraints on stellar masses at high redshift, which presents an ideal opportunity to investigate this relation further. 

We plot our bias corrected power law $\beta$ values against our {\tt Bagpipes} derived $\log_{10}M_{\star}$ in \autoref{fig:UV_slope_vs_mass_z_split}, fitting a power law function of the form 
\begin{equation}
    \beta = \beta_0 + \frac{\mathrm{d}\beta}{\mathrm{d}\log_{10}(M_{\star} / \mathrm{M}_{\odot})}\log_{10}(M_{\star} / \mathrm{M}_{\odot})
    \label{eq:Beta_mass_fit_func}
\end{equation}
where $\beta_0=\beta(\log_{10}(M_{\star} / \mathrm{M}_{\odot})=0)$ is a physically meaningless normalization factor. 20\% completeness contours derived from 5 realizations of the mock {\tt JAGUAR} catalogue (see \autoref{sec:Completeness_contamination} for details) are shown as red lines, with the limits of the {\tt JAGUAR} catalogue shown in black. Observational \citep{Finkelstein2012, Bhatawdekar2021} and simulated results from DELPHI \citep{Mauerhofer2023-DELPHI}, SC SAM GUREFT \citep{Yung2023-SC-SAM-GUREFT, Yung2024-SC-SAM-GUREFT}, DREaM \citep{Drakos2022-DREaM}, and FLARES \citep{Lovell2021_FLARES_I, Vijayan2021_FLARES_II, Wilkins2023_FLARES_V} are shown in the left hand panel. Bootstrapped median $\langle\beta\rangle$ and $\langle \log_{10}M_{\star} \rangle$ values in stellar mass bins are given in \autoref{tab:beta_mass_beta_corr} and $\mathrm{d}\beta/\mathrm{d}\log_{10}(M_{\star} / \mathrm{M}_{\odot})$ for bias corrected/uncorrected power law $\beta_{\mathrm{PL}}$ and {\tt Bagpipes} $\beta_{\mathrm{SED}}$ are tabulated in \autoref{tab:beta_MUV_beta_mass_fit_params} and plotted in comparison to observational results from \citet{Finkelstein2012} and \citet{Bhatawdekar2021} in \autoref{fig:UV_slope_vs_mass_slope}.

In our lowest redshift bin in \autoref{fig:UV_slope_vs_mass_z_split}, we see that the bootstrapped median points do not match the best-fit power law to the individual galaxies particularly well. While the choice of bin size plays a role here, we note that a power law is probably not the best fitting function to use in the future, although we use it here to provide direct comparisons to previous HST studies. In an attempt to quantify the threshold mass, $M_{\mathrm{thresh}}$, that our bootstrapped median data points galaxies become ``dust-free'', we fit a piecewise power law function that is flat below $M_{\mathrm{thresh}}$ and a power law at $M_{\star}>M_{\mathrm{thresh}}$ although without much success due to the large intrinsic scatter in the data. We note also that our results at $11<z<13$ suggest less dust attenuation \textit{on average} than predicted by the simulations, which was explored in more detail in \autoref{sec:beta_MUV_blue_beta_highz}.

The two most important takeaways from \autoref{fig:UV_slope_vs_mass_z_split}, however, are that the slope at $6.5<z<8.5$ and $8.5<z<11$ is shallower than observed by both \citet{Finkelstein2012} and \citet{Bhatawdekar2021} due to the presence of a large sample of low mass ($M_{\star}\sim10^{7.5}~\mathrm{M}_{\odot}$) red ($-2\lesssim\beta\lesssim-1$) galaxies, and that the slope at $11<z<13$ is much steeper than at lower redshift due to the non-detection of these low mass red objects. 

\subsubsection{Shallow \texorpdfstring{$\mathrm{d}\beta/\mathrm{d}M_{\star}$ at $6.5<z<8.5$}{dBeta/dMstar at 6.5 \textless z \textless 8.5}}
\label{sec:beta_mass_shallow_slope_lowz}

\begin{figure*}[h!]
    \centering
    \includegraphics[width=0.92\textwidth]{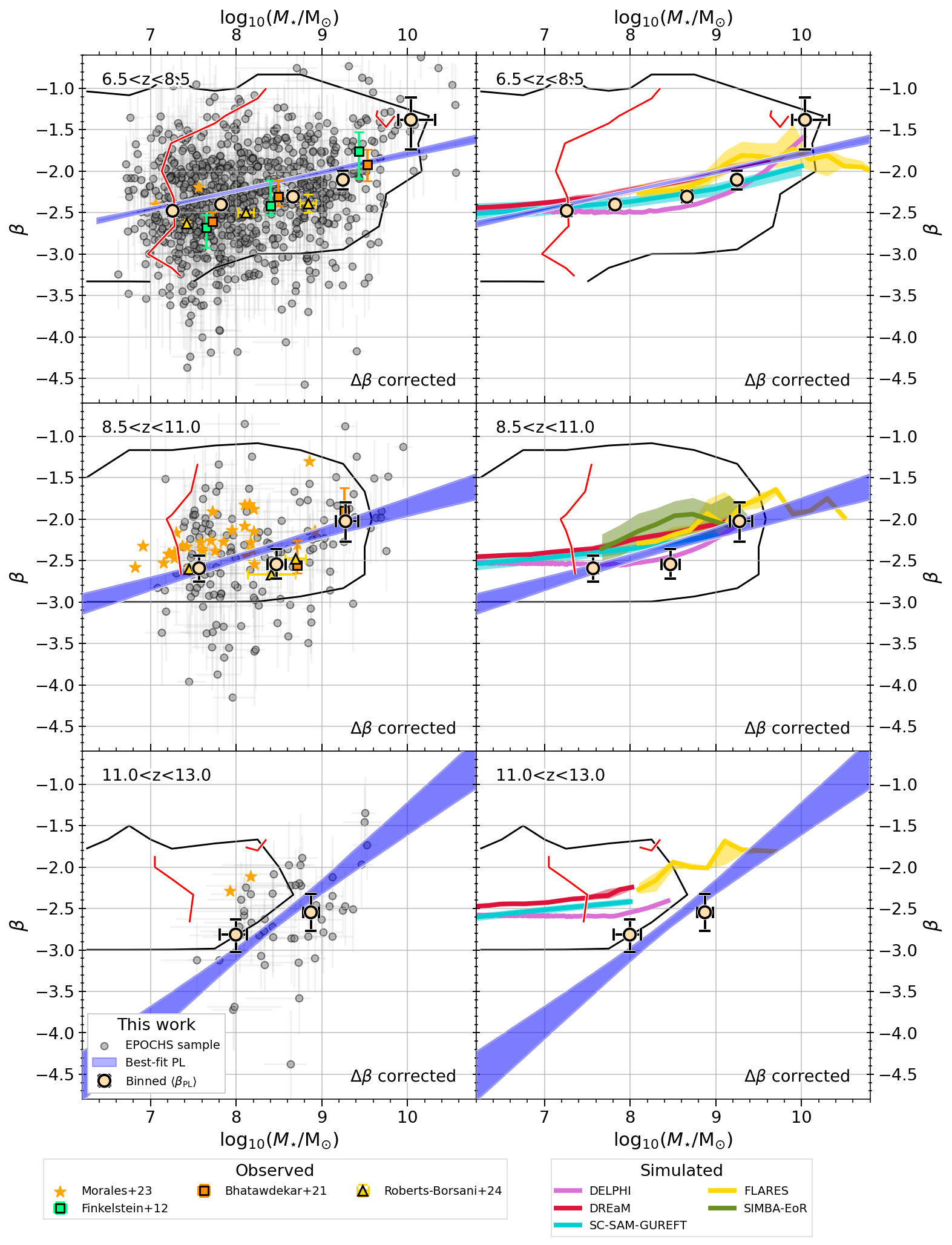}
    \caption{Redshift evolution of $\beta-M_{\star}$ for our EPOCHS-III sample, where the power law measured $\beta$ has been bias corrected. Bootstrapped median points are shown as beige circles, with individual galaxies in gray and the $16^{\mathrm{th}}-84^{\mathrm{th}}$ percentiles of the power law fit shown in blue. 20\% completeness contours from {\tt JAGUAR} are shown as red lines, with the {\tt JAGUAR} limit shown in black. We compare to observations \citep{Finkelstein2012, Bhatawdekar2021, Morales2023, Roberts-Borsani2024} and simulations (SC SAM GUREFT \citep[][$z=\{7, 9, 11\}$]{Yung2023-SC-SAM-GUREFT, Yung2024-SC-SAM-GUREFT}, DELPHI \citep[][$z=\{7.1, 9.1, 12.2\}$]{Mauerhofer2023-DELPHI}, FLARES \citep[][$z=\{7, 9, 12\}$]{Lovell2021_FLARES_I, Vijayan2021_FLARES_II, Wilkins2023_FLARES_V}, DREaM \citep[][$z=\{6.5, 8.5, 11\}$]{Drakos2022-DREaM} and SIMBA-EoR \citep[][$z=9$]{Dave2019-SIMBA-EoR, Wu2020-SIMBA-EoR}) in the left and right panels respectively. The black lines show limits of the {\tt{JAGUAR}} catalogue which mimics the CEERS observations, and the red lines show the 20\% completeness contours.}
    \label{fig:UV_slope_vs_mass_z_split}
\end{figure*}

While we do not go as deep as \citet{Bhatawdekar2021}, who utilize the strong gravitational lens of the MACS-0416 cluster to probe down to $M_{\mathrm{UV}}\lesssim-13.5$ at $z=6$, we notice that our selection criteria detect faint ($M_{\mathrm{UV}}\simeq-17.5$), low mass red objects more efficiently than their bluer counterparts. The reasoning behind this is somewhat unclear and depends on the difference in selection criteria used, although we note that these redder $\beta$ could potentially become accessible due to the increased rest-frame UV coverage up to $3000~\mathrm{\AA}$ provided by the F200W and F277W JWST/NIRCam wideband filters compared to HST/WFC3IR. The extent to which we are able to select these low mass red objects with JWST is apparent when comparing the 20\% completeness curves plotted in \autoref{fig:UV_slope_vs_mass_z_split} to those by \citet{Finkelstein2012} and \citet{Bhatawdekar2021} (see their Fig. 7 and Fig. 6 respectively). At $z\simeq7$, these pre-JWST completeness curves do not cover detections of galaxies redder than $\beta\simeq-2$ at $\log_{10}(M_{\star} / \mathrm{M}_{\odot})=7.5$, whereas our {\tt JAGUAR} completeness curves extend to $\beta\simeq-1.6$ at $6.5<z<8.5$.

\subsubsection{Steepening of \texorpdfstring{$\mathrm{d}\beta/\mathrm{d}M_{\star}$}{dBeta/dMstar} with increasing redshift}
\label{sec:beta_mass_steepening_slope}

\begin{figure}[h!]
    \centering
    \includegraphics[width=0.45\textwidth]{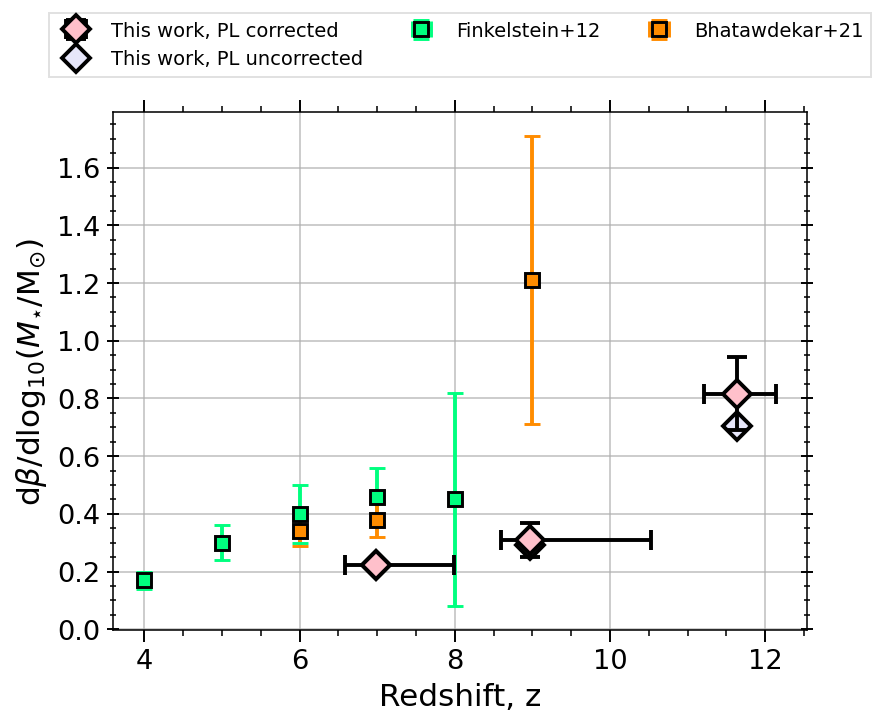}
    \caption{Evolution of the slope of the $\beta-M_{\star}$ relation, $\mathrm{d}\beta/\mathrm{d}\log(M_{\star}/\mathrm{M}_{\odot})$, for both our bias corrected (pink diamonds) and uncorrected (turquoise diamonds) power law $\beta$ measurements. We compare our results to those given in \citet{Finkelstein2012} (lime green) and \citet{Bhatawdekar2021} (orange), which use the SED fitting method to measure $\beta$. Consistently shallower results are measured at $z\simeq7$ and $z\simeq9$, and the first measurement at $z\simeq11.5$ is shown. We caution that we may be completeness dominated in the highest redshift bin.}
\label{fig:UV_slope_vs_mass_slope}
\end{figure}

From \autoref{fig:UV_slope_vs_mass_slope} we can see that our $\mathrm{d}\beta/\mathrm{d}\log(M_{\star}/\mathrm{M}_{\odot})$ increases significantly in our highest redshift bin from $0.31\pm0.06$ at $8.5<z<11$ to $0.81\pm0.13$ at $11<z<13$ due to the non-detection of low mass red galaxies that are seen in the lower redshift bins. At $11<z<13$, the most massive galaxies at $M_{\star}=10^{9.5}~\mathrm{M}_{\odot}$ are very red (with the reddest having $\beta=-1.3$), meaning that dust formation channels must already exist at these redshifts. Due to the young ages of these galaxies, Type II SNe are the most promising candidates as the major dust production mechanism. Since such red galaxies are not observed at lower masses, it is likely that the dust produced by these Type II SNe is lost in outflows instead of being retained by the large gravitational potential wells of the most massive galaxies \citep{Finkelstein2012}.

While our {\tt JAGUAR} 20\% completeness contours suggest that this parameter space should be detectable, we note that there are few galaxies in our {\tt JAGUAR} simulation at $11<z<13$. This is complemented by the fact that the {\tt JAGUAR} mock SEDs may not adequately reproduce the real Universe at high redshift, meaning that our completeness contours may not accurately represent the real completeness limits in this redshift bin. Galaxies with $\beta\gtrsim-2$, such as our sample of NIRCam selected red sources, lie outside of the HST 20\% completeness contours of \citet{Finkelstein2012} and \citet{Bhatawdekar2021}, meaning that these studies are likely completeness limited at these redshifts. In our highest redshift bin, we therefore cannot conclude that the trend is not induced by sample incompleteness since previous studies seem to have been.

\begin{table}[]
\centering
\caption{Bootstrapped binned results of median $\log_{10}(M_{\star}/\mathrm{M}_{\odot})$ and bias corrected power law $\beta$ for our EPOCHS-III sample as plotted in \autoref{fig:UV_slope_vs_mass_z_split}. We show results at $6.5<z<8.5$ in five $\log_{10}(M_{\star}/\mathrm{M}_{\odot})$ bins, as well as in three and two bins at the higher redshifts $8.5<z<11$ and $11<z<13$ respectively. The positive $\mathrm{d}\beta/\mathrm{d}M_{\star}$ slope is evident from these values in all three redshift bins.}
\label{tab:beta_mass_beta_corr}
\begin{tabular}{ccc}
\toprule
    $\log_{10}(M_{\star} / \mathrm{M}_{\odot})$ bin & $\langle \log_{10}(M_{\star} / \mathrm{M}_{\odot})\rangle$ & $\langle \beta \rangle$ \\ 
\hline 
\multicolumn{3}{c}{$6.5<z<8.5$} \\ 
\hline 
$\log_{10}(M_{\star} / \mathrm{M}_{\odot})<7.5$ & $7.26^{+0.02}_{-0.02}$ & $-2.48^{+0.07}_{-0.07}$ \\
$7.5<\log_{10}(M_{\star} / \mathrm{M}_{\odot})<8.25$ & $7.82^{+0.03}_{-0.03}$ & $-2.40^{+0.06}_{-0.06}$ \\
$8.25<\log_{10}(M_{\star} / \mathrm{M}_{\odot})<9.0$ & $8.66^{+0.03}_{-0.03}$ & $-2.31^{+0.07}_{-0.07}$ \\
$9.0<\log_{10}(M_{\star} / \mathrm{M}_{\odot})<9.75$ & $9.25^{+0.05}_{-0.05}$ & $-2.10^{+0.11}_{-0.11}$ \\
$\log_{10}(M_{\star} / \mathrm{M}_{\odot})>9.75$ & $10.04^{+0.28}_{-0.15}$ & $-1.38^{+0.27}_{-0.36}$ \\\hline 
\multicolumn{3}{c}{$8.5<z<11.0$} \\ 
\hline 
$\log_{10}(M_{\star} / \mathrm{M}_{\odot})<8.0$ & $7.57^{+0.05}_{-0.05}$ & $-2.59^{+0.16}_{-0.16}$ \\
$8.0<\log_{10}(M_{\star} / \mathrm{M}_{\odot})<9.0$ & $8.47^{+0.11}_{-0.10}$ & $-2.54^{+0.18}_{-0.18}$ \\
$\log_{10}(M_{\star} / \mathrm{M}_{\odot})>9.0$ & $9.28^{+0.15}_{-0.11}$ & $-2.02^{+0.23}_{-0.25}$ \\\hline 
\multicolumn{3}{c}{$11.0<z<13.0$} \\ 
\hline 
$\log_{10}(M_{\star} / \mathrm{M}_{\odot})<8.5$ & $8.00^{+0.13}_{-0.19}$ & $-2.81^{+0.18}_{-0.21}$ \\
$\log_{10}(M_{\star} / \mathrm{M}_{\odot})>8.5$ & $8.87^{+0.09}_{-0.09}$ & $-2.55^{+0.22}_{-0.23}$ \\
\botrule
\end{tabular}
\end{table}

\begin{table*}[]
\caption{Median redshifts, amplitudes and slopes for our power law fits to the $\beta-M_{\mathrm{UV}}$ and $\beta-M_{\star}$ relations in our 3 redshift bins ranging $6.5<z<13$. We show results using both the power law ($\beta_{\mathrm{PL}}$) and SED fitting ($\beta_{\mathrm{SED}}$) methods to measure $\beta$. In the upper panel results for the bias corrected power law $\beta$ are shown in brackets, and have the same error as the uncorrected values. In our highest redshift bin ($11<z<13$) we fix the slope of our $\beta-M_{\mathrm{UV}}$ to that of our $8.5<z<11$ bin (shown by $^\dagger$).}
\centering
\begin{tabular}{cccccc}
    \toprule
    z bin & $N_{\mathrm{gals}}$ & $\langle z \rangle$ & $\beta(M_{\mathrm{UV}}=-19)$ & $\mathrm{d}\beta/\mathrm{d}M_{\mathrm{UV}}$ & $\mathrm{d}\beta/\mathrm{d}\log(M_{\star}/\mathrm{M}_{\odot})$ \\
    \hline
    \multicolumn{6}{c}{$\beta_{\mathrm{PL}}$ ($\beta_{\mathrm{PL, corr}}$)} \\
    \hline
    $6.5<z<8.5$ & $823$ & $6.98^{+1.01}_{-0.39}$ & $-2.28(-2.19)\pm0.01$ & $-0.01(0.03)\pm0.02$ & $0.22(0.22)\pm0.02$ \\ 
    $8.5<z<11$ & $138$ & $8.97^{+1.56}_{-0.38}$ & $-2.59(-2.49)\pm0.06$ & $-0.19(-0.17)\pm0.06$ & $0.29(0.31)\pm0.06$ \\ 
    $11<z<13$ & $50$ & $11.64^{+0.49}_{-0.43}$ & $-2.83(-2.73)\pm0.06$ & $-0.19(-0.17)^\dagger$ & $0.70(0.81)\pm0.13$ \\
    \multicolumn{6}{c}{$\beta_{\mathrm{SED}}$} \\
    \hline
    $6.5<z<8.5$ & $823$ & $6.98^{+1.01}_{-0.39}$ & $-2.40\pm0.01$ & $-0.03\pm0.02$ & $0.09\pm0.01$ \\ 
    $8.5<z<11$ & $138$ & $8.97^{+1.56}_{-0.38}$ & $-2.48\pm0.02$ & $-0.05\pm0.03$ & $0.09\pm0.04$ \\ 
    $11<z<13$ & $50$ & $11.64^{+0.49}_{-0.43}$ & $-2.53\pm0.02$ & $-0.05^\dagger$ & $0.18\pm0.07$ \\
    \botrule
\end{tabular}
\label{tab:beta_MUV_beta_mass_fit_params}
\end{table*}
 
\subsection{The impact of sample contamination}
\label{sec:contamination_impact}

As in all galaxy samples, there will be some contaminant objects in our sample; these most likely arise as lower redshift galaxy interlopers due to Balmer-Lyman break degeneracy. In this work, we predict the number of these interlopers we expect to find using the {\tt JAGUAR} simulation \citep{Williams2018}, as explained in more detail in \autoref{sec:Completeness_contamination}. We assign a contamination percentage likelihood to each galaxy dependent on its origin survey and position in the ($M_{\mathrm{UV}}, \beta$) and ($M_{\star}, \beta$) parameter spaces. Adding these percentages together, we expect our sample to have $\{$40-90/823, 6-15/138, 2-3/50$\}$ contaminants in the respective $6.5<z<8.5$, $8.5<z<11$ and $11<z<13$ redshift bins (which are likely to be the bluest in our sample) corresponding to 5-10\%. This, of course, assumes that the distribution of galaxy colours in {\tt{JAGUAR}} matches the real Universe, and may not accurately represent the real contamination in our sample. A thorough comparison of the colours of simulations, including {\tt{JAGUAR}}, with NIRCam wide band photometric data from the CEERS survey is presented in \citet{Wilkins2023-colors}. Consequently, our simulation work identifies which regions of parameter space may be subject to contamination, but the precise contamination level may not be reliable. The reconciliation of this potential systematic is deemed beyond the scope of this work.

We exclude objects from our relationship fits that lie in the regions of parameter space found to have 10\%, 25\%, and 50\% contamination likelihood according to our {\tt JAGUAR} based simulations and re-fit our $\beta-M_{\mathrm{UV}}$ and $\beta-M_{\star}$ scaling relations to observe the difference that this has on our results. We perform the power law fitting of \autoref{eq:Beta_M_UV_fit_func} in the same fashion as in \autoref{fig:Beta_vs_z}, except this time we weight each galaxy, $i$, by a factor $w_i = 1 - \mathrm{Cont}_i(M_{\mathrm{UV}},\beta)$, where $\mathrm{Cont}_i(M_{\mathrm{UV}},\beta)$ is calculated for each galaxy as explained in \autoref{sec:Completeness_contamination}. 

\begin{figure}
    \centering
    \includegraphics[width=0.45\textwidth]{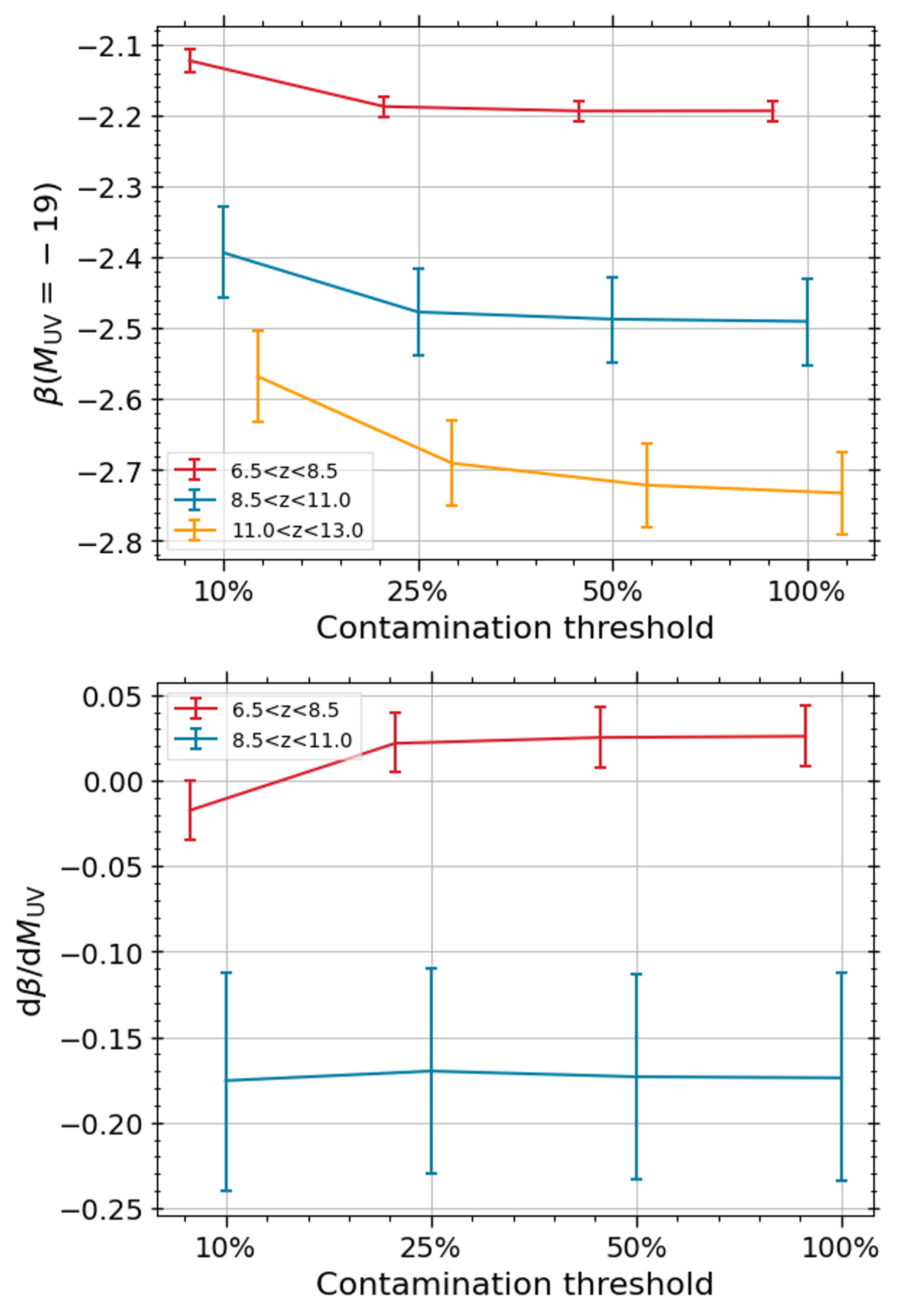}
    \caption{The impact of {\tt JAGUAR} estimated contamination on the amplitude (at $M_{\mathrm{UV}}=-19$; top panel) and slope (bottom panel) on our power law $\beta$ bias corrected $\beta-M_{\mathrm{UV}}$ relation measured in our 3 redshift bins. Shown on the right are the results when no contaminant objects are removed and moving leftwards we remove objects with $\mathrm{Cont}(M_{\mathrm{UV}}, \beta)<\{0.5, 0.25, 0.1\}$. Since we fix the $11<z<13$ slope, the trend of $\mathrm{d}\beta/\mathrm{d}M_{\mathrm{UV}}$ with contamination removal percentage in the lower panel matches the $8.5<z<11$ bin.}
    \label{fig:beta_M_UV_contamination}
\end{figure}

The impact of this on our $\beta-M_{\mathrm{UV}}$ fits is plotted in \autoref{fig:beta_M_UV_contamination}. We find that the largest impact on the power law slope of our $\beta-M_{\star}$ relation occurs at $6.5<z<8.5$, where we observe a decrease of $\simeq0.043$ from $\mathrm{d}\beta/\mathrm{d}M_{\mathrm{UV}}=0.026\pm0.017$ to $\mathrm{d}\beta/\mathrm{d}M_{\mathrm{UV}}=-0.017\pm0.017$ when removing all galaxies with contamination likelihood $>10\%$. Negligible evolution in $\mathrm{d}\beta/\mathrm{d}M_{\mathrm{UV}}$ is observed at $z>8.5$. We also observe a significant reddening of $\beta(M_{\mathrm{UV}}=-19)$ by $\simeq0.07$ ($6.5<z<8.5$), $\simeq0.10$ ($8.5<z<11$), and $\simeq0.16$ ($11<z<13$) when removing contaminant galaxies with contamination likelihood $>10\%$. This somewhat dampens our discussion on blue $\beta$ slopes should this high level of contamination be accurate, although we note that even with this reddening we still observe $\beta(M_{\mathrm{UV}}=-19)=-2.57\pm0.06$ in our $11<z<13$ bin which is still significantly bluer than the most up-to-date simulated results. Due to the small dependence on $\beta$ when binning the contamination in $\Theta=(M_{\star}, \beta)$, we find very little impact on the slope of $\beta-M_{\star}$ at any redshift. 

\section{Discussion}
\label{sec:Discussion}

\subsection{An abundance of faint, low mass, red galaxies at \texorpdfstring{$6.5<z<11$}{6.5 \textless z \textless 11}}

In the two lowest redshift bins of \autoref{fig:UV_slope_vs_mass_z_split} we observe an abundance of $\log_{10}(M_{\star} / \mathrm{M}_{\odot})<8.0$, $\beta>-2$ galaxies which are missed by studies of \citet{Finkelstein2012} and \citet{Bhatawdekar2021}. These objects are also observed with $M_{\mathrm{UV}}\gtrsim-19$ at $6.5<z<8.5$, which flatten the $\mathrm{d}\beta/\mathrm{d}M_{\mathrm{UV}}$. These could be a result of our failure to accurately trace the UV continuum; we may be biased red by high HI column density DLAs (already observed by JWST \citep[see][]{Heintz2023}) or strong resonant {\MgII}$-\lambda\lambda2796,2803$ line emission (henceforth {\MgII}), which may help trace $f_{\mathrm{esc, LyC}}$ \citep{Chisholm2020, Katz2022}. 

In addition, \citet{Schaerer2002} note that nebular dominated galaxies with hard ionizing fields (large $\log U$ values) can be biased red should they exist in the real Universe. Candidate nebular dominated galaxies, such as those presented in \citet{Cameron2023-nebular}, would also provide evidence for a long hypothesized top-heavy IMF in the early Universe should they exist. We use the {\tt synthesizer}\footnote{\url{https://github.com/flaresimulations/synthesizer}} \citep{Vijayan2021_FLARES_II} python code to measure the expected $\beta$ slope of a pure nebular dominated galaxy using a template with assumed $\log U = -2$, BPASS v2.2 SPS model, and \citet{Chabrier2003} IMF. We find $\beta_{\mathrm{neb, C94}}\simeq-1.0$ in the C94 filters (which have been seen to be biased red due to the increased wavelength coverage approaching two-photon continuum emission turnover) and $\beta_{\mathrm{neb, 2-window}}\simeq-1.4$ in two rest wavelength windows covering $\lambda_{\mathrm{rest}}=\{1250-1750,2250-2750\}~\mathrm{\AA}$. This two-window $\beta$ more accurately reflects our power law measurement technique, and therefore any nebular dominated galaxies, should they exist in our sample, would appear with $\beta\simeq-1.4$.

We do not attempt to quantify the number density of DLAs and nebular dominated galaxies in our EPOCHS-III sample as the defining features of these SEDs are not traceable with the JWST/NIRCam wide-bands used in these early blank-field surveys. Increased medium band coverage utilizing JWST/NIRCam F210M, F250M, F300M, and F335M probing $2000<\lambda_{\mathrm{rest}}~/~\mathrm{\AA}<3000$ at $6.5<z<8.5$ can be used to both estimate {\MgII} EWs and detect the Balmer jump at $3646~\mathrm{\AA}$ in nebular dominated galaxies. The detection of these features in unbiased medium-band photometric surveys, and subsequent modelling of the expected number densities of these systems, will allow us to distinguish whether these red galaxies are indeed as dusty as their red $\beta$ suggests.

\subsection{Dust implications}
\label{sec:Dust_implications}

From the results of our $\beta-M_{\star}$ relations, we propose possible scenarios for the average build up of galactic dust. We start in our highest redshift bin at $11<z<13$ where we observe a steep $\mathrm{d}\beta/\mathrm{d}\log_{10}(M_{\star}/\mathrm{M}_{\odot})=0.81\pm0.13$. The most massive galaxies in this redshift bin have $M_{\star}\simeq10^{9.5}~\mathrm{M}_{\odot}$ and $\beta\simeq-1.5$, meaning galactic dust must have been formed at these epochs. Due to the relatively young ages of galaxies at these redshifts, we attribute this to dust formed in Type II SNe on the smallest $\sim10$~Myr timescales (see also \citet{Finkelstein2012}). Since these SNe must also occur in the lowest mass systems, the lack of observed red, $M_{\star}<10^8~\mathrm{M}_{\odot}$ galaxies means that this dust is not retained, most likely via SNe feedback induced outflows which remove dust from the small gravitational potential wells of the low mass host galaxies. As well as this, low mass galaxies may well be prefentially bursty, meaning radiation pressure from O/B-type main sequence stars may contribute here. 

Following the initial phase of Type II SNe dust production domination at $z>11$, we observe an average reddening in galaxies at all stellar masses. This reddening is more prominent for the lower mass galaxies which flattens the slope of the $\beta-M_{\star}$ relation. One plausible explanation for this reddening is by dust production on slightly longer timescales than Type II SNe from short-lived carbon-rich Wolf-Rayet (WC) stars, which have been seen to produce copious amounts of dust ($10^{-10}-10^{-6}~\mathrm{M}_{\odot}\mathrm{yr}^{-1}$ per WC) in the local Universe \citep{Lau2020}. This production mechanism has been somewhat overlooked in the past due to the requirement of an O/B main-sequence companion star for efficient dust production, however with newer BPASS SPS models this can now be studied more thoroughly \citep[see][]{Lau2020}.

While WR stars are quite rare in typical local IMFs, more top-heavy IMFs which may be naively expected in the early Universe \citep[proposed by e.g.][]{Rasmussen2023, Steinhardt2023-HOT-IMFs} may produce enough WC stars (if sufficiently carbon-enriched) to fully account for the reddening observed between $z\simeq12$ and $z\simeq9$ should dust destruction processes not be prevalent in the reverse shock of the resulting SNe. Quantitatively, the IMF required for the two-photon nebular continuum to dominate in GS-NDG-9422 from \citet{Cameron2023-nebular} would produce a WR star for every $\sim140~\mathrm{M}_{\odot}$ of normal stellar population compared to every $\sim1300~\mathrm{M}_{\odot}$ in a typical local IMF. Evidence for dust produced via WC binaries, for instance from the $\lambda_{\mathrm{rest}}=2175~\mathrm{\AA}$ carbonaceous signature found at $z=6.71$ by \citet{Witstok2023}, would likely point towards a more top-heavy IMF.

By $z=6.5$ the Universe is approximately $825$~Myr old, with $500$~Myr of time having passed since the upper redshift limit of this study at $z=13$. Should all galaxies in our $6.5<z<8.5$ bin be $\sim500$~Myr old, we expect the stellar winds of asymptotic giant branch (AGB) stars to dominate the production of dust. Although these stars are less massive ($0.5-8~\mathrm{M}_{\odot}$), longer lived ($\gtrsim100$~Myr timescales) and produce less dust than WR stars, they are also more common. In addition, ISM dust grain growth may also be prevalent in this stage of galaxy evolution, although little is known regarding the size, shape, and chemical composition of dust grains at $z=6.5$ beyond theoretical predictions by, for example, \citet{Hensley2023}.

As well as the average reddening, we also observe an increasing scatter in $\beta-M_{\star}$ with decreasing redshift. Even though at lower redshift, there is more diversity in ages, metallicities, and environment at a given mass, \citet{Bouwens2012} show (in their Fig. 18) that the scatter in dust attenuation provides the largest impact on the scatter in $\beta$. We conclude, therefore, that the increase in scatter between $z=13$ and $z=6.5$ is due to the variety of dust production methods available in these galaxies (from Type II SNe, ISM dust grain growth, or WC/RSG/AGB stars), which are on average older, and may lead to a diversity of dust attenuation laws diverging from the standard \citet{Calzetti2000} or SMC/LMC laws adopted in the local Universe.

\subsection{A robust sample of blue \texorpdfstring{$\beta<-2.8$}{beta < 2.8} galaxies}
\label{sec:Blue_galaxies}

Our fits to $\beta-M_{\mathrm{UV}}$ in \autoref{fig:Beta_vs_M_UV_fit_params} show that on average the galaxy population is uniformly blue at $z\gtrsim11.0$, most likely due to the lack of dust at these high redshifts. Much work has been done in the past to investigate galaxies with ultra-blue $\beta$ slopes with minimal $A_{\mathrm{UV}}$ dust extinction, including the possibility of Pop.~III stars and an increasingly top-heavy IMF \citep[e.g.][]{Schaerer2002, Schaerer2003, Raiter2010b, Zackrisson2011}.

\begin{figure}
    \centering
    \includegraphics[width=0.45\textwidth]{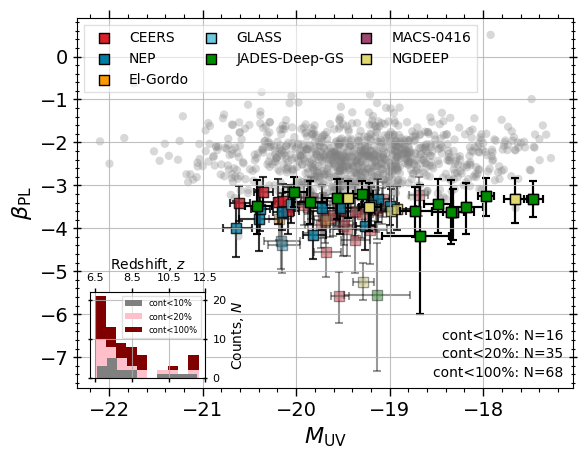}
    \caption{Sub-sample of 68 ultra-blue $\beta+\sigma_{\beta}<-2.8$ candidates with 33 and 16 having less than 20\% and 10\% contamination likelihood plotted as coloured points with more solid colours (less transparency). The rest of the EPOCHS-III sample within our redshift range of interest is plotted in black. Redshift distributions of the 3 sub-samples of ultra-blue candidate galaxies are shown in the lower left plot inset.}
    \label{fig:Blue_sample}
\end{figure}

We produce a tiered sample of 68 ultra-blue galaxies with $\beta+\sigma_{\beta}<-2.8$, 35 of which have $\mathrm{Cont}(M_{\mathrm{UV}}, \beta)<0.2$ with 16 galaxies from NGDEEP and JADES-Deep-GS having $\mathrm{Cont}(M_{\mathrm{UV}}, \beta)<0.1$. We plot this tiered sample in \autoref{fig:Blue_sample} with transparency defined by $1-\mathrm{Cont}(M_{\mathrm{UV}}, \beta)$.

To analyze the possible scenarios that could give rise to these extreme blue average $\beta$ values, we compare our results to the Pop.~I, II, and III instantaneous burst SED models produced by the Yggdrasil SPS code \citep{Zackrisson2011}. The Pop.~I and Pop.~II models are plotted with metallicities $Z_{\star}=\{0.02,0.2,1\}~\mathrm{Z}_{\odot}$, a Starburst99 single stellar population (SSP) based on Padova-AGB tracks \citep{Leitherer1999,Vazquez2005}, a gas density $n_{\mathrm{H}}=100~\mathrm{cm}^{-2}$, and a universal \citet{Kroupa2001} IMF in the range $0.1-100~\mathrm{M}_{\odot}$. The ``PopIII'' model uses the same \citet{Kroupa2001} IMF as the Pop.~I and Pop.~II models, although with a rescaled SSP from \citet{Schaerer2002}. The ``PopIII.2'' model uses a moderately top-heavy IMF (lognormal with $10~\mathrm{M}_{\odot}$ characteristic mass, dispersion $\sigma=1.0$, and $1-500~\mathrm{M}_{\odot}$ wings) from \citet{Raiter2010b}, while the ``PopIII.1'' model uses the most top-heavy IMF ($50-500~\mathrm{M}_{\odot}$ with \citet{Salpeter1955} IMF slope) from \citet{Schaerer2002}. All Pop.~III models assume zero metallicity by definition.

\begin{figure}
    \centering
    \includegraphics[width=0.45\textwidth]{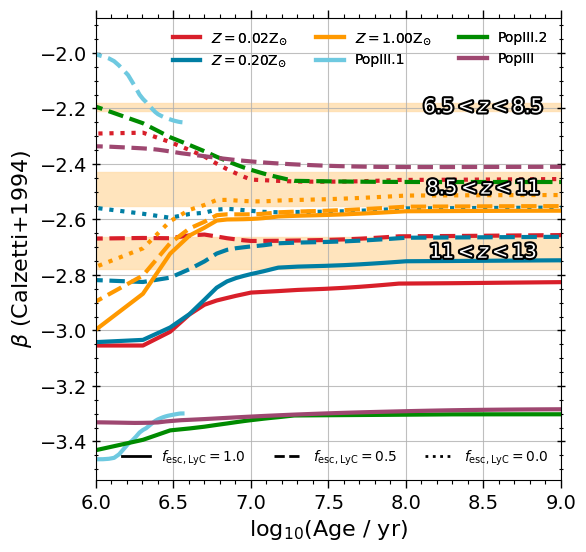}
    \caption{Power law measured UV $\beta$ slopes as a function of galaxy age for the instantaneous burst Yggdrasil population I, II, and III SEDs \citep{Zackrisson2011} measured in the 10 C94 filters to avoid bias from nebular rest-UV line emission prevalent at young ages. Bias corrected power law $\beta$ constraints for our three redshift bins are highlighted in light orange showing the most notable new $11<z<13$ ultra-blue $\langle \beta \rangle$ measurements.}
    \label{fig:yggdrasil_beta}
\end{figure}

In \autoref{fig:yggdrasil_beta}, we calculate $\beta$ in the 10 C94 filters for a range of ages, $10^6-10^9$~yr, and plot against our average $\beta$ results at $M_{\mathrm{UV}}=-19$ given in \autoref{tab:beta_MUV_beta_mass_fit_params} from our best-fitting power law to the $\beta-M_{\mathrm{UV}}$ relation in our 3 redshift bins. First of all, we conclude that since none of these dust-free models can reproduce our results at $6.5<z<8.5$, there must have been dust built up in the majority of these galaxies already at this epoch. We now focus our attention towards our highest redshift bin, where the bluest and most extreme $\langle\beta\rangle$ are observed. At these redshifts, we cannot rule out the possibility that there are Pop.~III sources within our sample and propose 2 plausible scenarios which could explain our observational results: 
\begin{enumerate}
    \item We are dominated by low-metallicity, $Z_{\star}<\mathrm{Z}_{\odot}$, stellar populations with moderate to extreme Lyman continuum leakage, $f_{\mathrm{esc, LyC}}>0.5$, in dust-free environments. It has been shown, however, that both FSPS \citep{Conroy2009,Conroy2010,Byler2017} and $\alpha$-enhanced BPASS \citep{Byrne2022-BPASS} models reduce the dependence of $\beta$ of $f_{\mathrm{esc, LyC}}$ and hence allow for lower values of $f_{\mathrm{esc, LyC}}$ to coincide with our observations \citep[see][Fig. 9]{Cullen2023b}.
    \item There exists a non-negligible number of galaxies in the sample with a more top-heavy (potentially Pop.~III) IMF which either must have $f_{\mathrm{esc, LyC}}\simeq1$, if enshrouded in dust, or moderate $f_{\mathrm{esc, LyC}}\gtrsim0.5$ if dust-free. We note that we have already seen evidence of strong metal lines in spectra at $z>11$ (e.g. GN-z11), making it unlikely that many of our high redshift galaxy sample host entirely metal-free Pop.~III stellar populations. There remains the possibility, however, that Pop.~III stars contribute a non-negligible amount to the stellar SED due to the co-existence and incomplete mixing of these with more metal-enriched stars \citep{Sarmento2018, Sarmento2019}.
\end{enumerate}

Aside from these two scenarios that would explain these blue $\beta$ mentioned above, there still remains the possibility that our subsample is biased blue by Ly$\alpha$ emission at (as much as $\Delta \beta = -0.6$ for $\mathrm{EW}_{\mathrm{Ly} \alpha}=300~\mathrm{\AA}$ at $z \sim 7$), is dominated by contamination from Balmer break galaxies, or is simply produced as a result of photometric scatter in our wide-band photometric NIRCam surveys.

\section{Conclusions} \label{sec:conclusions}

In this paper, we have calculated the UV continuum slopes, $\beta$, absolute UV magnitudes, $M_{\mathrm{UV}}$, and stellar masses, $M_{\star}$, for 1011 high redshift galaxies at $6.5<z<13$ taken from the EPOCHS v1 sample across $178.9~\mathrm{arcmin}^2$ of unmasked blank sky area. The main aims of this study are to both trace the build up of dust from the \textit{average} $\beta$ values as well as search for extremely blue $\beta<-3$ in \textit{individual} galaxies in the EoR. We favor the power law method to measure $\beta$ as it is not biased red by the Bayesian prior on $\beta$ and it better matches 41 cross-matched NIRSpec PRISM derived $\beta_{\mathrm{spec}}$ collated from the DJA. We summarize the main results of this paper below:

\begin{enumerate}
    \item We quantify the potential impact of rest-frame UV line emission, Lyman alpha emitters (LAEs) and damped Lyman alpha systems (DLAs) on the measured $\beta$ as a function of redshift in the range $6.5<z<13$. Biases can be seen as large as $\lvert \Delta\beta \rvert \simeq 0.1$ for UV-line emitters per $10~\mathrm{\AA}$ EW, $\Delta\beta=-0.6$ for the strongest LAEs with EW=$300~\mathrm{\AA}$, and $\Delta\beta=0.5$ for proximate DLAs with $N_{\mathrm{HI}}=10^{23.5}~\mathrm{cm}^{-2}$. Though some JWST/NIRCam medium band filters are present in several wide-field surveys, expanding the number of filters used in deep fields can help in reducing these biases, for instance in the JADES Origin Field (PID: 3215).
    
    \item Our $\beta$ bias simulations show that the faintest galaxies in our sample are biased blue, with maximum $\Delta\beta\simeq-0.55$. This bias is larger for redder galaxies due to the increased selectability of blue objects, and is especially poor ($\Delta\beta\sim-0.3$ even in the brightest sources) for red objects at $6.5<z<7.5$ in fields where blue supplementary HST/ACS data is not included and at $7.5<z<9.5$ in CEERS and NGDEEP where there is no F090W filter even when including F814W data from HST/ACS.
    
    \item We measure a decreasing trend of $\beta$ with redshift, $\beta=-1.51\pm0.08-(0.097\pm0.010)\times z$. This is corroborated by our $\beta(M_{\mathrm{UV}}=-19)$ decreasing from to $\beta(M_{\mathrm{UV}}=-19)=-2.19 \pm 0.06$ at $z\simeq7$ to $\beta(M_{\mathrm{UV}}=-19)=-2.73 \pm 0.06$ at $z\simeq11.6$, implying minimal \textit{average} dust attenuation at the highest redshifts and  deviating from the {\tt FLARES}, {\tt DELPHI}, {\tt SC SAM GUREFT}, and {\tt DREaM} simulations. Our $\beta-z$ relation is also discrepant with recent JWST observations by \citet{Topping2023} and falls between those by \citet{Cullen2023b} and \citet{Roberts-Borsani2024} due to differing sample $M_{\mathrm{UV}}$ distributions and selection procedures. 
    
    \item We measure a flatter $\mathrm{d}\beta/\mathrm{d}M_{\mathrm{UV}}=0.03\pm0.02$, leading an $M_{\mathrm{UV}}$ independent $\beta-z$ relation, and a shallower $\mathrm{d}\beta/\mathrm{d}\log_{10}(M_{\star} / \mathrm{M}_{\odot})=0.24\pm0.01$ at $z\simeq7$ than seen in previous HST studies \citep[e.g.][]{Finkelstein2012, Bhatawdekar2021}, revealing a large population of low mass, faint, red galaxies. These could be DLAs or nebular dominated galaxies, but if indeed reddened by dust this would imply either early dust production in the stellar winds of asymptotic giant branch (AGB) or carbon-rich Wolf-Rayet binaries coupled with reduced dust destruction in subsequent SNe reverse shocks.

    \item The observed steepening of $\mathrm{d}\beta/\mathrm{d}\log_{10}(M_{\star} / \mathrm{M}_{\odot})$ towards high redshift implies that dust produced by core-collapse SNe at the earliest times is ejected by SNe-induced outflows and retained by the large gravitational potential wells of high mass galaxies.

    \item We identify 68 $\beta+\sigma_{\beta}<-2.8$ ultra-blue galaxy candidates that are potential LyC leakers and may host Pop.~III or top-heavy IMFs, although comparison to spectroscopy shows the $\beta$ of individual objects is difficult to accurately constrain.
\end{enumerate}

We have collated one of the largest samples of high-redshift galaxies at $z>6.5$ and performed a comprehensive analysis of the $\beta$ biases associated with this sample. Whilst we speculate on the implications of our results regarding the dust content and production channels, we note that many scenarios here are degenerate and indistinguishable with photometric data alone. Our candidate ultra-blue ($\beta+\sigma_{\beta}<-2.8$) galaxies may provide evidence for Pop.~III stellar populations and/or significant Lyman continuum leakage, although high UV continuum SNR spectra from NIRSpec remain crucial to confirm these theories.

\section{Acknowledgments}

\vspace{10pt}

We acknowledge support from the ERC Advanced Investigator Grant EPOCHS (788113), as well as two studentships from STFC for DA and TH. LW acknowledges funding from the Faculty of Science \& Engineering at the University of Manchester. LF acknowledges financial support from Coordenação de Aperfeiçoamento de Pessoal de Nível Superior - Brazil (CAPES) in the form of a PhD studentship. RW, SHC, and RAJ acknowledge support from NASA JWST Interdisciplinary Scientist grants NAG5 12460, NNX14AN10G and 80NSSC18K0200 from GSFC. CC is supported by the National Natural Science Foundation of China, No. 11803044, 11933003, 12173045. This work is sponsored (in part) by the Chinese Academy of Sciences (CAS), through a grant to the CAS South America Center for Astronomy (CASSACA). We acknowledge the science research grants from the China Manned Space Project with No. CMS-CSST-2021-A05. MAM acknowledges the support of a National Research Council of Canada Plaskett Fellowship, and the Australian Research Council Centre of Excellence for All Sky Astrophysics in 3 Dimensions (ASTRO 3D), through project number CE17010001. MN acknowledges INAF-Mainstreams 1.05.01.86.20. CNAW acknowledges support from the NIRCam Science Team contract to the
University of Arizona, NAS5-02015.

This work is based on observations made with the NASA/ESA \textit{Hubble Space Telescope} (HST) and NASA/ESA/CSA \textit{James Webb Space Telescope} (JWST) obtained from the \texttt{Mikulski Archive for Space Telescopes} (\texttt{MAST}) at the \textit{Space Telescope Science Institute} (STScI), which is operated by the Association of Universities for Research in Astronomy, Inc., under NASA contract NAS 5-03127 for JWST, and NAS 5–26555 for HST. The PEARLS observations used in this work are associated with JWST programs 1176 and 2738. In addition, public datasets from JWST programs 1180, 1210, 1895, 1963 (JADES), 1324 (GLASS), 1345 (CEERS) and 2079 (NGDEEP) are also used within the work presented. Some of the data products presented herein were retrieved from the Dawn JWST Archive (DJA). DJA is an initiative of the Cosmic Dawn Center, which is funded by the Danish National Research Foundation under grant No. 140. The authors thank all involved in the construction and operations of the telescope as well as those who designed and executed these observations, their number are too large to list here and without each of their continued efforts such work would not be possible. The authors also thank Adam Carnall for their prompt help with {\tt Bagpipes} via email, as well as helpful discussions with Rebecca Bowler, Fergus Cullen, and Albert Zijlstra which significantly improved the discussion of results. This work is dedicated to the memory of our dedicated colleague and co-author Mario Nonino, who sadly passed during the completion of this work.

The authors thank Anthony Holloway and Sotirios Sanidas for their providing their expertise in high performance computing and other IT support throughout this work. Data used in this publication (including catalogues and imaging) will be made publicly available once initial works are completed with students involved in its reduction and analysis. The anticipated timescale is early Summer 2024. Products using GTO data will be made available as and when exclusive access periods lapse. This work makes use of {\tt astropy} \citep{Astropy2013,Astropy2018,Astropy2022}, {\tt matplotlib} \citep{Hunter2007}, {\tt reproject}, {\tt DrizzlePac} \citep{Hoffmann2021}, {\tt SciPy} \citep{2020SciPy-NMeth}, {\tt photutils} \citep{Bradley2022}, and the currently private {\tt galfind}. v1 of the {\tt galfind} code is expected to be released to the public within the next year.


%






\appendix

\section{The impact of little red dots}
\label{sec:little_red_dots}

\begin{figure*}
    \centering
    \includegraphics[width=0.95\textwidth]{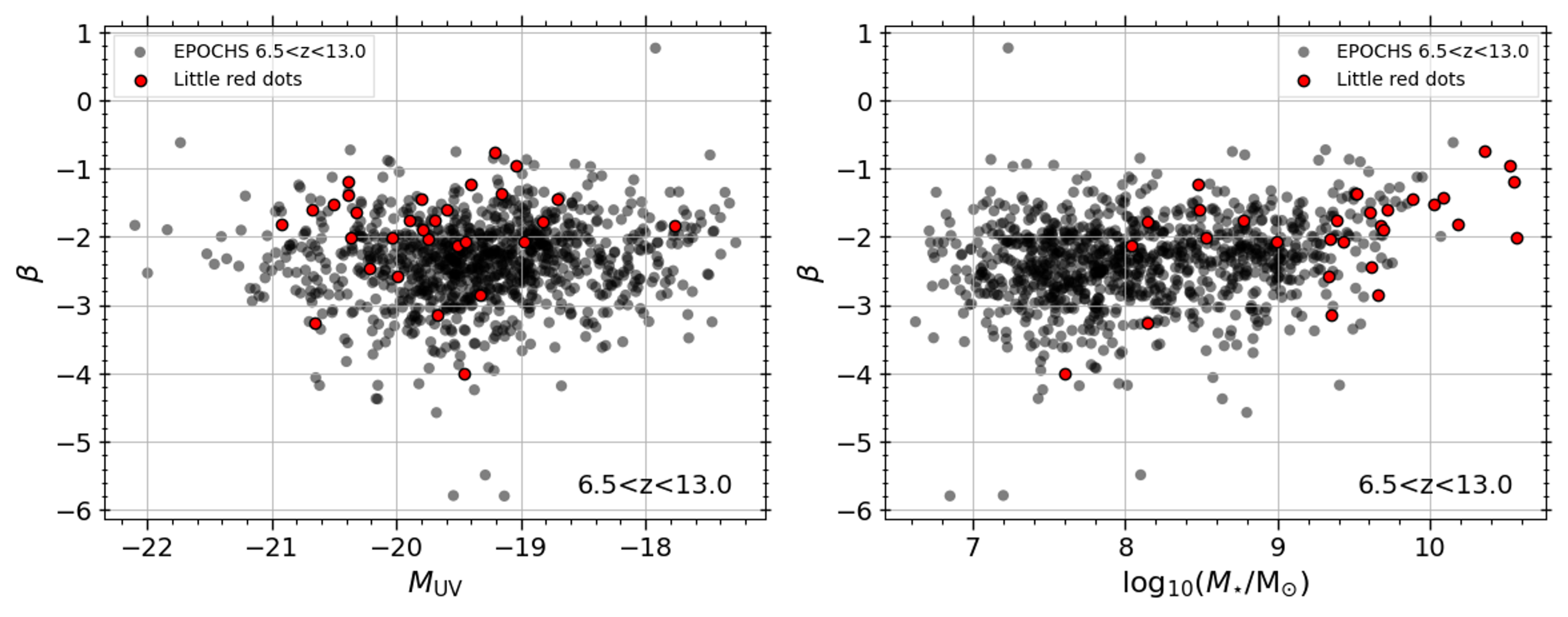}
    \caption{The location of the 34 ``little red dots'' identified in our EPOCHS-III sample using the \citet{Kokorev2024} selection criteria. The majority host high observed stellar masses, many of which are likely overestimated, and red rest-frame UV colours, although we note the large scatter in the $\beta$ distribution of these sources.}
    \label{fig:little_red_dots}
\end{figure*}

Inital JWST images have uncovered a large sample of ``little red dots'' (LRDs) that have been catalogued from photometric \citep[e.g.][]{Labbe2023, Kokorev2024} and spectroscopic \citep[e.g.][]{Matthee2023, Kocevski2024} data, some of which have been confirmed to be partially dust obscured AGN via their broadened H$\alpha$ lines. We identify 34 LRDs when applying the `\textit{red2}' color selection criteria, compactness criterion, and SNR requirements from \citet{Kokorev2024} to our EPOCHS-III sample, representing the same sub-sample as in EPOCHS-IV. The location of these systems in the $(\beta, M_{\mathrm{UV}})$ and $(\beta, M_{\star})$ parameter spaces is plotted in \autoref{fig:little_red_dots}.

These LRDs in general fall within the lowest redshift $6.5<z<8.5$ bin and are mainly intermediate to bright in the rest-frame UV compared with the rest of our EPOCHS-III sample. They are also known to exhibit high stellar masses due to the confusion of the red AGN SED component with an aged stellar population. The $\beta$ slopes of these LRDs exhibit a wide range of values, but on average are redder than the general SFG population, most likely due to dust production in enriched quasar-driven winds \citep[e.g.][]{Valiante2014, Sarangi2019}. While we do not re-compute our $\beta-M_{\mathrm{UV}}$ and $\beta-M_{\star}$ fits excluding these sources, we note that this would likely have minimal impact due to the small relative sample size compared to our full EPOCHS-III sample($34/1011$).

\section{NIRSpec PRISM cross-matches from the DJA}
\label{appendix:spectroscopy_tab}

A total of 24 of the 41 spectroscopically confirmed galaxies in our sample have been found by previous studies \citep{ArrabalHaro2023a, Bunker2023-NIRSpec, Curtis-Lake2023, Hainline2023a, Heintz2023b, Heintz2023, Nakajima2023, Tang2023, Sanders2024}, with 3 found in photometric surveys by \citet{Donnan2023} and \citet{Hainline2023a}. We provide results of these cross-matches in \autoref{tab:spec_beta}, where we calculate $\beta$ both from the NIRSpec PRISM data as well as the NIRCam photometry using the two techniques discussed in this work. The spectroscopic $\beta$ are measured in the C94 filters and the $1250<\lambda_{\mathrm{rest}}~/~\mathrm{\AA}<3000$ wavelength range showing the wavelength dependence of these $\beta$ measurements especially in low cases of low UV continuum signal-to-noise ratio (SNR).

\begin{table*}[]
    \centering
    \caption{Spectroscopically confirmed galaxies from a cross match of our EPOCHS-III sample with the DJA. We provide spectroscopic $\beta$ measurements in the C94 filters and the $1250-3000~\mathrm{\AA}$ rest-frame wavelength range as well as photometric $\beta$ from beta bias corrected power law fits to the photometry and {\tt{Bagpipes}} SED fitting. $\mathrm{SNR}_{\mathrm{UV}}$ indicates the SNR obtained from the PRISM spectrum averaged over $1250<\lambda_{\mathrm{rest}}~/~\mathrm{\AA}<3000$. We reference works that have previously reported these galaxies following the numbering system: 1) \citet{Donnan2023}, 2) \citet{Heintz2023b}, 3) \citet{Curtis-Lake2023}, 4) \citet{Bunker2023-NIRSpec}, 5) \citet{Hainline2023a}, 6) \citet{ArrabalHaro2023a}, 7) \citet{Heintz2023}, 8) \citet{Nakajima2023}, 9) \citet{Sanders2024}, 10) \citet{Tang2023}.}
    \label{tab:spec_beta}
    \setlength{\tabcolsep}{3.5pt}
    \begin{tabular}{ccccccccccc}
    \toprule
    \multirow{2}{*}{PID} & \multirow{2}{*}{RA} & \multirow{2}{*}{Dec} & \multirow{2}{*}{$z_{\mathrm{spec}}$} & $\beta_{\mathrm{spec}}$ & $\beta_{\mathrm{spec}}$ & \multirow{2}{*}{SNR$_{\mathrm{UV}}$} &  \multirow{2}{*}{$z_{\mathrm{phot}}$} & \multirow{2}{*}{$\beta_{\mathrm{phot, PL}}$} & \multirow{2}{*}{$\beta_{\mathrm{phot, SED}}$} & \multirow{2}{*}{Ref.} \\
    & & & & (C94) & (1250-3000{\AA}) & & & & & \\
    \hline
1180 & $53.155144$ & $-27.760742$ & $6.3139$ & $-2.26 \pm 0.03$ & $-2.27 \pm 0.08$ & $9.51$ & $6.52_{-0.07}^{+0.05}$ & $-2.53_{-0.32}^{+0.34}$ & $-2.44_{-0.05}^{+0.07}$ & - \\
1180 & $53.127316$ & $-27.788040$ & $6.3845$ & $-3.00 \pm 0.05$ & $-2.58 \pm 0.16$ & $3.81$ & $6.57_{-0.07}^{+0.04}$ & $-2.62_{-0.33}^{+0.35}$ & $-2.52_{-0.04}^{+0.05}$ & - \\
1180 & $53.137429$ & $-27.765207$ & $6.6223$ & $-2.30 \pm 0.04$ & $-2.40 \pm 0.15$ & $3.53$ & $6.59_{-0.06}^{+0.03}$ & $-2.32_{-0.34}^{+0.30}$ & $-2.46_{-0.04}^{+0.09}$ & - \\
1180 & $53.169516$ & $-27.753317$ & $6.6284$ & $-2.53 \pm 0.07$ & $-2.68 \pm 0.15$ & $4.66$ & $6.60_{-0.02}^{+0.02}$ & $-2.58_{-0.32}^{+0.32}$ & $-2.27_{-0.09}^{+0.10}$ & - \\
1180 & $53.118187$ & $-27.793008$ & $6.7895$ & $-1.96 \pm 0.06$ & $-2.38 \pm 0.18$ & $3.12$ & $6.89_{-0.04}^{+0.03}$ & $-2.31_{-0.32}^{+0.32}$ & $-2.13_{-0.16}^{+0.12}$ & - \\
1180 & $53.138054$ & $-27.781863$ & $7.1391$ & $-2.66 \pm 0.06$ & $-2.24 \pm 0.25$ & $2.12$ & $7.40_{-0.24}^{+0.05}$ & $-2.73_{-0.51}^{+0.48}$ & $-2.40_{-0.06}^{+0.10}$ & - \\
1180 & $53.161709$ & $-27.785391$ & $7.2349$ & $-1.45 \pm 0.04$ & $-1.96 \pm 0.13$ & $4.05$ & $7.45_{-0.11}^{+0.07}$ & $-2.27_{-0.48}^{+0.52}$ & $-1.73_{-0.14}^{+0.14}$ & - \\
1180 & $53.164824$ & $-27.788258$ & $7.2406$ & $-1.87 \pm 0.06$ & $-1.86 \pm 0.13$ & $3.47$ & $7.39_{-0.08}^{+0.09}$ & $-2.31_{-0.48}^{+0.50}$ & $-1.94_{-0.11}^{+0.12}$ & - \\
1210 & $53.162377$ & $-27.803300$ & $6.2942$ & $1.88 \pm 0.25$ & $-2.27 \pm 0.88$ & $0.88$ & $6.51_{-0.28}^{+0.04}$ & $-1.92_{-0.40}^{+0.46}$ & $-2.32_{-0.07}^{+0.07}$ & 4 \\
1210 & $53.175819$ & $-27.774465$ & $6.3351$ & $-1.70 \pm 0.03$ & $-1.66 \pm 0.10$ & $8.22$ & $6.58_{-0.11}^{+0.04}$ & $-2.08_{-0.28}^{+0.32}$ & $-2.37_{-0.05}^{+0.07}$ & 4 \\
1210 & $53.169041$ & $-27.778842$ & $6.6322$ & $-2.68 \pm 0.02$ & $-2.59 \pm 0.06$ & $13.86$ & $6.59_{-0.02}^{+0.03}$ & $-2.70_{-0.36}^{+0.34}$ & $-2.45_{-0.04}^{+0.06}$ & 4 \\
1210 & $53.151385$ & $-27.819159$ & $6.7074$ & $-2.03 \pm 0.05$ & $-1.87 \pm 0.17$ & $4.58$ & $6.63_{-0.04}^{+0.04}$ & $-2.17_{-0.34}^{+0.28}$ & $-2.39_{-0.07}^{+0.15}$ & 4 \\
1210 & $53.155794$ & $-27.815199$ & $6.7186$ & $-2.33 \pm 0.05$ & $-2.30 \pm 0.13$ & $6.18$ & $6.65_{-0.03}^{+0.03}$ & $-2.50_{-0.33}^{+0.33}$ & $-2.30_{-0.12}^{+0.08}$ & 4 \\
1210 & $53.152839$ & $-27.801940$ & $7.2621$ & $-2.26 \pm 0.03$ & $-2.35 \pm 0.10$ & $5.59$ & $7.56_{-0.29}^{+0.11}$ & $-2.36_{-0.49}^{+0.49}$ & $-2.48_{-0.03}^{+0.04}$ & 4 \\
1210 & $53.155083$ & $-27.801769$ & $7.2697$ & $-2.31 \pm 0.02$ & $-2.34 \pm 0.04$ & $17.63$ & $7.69_{-0.39}^{+0.14}$ & $-2.42_{-0.49}^{+0.51}$ & $-2.48_{-0.03}^{+0.03}$ & 4 \\
1210 & $53.156825$ & $-27.767159$ & $7.9807$ & $-2.17 \pm 0.02$ & $-2.21 \pm 0.06$ & $10.22$ & $7.95_{-0.08}^{+0.14}$ & $-2.71_{-0.50}^{+0.53}$ & $-2.15_{-0.22}^{+0.13}$ & 4 \\
1210 & $53.164468$ & $-27.802181$ & $8.4790$ & $-2.08 \pm 0.07$ & $-1.93 \pm 0.15$ & $3.15$ & $8.89_{-0.20}^{+0.07}$ & $-2.10_{-0.53}^{+0.51}$ & $-2.29_{-0.13}^{+0.13}$ & 4,5 \\
1210 & $53.167357$ & $-27.807502$ & $9.6886$ & $-2.71 \pm 0.03$ & $-2.75 \pm 0.07$ & $7.19$ & $9.44_{-0.04}^{+0.72}$ & $-2.23_{-0.32}^{+0.30}$ & $-2.36_{-0.13}^{+0.10}$ & - \\
1210 & $53.164763$ & $-27.774625$ & $11.5922$ & $-2.48 \pm 0.03$ & $-2.52 \pm 0.06$ & $8.25$ & $11.94_{-0.21}^{+0.19}$ & $-1.92_{-0.29}^{+0.27}$ & $-2.40_{-0.11}^{+0.10}$ & 3,4,5 \\
1345 & $214.731462$ & $52.736427$ & $5.3535$ & $-1.73 \pm 0.16$ & $-3.70 \pm 0.96$ & $0.69$ & $7.08_{-1.16}^{+0.23}$ & $-2.72_{-0.48}^{+0.42}$ & $-2.37_{-0.15}^{+0.11}$ & - \\
1345 & $214.806478$ & $52.878827$ & $6.1086$ & $-2.25 \pm 0.04$ & $-2.08 \pm 0.12$ & $6.36$ & $6.51_{-0.30}^{+0.04}$ & $-2.06_{-0.33}^{+0.33}$ & $-2.06_{-0.10}^{+0.11}$ & 8 \\
1345 & $214.832181$ & $52.885089$ & $6.6203$ & $-2.30 \pm 0.06$ & $-2.60 \pm 0.23$ & $2.42$ & $6.61_{-0.17}^{+0.04}$ & $-1.58_{-0.37}^{+0.35}$ & $-2.13_{-0.19}^{+0.15}$ & 8 \\
1345 & $215.128019$ & $52.984951$ & $6.6815$ & $-3.74 \pm 0.20$ & $-2.39 \pm 0.28$ & $6.40$ & $6.69_{-0.03}^{+0.02}$ & $-2.34_{-0.35}^{+0.33}$ & $-1.64_{-0.07}^{+0.06}$ & - \\
1345 & $214.789828$ & $52.730794$ & $6.7370$ & $-2.25 \pm 0.04$ & $-2.19 \pm 0.09$ & $7.17$ & $6.67_{-0.04}^{+0.03}$ & $-2.67_{-0.33}^{+0.35}$ & $-2.39_{-0.05}^{+0.05}$ & 8 \\
1345 & $214.948681$ & $52.853273$ & $6.7501$ & $0.56 \pm 0.30$ & $-1.00 \pm 0.55$ & $1.32$ & $6.77_{-0.04}^{+0.03}$ & $-2.18_{-0.32}^{+0.28}$ & $-2.41_{-0.05}^{+0.09}$ & - \\
1345 & $215.001120$ & $53.011273$ & $7.1028$ & $-2.17 \pm 0.04$ & $-2.44 \pm 0.16$ & $3.33$ & $7.12_{-0.06}^{+0.23}$ & $-2.67_{-0.38}^{+0.35}$ & $-2.42_{-0.07}^{+0.12}$ & 2,8,10 \\
1345 & $214.859185$ & $52.853595$ & $7.1135$ & $-1.38 \pm 0.05$ & $-2.23 \pm 0.15$ & $4.08$ & $7.03_{-0.03}^{+0.33}$ & $-2.34_{-0.32}^{+0.31}$ & $-2.43_{-0.06}^{+0.10}$ & 8 \\
1345 & $214.813057$ & $52.834241$ & $7.1785$ & $-2.54 \pm 0.05$ & $-2.48 \pm 0.14$ & $3.98$ & $7.37_{-0.19}^{+0.08}$ & $-3.11_{-0.51}^{+0.49}$ & $-2.46_{-0.05}^{+0.06}$ & 8,10 \\
1345 & $214.812062$ & $52.746747$ & $7.4757$ & $-2.74 \pm 0.04$ & $-2.43 \pm 0.12$ & $4.34$ & $7.38_{-0.24}^{+0.06}$ & $-2.38_{-0.48}^{+0.52}$ & $-2.32_{-0.12}^{+0.14}$ & 8 \\
1345 & $214.806079$ & $52.750868$ & $7.6487$ & $-2.16 \pm 0.05$ & $-2.15 \pm 0.14$ & $3.86$ & $8.20_{-0.38}^{+0.15}$ & $-2.70_{-0.47}^{+0.54}$ & $-2.40_{-0.05}^{+0.06}$ & 6 \\
1345 & $214.882999$ & $52.840419$ & $7.8314$ & $-1.55 \pm 0.04$ & $-1.66 \pm 0.12$ & $4.34$ & $8.06_{-0.12}^{+0.12}$ & $-2.54_{-0.52}^{+0.48}$ & $-2.29_{-0.13}^{+0.11}$ & 2,8,9,10 \\
1345 & $214.961271$ & $52.842360$ & $8.6351$ & $-2.96 \pm 0.16$ & $-1.73 \pm 0.30$ & $1.74$ & $8.64_{-0.27}^{+0.25}$ & $-2.82_{-0.66}^{+0.63}$ & $-2.02_{-0.16}^{+0.16}$ & 6,8 \\
1345 & $214.811853$ & $52.737113$ & $9.5635$ & $-2.64 \pm 0.05$ & $-2.58 \pm 0.13$ & $3.86$ & $10.42_{-0.65}^{+0.20}$ & $-1.98_{-0.42}^{+0.40}$ & $-2.18_{-0.08}^{+0.13}$ & 6 \\
1345 & $214.732527$ & $52.758097$ & $9.8505$ & $-3.07 \pm 0.07$ & $-2.90 \pm 0.17$ & $2.95$ & $9.61_{-0.04}^{+0.79}$ & $-2.18_{-0.30}^{+0.28}$ & $-2.30_{-0.13}^{+0.10}$ & 6 \\
2750 & $214.878972$ & $52.896751$ & $6.5357$ & $-2.57 \pm 0.03$ & $-2.59 \pm 0.07$ & $10.34$ & $6.58_{-0.13}^{+0.03}$ & $-2.38_{-0.33}^{+0.33}$ & $-2.48_{-0.04}^{+0.07}$ & - \\
2750 & $214.877890$ & $52.897677$ & $6.5361$ & $-2.63 \pm 0.05$ & $-2.56 \pm 0.12$ & $5.22$ & $6.54_{-0.32}^{+0.03}$ & $-2.08_{-0.31}^{+0.29}$ & $-2.47_{-0.05}^{+0.09}$ & - \\
2750 & $214.941618$ & $52.929132$ & $6.9806$ & $-1.99 \pm 0.03$ & $-2.06 \pm 0.08$ & $6.43$ & $7.04_{-0.01}^{+0.38}$ & $-2.05_{-0.37}^{+0.34}$ & $-2.21_{-0.06}^{+0.08}$ & - \\
2750 & $214.940489$ & $52.932559$ & $7.5524$ & $-2.48 \pm 0.04$ & $-2.44 \pm 0.13$ & $4.41$ & $7.47_{-0.45}^{+0.53}$ & $-2.62_{-0.75}^{+0.73}$ & $-2.46_{-0.05}^{+0.07}$ & - \\
2750 & $214.906639$ & $52.945504$ & $11.0419$ & $-1.82 \pm 0.05$ & $-1.85 \pm 0.13$ & $5.48$ & $11.54_{-0.34}^{+0.23}$ & $-1.88_{-0.41}^{+0.44}$ & $-2.17_{-0.11}^{+0.11}$ & - \\
2750 & $214.922777$ & $52.911524$ & $11.1168$ & $-1.65 \pm 0.06$ & $-2.08 \pm 0.21$ & $3.62$ & $11.22_{-0.85}^{+0.37}$ & $-2.61_{-0.50}^{+0.45}$ & $-2.40_{-0.09}^{+0.12}$ & 1,6 \\
2750 & $214.943146$ & $52.942446$ & $11.4111$ & $-1.88 \pm 0.04$ & $-2.10 \pm 0.11$ & $5.14$ & $11.90_{-0.32}^{+0.17}$ & $-3.13_{-0.42}^{+0.43}$ & $-2.53_{-0.03}^{+0.06}$ & 6,7 \\
    \botrule
    \end{tabular}
\end{table*}


\bibliography{main}{}

\begin{thebibliography}{}
\expandafter\ifx\csname natexlab\endcsname\relax\def\natexlab#1{#1}\fi
\providecommand{\url}[1]{\href{#1}{#1}}
\providecommand{\dodoi}[1]{doi:~\href{http://doi.org/#1}{\nolinkurl{#1}}}
\providecommand{\doeprint}[1]{\href{http://ascl.net/#1}{\nolinkurl{http://ascl.net/#1}}}
\providecommand{\doarXiv}[1]{\href{https://arxiv.org/abs/#1}{\nolinkurl{https://arxiv.org/abs/#1}}}

\bibitem[{{Adams} {et~al.}(2023{\natexlab{a}}){Adams}, {Conselice}, {Ferreira},
  {Austin}, {Trussler}, {Juod{\v{z}}balis}, {Wilkins}, {Caruana}, {Dayal},
  {Verma}, \& {Vijayan}}]{Adams2023}
{Adams}, N.~J., {Conselice}, C.~J., {Ferreira}, L., {et~al.}
  2023{\natexlab{a}}, \mnras, 518, 4755, \dodoi{10.1093/mnras/stac3347}

\bibitem[{{Adams} {et~al.}(2023{\natexlab{b}}){Adams}, {Conselice}, {Austin},
  {Harvey}, {Ferreira}, {Trussler}, {Juodzbalis}, {Li}, {Windhorst}, {Cohen},
  {Jansen}, {Summers}, {Tompkins}, {Driver}, {Robotham}, {D'Silva}, {Yan},
  {Coe}, {Frye}, {Grogin}, {Koekemoer}, {Marshall}, {Pirzkal}, {Ryan},
  {Maksym}, {Rutkowski}, {Willmer}, {Hammel}, {Nonino}, {Bhatawdekar},
  {Wilkins}, {Bradley}, {Broadhurst}, {Cheng}, {Dole}, {Hathi}, \&
  {Zitrin}}]{Adams2023-EPOCHS-II}
{Adams}, N.~J., {Conselice}, C.~J., {Austin}, D., {et~al.} 2023{\natexlab{b}},
  arXiv e-prints, arXiv:2304.13721, \dodoi{10.48550/arXiv.2304.13721}

\bibitem[{{Arrabal Haro} {et~al.}(2023{\natexlab{a}}){Arrabal Haro},
  {Dickinson}, {Finkelstein}, {Fujimoto}, {Fern{\'a}ndez}, {Kartaltepe},
  {Jung}, {Cole}, {Burgarella}, {Chworowsky}, {Hutchison}, {Morales},
  {Papovich}, {Simons}, {Amor{\'\i}n}, {Backhaus}, {Bagley}, {Bisigello},
  {Calabr{\`o}}, {Castellano}, {Cleri}, {Dav{\'e}}, {Dekel}, {Ferguson},
  {Fontana}, {Gawiser}, {Giavalisco}, {Harish}, {Hathi}, {Hirschmann},
  {Holwerda}, {Huertas-Company}, {Koekemoer}, {Larson}, {Lucas}, {Mobasher},
  {P{\'e}rez-Gonz{\'a}lez}, {Pirzkal}, {Rose}, {Santini}, {Trump}, {de la
  Vega}, {Wang}, {Weiner}, {Wilkins}, {Yang}, {Yung}, \&
  {Zavala}}]{ArrabalHaro2023a}
{Arrabal Haro}, P., {Dickinson}, M., {Finkelstein}, S.~L., {et~al.}
  2023{\natexlab{a}}, \apjl, 951, L22, \dodoi{10.3847/2041-8213/acdd54}

\bibitem[{{Arrabal Haro} {et~al.}(2023{\natexlab{b}}){Arrabal Haro},
  {Dickinson}, {Finkelstein}, {Kartaltepe}, {Donnan}, {Burgarella}, {Carnall},
  {Cullen}, {Dunlop}, {Fern{\'a}ndez}, {Fujimoto}, {Jung}, {Krips}, {Larson},
  {Papovich}, {P{\'e}rez-Gonz{\'a}lez}, {Amor{\'\i}n}, {Bagley}, {Buat},
  {Casey}, {Chworowsky}, {Cohen}, {Ferguson}, {Giavalisco}, {Huertas-Company},
  {Hutchison}, {Kocevski}, {Koekemoer}, {Lucas}, {McLeod}, {McLure}, {Pirzkal},
  {Seill{\'e}}, {Trump}, {Weiner}, {Wilkins}, \& {Zavala}}]{ArrabalHaro2023b}
---. 2023{\natexlab{b}}, \nat, 622, 707, \dodoi{10.1038/s41586-023-06521-7}

\bibitem[{{Astropy Collaboration} {et~al.}(2013){Astropy Collaboration},
  {Robitaille}, {Tollerud}, {Greenfield}, {Droettboom}, {Bray}, {Aldcroft},
  {Davis}, {Ginsburg}, {Price-Whelan}, {Kerzendorf}, {Conley}, {Crighton},
  {Barbary}, {Muna}, {Ferguson}, {Grollier}, {Parikh}, {Nair}, {Unther},
  {Deil}, {Woillez}, {Conseil}, {Kramer}, {Turner}, {Singer}, {Fox}, {Weaver},
  {Zabalza}, {Edwards}, {Azalee Bostroem}, {Burke}, {Casey}, {Crawford},
  {Dencheva}, {Ely}, {Jenness}, {Labrie}, {Lim}, {Pierfederici}, {Pontzen},
  {Ptak}, {Refsdal}, {Servillat}, \& {Streicher}}]{Astropy2013}
{Astropy Collaboration}, {Robitaille}, T.~P., {Tollerud}, E.~J., {et~al.} 2013,
  \aap, 558, A33, \dodoi{10.1051/0004-6361/201322068}

\bibitem[{{Astropy Collaboration} {et~al.}(2018){Astropy Collaboration},
  {Price-Whelan}, {Sip{\H{o}}cz}, {G{\"u}nther}, {Lim}, {Crawford}, {Conseil},
  {Shupe}, {Craig}, {Dencheva}, {Ginsburg}, {VanderPlas}, {Bradley},
  {P{\'e}rez-Su{\'a}rez}, {de Val-Borro}, {Aldcroft}, {Cruz}, {Robitaille},
  {Tollerud}, {Ardelean}, {Babej}, {Bach}, {Bachetti}, {Bakanov}, {Bamford},
  {Barentsen}, {Barmby}, {Baumbach}, {Berry}, {Biscani}, {Boquien}, {Bostroem},
  {Bouma}, {Brammer}, {Bray}, {Breytenbach}, {Buddelmeijer}, {Burke},
  {Calderone}, {Cano Rodr{\'\i}guez}, {Cara}, {Cardoso}, {Cheedella}, {Copin},
  {Corrales}, {Crichton}, {D'Avella}, {Deil}, {Depagne}, {Dietrich}, {Donath},
  {Droettboom}, {Earl}, {Erben}, {Fabbro}, {Ferreira}, {Finethy}, {Fox},
  {Garrison}, {Gibbons}, {Goldstein}, {Gommers}, {Greco}, {Greenfield},
  {Groener}, {Grollier}, {Hagen}, {Hirst}, {Homeier}, {Horton}, {Hosseinzadeh},
  {Hu}, {Hunkeler}, {Ivezi{\'c}}, {Jain}, {Jenness}, {Kanarek}, {Kendrew},
  {Kern}, {Kerzendorf}, {Khvalko}, {King}, {Kirkby}, {Kulkarni}, {Kumar},
  {Lee}, {Lenz}, {Littlefair}, {Ma}, {Macleod}, {Mastropietro}, {McCully},
  {Montagnac}, {Morris}, {Mueller}, {Mumford}, {Muna}, {Murphy}, {Nelson},
  {Nguyen}, {Ninan}, {N{\"o}the}, {Ogaz}, {Oh}, {Parejko}, {Parley}, {Pascual},
  {Patil}, {Patil}, {Plunkett}, {Prochaska}, {Rastogi}, {Reddy Janga},
  {Sabater}, {Sakurikar}, {Seifert}, {Sherbert}, {Sherwood-Taylor}, {Shih},
  {Sick}, {Silbiger}, {Singanamalla}, {Singer}, {Sladen}, {Sooley},
  {Sornarajah}, {Streicher}, {Teuben}, {Thomas}, {Tremblay}, {Turner},
  {Terr{\'o}n}, {van Kerkwijk}, {de la Vega}, {Watkins}, {Weaver}, {Whitmore},
  {Woillez}, {Zabalza}, \& {Astropy Contributors}}]{Astropy2018}
{Astropy Collaboration}, {Price-Whelan}, A.~M., {Sip{\H{o}}cz}, B.~M., {et~al.}
  2018, \aj, 156, 123, \dodoi{10.3847/1538-3881/aabc4f}

\bibitem[{{Astropy Collaboration} {et~al.}(2022){Astropy Collaboration},
  {Price-Whelan}, {Lim}, {Earl}, {Starkman}, {Bradley}, {Shupe}, {Patil},
  {Corrales}, {Brasseur}, {N{\"o}the}, {Donath}, {Tollerud}, {Morris},
  {Ginsburg}, {Vaher}, {Weaver}, {Tocknell}, {Jamieson}, {van Kerkwijk},
  {Robitaille}, {Merry}, {Bachetti}, {G{\"u}nther}, {Aldcroft},
  {Alvarado-Montes}, {Archibald}, {B{\'o}di}, {Bapat}, {Barentsen},
  {Baz{\'a}n}, {Biswas}, {Boquien}, {Burke}, {Cara}, {Cara}, {Conroy},
  {Conseil}, {Craig}, {Cross}, {Cruz}, {D'Eugenio}, {Dencheva}, {Devillepoix},
  {Dietrich}, {Eigenbrot}, {Erben}, {Ferreira}, {Foreman-Mackey}, {Fox},
  {Freij}, {Garg}, {Geda}, {Glattly}, {Gondhalekar}, {Gordon}, {Grant},
  {Greenfield}, {Groener}, {Guest}, {Gurovich}, {Handberg}, {Hart},
  {Hatfield-Dodds}, {Homeier}, {Hosseinzadeh}, {Jenness}, {Jones}, {Joseph},
  {Kalmbach}, {Karamehmetoglu}, {Ka{\l}uszy{\'n}ski}, {Kelley}, {Kern},
  {Kerzendorf}, {Koch}, {Kulumani}, {Lee}, {Ly}, {Ma}, {MacBride}, {Maljaars},
  {Muna}, {Murphy}, {Norman}, {O'Steen}, {Oman}, {Pacifici}, {Pascual},
  {Pascual-Granado}, {Patil}, {Perren}, {Pickering}, {Rastogi}, {Roulston},
  {Ryan}, {Rykoff}, {Sabater}, {Sakurikar}, {Salgado}, {Sanghi}, {Saunders},
  {Savchenko}, {Schwardt}, {Seifert-Eckert}, {Shih}, {Jain}, {Shukla}, {Sick},
  {Simpson}, {Singanamalla}, {Singer}, {Singhal}, {Sinha}, {Sip{\H{o}}cz},
  {Spitler}, {Stansby}, {Streicher}, {{\v{S}}umak}, {Swinbank}, {Taranu},
  {Tewary}, {Tremblay}, {de Val-Borro}, {Van Kooten}, {Vasovi{\'c}}, {Verma},
  {de Miranda Cardoso}, {Williams}, {Wilson}, {Winkel}, {Wood-Vasey}, {Xue},
  {Yoachim}, {Zhang}, {Zonca}, \& {Astropy Project Contributors}}]{Astropy2022}
{Astropy Collaboration}, {Price-Whelan}, A.~M., {Lim}, P.~L., {et~al.} 2022,
  \apj, 935, 167, \dodoi{10.3847/1538-4357/ac7c74}

\bibitem[{{Atek} {et~al.}(2023){Atek}, {Shuntov}, {Furtak}, {Richard}, {Kneib},
  {Mahler}, {Zitrin}, {McCracken}, {Charlot}, {Chevallard}, \&
  {Chemerynska}}]{Atek2023-SMACS}
{Atek}, H., {Shuntov}, M., {Furtak}, L.~J., {et~al.} 2023, \mnras, 519, 1201,
  \dodoi{10.1093/mnras/stac3144}

\bibitem[{{Austin} {et~al.}(2023){Austin}, {Adams}, {Conselice}, {Harvey},
  {Ormerod}, {Trussler}, {Li}, {Ferreira}, {Dayal}, \&
  {Juod{\v{z}}balis}}]{Austin2023}
{Austin}, D., {Adams}, N., {Conselice}, C.~J., {et~al.} 2023, \apjl, 952, L7,
  \dodoi{10.3847/2041-8213/ace18d}

\bibitem[{{Bagley} {et~al.}(2023{\natexlab{a}}){Bagley}, {Finkelstein},
  {Koekemoer}, {Ferguson}, {Arrabal Haro}, {Dickinson}, {Kartaltepe},
  {Papovich}, {P{\'e}rez-Gonz{\'a}lez}, {Pirzkal}, {Somerville}, {Willmer},
  {Yang}, {Yung}, {Fontana}, {Grazian}, {Grogin}, {Hirschmann}, {Kewley},
  {Kirkpatrick}, {Kocevski}, {Lotz}, {Medrano}, {Morales}, {Pentericci},
  {Ravindranath}, {Trump}, {Wilkins}, {Calabr{\`o}}, {Cooper}, {Costantin}, {de
  la Vega}, {Hilbert}, {Hutchison}, {Larson}, {Lucas}, {McGrath}, {Ryan},
  {Wang}, \& {Wuyts}}]{Bagley2023-CEERS}
{Bagley}, M.~B., {Finkelstein}, S.~L., {Koekemoer}, A.~M., {et~al.}
  2023{\natexlab{a}}, \apjl, 946, L12, \dodoi{10.3847/2041-8213/acbb08}

\bibitem[{{Bagley} {et~al.}(2023{\natexlab{b}}){Bagley}, {Pirzkal},
  {Finkelstein}, {Papovich}, {Berg}, {Lotz}, {Leung}, {Ferguson}, {Koekemoer},
  {Dickinson}, {Kartaltepe}, {Kocevski}, {Somerville}, {Yung}, {Backhaus},
  {Casey}, {Castellano}, {Ch{\'a}vez Ortiz}, {Chworowsky}, {Cox}, {Dav{\'e}},
  {Davis}, {Estrada-Carpenter}, {Fontana}, {Fujimoto}, {Gardner}, {Giavalisco},
  {Grazian}, {Grogin}, {Hathi}, {Hutchison}, {Jaskot}, {Jung}, {Kewley},
  {Kirkpatrick}, {Larson}, {Matharu}, {Natarajan}, {Pentericci},
  {P{\'e}rez-Gonz{\'a}lez}, {Ravindranath}, {Rothberg}, {Ryan}, {Shen},
  {Simons}, {Snyder}, {Trump}, \& {Wilkins}}]{Bagley2023-NGDEEP}
{Bagley}, M.~B., {Pirzkal}, N., {Finkelstein}, S.~L., {et~al.}
  2023{\natexlab{b}}, arXiv e-prints, arXiv:2302.05466,
  \dodoi{10.48550/arXiv.2302.05466}

\bibitem[{Barbary(2016)}]{Barbary2016-sep}
Barbary, K. 2016, Journal of Open Source Software, 1, 58,
  \dodoi{10.21105/joss.00058}

\bibitem[{{Bertin} \& {Arnouts}(1996)}]{Bertin1996-SExtractor}
{Bertin}, E., \& {Arnouts}, S. 1996, \aaps, 117, 393,
  \dodoi{10.1051/aas:1996164}

\bibitem[{{Bezanson} {et~al.}(2022){Bezanson}, {Labbe}, {Whitaker}, {Leja},
  {Price}, {Franx}, {Brammer}, {Marchesini}, {Zitrin}, {Wang}, {Weaver},
  {Furtak}, {Atek}, {Coe}, {Cutler}, {Dayal}, {van Dokkum}, {Feldmann},
  {Forster Schreiber}, {Fujimoto}, {Geha}, {Glazebrook}, {de Graaff}, {Greene},
  {Juneau}, {Kassin}, {Kriek}, {Khullar}, {Maseda}, {Mowla}, {Muzzin},
  {Nanayakkara}, {Nelson}, {Oesch}, {Pacifici}, {Pan}, {Papovich}, {Setton},
  {Shapley}, {Smit}, {Stefanon}, {Taylor}, \& {Williams}}]{Bezanson2022}
{Bezanson}, R., {Labbe}, I., {Whitaker}, K.~E., {et~al.} 2022, arXiv e-prints,
  arXiv:2212.04026, \dodoi{10.48550/arXiv.2212.04026}

\bibitem[{{Bhatawdekar} \& {Conselice}(2021)}]{Bhatawdekar2021}
{Bhatawdekar}, R., \& {Conselice}, C.~J. 2021, \apj, 909, 144,
  \dodoi{10.3847/1538-4357/abdd3f}

\bibitem[{{Bianchi} {et~al.}(2009){Bianchi}, {Schneider}, \&
  {Valiante}}]{Bianchi2009}
{Bianchi}, S., {Schneider}, R., \& {Valiante}, R. 2009, in Astronomical Society
  of the Pacific Conference Series, Vol. 414, Cosmic Dust - Near and Far, ed.
  T.~{Henning}, E.~{Gr{\"u}n}, \& J.~{Steinacker}, 65

\bibitem[{{Blaizot} {et~al.}(2023){Blaizot}, {Garel}, {Verhamme}, {Katz},
  {Kimm}, {Michel-Dansac}, {Mitchell}, {Rosdahl}, \& {Trebitsch}}]{Blaizot2023}
{Blaizot}, J., {Garel}, T., {Verhamme}, A., {et~al.} 2023, \mnras, 523, 3749,
  \dodoi{10.1093/mnras/stad1523}

\bibitem[{{Bocchio} {et~al.}(2016){Bocchio}, {Marassi}, {Schneider}, {Bianchi},
  {Limongi}, \& {Chieffi}}]{Bocchio2016}
{Bocchio}, M., {Marassi}, S., {Schneider}, R., {et~al.} 2016, \aap, 587, A157,
  \dodoi{10.1051/0004-6361/201527432}

\bibitem[{{Bohlin}(2016)}]{Bohlin2016}
{Bohlin}, R.~C. 2016, \aj, 152, 60, \dodoi{10.3847/0004-6256/152/3/60}

\bibitem[{{B{\"o}ker} {et~al.}(2023){B{\"o}ker}, {Beck}, {Birkmann},
  {Giardino}, {Keyes}, {Kumari}, {Muzerolle}, {Rawle}, {Zeidler}, {Abul-Huda},
  {Alves de Oliveira}, {Arribas}, {Bechtold}, {Bhatawdekar}, {Bonaventura},
  {Bunker}, {Cameron}, {Carniani}, {Charlot}, {Curti}, {Espinoza}, {Ferruit},
  {Franx}, {Jakobsen}, {Karakla}, {L{\'o}pez-Caniego}, {L{\"u}tzgendorf},
  {Maiolino}, {Manjavacas}, {Marston}, {Moseley}, {Ogle}, {Perna},
  {Pe{\~n}a-Guerrero}, {Pirzkal}, {Plesha}, {Proffitt}, {Rauscher}, {Rix},
  {Rodr{\'\i}guez del Pino}, {Rustamkulov}, {Sabbi}, {Sing}, {Sirianni}, {te
  Plate}, {{\'U}beda}, {Wahlgren}, {Wislowski}, {Wu}, \& {Willott}}]{Boker2023}
{B{\"o}ker}, T., {Beck}, T.~L., {Birkmann}, S.~M., {et~al.} 2023, \pasp, 135,
  038001, \dodoi{10.1088/1538-3873/acb846}

\bibitem[{{Bouwens} {et~al.}(2023{\natexlab{a}}){Bouwens}, {Illingworth},
  {Oesch}, {Stefanon}, {Naidu}, {van Leeuwen}, \& {Magee}}]{Bouwens2023}
{Bouwens}, R., {Illingworth}, G., {Oesch}, P., {et~al.} 2023{\natexlab{a}},
  \mnras, 523, 1009, \dodoi{10.1093/mnras/stad1014}

\bibitem[{{Bouwens} {et~al.}(2023{\natexlab{b}}){Bouwens}, {Illingworth},
  {Oesch}, {Stefanon}, {Naidu}, {van Leeuwen}, \& {Magee}}]{Bouwens2022c}
---. 2023{\natexlab{b}}, \mnras, 523, 1009, \dodoi{10.1093/mnras/stad1014}

\bibitem[{{Bouwens} {et~al.}(2015{\natexlab{a}}){Bouwens}, {Illingworth},
  {Oesch}, {Caruana}, {Holwerda}, {Smit}, \& {Wilkins}}]{Bouwens2015-EoR}
{Bouwens}, R.~J., {Illingworth}, G.~D., {Oesch}, P.~A., {et~al.}
  2015{\natexlab{a}}, \apj, 811, 140, \dodoi{10.1088/0004-637X/811/2/140}

\bibitem[{{Bouwens} {et~al.}(2009){Bouwens}, {Illingworth}, {Franx}, {Chary},
  {Meurer}, {Conselice}, {Ford}, {Giavalisco}, \& {van Dokkum}}]{Bouwens2009}
{Bouwens}, R.~J., {Illingworth}, G.~D., {Franx}, M., {et~al.} 2009, \apj, 705,
  936, \dodoi{10.1088/0004-637X/705/1/936}

\bibitem[{{Bouwens} {et~al.}(2010){Bouwens}, {Illingworth}, {Oesch}, {Trenti},
  {Stiavelli}, {Carollo}, {Franx}, {van Dokkum}, {Labb{\'e}}, \&
  {Magee}}]{Bouwens2010}
{Bouwens}, R.~J., {Illingworth}, G.~D., {Oesch}, P.~A., {et~al.} 2010, \apjl,
  708, L69, \dodoi{10.1088/2041-8205/708/2/L69}

\bibitem[{{Bouwens} {et~al.}(2011){Bouwens}, {Illingworth}, {Oesch},
  {Labb{\'e}}, {Trenti}, {van Dokkum}, {Franx}, {Stiavelli}, {Carollo},
  {Magee}, \& {Gonzalez}}]{Bouwens2011-UVLF}
---. 2011, \apj, 737, 90, \dodoi{10.1088/0004-637X/737/2/90}

\bibitem[{{Bouwens} {et~al.}(2012){Bouwens}, {Illingworth}, {Oesch}, {Franx},
  {Labb{\'e}}, {Trenti}, {van Dokkum}, {Carollo}, {Gonz{\'a}lez}, {Smit}, \&
  {Magee}}]{Bouwens2012}
---. 2012, \apj, 754, 83, \dodoi{10.1088/0004-637X/754/2/83}

\bibitem[{{Bouwens} {et~al.}(2014){Bouwens}, {Illingworth}, {Oesch},
  {Labb{\'e}}, {van Dokkum}, {Trenti}, {Franx}, {Smit}, {Gonzalez}, \&
  {Magee}}]{Bouwens2014a}
---. 2014, \apj, 793, 115, \dodoi{10.1088/0004-637X/793/2/115}

\bibitem[{{Bouwens} {et~al.}(2015{\natexlab{b}}){Bouwens}, {Illingworth},
  {Oesch}, {Trenti}, {Labb{\'e}}, {Bradley}, {Carollo}, {van Dokkum},
  {Gonzalez}, {Holwerda}, {Franx}, {Spitler}, {Smit}, \& {Magee}}]{Bouwens2015}
---. 2015{\natexlab{b}}, \apj, 803, 34, \dodoi{10.1088/0004-637X/803/1/34}

\bibitem[{{Bouwens} {et~al.}(2016){Bouwens}, {Oesch}, {Labb{\'e}},
  {Illingworth}, {Fazio}, {Coe}, {Holwerda}, {Smit}, {Stefanon}, {van Dokkum},
  {Trenti}, {Ashby}, {Huang}, {Spitler}, {Straatman}, {Bradley}, \&
  {Magee}}]{Bouwens2016}
{Bouwens}, R.~J., {Oesch}, P.~A., {Labb{\'e}}, I., {et~al.} 2016, \apj, 830,
  67, \dodoi{10.3847/0004-637X/830/2/67}

\bibitem[{{Bouwens} {et~al.}(2022){Bouwens}, {Smit}, {Schouws}, {Stefanon},
  {Bowler}, {Endsley}, {Gonzalez}, {Inami}, {Stark}, {Oesch}, {Hodge},
  {Aravena}, {da Cunha}, {Dayal}, {de Looze}, {Ferrara}, {Fudamoto},
  {Graziani}, {Li}, {Nanayakkara}, {Pallottini}, {Schneider}, {Sommovigo},
  {Topping}, {van der Werf}, {Algera}, {Barrufet}, {Hygate}, {Labb{\'e}},
  {Riechers}, \& {Witstok}}]{Bouwens2022}
{Bouwens}, R.~J., {Smit}, R., {Schouws}, S., {et~al.} 2022, \apj, 931, 160,
  \dodoi{10.3847/1538-4357/ac5a4a}

\bibitem[{{Bowler} {et~al.}(2024){Bowler}, {Inami}, {Sommovigo}, {Smit},
  {Algera}, {Aravena}, {Barrufet}, {Bouwens}, {da Cunha}, {Cullen}, {Dayal},
  {De Looze}, {Dunlop}, {Fudamoto}, {Mauerhofer}, {McLure}, {Stefanon},
  {Schneider}, {Ferrara}, {Graziani}, {Hodge}, {Nanayakkara}, {Palla},
  {Schouws}, {Stark}, \& {van der Werf}}]{Bowler2024}
{Bowler}, R.~A.~A., {Inami}, H., {Sommovigo}, L., {et~al.} 2024, \mnras, 527,
  5808, \dodoi{10.1093/mnras/stad3578}

\bibitem[{Bradley {et~al.}(2022)Bradley, Sipőcz, Robitaille, Tollerud,
  Vinícius, Deil, Barbary, Wilson, Busko, Donath, Günther, Cara, Lim,
  Meßlinger, Conseil, Bostroem, Droettboom, Bray, Bratholm, Barentsen, Craig,
  Rathi, Pascual, Perren, Georgiev, de~Val-Borro, Kerzendorf, Bach, Quint, \&
  Souchereau}]{Bradley2022}
Bradley, L., Sipőcz, B., Robitaille, T., {et~al.} 2022, Zenodo,
  \dodoi{10.5281/zenodo.6825092}

\bibitem[{{Brammer} {et~al.}(2008){Brammer}, {van Dokkum}, \&
  {Coppi}}]{Brammer2008-EAZY}
{Brammer}, G.~B., {van Dokkum}, P.~G., \& {Coppi}, P. 2008, \apj, 686, 1503,
  \dodoi{10.1086/591786}

\bibitem[{{Bruzual} \& {Charlot}(2003)}]{BC03}
{Bruzual}, G., \& {Charlot}, S. 2003, \mnras, 344, 1000,
  \dodoi{10.1046/j.1365-8711.2003.06897.x}

\bibitem[{{Bunker} {et~al.}(2023){Bunker}, {Cameron}, {Curtis-Lake},
  {Jakobsen}, {Carniani}, {Curti}, {Witstok}, {Maiolino}, {D'Eugenio},
  {Looser}, {Willott}, {Bonaventura}, {Hainline}, {Uebler}, {Willmer},
  {Saxena}, {Smit}, {Alberts}, {Arribas}, {Baker}, {Baum}, {Bhatawdekar},
  {Bowler}, {Boyett}, {Charlot}, {Chen}, {Chevallard}, {Circosta}, {DeCoursey},
  {de Graaff}, {Egami}, {Eisenstein}, {Endsley}, {Ferruit}, {Giardino},
  {Hausen}, {Helton}, {Hviding}, {Ji}, {Johnson}, {Jones}, {Kumari}, {Laseter},
  {Luetzgendorf}, {Maseda}, {Nelson}, {Parlanti}, {Perna}, {Rawle}, {Rix},
  {Rieke}, {Robertson}, {Rodriguez Del Pino}, {Sandles}, {Scholtz}, {Sharpe},
  {Skarbinski}, {Stark}, {Sun}, {Tacchella}, {Topping}, {Villanueva},
  {Wallace}, {Williams}, \& {Woodrum}}]{Bunker2023-NIRSpec}
{Bunker}, A.~J., {Cameron}, A.~J., {Curtis-Lake}, E., {et~al.} 2023, arXiv
  e-prints, arXiv:2306.02467, \dodoi{10.48550/arXiv.2306.02467}

\bibitem[{{Bushouse} {et~al.}(2022){Bushouse}, {Eisenhamer}, {Dencheva},
  {Davies}, {Greenfield}, {Morrison}, {Hodge}, {Simon}, {Grumm}, {Droettboom},
  {Slavich}, {Sosey}, {Pauly}, {Miller}, {Jedrzejewski}, {Hack}, {Davis},
  {Crawford}, {Law}, {Gordon}, {Regan}, {Cara}, {MacDonald}, {Bradley},
  {Shanahan}, {Jamieson}, {Teodoro}, \& {Williams}}]{Bushouse2022}
{Bushouse}, H., {Eisenhamer}, J., {Dencheva}, N., {et~al.} 2022, {JWST
  Calibration Pipeline}, 1.8.2,  Zenodo, \dodoi{10.5281/zenodo.7325378}

\bibitem[{{Byler} {et~al.}(2017){Byler}, {Dalcanton}, {Conroy}, \&
  {Johnson}}]{Byler2017}
{Byler}, N., {Dalcanton}, J.~J., {Conroy}, C., \& {Johnson}, B.~D. 2017, \apj,
  840, 44, \dodoi{10.3847/1538-4357/aa6c66}

\bibitem[{{Byrne} {et~al.}(2022){Byrne}, {Stanway}, {Eldridge}, {McSwiney}, \&
  {Townsend}}]{Byrne2022-BPASS}
{Byrne}, C.~M., {Stanway}, E.~R., {Eldridge}, J.~J., {McSwiney}, L., \&
  {Townsend}, O.~T. 2022, \mnras, 512, 5329, \dodoi{10.1093/mnras/stac807}

\bibitem[{{Calvi} {et~al.}(2016){Calvi}, {Trenti}, {Stiavelli}, {Oesch},
  {Bradley}, {Schmidt}, {Coe}, {Brammer}, {Bernard}, {Bouwens}, {Carrasco},
  {Carollo}, {Holwerda}, {MacKenty}, {Mason}, {Shull}, \& {Treu}}]{Calvi2016}
{Calvi}, V., {Trenti}, M., {Stiavelli}, M., {et~al.} 2016, \apj, 817, 120,
  \dodoi{10.3847/0004-637X/817/2/120}

\bibitem[{{Calzetti} {et~al.}(2000){Calzetti}, {Armus}, {Bohlin}, {Kinney},
  {Koornneef}, \& {Storchi-Bergmann}}]{Calzetti2000}
{Calzetti}, D., {Armus}, L., {Bohlin}, R.~C., {et~al.} 2000, \apj, 533, 682,
  \dodoi{10.1086/308692}

\bibitem[{{Calzetti} {et~al.}(1994){Calzetti}, {Kinney}, \&
  {Storchi-Bergmann}}]{Calzetti1994}
{Calzetti}, D., {Kinney}, A.~L., \& {Storchi-Bergmann}, T. 1994, \apj, 429,
  582, \dodoi{10.1086/174346}

\bibitem[{{Cameron} {et~al.}(2023{\natexlab{a}}){Cameron}, {Katz}, {Rey}, \&
  {Saxena}}]{Cameron2023-GNz11}
{Cameron}, A.~J., {Katz}, H., {Rey}, M.~P., \& {Saxena}, A. 2023{\natexlab{a}},
  \mnras, 523, 3516, \dodoi{10.1093/mnras/stad1579}

\bibitem[{{Cameron} {et~al.}(2023{\natexlab{b}}){Cameron}, {Katz}, {Witten},
  {Saxena}, {Laporte}, \& {Bunker}}]{Cameron2023-nebular}
{Cameron}, A.~J., {Katz}, H., {Witten}, C., {et~al.} 2023{\natexlab{b}}, arXiv
  e-prints, arXiv:2311.02051, \dodoi{10.48550/arXiv.2311.02051}

\bibitem[{{Carnall} {et~al.}(2018){Carnall}, {McLure}, {Dunlop}, \&
  {Dav{\'e}}}]{Carnall2018}
{Carnall}, A.~C., {McLure}, R.~J., {Dunlop}, J.~S., \& {Dav{\'e}}, R. 2018,
  \mnras, 480, 4379, \dodoi{10.1093/mnras/sty2169}

\bibitem[{{Castellano} {et~al.}(2022){Castellano}, {Fontana}, {Treu},
  {Santini}, {Merlin}, {Leethochawalit}, {Trenti}, {Mestric}, {Vanzella},
  {Bonchi}, {Belfiori}, {Nonino}, {Paris}, {Polenta}, {Roberts-Borsani},
  {Boyett}, {Calabro}, {Glazebrook}, {Grillo}, {Mascia}, {Mason}, {Mercurio},
  {Morishita}, {Nanayakkara}, {Pentericci}, {Rosati}, {Vulcani}, {Wang}, \&
  {Yang}}]{Castellano2022}
{Castellano}, M., {Fontana}, A., {Treu}, T., {et~al.} 2022, arXiv e-prints,
  arXiv:2207.09436.
\newblock \doarXiv{2207.09436}

\bibitem[{{Castellano} {et~al.}(2023){Castellano}, {Fontana}, {Treu}, {Merlin},
  {Santini}, {Bergamini}, {Grillo}, {Rosati}, {Acebron}, {Leethochawalit},
  {Paris}, {Bonchi}, {Belfiori}, {Calabr{\`o}}, {Correnti}, {Nonino},
  {Polenta}, {Trenti}, {Boyett}, {Brammer}, {Broadhurst}, {Caminha}, {Chen},
  {Filippenko}, {Fortuni}, {Glazebrook}, {Mascia}, {Mason}, {Menci},
  {Meneghetti}, {Mercurio}, {Metha}, {Morishita}, {Nanayakkara}, {Pentericci},
  {Roberts-Borsani}, {Roy}, {Vanzella}, {Vulcani}, {Yang}, \&
  {Wang}}]{Castellano2023-GLASS-XIX}
---. 2023, \apjl, 948, L14, \dodoi{10.3847/2041-8213/accea5}

\bibitem[{{Castellano} {et~al.}(2024){Castellano}, {Napolitano}, {Fontana},
  {Roberts-Borsani}, {Treu}, {Vanzella}, {Zavala}, {Arrabal Haro},
  {Calabr{\`o}}, {Llerena}, {Mascia}, {Merlin}, {Paris}, {Pentericci},
  {Santini}, {Bakx}, {Bergamini}, {Cupani}, {Dickinson}, {Filippenko},
  {Glazebrook}, {Grillo}, {Kelly}, {Malkan}, {Mason}, {Morishita},
  {Nanayakkara}, {Rosati}, {Sani}, {Wang}, \& {Yoon}}]{Castellano2024}
{Castellano}, M., {Napolitano}, L., {Fontana}, A., {et~al.} 2024, arXiv
  e-prints, arXiv:2403.10238, \dodoi{10.48550/arXiv.2403.10238}

\bibitem[{{Chabrier}(2003)}]{Chabrier2003}
{Chabrier}, G. 2003, \pasp, 115, 763, \dodoi{10.1086/376392}

\bibitem[{{Charbonnel} {et~al.}(2023){Charbonnel}, {Schaerer}, {Prantzos},
  {Ram{\'\i}rez-Galeano}, {Fragos}, {Kuruvanthodi}, {Marques-Chaves}, \&
  {Gieles}}]{Charbonnel2023}
{Charbonnel}, C., {Schaerer}, D., {Prantzos}, N., {et~al.} 2023, \aap, 673, L7,
  \dodoi{10.1051/0004-6361/202346410}

\bibitem[{{Charlot} \& {Fall}(2000)}]{CharlotFall2000}
{Charlot}, S., \& {Fall}, S.~M. 2000, \apj, 539, 718, \dodoi{10.1086/309250}

\bibitem[{{Chen} {et~al.}(2024){Chen}, {Stark}, {Mason}, {Topping}, {Whitler},
  {Tang}, {Endsley}, \& {Charlot}}]{Chen2024}
{Chen}, Z., {Stark}, D.~P., {Mason}, C., {et~al.} 2024, \mnras,
  \dodoi{10.1093/mnras/stae455}

\bibitem[{{Chevallard} \& {Charlot}(2016)}]{Chevallard2016-BEAGLE}
{Chevallard}, J., \& {Charlot}, S. 2016, \mnras, 462, 1415,
  \dodoi{10.1093/mnras/stw1756}

\bibitem[{{Chisholm} {et~al.}(2020){Chisholm}, {Prochaska}, {Schaerer},
  {Gazagnes}, \& {Henry}}]{Chisholm2020}
{Chisholm}, J., {Prochaska}, J.~X., {Schaerer}, D., {Gazagnes}, S., \& {Henry},
  A. 2020, \mnras, 498, 2554, \dodoi{10.1093/mnras/staa2470}

\bibitem[{{Chisholm} {et~al.}(2022){Chisholm}, {Saldana-Lopez}, {Flury},
  {Schaerer}, {Jaskot}, {Amor{\'\i}n}, {Atek}, {Finkelstein}, {Fleming},
  {Ferguson}, {Fern{\'a}ndez}, {Giavalisco}, {Hayes}, {Heckman}, {Henry}, {Ji},
  {Marques-Chaves}, {Mauerhofer}, {McCandliss}, {Oey}, {{\"O}stlin},
  {Rutkowski}, {Scarlata}, {Thuan}, {Trebitsch}, {Wang}, {Worseck}, \&
  {Xu}}]{Chisholm2022}
{Chisholm}, J., {Saldana-Lopez}, A., {Flury}, S., {et~al.} 2022, \mnras, 517,
  5104, \dodoi{10.1093/mnras/stac2874}

\bibitem[{{Choustikov} {et~al.}(2023){Choustikov}, {Katz}, {Saxena}, {Cameron},
  {Devriendt}, {Slyz}, {Rosdahl}, {Blaizot}, \&
  {Michel-Dansac}}]{Choustikov2023}
{Choustikov}, N., {Katz}, H., {Saxena}, A., {et~al.} 2023, arXiv e-prints,
  arXiv:2304.08526, \dodoi{10.48550/arXiv.2304.08526}

\bibitem[{{Conroy} \& {Gunn}(2010)}]{Conroy2010}
{Conroy}, C., \& {Gunn}, J.~E. 2010, {FSPS: Flexible Stellar Population
  Synthesis}, Astrophysics Source Code Library, record ascl:1010.043

\bibitem[{{Conroy} {et~al.}(2009){Conroy}, {Gunn}, \& {White}}]{Conroy2009}
{Conroy}, C., {Gunn}, J.~E., \& {White}, M. 2009, \apj, 699, 486,
  \dodoi{10.1088/0004-637X/699/1/486}

\bibitem[{{Cullen} {et~al.}(2023{\natexlab{a}}){Cullen}, {McLure}, {McLeod},
  {Dunlop}, {Donnan}, {Carnall}, {Bowler}, {Begley}, {Hamadouche}, \&
  {Stanton}}]{Cullen2023a}
{Cullen}, F., {McLure}, R.~J., {McLeod}, D.~J., {et~al.} 2023{\natexlab{a}},
  \mnras, 520, 14, \dodoi{10.1093/mnras/stad073}

\bibitem[{{Cullen} {et~al.}(2023{\natexlab{b}}){Cullen}, {McLeod}, {McLure},
  {Dunlop}, {Donnan}, {Carnall}, {Keating}, {Magee}, {Arellano-Cordova},
  {Bowler}, {Begley}, {Flury}, {Hamadouche}, \& {Stanton}}]{Cullen2023b}
{Cullen}, F., {McLeod}, D.~J., {McLure}, R.~J., {et~al.} 2023{\natexlab{b}},
  arXiv e-prints, arXiv:2311.06209, \dodoi{10.48550/arXiv.2311.06209}

\bibitem[{{Curtis-Lake} {et~al.}(2023){Curtis-Lake}, {Carniani}, {Cameron},
  {Charlot}, {Jakobsen}, {Maiolino}, {Bunker}, {Witstok}, {Smit}, {Chevallard},
  {Willott}, {Ferruit}, {Arribas}, {Bonaventura}, {Curti}, {D'Eugenio},
  {Franx}, {Giardino}, {Looser}, {L{\"u}tzgendorf}, {Maseda}, {Rawle}, {Rix},
  {Rodr{\'\i}guez del Pino}, {{\"U}bler}, {Sirianni}, {Dressler}, {Egami},
  {Eisenstein}, {Endsley}, {Hainline}, {Hausen}, {Johnson}, {Rieke},
  {Robertson}, {Shivaei}, {Stark}, {Tacchella}, {Williams}, {Willmer},
  {Bhatawdekar}, {Bowler}, {Boyett}, {Chen}, {de Graaff}, {Helton}, {Hviding},
  {Jones}, {Kumari}, {Lyu}, {Nelson}, {Perna}, {Sandles}, {Saxena}, {Suess},
  {Sun}, {Topping}, {Wallace}, \& {Whitler}}]{Curtis-Lake2023}
{Curtis-Lake}, E., {Carniani}, S., {Cameron}, A., {et~al.} 2023, Nature
  Astronomy, 7, 622, \dodoi{10.1038/s41550-023-01918-w}

\bibitem[{{Dav{\'e}} {et~al.}(2019){Dav{\'e}}, {Angl{\'e}s-Alc{\'a}zar},
  {Narayanan}, {Li}, {Rafieferantsoa}, \& {Appleby}}]{Dave2019-SIMBA-EoR}
{Dav{\'e}}, R., {Angl{\'e}s-Alc{\'a}zar}, D., {Narayanan}, D., {et~al.} 2019,
  \mnras, 486, 2827, \dodoi{10.1093/mnras/stz937}

\bibitem[{{De Barros} {et~al.}(2019){De Barros}, {Oesch}, {Labb{\'e}},
  {Stefanon}, {Gonz{\'a}lez}, {Smit}, {Bouwens}, \&
  {Illingworth}}]{DeBarros2019}
{De Barros}, S., {Oesch}, P.~A., {Labb{\'e}}, I., {et~al.} 2019, \mnras, 489,
  2355, \dodoi{10.1093/mnras/stz940}

\bibitem[{{D'Eugenio} {et~al.}(2023){D'Eugenio}, {Maiolino}, {Carniani},
  {Curtis-Lake}, {Witstok}, {Chevallard}, {Charlot}, {Baker}, {Arribas},
  {Boyett}, {Bunker}, {Curti}, {Eisenstein}, {Hainline}, {Ji}, {Johnson},
  {Looser}, {Nakajima}, {Nelson}, {Rieke}, {Robertson}, {Scholtz}, {Smit},
  {Venturi}, {Tacchella}, {Uebler}, {Willmer}, \& {Willott}}]{D'Eugenio2023}
{D'Eugenio}, F., {Maiolino}, R., {Carniani}, S., {et~al.} 2023, arXiv e-prints,
  arXiv:2311.09908, \dodoi{10.48550/arXiv.2311.09908}

\bibitem[{{Donnan} {et~al.}(2023){Donnan}, {McLeod}, {Dunlop}, {McLure},
  {Carnall}, {Begley}, {Cullen}, {Hamadouche}, {Bowler}, {Magee}, {McCracken},
  {Milvang-Jensen}, {Moneti}, \& {Targett}}]{Donnan2023}
{Donnan}, C.~T., {McLeod}, D.~J., {Dunlop}, J.~S., {et~al.} 2023, \mnras, 518,
  6011, \dodoi{10.1093/mnras/stac3472}

\bibitem[{{Donnan} {et~al.}(2024){Donnan}, {McLure}, {Dunlop}, {McLeod},
  {Magee}, {Arellano-C{\'o}rdova}, {Barrufet}, {Begley}, {Bowler}, {Carnall},
  {Cullen}, {Ellis}, {Fontana}, {Illingworth}, {Grogin}, {Hamadouche},
  {Koekemoer}, {Liu}, {Mason}, {Santini}, \& {Stanton}}]{Donnan2024}
{Donnan}, C.~T., {McLure}, R.~J., {Dunlop}, J.~S., {et~al.} 2024, arXiv
  e-prints, arXiv:2403.03171, \dodoi{10.48550/arXiv.2403.03171}

\bibitem[{{Draine}(2009)}]{Draine2009}
{Draine}, B.~T. 2009, in Astronomical Society of the Pacific Conference Series,
  Vol. 414, Cosmic Dust - Near and Far, ed. T.~{Henning}, E.~{Gr{\"u}n}, \&
  J.~{Steinacker}, 453, \dodoi{10.48550/arXiv.0903.1658}

\bibitem[{{Draine}(2011)}]{Draine2011}
{Draine}, B.~T. 2011, {Physics of the Interstellar and Intergalactic Medium}

\bibitem[{{Draine} \& {Salpeter}(1979)}]{Draine1979}
{Draine}, B.~T., \& {Salpeter}, E.~E. 1979, \apj, 231, 77,
  \dodoi{10.1086/157165}

\bibitem[{{Drakos} {et~al.}(2022){Drakos}, {Villasenor}, {Robertson}, {Hausen},
  {Dickinson}, {Ferguson}, {Furlanetto}, {Greene}, {Madau}, {Shapley}, {Stark},
  \& {Wechsler}}]{Drakos2022-DREaM}
{Drakos}, N.~E., {Villasenor}, B., {Robertson}, B.~E., {et~al.} 2022, \apj,
  926, 194, \dodoi{10.3847/1538-4357/ac46fb}

\bibitem[{{Driver} {et~al.}(2011){Driver}, {Hill}, {Kelvin}, {Robotham},
  {Liske}, {Norberg}, {Baldry}, {Bamford}, {Hopkins}, {Loveday}, {Peacock},
  {Andrae}, {Bland-Hawthorn}, {Brough}, {Brown}, {Cameron}, {Ching}, {Colless},
  {Conselice}, {Croom}, {Cross}, {de Propris}, {Dye}, {Drinkwater}, {Ellis},
  {Graham}, {Grootes}, {Gunawardhana}, {Jones}, {van Kampen}, {Maraston},
  {Nichol}, {Parkinson}, {Phillipps}, {Pimbblet}, {Popescu}, {Prescott},
  {Roseboom}, {Sadler}, {Sansom}, {Sharp}, {Smith}, {Taylor}, {Thomas},
  {Tuffs}, {Wijesinghe}, {Dunne}, {Frenk}, {Jarvis}, {Madore}, {Meyer},
  {Seibert}, {Staveley-Smith}, {Sutherland}, \& {Warren}}]{Driver2011}
{Driver}, S.~P., {Hill}, D.~T., {Kelvin}, L.~S., {et~al.} 2011, \mnras, 413,
  971, \dodoi{10.1111/j.1365-2966.2010.18188.x}

\bibitem[{{Dunlop} {et~al.}(2012){Dunlop}, {McLure}, {Robertson}, {Ellis},
  {Stark}, {Cirasuolo}, \& {de Ravel}}]{Dunlop2012}
{Dunlop}, J.~S., {McLure}, R.~J., {Robertson}, B.~E., {et~al.} 2012, \mnras,
  420, 901, \dodoi{10.1111/j.1365-2966.2011.20102.x}

\bibitem[{{Dunlop} {et~al.}(2013){Dunlop}, {Rogers}, {McLure}, {Ellis},
  {Robertson}, {Koekemoer}, {Dayal}, {Curtis-Lake}, {Wild}, {Charlot},
  {Bowler}, {Schenker}, {Ouchi}, {Ono}, {Cirasuolo}, {Furlanetto}, {Stark},
  {Targett}, \& {Schneider}}]{Dunlop2013}
{Dunlop}, J.~S., {Rogers}, A.~B., {McLure}, R.~J., {et~al.} 2013, \mnras, 432,
  3520, \dodoi{10.1093/mnras/stt702}

\bibitem[{{Eisenstein} {et~al.}(2023){Eisenstein}, {Willott}, {Alberts},
  {Arribas}, {Bonaventura}, {Bunker}, {Cameron}, {Carniani}, {Charlot},
  {Curtis-Lake}, {D'Eugenio}, {Endsley}, {Ferruit}, {Giardino}, {Hainline},
  {Hausen}, {Jakobsen}, {Johnson}, {Maiolino}, {Rieke}, {Rieke}, {Rix},
  {Robertson}, {Stark}, {Tacchella}, {Williams}, {Willmer}, {Baker}, {Baum},
  {Bhatawdekar}, {Boyett}, {Chen}, {Chevallard}, {Circosta}, {Curti},
  {Danhaive}, {DeCoursey}, {de Graaff}, {Dressler}, {Egami}, {Helton},
  {Hviding}, {Ji}, {Jones}, {Kumari}, {L{\"u}tzgendorf}, {Laseter}, {Looser},
  {Lyu}, {Maseda}, {Nelson}, {Parlanti}, {Perna}, {Pusk{\'a}s}, {Rawle},
  {Rodr{\'\i}guez Del Pino}, {Sandles}, {Saxena}, {Scholtz}, {Sharpe},
  {Shivaei}, {Silcock}, {Simmonds}, {Skarbinski}, {Smit}, {Stone}, {Suess},
  {Sun}, {Tang}, {Topping}, {{\"U}bler}, {Villanueva}, {Wallace}, {Whitler},
  {Witstok}, \& {Woodrum}}]{Eisenstein2023}
{Eisenstein}, D.~J., {Willott}, C., {Alberts}, S., {et~al.} 2023, arXiv
  e-prints, arXiv:2306.02465, \dodoi{10.48550/arXiv.2306.02465}

\bibitem[{{Eldridge} {et~al.}(2017){Eldridge}, {Stanway}, {Xiao}, {McClelland},
  {Taylor}, {Ng}, {Greis}, \& {Bray}}]{Eldridge2017-BPASS}
{Eldridge}, J.~J., {Stanway}, E.~R., {Xiao}, L., {et~al.} 2017, \pasa, 34,
  e058, \dodoi{10.1017/pasa.2017.51}

\bibitem[{{Ellis} {et~al.}(2013){Ellis}, {McLure}, {Dunlop}, {Robertson},
  {Ono}, {Schenker}, {Koekemoer}, {Bowler}, {Ouchi}, {Rogers}, {Curtis-Lake},
  {Schneider}, {Charlot}, {Stark}, {Furlanetto}, \&
  {Cirasuolo}}]{Ellis2013-HUDF12}
{Ellis}, R.~S., {McLure}, R.~J., {Dunlop}, J.~S., {et~al.} 2013, \apjl, 763,
  L7, \dodoi{10.1088/2041-8205/763/1/L7}

\bibitem[{{Endsley} {et~al.}(2023){Endsley}, {Stark}, {Whitler}, {Topping},
  {Chen}, {Plat}, {Chisholm}, \& {Charlot}}]{Endsley2023}
{Endsley}, R., {Stark}, D.~P., {Whitler}, L., {et~al.} 2023, \mnras, 524, 2312,
  \dodoi{10.1093/mnras/stad1919}

\bibitem[{{Feltre} {et~al.}(2016){Feltre}, {Charlot}, \& {Gutkin}}]{Feltre2016}
{Feltre}, A., {Charlot}, S., \& {Gutkin}, J. 2016, \mnras, 456, 3354,
  \dodoi{10.1093/mnras/stv2794}

\bibitem[{{Ferland} {et~al.}(2017){Ferland}, {Chatzikos}, {Guzm{\'a}n},
  {Lykins}, {van Hoof}, {Williams}, {Abel}, {Badnell}, {Keenan}, {Porter}, \&
  {Stancil}}]{Ferland2017-CLOUDY}
{Ferland}, G.~J., {Chatzikos}, M., {Guzm{\'a}n}, F., {et~al.} 2017, \rmxaa, 53,
  385, \dodoi{10.48550/arXiv.1705.10877}

\bibitem[{{Ferrara} {et~al.}(2023){Ferrara}, {Pallottini}, \&
  {Dayal}}]{Ferrara2023}
{Ferrara}, A., {Pallottini}, A., \& {Dayal}, P. 2023, \mnras, 522, 3986,
  \dodoi{10.1093/mnras/stad1095}

\bibitem[{{Ferreira} {et~al.}(2022){Ferreira}, {Adams}, {Conselice},
  {Sazonova}, {Austin}, {Caruana}, {Ferrari}, {Verma}, {Trussler},
  {Broadhurst}, {Diego}, {Frye}, {Pascale}, {Wilkins}, {Windhorst}, \&
  {Zitrin}}]{Ferreira2022}
{Ferreira}, L., {Adams}, N., {Conselice}, C.~J., {et~al.} 2022, \apjl, 938, L2,
  \dodoi{10.3847/2041-8213/ac947c}

\bibitem[{{Ferruit} {et~al.}(2022){Ferruit}, {Jakobsen}, {Giardino}, {Rawle},
  {Alves de Oliveira}, {Arribas}, {Beck}, {Birkmann}, {B{\"o}ker}, {Bunker},
  {Charlot}, {de Marchi}, {Franx}, {Henry}, {Karakla}, {Kassin}, {Kumari},
  {L{\'o}pez-Caniego}, {L{\"u}tzgendorf}, {Maiolino}, {Manjavacas}, {Marston},
  {Moseley}, {Muzerolle}, {Pirzkal}, {Rauscher}, {Rix}, {Sabbi}, {Sirianni},
  {te Plate}, {Valenti}, {Willott}, \& {Zeidler}}]{Ferruit2022}
{Ferruit}, P., {Jakobsen}, P., {Giardino}, G., {et~al.} 2022, \aap, 661, A81,
  \dodoi{10.1051/0004-6361/202142673}

\bibitem[{{Finkelstein} {et~al.}(2010){Finkelstein}, {Papovich}, {Giavalisco},
  {Reddy}, {Ferguson}, {Koekemoer}, \& {Dickinson}}]{Finkelstein2010}
{Finkelstein}, S.~L., {Papovich}, C., {Giavalisco}, M., {et~al.} 2010, \apj,
  719, 1250, \dodoi{10.1088/0004-637X/719/2/1250}

\bibitem[{{Finkelstein} {et~al.}(2012){Finkelstein}, {Papovich}, {Salmon},
  {Finlator}, {Dickinson}, {Ferguson}, {Giavalisco}, {Koekemoer}, {Reddy},
  {Bassett}, {Conselice}, {Dunlop}, {Faber}, {Grogin}, {Hathi}, {Kocevski},
  {Lai}, {Lee}, {McLure}, {Mobasher}, \& {Newman}}]{Finkelstein2012}
{Finkelstein}, S.~L., {Papovich}, C., {Salmon}, B., {et~al.} 2012, \apj, 756,
  164, \dodoi{10.1088/0004-637X/756/2/164}

\bibitem[{{Finkelstein} {et~al.}(2022){Finkelstein}, {Bagley}, {Arrabal Haro},
  {Dickinson}, {Ferguson}, {Kartaltepe}, {Papovich}, {Burgarella}, {Kocevski},
  {Huertas-Company}, {Iyer}, {Larson}, {P{\'e}rez-Gonz{\'a}lez}, {Rose},
  {Tacchella}, {Wilkins}, {Chworowsky}, {Medrano}, {Morales}, {Somerville},
  {Yung}, {Fontana}, {Giavalisco}, {Grazian}, {Grogin}, {Kewley}, {Koekemoer},
  {Kirkpatrick}, {Kurczynski}, {Lotz}, {Pentericci}, {Pirzkal}, {Ravindranath},
  {Ryan}, {Trump}, {Yang}, {Almaini}, {Amor{\'\i}n}, {Annunziatella},
  {Backhaus}, {Barro}, {Behroozi}, {Bell}, {Bhatawdekar}, {Bisigello}, {Bromm},
  {Buat}, {Buitrago}, {Calabr{\'o}}, {Casey}, {Castellano}, {Ch{\'a}vez Ortiz},
  {Ciesla}, {Cleri}, {Cohen}, {Cole}, {Cooke}, {Cooper}, {Cooray}, {Costantin},
  {Cox}, {Croton}, {Daddi}, {Dav{\'e}}, {de la Vega}, {Dekel}, {Elbaz},
  {Estrada-Carpenter}, {Faber}, {Fern{\'a}ndez}, {Finkelstein}, {Freundlich},
  {Fujimoto}, {Garc{\'\i}a-Argum{\'a}nez}, {Gardner}, {Gawiser},
  {G{\'o}mez-Guijarro}, {Guo}, {Hamilton}, {Hathi}, {Holwerda}, {Hirschmann},
  {Hutchison}, {Jha}, {Jogee}, {Juneau}, {Jung}, {Kassin}, {Le Bail}, {Leung},
  {Lucas}, {Magnelli}, {Mantha}, {Matharu}, {McGrath}, {McIntosh}, {Merlin},
  {Mobasher}, {Newman}, {Nicholls}, {Pandya}, {Rafelski}, {Ronayne}, {Santini},
  {Seill{\'e}}, {Shah}, {Shen}, {Simons}, {Snyder}, {Stanway}, {Straughn},
  {Teplitz}, {Vanderhoof}, {Vega-Ferrero}, {Wang}, {Weiner}, {Willmer},
  {Wuyts}, \& {Zavala}}]{Finkelstein2022}
{Finkelstein}, S.~L., {Bagley}, M.~B., {Arrabal Haro}, P., {et~al.} 2022, arXiv
  e-prints, arXiv:2207.12474.
\newblock \doarXiv{2207.12474}

\bibitem[{{Finkelstein} {et~al.}(2023{\natexlab{a}}){Finkelstein}, {Bagley},
  {Ferguson}, {Wilkins}, {Kartaltepe}, {Papovich}, {Yung}, {Arrabal Haro},
  {Behroozi}, {Dickinson}, {Kocevski}, {Koekemoer}, {Larson}, {Le Bail},
  {Morales}, {P{\'e}rez-Gonz{\'a}lez}, {Burgarella}, {Dav{\'e}}, {Hirschmann},
  {Somerville}, {Wuyts}, {Bromm}, {Casey}, {Fontana}, {Fujimoto}, {Gardner},
  {Giavalisco}, {Grazian}, {Grogin}, {Hathi}, {Hutchison}, {Jha}, {Jogee},
  {Kewley}, {Kirkpatrick}, {Long}, {Lotz}, {Pentericci}, {Pierel}, {Pirzkal},
  {Ravindranath}, {Ryan}, {Trump}, {Yang}, {Bhatawdekar}, {Bisigello}, {Buat},
  {Calabr{\`o}}, {Castellano}, {Cleri}, {Cooper}, {Croton}, {Daddi}, {Dekel},
  {Elbaz}, {Franco}, {Gawiser}, {Holwerda}, {Huertas-Company}, {Jaskot},
  {Leung}, {Lucas}, {Mobasher}, {Pandya}, {Tacchella}, {Weiner}, \&
  {Zavala}}]{Finkelstein2022-CEERSKP1}
{Finkelstein}, S.~L., {Bagley}, M.~B., {Ferguson}, H.~C., {et~al.}
  2023{\natexlab{a}}, \apjl, 946, L13, \dodoi{10.3847/2041-8213/acade4}

\bibitem[{{Finkelstein} {et~al.}(2023{\natexlab{b}}){Finkelstein}, {Leung},
  {Bagley}, {Dickinson}, {Ferguson}, {Papovich}, {Akins}, {Arrabal Haro},
  {Dave}, {Dekel}, {Kartaltepe}, {Kocevski}, {Koekemoer}, {Pirzkal},
  {Somerville}, {Yung}, {Amorin}, {Backhaus}, {Behroozi}, {Bisigello}, {Bromm},
  {Casey}, {Chavez Ortiz}, {Cheng}, {Chworowsky}, {Cleri}, {Cooper}, {Davis},
  {de la Vega}, {Elbaz}, {Franco}, {Fontana}, {Fujimoto}, {Giavalisco},
  {Grogin}, {Holwerda}, {Huertas-Company}, {Hirschmann}, {Iyer}, {Jogee},
  {Jung}, {Larson}, {Lucas}, {Mobasher}, {Morales}, {Morley}, {Mukherjee},
  {Perez-Gonzalez}, {Ravindranath}, {Rodighiero}, {Rowland}, {Tacchella},
  {Taylor}, {Trump}, \& {Wilkins}}]{Finkelstein2023}
{Finkelstein}, S.~L., {Leung}, G. C.~K., {Bagley}, M.~B., {et~al.}
  2023{\natexlab{b}}, arXiv e-prints, arXiv:2311.04279,
  \dodoi{10.48550/arXiv.2311.04279}

\bibitem[{{Fujimoto} {et~al.}(2023){Fujimoto}, {Arrabal Haro}, {Dickinson},
  {Finkelstein}, {Kartaltepe}, {Larson}, {Burgarella}, {Bagley}, {Behroozi},
  {Chworowsky}, {Hirschmann}, {Trump}, {Wilkins}, {Yung}, {Koekemoer},
  {Papovich}, {Pirzkal}, {Ferguson}, {Fontana}, {Grogin}, {Grazian}, {Kewley},
  {Kocevski}, {Lotz}, {Pentericci}, {Ravindranath}, {Somerville}, {Wilkins},
  {Amor{\'\i}n}, {Backhaus}, {Calabr{\`o}}, {Casey}, {Cooper}, {Fern{\'a}ndez},
  {Franco}, {Giavalisco}, {Hathi}, {Harish}, {Hutchison}, {Iyer}, {Jung},
  {Lucas}, \& {Zavala}}]{Fujimoto2023}
{Fujimoto}, S., {Arrabal Haro}, P., {Dickinson}, M., {et~al.} 2023, \apjl, 949,
  L25, \dodoi{10.3847/2041-8213/acd2d9}

\bibitem[{{Furtak} {et~al.}(2023){Furtak}, {Shuntov}, {Atek}, {Zitrin},
  {Richard}, {Lehnert}, \& {Chevallard}}]{Furtak2023-SMACS}
{Furtak}, L.~J., {Shuntov}, M., {Atek}, H., {et~al.} 2023, \mnras, 519, 3064,
  \dodoi{10.1093/mnras/stac3717}

\bibitem[{{Gaia Collaboration} {et~al.}(2018){Gaia Collaboration}, {Brown},
  {Vallenari}, {Prusti}, {de Bruijne}, {Babusiaux}, {Bailer-Jones}, {Biermann},
  {Evans}, {Eyer}, {Jansen}, {Jordi}, {Klioner}, {Lammers}, {Lindegren},
  {Luri}, {Mignard}, {Panem}, {Pourbaix}, {Randich}, {Sartoretti}, {Siddiqui},
  {Soubiran}, {van Leeuwen}, {Walton}, {Arenou}, {Bastian}, {Cropper},
  {Drimmel}, {Katz}, {Lattanzi}, {Bakker}, {Cacciari}, {Casta{\~n}eda},
  {Chaoul}, {Cheek}, {De Angeli}, {Fabricius}, {Guerra}, {Holl}, {Masana},
  {Messineo}, {Mowlavi}, {Nienartowicz}, {Panuzzo}, {Portell}, {Riello},
  {Seabroke}, {Tanga}, {Th{\'e}venin}, {Gracia-Abril}, {Comoretto},
  {Garcia-Reinaldos}, {Teyssier}, {Altmann}, {Andrae}, {Audard},
  {Bellas-Velidis}, {Benson}, {Berthier}, {Blomme}, {Burgess}, {Busso},
  {Carry}, {Cellino}, {Clementini}, {Clotet}, {Creevey}, {Davidson}, {De
  Ridder}, {Delchambre}, {Dell'Oro}, {Ducourant},
  {Fern{\'a}ndez-Hern{\'a}ndez}, {Fouesneau}, {Fr{\'e}mat}, {Galluccio},
  {Garc{\'\i}a-Torres}, {Gonz{\'a}lez-N{\'u}{\~n}ez}, {Gonz{\'a}lez-Vidal},
  {Gosset}, {Guy}, {Halbwachs}, {Hambly}, {Harrison}, {Hern{\'a}ndez},
  {Hestroffer}, {Hodgkin}, {Hutton}, {Jasniewicz}, {Jean-Antoine-Piccolo},
  {Jordan}, {Korn}, {Krone-Martins}, {Lanzafame}, {Lebzelter}, {L{\"o}ffler},
  {Manteiga}, {Marrese}, {Mart{\'\i}n-Fleitas}, {Moitinho}, {Mora}, {Muinonen},
  {Osinde}, {Pancino}, {Pauwels}, {Petit}, {Recio-Blanco}, {Richards},
  {Rimoldini}, {Robin}, {Sarro}, {Siopis}, {Smith}, {Sozzetti}, {S{\"u}veges},
  {Torra}, {van Reeven}, {Abbas}, {Abreu Aramburu}, {Accart}, {Aerts},
  {Altavilla}, {{\'A}lvarez}, {Alvarez}, {Alves}, {Anderson}, {Andrei},
  {Anglada Varela}, {Antiche}, {Antoja}, {Arcay}, {Astraatmadja}, {Bach},
  {Baker}, {Balaguer-N{\'u}{\~n}ez}, {Balm}, {Barache}, {Barata}, {Barbato},
  {Barblan}, {Barklem}, {Barrado}, {Barros}, {Barstow}, {Bartholom{\'e}
  Mu{\~n}oz}, {Bassilana}, {Becciani}, {Bellazzini}, {Berihuete}, {Bertone},
  {Bianchi}, {Bienaym{\'e}}, {Blanco-Cuaresma}, {Boch}, {Boeche}, {Bombrun},
  {Borrachero}, {Bossini}, {Bouquillon}, {Bourda}, {Bragaglia}, {Bramante},
  {Breddels}, {Bressan}, {Brouillet}, {Br{\"u}semeister}, {Brugaletta},
  {Bucciarelli}, {Burlacu}, {Busonero}, {Butkevich}, {Buzzi}, {Caffau},
  {Cancelliere}, {Cannizzaro}, {Cantat-Gaudin}, {Carballo}, {Carlucci},
  {Carrasco}, {Casamiquela}, {Castellani}, {Castro-Ginard}, {Charlot},
  {Chemin}, {Chiavassa}, {Cocozza}, {Costigan}, {Cowell}, {Crifo}, {Crosta},
  {Crowley}, {Cuypers}, {Dafonte}, {Damerdji}, {Dapergolas}, {David}, {David},
  {de Laverny}, {De Luise}, {De March}, {de Martino}, {de Souza}, {de Torres},
  {Debosscher}, {del Pozo}, {Delbo}, {Delgado}, {Delgado}, {Di Matteo},
  {Diakite}, {Diener}, {Distefano}, {Dolding}, {Drazinos}, {Dur{\'a}n},
  {Edvardsson}, {Enke}, {Eriksson}, {Esquej}, {Eynard Bontemps}, {Fabre},
  {Fabrizio}, {Faigler}, {Falc{\~a}o}, {Farr{\`a}s Casas}, {Federici},
  {Fedorets}, {Fernique}, {Figueras}, {Filippi}, {Findeisen}, {Fonti},
  {Fraile}, {Fraser}, {Fr{\'e}zouls}, {Gai}, {Galleti}, {Garabato},
  {Garc{\'\i}a-Sedano}, {Garofalo}, {Garralda}, {Gavel}, {Gavras}, {Gerssen},
  {Geyer}, {Giacobbe}, {Gilmore}, {Girona}, {Giuffrida}, {Glass}, {Gomes},
  {Granvik}, {Gueguen}, {Guerrier}, {Guiraud}, {Guti{\'e}rrez-S{\'a}nchez},
  {Haigron}, {Hatzidimitriou}, {Hauser}, {Haywood}, {Heiter}, {Helmi}, {Heu},
  {Hilger}, {Hobbs}, {Hofmann}, {Holland}, {Huckle}, {Hypki}, {Icardi},
  {Jan{\ss}en}, {Jevardat de Fombelle}, {Jonker}, {Juh{\'a}sz}, {Julbe},
  {Karampelas}, {Kewley}, {Klar}, {Kochoska}, {Kohley}, {Kolenberg},
  {Kontizas}, {Kontizas}, {Koposov}, {Kordopatis}, {Kostrzewa-Rutkowska},
  {Koubsky}, {Lambert}, {Lanza}, {Lasne}, {Lavigne}, {Le Fustec}, {Le
  Poncin-Lafitte}, {Lebreton}, {Leccia}, {Leclerc}, {Lecoeur-Taibi},
  {Lenhardt}, {Leroux}, {Liao}, {Licata}, {Lindstr{\o}m}, {Lister}, {Livanou},
  {Lobel}, {L{\'o}pez}, {Managau}, {Mann}, {Mantelet}, {Marchal}, {Marchant},
  {Marconi}, {Marinoni}, {Marschalk{\'o}}, {Marshall}, {Martino}, {Marton},
  {Mary}, {Massari}, {Matijevi{\v{c}}}, {Mazeh}, {McMillan}, {Messina},
  {Michalik}, {Millar}, {Molina}, {Molinaro}, {Moln{\'a}r}, {Montegriffo},
  {Mor}, {Morbidelli}, {Morel}, {Morris}, {Mulone}, {Muraveva}, {Musella},
  {Nelemans}, {Nicastro}, {Noval}, {O'Mullane}, {Ord{\'e}novic},
  {Ord{\'o}{\~n}ez-Blanco}, {Osborne}, {Pagani}, {Pagano}, {Pailler},
  {Palacin}, {Palaversa}, {Panahi}, {Pawlak}, {Piersimoni}, {Pineau}, {Plachy},
  {Plum}, {Poggio}, {Poujoulet}, {Pr{\v{s}}a}, {Pulone}, {Racero}, {Ragaini},
  {Rambaux}, {Ramos-Lerate}, {Regibo}, {Reyl{\'e}}, {Riclet}, {Ripepi}, {Riva},
  {Rivard}, {Rixon}, {Roegiers}, {Roelens}, {Romero-G{\'o}mez}, {Rowell},
  {Royer}, {Ruiz-Dern}, {Sadowski}, {Sagrist{\`a} Sell{\'e}s}, {Sahlmann},
  {Salgado}, {Salguero}, {Sanna}, {Santana-Ros}, {Sarasso}, {Savietto},
  {Schultheis}, {Sciacca}, {Segol}, {Segovia}, {S{\'e}gransan}, {Shih},
  {Siltala}, {Silva}, {Smart}, {Smith}, {Solano}, {Solitro}, {Sordo}, {Soria
  Nieto}, {Souchay}, {Spagna}, {Spoto}, {Stampa}, {Steele},
  {Steidelm{\"u}ller}, {Stephenson}, {Stoev}, {Suess}, {Surdej}, {Szabados},
  {Szegedi-Elek}, {Tapiador}, {Taris}, {Tauran}, {Taylor}, {Teixeira},
  {Terrett}, {Teyssandier}, {Thuillot}, {Titarenko}, {Torra Clotet}, {Turon},
  {Ulla}, {Utrilla}, {Uzzi}, {Vaillant}, {Valentini}, {Valette}, {van Elteren},
  {Van Hemelryck}, {van Leeuwen}, {Vaschetto}, {Vecchiato}, {Veljanoski},
  {Viala}, {Vicente}, {Vogt}, {von Essen}, {Voss}, {Votruba}, {Voutsinas},
  {Walmsley}, {Weiler}, {Wertz}, {Wevers}, {Wyrzykowski}, {Yoldas},
  {{\v{Z}}erjal}, {Ziaeepour}, {Zorec}, {Zschocke}, {Zucker}, {Zurbach}, \&
  {Zwitter}}]{GAIA-DR2}
{Gaia Collaboration}, {Brown}, A.~G.~A., {Vallenari}, A., {et~al.} 2018, \aap,
  616, A1, \dodoi{10.1051/0004-6361/201833051}

\bibitem[{{Gail} {et~al.}(2009){Gail}, {Zhukovska}, {Hoppe}, \&
  {Trieloff}}]{Gail2009}
{Gail}, H.~P., {Zhukovska}, S.~V., {Hoppe}, P., \& {Trieloff}, M. 2009, \apj,
  698, 1136, \dodoi{10.1088/0004-637X/698/2/1136}

\bibitem[{{Gardner} {et~al.}(2023){Gardner}, {Mather}, {Abbott}, {Abell},
  {Abernathy}, {Abney}, {Abraham}, {Abraham}, {Abul-Huda}, {Acton}, {Adams},
  {Adams}, {Adler}, {Adriaensen}, {Aguilar}, {Ahmed}, {Ahmed}, {Ahmed},
  {Albat}, {Albert}, {Alberts}, {Aldridge}, {Allen}, {Allen}, {Altenburg},
  {Altunc}, {Alvarez}, {{\'A}lvarez-M{\'a}rquez}, {Alves de Oliveira},
  {Ambrose}, {Anandakrishnan}, {Andersen}, {Anderson}, {Anderson}, {Anderson},
  {Anderson}, {Aprea}, {Archer}, {Arenberg}, {Argyriou}, {Arribas}, {Artigau},
  {Arvai}, {Atcheson}, {Atkinson}, {Averbukh}, {Aymergen}, {Bacinski},
  {Baggett}, {Bagnasco}, {Baker}, {Balzano}, {Banks}, {Baran}, {Barker},
  {Barrett}, {Barringer}, {Barto}, {Bast}, {Baudoz}, {Baum}, {Beatty},
  {Beaulieu}, {Bechtold}, {Beck}, {Beddard}, {Beichman}, {Bellagama}, {Bely},
  {Berger}, {Bergeron}, {Bernier}, {Bertch}, {Beskow}, {Betz}, {Biagetti},
  {Birkmann}, {Bjorklund}, {Blackwood}, {Blazek}, {Blossfeld}, {Bluth},
  {Boccaletti}, {Boegner}, {Bohlin}, {Boia}, {B{\"o}ker}, {Bonaventura},
  {Bond}, {Bosley}, {Boucarut}, {Bouchet}, {Bouwman}, {Bower}, {Bowers},
  {Bowers}, {Boyce}, {Boyer}, {Boyer}, {Boyer}, {Boyer}, {Bradley}, {Brady},
  {Brandl}, {Brannen}, {Breda}, {Bremmer}, {Brennan}, {Bresnahan}, {Bright},
  {Broiles}, {Bromenschenkel}, {Brooks}, {Brooks}, {Brown}, {Brown}, {Brown},
  {Bruce}, {Bryson}, {Bujanda}, {Bullock}, {Bunker}, {Bureo}, {Burt}, {Bush},
  {Bushouse}, {Bussman}, {Cabaud}, {Cale}, {Calhoon}, {Calvani}, {Canipe},
  {Caputo}, {Cara}, {Carey}, {Case}, {Cesari}, {Cetorelli}, {Chance},
  {Chandler}, {Chaney}, {Chapman}, {Charlot}, {Chayer}, {Cheezum}, {Chen},
  {Chen}, {Cherinka}, {Chichester}, {Chilton}, {Chittiraibalan}, {Clampin},
  {Clark}, {Clark}, {Clark}, {Claybrooks}, {Cleveland}, {Cohen}, {Cohen},
  {Col{\'o}n}, {Coleman}, {Colina}, {Comber}, {Comeau}, {Comer}, {Conde Reis},
  {Connolly}, {Conroy}, {Contos}, {Contreras}, {Cook}, {Cooper}, {Cooper},
  {Correia}, {Correnti}, {Cossou}, {Costanza}, {Coulais}, {Cox}, {Coyle},
  {Cracraft}, {Crew}, {Curtis}, {Cusveller}, {Da Costa Maciel}, {Dailey},
  {Daugeron}, {Davidson}, {Davies}, {Davis}, {Davis}, {Day}, {de Chambure}, {de
  Jong}, {De Marchi}, {Dean}, {Decker}, {Delisa}, {Dell}, {Dellagatta},
  {Dembinska}, {Demosthenes}, {Dencheva}, {Deneu}, {DePriest}, {Deschenes},
  {Dethienne}, {Detre}, {Diaz}, {Dicken}, {DiFelice}, {Dillman}, {Disharoon},
  {Dixon}, {Doggett}, {Dominguez}, {Donaldson}, {Doria-Warner}, {Santos},
  {Doty}, {Douglas}, {Doyon}, {Dressler}, {Driggers}, {Driggers}, {Dunn},
  {DuPrie}, {Dupuis}, {Durning}, {Dutta}, {Earl}, {Eccleston}, {Ecobichon},
  {Egami}, {Ehrenwinkler}, {Eisenhamer}, {Eisenhower}, {Eisenstein}, {El
  Hamel}, {Elie}, {Elliott}, {Elliott}, {Engesser}, {Espinoza}, {Etienne},
  {Etxaluze}, {Evans}, {Fabreguettes}, {Falcolini}, {Falini}, {Fatig},
  {Feeney}, {Feinberg}, {Fels}, {Ferdous}, {Ferguson}, {Ferrarese}, {Ferreira},
  {Ferruit}, {Ferry}, {Filippazzo}, {Firre}, {Fix}, {Flagey}, {Flanagan},
  {Fleming}, {Florian}, {Flynn}, {Foiadelli}, {Fontaine}, {Fontanella},
  {Forshay}, {Fortner}, {Fox}, {Framarini}, {Francisco}, {Franck}, {Franx},
  {Franz}, {Friedman}, {Friend}, {Frost}, {Fu}, {Fullerton}, {Gaillard},
  {Galkin}, {Gallagher}, {Galyer}, {Garc{\'\i}a Mar{\'\i}n}, {Gardner},
  {Garland}, {Garrett}, {Gasman}, {G{\'a}sp{\'a}r}, {Gastaud}, {Gaudreau},
  {Gauthier}, {Geers}, {Geithner}, {Gennaro}, {Gerber}, {Gereau}, {Giampaoli},
  {Giardino}, {Gibbons}, {Gilbert}, {Gilman}, {Girard}, {Giuliano}, {Gkountis},
  {Glasse}, {Glassmire}, {Glauser}, {Glazer}, {Goldberg}, {Golimowski},
  {Gonzaga}, {Gordon}, {Gordon}, {Goudfrooij}, {Gough}, {Graham}, {Grau},
  {Green}, {Greene}, {Greene}, {Greenfield}, {Greenhouse}, {Greve}, {Greville},
  {Grimaldi}, {Groe}, {Groebner}, {Grumm}, {Grundy}, {G{\"u}del}, {Guillard},
  {Guldalian}, {Gunn}, {Gurule}, {Gutman}, {Guy}, {Guyot}, {Hack}, {Haderlein},
  {Hagan}, {Hagedorn}, {Hainline}, {Haley}, {Hami}, {Hamilton}, {Hammann},
  {Hammel}, {Hanley}, {Hansen}, {Hardy}, {Harnisch}, {Harr}, {Harris}, {Hart},
  {Hartig}, {Hasan}, {Hashim}, {Hashimoto}, {Haskins}, {Hawkins}, {Hayden},
  {Hayden}, {Healy}, {Hecht}, {Heeg}, {Hejal}, {Helm}, {Hengemihle}, {Henning},
  {Henry}, {Henry}, {Henshaw}, {Hernandez}, {Herrington}, {Heske}, {Hesman},
  {Hickey}, {Hilbert}, {Hines}, {Hinz}, {Hirsch}, {Hitcho}, {Hodapp}, {Hodge},
  {Hoffman}, {Holfeltz}, {Holler}, {Hoppa}, {Horner}, {Howard}, {Howard},
  {Huber}, {Hunkeler}, {Hunter}, {Hunter}, {Hurd}, {Hurst}, {Hutchings},
  {Hylan}, {Ignat}, {Illingworth}, {Irish}, {Isaacs}, {Jackson}, {Jaffe},
  {Jahic}, {Jahromi}, {Jakobsen}, {James}, {James}, {James}, {Jamieson},
  {Jandra}, {Jayawardhana}, {Jedrzejewski}, {Jeffers}, {Jensen}, {Joanne},
  {Johns}, {Johnson}, {Johnson}, {Johnson}, {Johnson}, {Johnson}, {Johnson},
  {Johnstone}, {Jollet}, {Jones}, {Jones}, {Jones}, {Jones}, {Jones}, {Jordan},
  {Jordan}, {Jue}, {Jurkowski}, {Justis}, {Justtanont}, {Kaleida}, {Kalirai},
  {Kalmanson}, {Kaltenegger}, {Kammerer}, {Kan}, {Kanarek}, {Kao}, {Karakla},
  {Karl}, {Kassin}, {Kauffman}, {Kavanagh}, {Kelley}, {Kelly}, {Kendrew},
  {Kennedy}, {Kenny}, {Keski-Kuha}, {Keyes}, {Khan}, {Kidwell}, {Kimble},
  {King}, {King}, {Kinzel}, {Kirk}, {Kirkpatrick}, {Klaassen}, {Klingemann},
  {Klintworth}, {Knapp}, {Knight}, {Knollenberg}, {Knutsen}, {Koehler},
  {Koekemoer}, {Kofler}, {Kontson}, {Kovacs}, {Kozhurina-Platais}, {Krause},
  {Kriss}, {Krist}, {Kristoffersen}, {Krogel}, {Krueger}, {Kulp}, {Kumari},
  {Kwan}, {Kyprianou}, {Labador}, {Labiano}, {Lafreni{\`e}re}, {Lagage},
  {Laidler}, {Laine}, {Laird}, {Lajoie}, {Lallo}, {Lam}, {LaMassa}, {Lambros},
  {Lampenfield}, {Lander}, {Langston}, {Larson}, {Larson}, {LaVerghetta},
  {Law}, {Lawrence}, {Lee}, {Lee}, {Lee}, {Leisenring}, {Leveille}, {Levenson},
  {Levi}, {Levine}, {Lewis}, {Lewis}, {Lewis}, {Libralato}, {Lidon},
  {Liebrecht}, {Lightsey}, {Lilly}, {Lim}, {Lim}, {Ling}, {Link}, {Link},
  {Lipinski}, {Liu}, {Lo}, {Lobmeyer}, {Logue}, {Long}, {Long}, {Long}, {Long},
  {L{\'o}pez-Caniego}, {Lotz}, {Love-Pruitt}, {Lubskiy}, {Luers}, {Luetgens},
  {Luevano}, {Lui}, {Lund}, {Lundquist}, {Lunine}, {L{\"u}tzgendorf}, {Lynch},
  {MacDonald}, {MacDonald}, {Macias}, {Macklis}, {Maghami}, {Maharaja},
  {Maiolino}, {Makrygiannis}, {Malla}, {Malumuth}, {Manjavacas}, {Marini},
  {Marrione}, {Marston}, {Martel}, {Martin}, {Martin}, {Martinez}, {Maschmann},
  {Masci}, {Masetti}, {Maszkiewicz}, {Matthews}, {Matuskey}, {McBrayer},
  {McCarthy}, {McCaughrean}, {McClare}, {McClare}, {McCloskey}, {McClurg},
  {McCoy}, {McElwain}, {McGregor}, {McGuffey}, {McKay}, {McKenzie}, {McLean},
  {McMaster}, {McNeil}, {De Meester}, {Mehalick}, {Meixner}, {Mel{\'e}ndez},
  {Menzel}, {Menzel}, {Merz}, {Mesterharm}, {Meyer}, {Meyett}, {Meza},
  {Midwinter}, {Milam}, {Miller}, {Miller}, {Miskey}, {Misselt}, {Mitchell},
  {Mohan}, {Montoya}, {Moran}, {Morishita}, {Moro-Mart{\'\i}n}, {Morrison},
  {Morrison}, {Morse}, {Moschos}, {Moseley}, {Mosier}, {Mosner}, {Mountain},
  {Muckenthaler}, {Mueller}, {Mueller}, {Muhiem}, {M{\"u}hlmann}, {Mullally},
  {Mullen}, {Munger}, {Murphy}, {Murray}, {Muzerolle}, {Mycroft}, {Myers},
  {Myers}, {Myers}, {Myers}, {Myrick}, {Nagle}, {Nayak}, {Naylor}, {Neff},
  {Nelan}, {Nella}, {Nguyen}, {Nguyen}, {Nickson}, {Nidhiry}, {Niedner},
  {Nieto-Santisteban}, {Nikolov}, {Nishisaka}, {Noriega-Crespo}, {Nota},
  {O'Mara}, {Oboryshko}, {O'Brien}, {Ochs}, {Offenberg}, {Ogle}, {Ohl},
  {Olmsted}, {Osborne}, {O'Shaughnessy}, {{\"O}stlin}, {O'Sullivan}, {Otor},
  {Ottens}, {Ouellette}, {Outlaw}, {Owens}, {Pacifici}, {Page}, {Paranilam},
  {Park}, {Parrish}, {Paschal}, {Patapis}, {Patel}, {Patrick}, {Pattishall},
  {Paul}, {Paul}, {Pauly}, {Pavlovsky}, {Pe{\~n}a-Guerrero}, {Pedder}, {Peek},
  {Pelham}, {Penanen}, {Perriello}, {Perrin}, {Perrine}, {Perrygo}, {Peslier},
  {Petach}, {Peterson}, {Pfarr}, {Pierson}, {Pietraszkiewicz}, {Pilchen},
  {Pipher}, {Pirzkal}, {Pitman}, {Player}, {Plesha}, {Plitzke}, {Pohner},
  {Poletis}, {Pollizzi}, {Polster}, {Pontius}, {Pontoppidan}, {Porges},
  {Potter}, {Prescott}, {Proffitt}, {Pueyo}, {Quispe Neira}, {Radich}, {Rager},
  {Rameau}, {Ramey}, {Ramos Alarcon}, {Rampini}, {Rapp}, {Rashford},
  {Rauscher}, {Ravindranath}, {Rawle}, {Rawlings}, {Ray}, {Regan}, {Rehm},
  {Rehm}, {Reid}, {Reis}, {Renk}, {Reoch}, {Ressler}, {Rest}, {Reynolds},
  {Richon}, {Richon}, {Ridgaway}, {Riedel}, {Rieke}, {Rieke}, {Rifelli},
  {Rigby}, {Riggs}, {Ringel}, {Ritchie}, {Rix}, {Robberto}, {Robinson},
  {Robinson}, {Robinson}, {Rock}, {Rodriguez}, {Rodr{\'\i}guez del Pino},
  {Roellig}, {Rohrbach}, {Roman}, {Romelfanger}, {Romo}, {Rosales}, {Rose},
  {Roteliuk}, {Roth}, {Rothwell}, {Rouzaud}, {Rowe}, {Rowlands}, {Roy},
  {Royer}, {Rui}, {Rumler}, {Rumpl}, {Russ}, {Ryan}, {Ryan}, {Saad}, {Sabata},
  {Sabatino}, {Sabbi}, {Sabelhaus}, {Sabia}, {Sahu}, {Saif}, {Salvignol},
  {Samara-Ratna}, {Samuelson}, {Sanders}, {Sappington}, {Sargent}, {Sauer},
  {Savadkin}, {Sawicki}, {Schappell}, {Scheffer}, {Scheithauer}, {Scherer},
  {Schiff}, {Schlawin}, {Schmeitzky}, {Schmitz}, {Schmude}, {Schneider},
  {Schreiber}, {Schroeven-Deceuninck}, {Schultz}, {Schwab}, {Schwartz},
  {Scoccimarro}, {Scott}, {Scott}, {Seaton}, {Seely}, {Seery}, {Seidleck},
  {Sembach}, {Shanahan}, {Shaughnessy}, {Shaw}, {Shay}, {Sheehan}, {Sheth},
  {Shih}, {Shivaei}, {Siegel}, {Sienkiewicz}, {Simmons}, {Simon}, {Sirianni},
  {Sivaramakrishnan}, {Slade}, {Sloan}, {Slocum}, {Slowinski}, {Smith},
  {Smith}, {Smith}, {Smith}, {Smith}, {Smith}, {Smolik}, {Soderblom}, {Sohn},
  {Sokol}, {Sonneborn}, {Sontag}, {Sooy}, {Soummer}, {Southwood}, {Spain},
  {Sparmo}, {Speer}, {Spencer}, {Sprofera}, {Stallcup}, {Stanley},
  {Stansberry}, {Stark}, {Starr}, {Stassi}, {Steck}, {Steeley}, {Stephens},
  {Stephenson}, {Stewart}, {Stiavelli}, {}, {Strada}, {Straughn}, {Streetman},
  {Strickland}, {Strobele}, {Stuhlinger}, {Stys}, {Such}, {Sukhatme},
  {Sullivan}, {Sullivan}, {Sumner}, {Sun}, {Sunnquist}, {Swade}, {Swam},
  {Swenton}, {Swoish}, {Tam Litten}, {Tamas}, {Tao}, {Taylor}, {Taylor}, {te
  Plate}, {Van Tea}, {Teague}, {Telfer}, {Temim}, {Texter}, {Thatte},
  {Thompson}, {Thompson}, {Thomson}, {Thronson}, {Tierney}, {Tikkanen},
  {Tinnin}, {Tippet}, {Todd}, {Tran}, {Trauger}, {Trejo}, {Vinh Truong},
  {Tsukamoto}, {Tufail}, {Tumlinson}, {Tustain}, {Tyra}, {Ubeda}, {Underwood},
  {Uzzo}, {Vaclavik}, {Valenduc}, {Valenti}, {Van Campen}, {van de Wetering},
  {Van Der Marel}, {van Haarlem}, {Vandenbussche}, {van Dishoeck},
  {Vanterpool}, {Vernoy}, {Vila Costas}, {Volk}, {Voorzaat}, {Voyton}, {Vydra},
  {Waddy}, {Waelkens}, {Wahlgren}, {Walker}, {Wander}, {Warfield}, {Warner},
  {Wasiak}, {Wasiak}, {Wehner}, {Weiler}, {Weilert}, {Weiss}, {Wells}, {Welty},
  {Wheate}, {Wheeler}, {White}, {Whitehouse}, {Whiteleather}, {Whitman},
  {Williams}, {Willmer}, {Willott}, {Willoughby}, {Wilson}, {Wilson}, {Wilson},
  {Windhorst}, {Wislowski}, {Wolfe}, {Wolfe}, {Wolff}, {Wondel}, {Woo},
  {Woods}, {Worden}, {Workman}, {Wright}, {Wu}, {Wu}, {Wun}, {Wymer},
  {Yadetie}, {Yan}, {Yang}, {Yates}, {Yeager}, {Yerger}, {Young}, {Young},
  {Yu}, {Yu}, {Zak}, {Zeidler}, {Zepp}, {Zhou}, {Zincke}, {Zonak}, \&
  {Zondag}}]{Gardner2023}
{Gardner}, J.~P., {Mather}, J.~C., {Abbott}, R., {et~al.} 2023, \pasp, 135,
  068001, \dodoi{10.1088/1538-3873/acd1b5}

\bibitem[{{Gonzalez-Perez} {et~al.}(2013){Gonzalez-Perez}, {Lacey}, {Baugh},
  {Frenk}, \& {Wilkins}}]{GonzalezPerez2013}
{Gonzalez-Perez}, V., {Lacey}, C.~G., {Baugh}, C.~M., {Frenk}, C.~S., \&
  {Wilkins}, S.~M. 2013, \mnras, 429, 1609, \dodoi{10.1093/mnras/sts446}

\bibitem[{{Gordon} {et~al.}(2003){Gordon}, {Clayton}, {Misselt}, {Landolt}, \&
  {Wolff}}]{Gordon2003}
{Gordon}, K.~D., {Clayton}, G.~C., {Misselt}, K.~A., {Landolt}, A.~U., \&
  {Wolff}, M.~J. 2003, \apj, 594, 279, \dodoi{10.1086/376774}

\bibitem[{{Grogin} {et~al.}(2011){Grogin}, {Kocevski}, {Faber}, {Ferguson},
  {Koekemoer}, {Riess}, {Acquaviva}, {Alexander}, {Almaini}, {Ashby}, {Barden},
  {Bell}, {Bournaud}, {Brown}, {Caputi}, {Casertano}, {Cassata}, {Castellano},
  {Challis}, {Chary}, {Cheung}, {Cirasuolo}, {Conselice}, {Roshan Cooray},
  {Croton}, {Daddi}, {Dahlen}, {Dav{\'e}}, {de Mello}, {Dekel}, {Dickinson},
  {Dolch}, {Donley}, {Dunlop}, {Dutton}, {Elbaz}, {Fazio}, {Filippenko},
  {Finkelstein}, {Fontana}, {Gardner}, {Garnavich}, {Gawiser}, {Giavalisco},
  {Grazian}, {Guo}, {Hathi}, {H{\"a}ussler}, {Hopkins}, {Huang}, {Huang},
  {Jha}, {Kartaltepe}, {Kirshner}, {Koo}, {Lai}, {Lee}, {Li}, {Lotz}, {Lucas},
  {Madau}, {McCarthy}, {McGrath}, {McIntosh}, {McLure}, {Mobasher},
  {Moustakas}, {Mozena}, {Nandra}, {Newman}, {Niemi}, {Noeske}, {Papovich},
  {Pentericci}, {Pope}, {Primack}, {Rajan}, {Ravindranath}, {Reddy}, {Renzini},
  {Rix}, {Robaina}, {Rodney}, {Rosario}, {Rosati}, {Salimbeni}, {Scarlata},
  {Siana}, {Simard}, {Smidt}, {Somerville}, {Spinrad}, {Straughn}, {Strolger},
  {Telford}, {Teplitz}, {Trump}, {van der Wel}, {Villforth}, {Wechsler},
  {Weiner}, {Wiklind}, {Wild}, {Wilson}, {Wuyts}, {Yan}, \&
  {Yun}}]{Grogin2011-CANDELS}
{Grogin}, N.~A., {Kocevski}, D.~D., {Faber}, S.~M., {et~al.} 2011, \apjs, 197,
  35, \dodoi{10.1088/0067-0049/197/2/35}

\bibitem[{{Groth} {et~al.}(1994){Groth}, {Kristian}, {Lynds}, {O'Neil},
  {Balsano}, {Rhodes}, \& {WFPC-1 IDT}}]{Groth1994}
{Groth}, E.~J., {Kristian}, J.~A., {Lynds}, R., {et~al.} 1994, in American
  Astronomical Society Meeting Abstracts, Vol. 185, American Astronomical
  Society Meeting Abstracts, 53.09

\bibitem[{{Gunn} \& {Peterson}(1965)}]{Gunn-Peterson1965}
{Gunn}, J.~E., \& {Peterson}, B.~A. 1965, \apj, 142, 1633,
  \dodoi{10.1086/148444}

\bibitem[{{Gutkin} {et~al.}(2016){Gutkin}, {Charlot}, \&
  {Bruzual}}]{Gutkin2016}
{Gutkin}, J., {Charlot}, S., \& {Bruzual}, G. 2016, \mnras, 462, 1757,
  \dodoi{10.1093/mnras/stw1716}

\bibitem[{{Haiman}(2002)}]{Haiman2002}
{Haiman}, Z. 2002, \apjl, 576, L1, \dodoi{10.1086/343101}

\bibitem[{{Hainline} {et~al.}(2023{\natexlab{a}}){Hainline}, {Robertson},
  {Tacchella}, {Rieke}, {Eisenstein}, {Helton}, {Whitler}, {Topping}, {Sun},
  {Hviding}, \& {Jades Collaboration}}]{Hainline2023b}
{Hainline}, K., {Robertson}, B., {Tacchella}, S., {et~al.} 2023{\natexlab{a}},
  in American Astronomical Society Meeting Abstracts, Vol.~55, American
  Astronomical Society Meeting Abstracts, 212.02

\bibitem[{{Hainline} {et~al.}(2023{\natexlab{b}}){Hainline}, {Johnson},
  {Robertson}, {Tacchella}, {Helton}, {Sun}, {Eisenstein}, {Simmonds},
  {Topping}, {Whitler}, {Willmer}, {Rieke}, {Suess}, {Hviding}, {Cameron},
  {Alberts}, {Baker}, {Bhatawdekar}, {Boyett}, {Bunker}, {Carniani}, {Charlot},
  {Chen}, {Curti}, {Curtis-Lake}, {D'Eugenio}, {Egami}, {Endsley}, {Hausen},
  {Ji}, {Looser}, {Lyu}, {Maiolino}, {Nelson}, {Puskas}, {Rawle}, {Sandles},
  {Saxena}, {Smit}, {Stark}, {Williams}, {Willott}, \&
  {Witstok}}]{Hainline2023a}
{Hainline}, K.~N., {Johnson}, B.~D., {Robertson}, B., {et~al.}
  2023{\natexlab{b}}, arXiv e-prints, arXiv:2306.02468,
  \dodoi{10.48550/arXiv.2306.02468}

\bibitem[{{Harikane} {et~al.}(2022){Harikane}, {Ouchi}, {Oguri}, {Ono},
  {Nakajima}, {Isobe}, {Umeda}, {Mawatari}, \& {Zhang}}]{Harikane2022}
{Harikane}, Y., {Ouchi}, M., {Oguri}, M., {et~al.} 2022, arXiv e-prints,
  arXiv:2208.01612.
\newblock \doarXiv{2208.01612}

\bibitem[{{Harvey} {et~al.}(2024){Harvey}, {Conselice}, {Adams}, {Austin},
  {Juodzbalis}, {Trussler}, {Li}, {Ormerod}, {Ferreira}, {Duan}, {Westcott},
  {Harris}, {Bhatawdekar}, {Coe}, {Cohen}, {Caruana}, {Cheng}, {Driver},
  {Frye}, {Furtak}, {Grogin}, {Hathi}, {Holwerda}, {Jansen}, {Koekemoer},
  {Lovell}, {Marshall}, {Nonino}, {Smail}, {Vijayan}, {Wilkins}, {Windhorst},
  {Willmer}, {Yan}, \& {Zitrin}}]{harvey2024epochs}
{Harvey}, T., {Conselice}, C., {Adams}, N.~J., {et~al.} 2024, arXiv e-prints,
  arXiv:2403.03908, \dodoi{10.48550/arXiv.2403.03908}

\bibitem[{{Hathi} {et~al.}(2008){Hathi}, {Malhotra}, \& {Rhoads}}]{Hathi2008}
{Hathi}, N.~P., {Malhotra}, S., \& {Rhoads}, J.~E. 2008, \apj, 673, 686,
  \dodoi{10.1086/524836}

\bibitem[{{Hathi} {et~al.}(2013){Hathi}, {Cohen}, {Ryan}, {Finkelstein},
  {McCarthy}, {Windhorst}, {Yan}, {Koekemoer}, {Rutkowski}, {O'Connell},
  {Straughn}, {Balick}, {Bond}, {Calzetti}, {Disney}, {Dopita}, {Frogel},
  {Hall}, {Holtzman}, {Kimble}, {Paresce}, {Saha}, {Silk}, {Trauger}, {Walker},
  {Whitmore}, \& {Young}}]{Hathi2013}
{Hathi}, N.~P., {Cohen}, S.~H., {Ryan}, R.~E., J., {et~al.} 2013, \apj, 765,
  88, \dodoi{10.1088/0004-637X/765/2/88}

\bibitem[{{Heintz} {et~al.}(2023{\natexlab{a}}){Heintz}, {Watson}, {Brammer},
  {Vejlgaard}, {Hutter}, {Strait}, {Matthee}, {Oesch}, {Jakobsson}, {Tanvir},
  {Laursen}, {Naidu}, {Mason}, {Killi}, {Jung}, {Hsiao}, {Abdurro'uf}, {Coe},
  {Arrabal Haro}, {Finkelstein}, \& {Toft}}]{Heintz2023}
{Heintz}, K.~E., {Watson}, D., {Brammer}, G., {et~al.} 2023{\natexlab{a}},
  arXiv e-prints, arXiv:2306.00647, \dodoi{10.48550/arXiv.2306.00647}

\bibitem[{{Heintz} {et~al.}(2023{\natexlab{b}}){Heintz}, {Brammer},
  {Gim{\'e}nez-Arteaga}, {Strait}, {del P. Lagos}, {Vijayan}, {Matthee},
  {Watson}, {Mason}, {Hutter}, {Toft}, {Fynbo}, \& {Oesch}}]{Heintz2023b}
{Heintz}, K.~E., {Brammer}, G.~B., {Gim{\'e}nez-Arteaga}, C., {et~al.}
  2023{\natexlab{b}}, Nature Astronomy, 7, 1517,
  \dodoi{10.1038/s41550-023-02078-7}

\bibitem[{{Heintz} {et~al.}(2024){Heintz}, {Brammer}, {Watson}, {Oesch},
  {Keating}, {Hayes}, {Abdurro'uf}, {Arellano-C{\'o}rdova}, {Carnall},
  {Christiansen}, {Cullen}, {Dav{\'e}}, {Dayal}, {Ferrara}, {Finlator},
  {Fynbo}, {Flury}, {Gelli}, {Gillman}, {Gottumukkala}, {Gould}, {Greve},
  {Hardin}, {Y. -Y Hsiao}, {Hutter}, {Jakobsson}, {Killi}, {Khosravaninezhad},
  {Laursen}, {Lee}, {Magdis}, {Matthee}, {Naidu}, {Narayanan}, {Pollock},
  {Prescott}, {Rusakov}, {Shuntov}, {Sneppen}, {Smit}, {Tanvir}, {Terp},
  {Toft}, {Valentino}, {Vijayan}, {Weaver}, {Wise}, \& {Witstok}}]{Heintz2024}
{Heintz}, K.~E., {Brammer}, G.~B., {Watson}, D., {et~al.} 2024, arXiv e-prints,
  arXiv:2404.02211, \dodoi{10.48550/arXiv.2404.02211}

\bibitem[{{Hensley} \& {Draine}(2023)}]{Hensley2023}
{Hensley}, B.~S., \& {Draine}, B.~T. 2023, \apj, 948, 55,
  \dodoi{10.3847/1538-4357/acc4c2}

\bibitem[{{Hodges}(1958)}]{scipy.stats.ks2_samp}
{Hodges}, J.~L. 1958, Arkiv for Matematik, 3, 469, \dodoi{10.1007/BF02589501}

\bibitem[{{Hoffmann} {et~al.}(2021){Hoffmann}, {Mack}, {Avila}, {Martlin},
  {Cohen}, \& {Bajaj}}]{Hoffmann2021}
{Hoffmann}, S.~L., {Mack}, J., {Avila}, R., {et~al.} 2021, in American
  Astronomical Society Meeting Abstracts, Vol.~53, American Astronomical
  Society Meeting Abstracts, 216.02

\bibitem[{{H{\"o}fner} \& {Olofsson}(2018)}]{Hofner2018}
{H{\"o}fner}, S., \& {Olofsson}, H. 2018, \aapr, 26, 1,
  \dodoi{10.1007/s00159-017-0106-5}

\bibitem[{{Hsiao} {et~al.}(2023){Hsiao}, {Abdurro'uf}, {Coe}, {Larson}, {Jung},
  {Mingozzi}, {Dayal}, {Kumari}, {Kokorev}, {Vikaeus}, {Brammer}, {Furtak},
  {Adamo}, {Andrade-Santos}, {Antwi-Danso}, {Bradac}, {Bradley}, {Broadhurst},
  {Carnall}, {Conselice}, {Diego}, {Donahue}, {Eldridge}, {Fujimoto}, {Henry},
  {Hernandez}, {Hutchison}, {James}, {Norman}, {Park}, {Pirzkal}, {Postman},
  {Ricotti}, {Rigby}, {Vanzella}, {Welch}, {Wilkins}, {Windhorst}, {Xu},
  {Zackrisson}, \& {Zitrin}}]{Hsiao2023}
{Hsiao}, T. Y.-Y., {Abdurro'uf}, {Coe}, D., {et~al.} 2023, arXiv e-prints,
  arXiv:2305.03042, \dodoi{10.48550/arXiv.2305.03042}

\bibitem[{Hunter(2007)}]{Hunter2007}
Hunter, J.~D. 2007, Computing in Science \& Engineering, 9, 90,
  \dodoi{10.1109/MCSE.2007.55}

\bibitem[{{Inayoshi} {et~al.}(2022){Inayoshi}, {Harikane}, {Inoue}, {Li}, \&
  {Ho}}]{Inayoshi2022}
{Inayoshi}, K., {Harikane}, Y., {Inoue}, A.~K., {Li}, W., \& {Ho}, L.~C. 2022,
  \apjl, 938, L10, \dodoi{10.3847/2041-8213/ac9310}

\bibitem[{{Inoue} {et~al.}(2014){Inoue}, {Shimizu}, {Iwata}, \&
  {Tanaka}}]{Inoue2014}
{Inoue}, A.~K., {Shimizu}, I., {Iwata}, I., \& {Tanaka}, M. 2014, \mnras, 442,
  1805, \dodoi{10.1093/mnras/stu936}

\bibitem[{{Jaacks} {et~al.}(2018){Jaacks}, {Finkelstein}, \&
  {Bromm}}]{Jaacks2018}
{Jaacks}, J., {Finkelstein}, S.~L., \& {Bromm}, V. 2018, \mnras, 475, 3883,
  \dodoi{10.1093/mnras/sty049}

\bibitem[{{Jakobsen} {et~al.}(2022){Jakobsen}, {Ferruit}, {Alves de Oliveira},
  {Arribas}, {Bagnasco}, {Barho}, {Beck}, {Birkmann}, {B{\"o}ker}, {Bunker},
  {Charlot}, {de Jong}, {de Marchi}, {Ehrenwinkler}, {Falcolini}, {Fels},
  {Franx}, {Franz}, {Funke}, {Giardino}, {Gnata}, {Holota}, {Honnen}, {Jensen},
  {Jentsch}, {Johnson}, {Jollet}, {Karl}, {Kling}, {K{\"o}hler}, {Kolm},
  {Kumari}, {Lander}, {Lemke}, {L{\'o}pez-Caniego}, {L{\"u}tzgendorf},
  {Maiolino}, {Manjavacas}, {Marston}, {Maschmann}, {Maurer}, {Messerschmidt},
  {Moseley}, {Mosner}, {Mott}, {Muzerolle}, {Pirzkal}, {Pittet}, {Plitzke},
  {Posselt}, {Rapp}, {Rauscher}, {Rawle}, {Rix}, {R{\"o}del}, {Rumler},
  {Sabbi}, {Salvignol}, {Schmid}, {Sirianni}, {Smith}, {Strada}, {te Plate},
  {Valenti}, {Wettemann}, {Wiehe}, {Wiesmayer}, {Willott}, {Wright}, {Zeidler},
  \& {Zincke}}]{Jakobsen2022}
{Jakobsen}, P., {Ferruit}, P., {Alves de Oliveira}, C., {et~al.} 2022, \aap,
  661, A80, \dodoi{10.1051/0004-6361/202142663}

\bibitem[{{Jansen} \& {Windhorst}(2018)}]{Jansen2018}
{Jansen}, R.~A., \& {Windhorst}, R.~A. 2018, \pasp, 130, 124001,
  \dodoi{10.1088/1538-3873/aae476}

\bibitem[{{Jaskot} \& {Ravindranath}(2016)}]{Jaskot2016}
{Jaskot}, A.~E., \& {Ravindranath}, S. 2016, \apj, 833, 136,
  \dodoi{10.3847/1538-4357/833/2/136}

\bibitem[{{Johnson} {et~al.}(2021){Johnson}, {Leja}, {Conroy}, \&
  {Speagle}}]{Johnson2021}
{Johnson}, B.~D., {Leja}, J., {Conroy}, C., \& {Speagle}, J.~S. 2021, \apjs,
  254, 22, \dodoi{10.3847/1538-4365/abef67}

\bibitem[{{Kannan} {et~al.}(2022){Kannan}, {Smith}, {Garaldi}, {Shen},
  {Vogelsberger}, {Pakmor}, {Springel}, \& {Hernquist}}]{Kannan2022-THESAN}
{Kannan}, R., {Smith}, A., {Garaldi}, E., {et~al.} 2022, \mnras, 514, 3857,
  \dodoi{10.1093/mnras/stac1557}

\bibitem[{{Katz} {et~al.}(2022){Katz}, {Garel}, {Rosdahl}, {Mauerhofer},
  {Kimm}, {Blaizot}, {Michel-Dansac}, {Devriendt}, {Slyz}, \&
  {Haehnelt}}]{Katz2022}
{Katz}, H., {Garel}, T., {Rosdahl}, J., {et~al.} 2022, \mnras, 515, 4265,
  \dodoi{10.1093/mnras/stac1437}

\bibitem[{{Keating} {et~al.}(2023){Keating}, {Bolton}, {Cullen}, {Haehnelt},
  {Puchwein}, \& {Kulkarni}}]{Keating2023a}
{Keating}, L.~C., {Bolton}, J.~S., {Cullen}, F., {et~al.} 2023, arXiv e-prints,
  arXiv:2308.05800, \dodoi{10.48550/arXiv.2308.05800}

\bibitem[{{Kirchschlager} {et~al.}(2024){Kirchschlager}, {Mattsson}, \&
  {Gent}}]{Kirchschlager2024}
{Kirchschlager}, F., {Mattsson}, L., \& {Gent}, F.~A. 2024, arXiv e-prints,
  arXiv:2402.06543, \dodoi{10.48550/arXiv.2402.06543}

\bibitem[{{Kirchschlager} {et~al.}(2019){Kirchschlager}, {Schmidt}, {Barlow},
  {Fogerty}, {Bevan}, \& {Priestley}}]{Kirchschlager2019}
{Kirchschlager}, F., {Schmidt}, F.~D., {Barlow}, M.~J., {et~al.} 2019, \mnras,
  489, 4465, \dodoi{10.1093/mnras/stz2399}

\bibitem[{{Kocevski} {et~al.}(2024){Kocevski}, {Finkelstein}, {Barro},
  {Taylor}, {Calabr{\`o}}, {Laloux}, {Buchner}, {Trump}, {Leung}, {Yang},
  {Dickinson}, {P{\'e}rez-Gonz{\'a}lez}, {Pacucci}, {Inayoshi}, {Somerville},
  {McGrath}, {Akins}, {Arrabal Haro}, {Bagley}, {Bowler}, {Carnall}, {Casey},
  {Cheng}, {Cleri}, {Costantin}, {Cullen}, {Davis}, {Donnan}, {Dunlop},
  {Ellis}, {Ferguson}, {Fujimoto}, {Fontana}, {Giavalisco}, {Grazian},
  {Grogin}, {Hathi}, {Hirschmann}, {Huertas-Company}, {Holwerda},
  {Illingworth}, {Juneau}, {Kartaltepe}, {Koekemoer}, {Li}, {Lucas}, {Magee},
  {Mason}, {McLeod}, {McLure}, {Napolitano}, {Papovich}, {Pirzkal},
  {Rodighiero}, {Santini}, {Wilkins}, \& {Yung}}]{Kocevski2024}
{Kocevski}, D.~D., {Finkelstein}, S.~L., {Barro}, G., {et~al.} 2024, arXiv
  e-prints, arXiv:2404.03576, \dodoi{10.48550/arXiv.2404.03576}

\bibitem[{{Koekemoer} {et~al.}(2011){Koekemoer}, {Faber}, {Ferguson}, {Grogin},
  {Kocevski}, {Koo}, {Lai}, {Lotz}, {Lucas}, {McGrath}, {Ogaz}, {Rajan},
  {Riess}, {Rodney}, {Strolger}, {Casertano}, {Castellano}, {Dahlen},
  {Dickinson}, {Dolch}, {Fontana}, {Giavalisco}, {Grazian}, {Guo}, {Hathi},
  {Huang}, {van der Wel}, {Yan}, {Acquaviva}, {Alexander}, {Almaini}, {Ashby},
  {Barden}, {Bell}, {Bournaud}, {Brown}, {Caputi}, {Cassata}, {Challis},
  {Chary}, {Cheung}, {Cirasuolo}, {Conselice}, {Roshan Cooray}, {Croton},
  {Daddi}, {Dav{\'e}}, {de Mello}, {de Ravel}, {Dekel}, {Donley}, {Dunlop},
  {Dutton}, {Elbaz}, {Fazio}, {Filippenko}, {Finkelstein}, {Frazer}, {Gardner},
  {Garnavich}, {Gawiser}, {Gruetzbauch}, {Hartley}, {H{\"a}ussler},
  {Herrington}, {Hopkins}, {Huang}, {Jha}, {Johnson}, {Kartaltepe},
  {Khostovan}, {Kirshner}, {Lani}, {Lee}, {Li}, {Madau}, {McCarthy},
  {McIntosh}, {McLure}, {McPartland}, {Mobasher}, {Moreira}, {Mortlock},
  {Moustakas}, {Mozena}, {Nandra}, {Newman}, {Nielsen}, {Niemi}, {Noeske},
  {Papovich}, {Pentericci}, {Pope}, {Primack}, {Ravindranath}, {Reddy},
  {Renzini}, {Rix}, {Robaina}, {Rosario}, {Rosati}, {Salimbeni}, {Scarlata},
  {Siana}, {Simard}, {Smidt}, {Snyder}, {Somerville}, {Spinrad}, {Straughn},
  {Telford}, {Teplitz}, {Trump}, {Vargas}, {Villforth}, {Wagner}, {Wandro},
  {Wechsler}, {Weiner}, {Wiklind}, {Wild}, {Wilson}, {Wuyts}, \&
  {Yun}}]{Koekemoer2011-CANDELS}
{Koekemoer}, A.~M., {Faber}, S.~M., {Ferguson}, H.~C., {et~al.} 2011, \apjs,
  197, 36, \dodoi{10.1088/0067-0049/197/2/36}

\bibitem[{{Koekemoer} {et~al.}(2013){Koekemoer}, {Ellis}, {McLure}, {Dunlop},
  {Robertson}, {Ono}, {Schenker}, {Ouchi}, {Bowler}, {Rogers}, {Curtis-Lake},
  {Schneider}, {Charlot}, {Stark}, {Furlanetto}, {Cirasuolo}, {Wild}, \&
  {Targett}}]{Koekemoer2013-HUDF12}
{Koekemoer}, A.~M., {Ellis}, R.~S., {McLure}, R.~J., {et~al.} 2013, \apjs, 209,
  3, \dodoi{10.1088/0067-0049/209/1/3}

\bibitem[{{Kokorev} {et~al.}(2024){Kokorev}, {Caputi}, {Greene}, {Dayal},
  {Trebitsch}, {Cutler}, {Fujimoto}, {Labb{\'e}}, {Miller}, {Iani},
  {Navarro-Carrera}, \& {Rinaldi}}]{Kokorev2024}
{Kokorev}, V., {Caputi}, K.~I., {Greene}, J.~E., {et~al.} 2024, arXiv e-prints,
  arXiv:2401.09981, \dodoi{10.48550/arXiv.2401.09981}

\bibitem[{{Kron}(1980)}]{Kron1980}
{Kron}, R.~G. 1980, \apjs, 43, 305, \dodoi{10.1086/190669}

\bibitem[{{Kroupa}(2001)}]{Kroupa2001}
{Kroupa}, P. 2001, \mnras, 322, 231, \dodoi{10.1046/j.1365-8711.2001.04022.x}

\bibitem[{{Labbe} {et~al.}(2023){Labbe}, {Greene}, {Bezanson}, {Fujimoto},
  {Furtak}, {Goulding}, {Matthee}, {Naidu}, {Oesch}, {Atek}, {Brammer},
  {Chemerynska}, {Coe}, {Cutler}, {Dayal}, {Feldmann}, {Franx}, {Glazebrook},
  {Leja}, {Marchesini}, {Maseda}, {Nanayakkara}, {Nelson}, {Pan}, {Papovich},
  {Price}, {Suess}, {Wang}, {Whitaker}, {Williams}, \& {Zitrin}}]{Labbe2023}
{Labbe}, I., {Greene}, J.~E., {Bezanson}, R., {et~al.} 2023, arXiv e-prints,
  arXiv:2306.07320, \dodoi{10.48550/arXiv.2306.07320}

\bibitem[{{Lanzetta}(2000)}]{Lanzetta2000}
{Lanzetta}, K. 2000, in Encyclopedia of Astronomy and Astrophysics, ed.
  P.~{Murdin}, 2141, \dodoi{10.1888/0333750888/2141}

\bibitem[{{Larson} {et~al.}(2023){Larson}, {Hutchison}, {Bagley},
  {Finkelstein}, {Yung}, {Somerville}, {Hirschmann}, {Brammer}, {Holwerda},
  {Papovich}, {Morales}, \& {Wilkins}}]{Larson2023}
{Larson}, R.~L., {Hutchison}, T.~A., {Bagley}, M., {et~al.} 2023, \apj, 958,
  141, \dodoi{10.3847/1538-4357/acfed4}

\bibitem[{{Lau} {et~al.}(2020){Lau}, {Eldridge}, {Hankins}, {Lamberts},
  {Sakon}, \& {Williams}}]{Lau2020}
{Lau}, R.~M., {Eldridge}, J.~J., {Hankins}, M.~J., {et~al.} 2020, \apj, 898,
  74, \dodoi{10.3847/1538-4357/ab9cb5}

\bibitem[{{Lau} {et~al.}(2022){Lau}, {Hankins}, {Han}, {Argyriou}, {Corcoran},
  {Eldridge}, {Endo}, {Fox}, {Garcia Marin}, {Gull}, {Jones}, {Hamaguchi},
  {Lamberts}, {Law}, {Madura}, {Marchenko}, {Matsuhara}, {Moffat}, {Morris},
  {Morris}, {Onaka}, {Ressler}, {Richardson}, {Russell}, {Sanchez-Bermudez},
  {Smith}, {Soulain}, {Stevens}, {Tuthill}, {Weigelt}, {Williams}, \&
  {Yamaguchi}}]{Lau2022}
{Lau}, R.~M., {Hankins}, M.~J., {Han}, Y., {et~al.} 2022, Nature Astronomy, 6,
  1308, \dodoi{10.1038/s41550-022-01812-x}

\bibitem[{{Leitherer} {et~al.}(1999){Leitherer}, {Schaerer}, {Goldader},
  {Delgado}, {Robert}, {Kune}, {de Mello}, {Devost}, \&
  {Heckman}}]{Leitherer1999}
{Leitherer}, C., {Schaerer}, D., {Goldader}, J.~D., {et~al.} 1999, \apjs, 123,
  3, \dodoi{10.1086/313233}

\bibitem[{{Leja} {et~al.}(2019){Leja}, {Carnall}, {Johnson}, {Conroy}, \&
  {Speagle}}]{Leja2019}
{Leja}, J., {Carnall}, A.~C., {Johnson}, B.~D., {Conroy}, C., \& {Speagle},
  J.~S. 2019, \apj, 876, 3, \dodoi{10.3847/1538-4357/ab133c}

\bibitem[{{Leung} {et~al.}(2023){Leung}, {Bagley}, {Finkelstein}, {Ferguson},
  {Koekemoer}, {P{\'e}rez-Gonz{\'a}lez}, {Morales}, {Kocevski}, {Yang},
  {Somerville}, {Wilkins}, {Yung}, {Fujimoto}, {Larson}, {Papovich}, {Pirzkal},
  {Berg}, {Lotz}, {Castellano}, {Ch{\'a}vez Ortiz}, {Cheng}, {Dickinson},
  {Giavalisco}, {Hathi}, {Hutchison}, {Jung}, {Kartaltepe}, {Natarajan}, \&
  {Rothberg}}]{Leung2023}
{Leung}, G. C.~K., {Bagley}, M.~B., {Finkelstein}, S.~L., {et~al.} 2023, \apjl,
  954, L46, \dodoi{10.3847/2041-8213/acf365}

\bibitem[{{Levesque} {et~al.}(2006){Levesque}, {Massey}, {Olsen}, {Plez},
  {Meynet}, \& {Maeder}}]{Levesque2006}
{Levesque}, E.~M., {Massey}, P., {Olsen}, K.~A.~G., {et~al.} 2006, \apj, 645,
  1102, \dodoi{10.1086/504417}

\bibitem[{{Liske} {et~al.}(2015){Liske}, {Baldry}, {Driver}, {Tuffs},
  {Alpaslan}, {Andrae}, {Brough}, {Cluver}, {Grootes}, {Gunawardhana},
  {Kelvin}, {Loveday}, {Robotham}, {Taylor}, {Bamford}, {Bland-Hawthorn},
  {Brown}, {Drinkwater}, {Hopkins}, {Meyer}, {Norberg}, {Peacock}, {Agius},
  {Andrews}, {Bauer}, {Ching}, {Colless}, {Conselice}, {Croom}, {Davies}, {De
  Propris}, {Dunne}, {Eardley}, {Ellis}, {Foster}, {Frenk}, {H{\"a}u{\ss}ler},
  {Holwerda}, {Howlett}, {Ibarra}, {Jarvis}, {Jones}, {Kafle}, {Lacey},
  {Lange}, {Lara-L{\'o}pez}, {L{\'o}pez-S{\'a}nchez}, {Maddox}, {Madore},
  {McNaught-Roberts}, {Moffett}, {Nichol}, {Owers}, {Palamara}, {Penny},
  {Phillipps}, {Pimbblet}, {Popescu}, {Prescott}, {Proctor}, {Sadler},
  {Sansom}, {Seibert}, {Sharp}, {Sutherland}, {V{\'a}zquez-Mata}, {van Kampen},
  {Wilkins}, {Williams}, \& {Wright}}]{Liske2015}
{Liske}, J., {Baldry}, I.~K., {Driver}, S.~P., {et~al.} 2015, \mnras, 452,
  2087, \dodoi{10.1093/mnras/stv1436}

\bibitem[{{Looser} {et~al.}(2023){Looser}, {D'Eugenio}, {Maiolino}, {Witstok},
  {Sandles}, {Curtis-Lake}, {Chevallard}, {Tacchella}, {Johnson}, {Baker},
  {Suess}, {Carniani}, {Ferruit}, {Arribas}, {Bonaventura}, {Bunker},
  {Cameron}, {Charlot}, {Curti}, {de Graaff}, {Maseda}, {Rawle}, {Rix},
  {Rodriguez Del Pino}, {Smit}, {{\"U}bler}, {Willott}, {Alberts}, {Egami},
  {Eisenstein}, {Endsley}, {Hausen}, {Rieke}, {Robertson}, {Shivaei},
  {Williams}, {Boyett}, {Chen}, {Ji}, {Jones}, {Kumari}, {Nelson}, {Perna},
  {Saxena}, \& {Scholtz}}]{Looser2023}
{Looser}, T.~J., {D'Eugenio}, F., {Maiolino}, R., {et~al.} 2023, arXiv
  e-prints, arXiv:2302.14155, \dodoi{10.48550/arXiv.2302.14155}

\bibitem[{{Lotz} {et~al.}(2017){Lotz}, {Koekemoer}, {Coe}, {Grogin}, {Capak},
  {Mack}, {Anderson}, {Avila}, {Barker}, {Borncamp}, {Brammer}, {Durbin},
  {Gunning}, {Hilbert}, {Jenkner}, {Khandrika}, {Levay}, {Lucas}, {MacKenty},
  {Ogaz}, {Porterfield}, {Reid}, {Robberto}, {Royle}, {Smith},
  {Storrie-Lombardi}, {Sunnquist}, {Surace}, {Taylor}, {Williams}, {Bullock},
  {Dickinson}, {Finkelstein}, {Natarajan}, {Richard}, {Robertson}, {Tumlinson},
  {Zitrin}, {Flanagan}, {Sembach}, {Soifer}, \& {Mountain}}]{Lotz2017}
{Lotz}, J.~M., {Koekemoer}, A., {Coe}, D., {et~al.} 2017, \apj, 837, 97,
  \dodoi{10.3847/1538-4357/837/1/97}

\bibitem[{{Lovell} {et~al.}(2021){Lovell}, {Vijayan}, {Thomas}, {Wilkins},
  {Barnes}, {Irodotou}, \& {Roper}}]{Lovell2021_FLARES_I}
{Lovell}, C.~C., {Vijayan}, A.~P., {Thomas}, P.~A., {et~al.} 2021, \mnras, 500,
  2127, \dodoi{10.1093/mnras/staa3360}

\bibitem[{{Lu} {et~al.}(2024){Lu}, {Mason}, {Hutter}, {Mesinger}, {Qin},
  {Stark}, \& {Endsley}}]{Lu2024}
{Lu}, T.-Y., {Mason}, C.~A., {Hutter}, A., {et~al.} 2024, \mnras, 528, 4872,
  \dodoi{10.1093/mnras/stae266}

\bibitem[{{Madau} \& {Dickinson}(2014)}]{Madau2014}
{Madau}, P., \& {Dickinson}, M. 2014, \araa, 52, 415,
  \dodoi{10.1146/annurev-astro-081811-125615}

\bibitem[{{Mainali} {et~al.}(2017){Mainali}, {Kollmeier}, {Stark}, {Simcoe},
  {Walth}, {Newman}, \& {Miller}}]{Mainali2017}
{Mainali}, R., {Kollmeier}, J.~A., {Stark}, D.~P., {et~al.} 2017, \apjl, 836,
  L14, \dodoi{10.3847/2041-8213/836/1/L14}

\bibitem[{{Malhotra} \& {Rhoads}(2006)}]{Malhotra2006}
{Malhotra}, S., \& {Rhoads}, J.~E. 2006, \apjl, 647, L95,
  \dodoi{10.1086/506983}

\bibitem[{{Marassi} {et~al.}(2019){Marassi}, {Schneider}, {Limongi}, {Chieffi},
  {Graziani}, \& {Bianchi}}]{Marassi2019}
{Marassi}, S., {Schneider}, R., {Limongi}, M., {et~al.} 2019, \mnras, 484,
  2587, \dodoi{10.1093/mnras/sty3323}

\bibitem[{Marley {et~al.}(2021)Marley, Saumon, Morley, Fortney, Visscher,
  Freedman, \& Lupu}]{Marley2021-SonoraBobcat}
Marley, M., Saumon, D., Morley, C., {et~al.} 2021, Zenodo,
  \dodoi{10.5281/zenodo.5063476}

\bibitem[{{Mascia} {et~al.}(2023){Mascia}, {Pentericci}, {Calabr{\`o}}, {Treu},
  {Santini}, {Yang}, {Napolitano}, {Roberts-Borsani}, {Bergamini}, {Grillo},
  {Rosati}, {Vulcani}, {Castellano}, {Boyett}, {Fontana}, {Glazebrook},
  {Henry}, {Mason}, {Merlin}, {Morishita}, {Nanayakkara}, {Paris}, {Roy},
  {Williams}, {Wang}, {Brammer}, {Brada{\v{c}}}, {Chen}, {Kelly}, {Koekemoer},
  {Trenti}, \& {Windhorst}}]{Mascia2023-GLASS}
{Mascia}, S., {Pentericci}, L., {Calabr{\`o}}, A., {et~al.} 2023, \aap, 672,
  A155, \dodoi{10.1051/0004-6361/202345866}

\bibitem[{{Mason} {et~al.}(2023){Mason}, {Trenti}, \& {Treu}}]{Mason2023}
{Mason}, C.~A., {Trenti}, M., \& {Treu}, T. 2023, \mnras, 521, 497,
  \dodoi{10.1093/mnras/stad035}

\bibitem[{{Matthee} {et~al.}(2023){Matthee}, {Naidu}, {Brammer}, {Chisholm},
  {Eilers}, {Goulding}, {Greene}, {Kashino}, {Labbe}, {Lilly}, {Mackenzie},
  {Oesch}, {Weibel}, {Wuyts}, {Xiao}, {Bordoloi}, {Bouwens}, {van Dokkum},
  {Illingworth}, {Kramarenko}, {Maseda}, {Mason}, {Meyer}, {Nelson}, {Reddy},
  {Shivaei}, {Simcoe}, \& {Yue}}]{Matthee2023}
{Matthee}, J., {Naidu}, R.~P., {Brammer}, G., {et~al.} 2023, arXiv e-prints,
  arXiv:2306.05448, \dodoi{10.48550/arXiv.2306.05448}

\bibitem[{{Mauerhofer} \& {Dayal}(2023)}]{Mauerhofer2023-DELPHI}
{Mauerhofer}, V., \& {Dayal}, P. 2023, \mnras, 526, 2196,
  \dodoi{10.1093/mnras/stad2734}

\bibitem[{{McLeod} {et~al.}(2023){McLeod}, {Donnan}, {McLure}, {Dunlop},
  {Magee}, {Begley}, {Carnall}, {Cullen}, {Ellis}, {Hamadouche}, \&
  {Stanton}}]{McLeod2023}
{McLeod}, D.~J., {Donnan}, C.~T., {McLure}, R.~J., {et~al.} 2023, \mnras,
  \dodoi{10.1093/mnras/stad3471}

\bibitem[{{McLure} {et~al.}(2011){McLure}, {Dunlop}, {de Ravel}, {Cirasuolo},
  {Ellis}, {Schenker}, {Robertson}, {Koekemoer}, {Stark}, \&
  {Bowler}}]{McLure2011}
{McLure}, R.~J., {Dunlop}, J.~S., {de Ravel}, L., {et~al.} 2011, \mnras, 418,
  2074, \dodoi{10.1111/j.1365-2966.2011.19626.x}

\bibitem[{{McQuinn}(2016)}]{McQuinn2016}
{McQuinn}, M. 2016, \araa, 54, 313, \dodoi{10.1146/annurev-astro-082214-122355}

\bibitem[{{Menanteau} {et~al.}(2012){Menanteau}, {Hughes}, {Sif{\'o}n},
  {Hilton}, {Gonz{\'a}lez}, {Infante}, {Barrientos}, {Baker}, {Bond}, {Das},
  {Devlin}, {Dunkley}, {Hajian}, {Hincks}, {Kosowsky}, {Marsden}, {Marriage},
  {Moodley}, {Niemack}, {Nolta}, {Page}, {Reese}, {Sehgal}, {Sievers},
  {Spergel}, {Staggs}, \& {Wollack}}]{Menanteau2012}
{Menanteau}, F., {Hughes}, J.~P., {Sif{\'o}n}, C., {et~al.} 2012, \apj, 748, 7,
  \dodoi{10.1088/0004-637X/748/1/7}

\bibitem[{{Meurer} {et~al.}(1999){Meurer}, {Heckman}, \&
  {Calzetti}}]{Meurer1999}
{Meurer}, G.~R., {Heckman}, T.~M., \& {Calzetti}, D. 1999, \apj, 521, 64,
  \dodoi{10.1086/307523}

\bibitem[{{Micelotta} {et~al.}(2018){Micelotta}, {Matsuura}, \&
  {Sarangi}}]{Micelotta2018}
{Micelotta}, E.~R., {Matsuura}, M., \& {Sarangi}, A. 2018, \ssr, 214, 53,
  \dodoi{10.1007/s11214-018-0484-7}

\bibitem[{{Miralda-Escud{\'e}}(1998)}]{Miralda-Escude1998}
{Miralda-Escud{\'e}}, J. 1998, \apj, 501, 15, \dodoi{10.1086/305799}

\bibitem[{{Mirocha} \& {Furlanetto}(2023)}]{Mirocha2023}
{Mirocha}, J., \& {Furlanetto}, S.~R. 2023, \mnras, 519, 843,
  \dodoi{10.1093/mnras/stac3578}

\bibitem[{{Momcheva} {et~al.}(2016){Momcheva}, {Brammer}, {van Dokkum},
  {Skelton}, {Whitaker}, {Nelson}, {Fumagalli}, {Maseda}, {Leja}, {Franx},
  {Rix}, {Bezanson}, {Da Cunha}, {Dickey}, {F{\"o}rster Schreiber},
  {Illingworth}, {Kriek}, {Labb{\'e}}, {Ulf Lange}, {Lundgren}, {Magee},
  {Marchesini}, {Oesch}, {Pacifici}, {Patel}, {Price}, {Tal}, {Wake}, {van der
  Wel}, \& {Wuyts}}]{Momcheva2016}
{Momcheva}, I.~G., {Brammer}, G.~B., {van Dokkum}, P.~G., {et~al.} 2016, \apjs,
  225, 27, \dodoi{10.3847/0067-0049/225/2/27}

\bibitem[{{Morales} {et~al.}(2023){Morales}, {Finkelstein}, {Leung}, {Bagley},
  {Cleri}, {Dave}, {Dickinson}, {Ferguson}, {Hathi}, {Jones}, {Koekemoer},
  {Papovich}, {Perez-Gonzalez}, {Pirzkal}, {Smith}, {Wilkins}, \&
  {Yung}}]{Morales2023}
{Morales}, A.~M., {Finkelstein}, S.~L., {Leung}, G. C.~K., {et~al.} 2023, arXiv
  e-prints, arXiv:2311.04294, \dodoi{10.48550/arXiv.2311.04294}

\bibitem[{{Naidu} {et~al.}(2022){Naidu}, {Oesch}, {van Dokkum}, {Nelson},
  {Suess}, {Whitaker}, {Allen}, {Bezanson}, {Bouwens}, {Brammer}, {Conroy},
  {Illingworth}, {Labbe}, {Leja}, {Leonova}, {Matthee}, {Price}, {Setton},
  {Strait}, {Stefanon}, {Tacchella}, {Toft}, {Weaver}, \& {Weibel}}]{Naidu2022}
{Naidu}, R.~P., {Oesch}, P.~A., {van Dokkum}, P., {et~al.} 2022, arXiv
  e-prints, arXiv:2207.09434.
\newblock \doarXiv{2207.09434}

\bibitem[{{Nakajima} {et~al.}(2023){Nakajima}, {Ouchi}, {Isobe}, {Harikane},
  {Zhang}, {Ono}, {Umeda}, \& {Oguri}}]{Nakajima2023}
{Nakajima}, K., {Ouchi}, M., {Isobe}, Y., {et~al.} 2023, \apjs, 269, 33,
  \dodoi{10.3847/1538-4365/acd556}

\bibitem[{{Nakane} {et~al.}(2023){Nakane}, {Ouchi}, {Nakajima}, {Harikane},
  {Ono}, {Umeda}, {Isobe}, {Zhang}, \& {Xu}}]{Nakane2023}
{Nakane}, M., {Ouchi}, M., {Nakajima}, K., {et~al.} 2023, arXiv e-prints,
  arXiv:2312.06804, \dodoi{10.48550/arXiv.2312.06804}

\bibitem[{{Nanayakkara} {et~al.}(2019){Nanayakkara}, {Brinchmann}, {Boogaard},
  {Bouwens}, {Cantalupo}, {Feltre}, {Kollatschny}, {Marino}, {Maseda},
  {Matthee}, {Paalvast}, {Richard}, \& {Verhamme}}]{Nanayakkara2019}
{Nanayakkara}, T., {Brinchmann}, J., {Boogaard}, L., {et~al.} 2019, \aap, 624,
  A89, \dodoi{10.1051/0004-6361/201834565}

\bibitem[{{Nanayakkara} {et~al.}(2023){Nanayakkara}, {Glazebrook}, {Jacobs},
  {Bonchi}, {Castellano}, {Fontana}, {Mason}, {Merlin}, {Morishita}, {Paris},
  {Trenti}, {Treu}, {Calabr{\`o}}, {Boyett}, {Bradac}, {Leethochawalit},
  {Marchesini}, {Santini}, {Strait}, {Vanzella}, {Vulcani}, {Wang}, \&
  {Yang}}]{Nanayakkara2023}
{Nanayakkara}, T., {Glazebrook}, K., {Jacobs}, C., {et~al.} 2023, \apjl, 947,
  L26, \dodoi{10.3847/2041-8213/acbfb9}

\bibitem[{{Napolitano} {et~al.}(2024){Napolitano}, {Pentericci}, {Santini},
  {Calabr{\`o}}, {Mascia}, {Llerena}, {Castellano}, {Dickinson}, {Finkelstein},
  {Amorin}, {Arrabal Haro}, {Bagley}, {Bhatawdekar}, {Cleri}, {Davis},
  {Gardner}, {Gawiser}, {Giavalisco}, {Hathi}, {Hu}, {Jung}, {Kartaltepe},
  {Koekemoer}, {Merlin}, {Mobasher}, {Papovich}, {Park}, {Pirzkal}, {Trump},
  {Wilkins}, \& {Yung}}]{Napolitano2024}
{Napolitano}, L., {Pentericci}, L., {Santini}, P., {et~al.} 2024, arXiv
  e-prints, arXiv:2402.11220, \dodoi{10.48550/arXiv.2402.11220}

\bibitem[{{O'Brien} {et~al.}(2024){O'Brien}, {Jansen}, {Grogin}, {Cohen},
  {Smith}, {Silver}, {Maksym}, {Windhorst}, {Koekemoer}, {Hathi}, {Willmer},
  {Frye}, {Alpaslan}, {Ashby}, {Ashcraft}, {Bonoli}, {Brisken}, {Cappelluti},
  {Civano}, {Conselice}, {Dhillon}, {Driver}, {Duncan}, {Dupke}, {Elvis},
  {Fazio}, {Finkelstein}, {Gim}, {Griffiths}, {Hammel}, {Hyun}, {Im}, {Jones},
  {Kim}, {Ladjelate}, {Larson}, {Malhotra}, {Marshall}, {Milam}, {Pierel},
  {Rhoads}, {Rodney}, {R{\"o}ttgering}, {Rutkowski}, {Ryan}, {Ward}, {White},
  {van Weeren}, {Zhao}, {Summers}, {D'Silva}, {Ortiz}, {Robotham}, {Coe},
  {Nonino}, {Pirzkal}, {Yan}, \& {Acharya}}]{OBrien2024}
{O'Brien}, R., {Jansen}, R.~A., {Grogin}, N.~A., {et~al.} 2024, arXiv e-prints,
  arXiv:2401.04944, \dodoi{10.48550/arXiv.2401.04944}

\bibitem[{{Oesch} {et~al.}(2018){Oesch}, {Bouwens}, {Illingworth}, {Labb{\'e}},
  \& {Stefanon}}]{Oesch2018}
{Oesch}, P.~A., {Bouwens}, R.~J., {Illingworth}, G.~D., {Labb{\'e}}, I., \&
  {Stefanon}, M. 2018, \apj, 855, 105, \dodoi{10.3847/1538-4357/aab03f}

\bibitem[{{Oesch} {et~al.}(2013){Oesch}, {Bouwens}, {Illingworth}, {Labb{\'e}},
  {Franx}, {van Dokkum}, {Trenti}, {Stiavelli}, {Gonzalez}, \&
  {Magee}}]{Oesch2013}
{Oesch}, P.~A., {Bouwens}, R.~J., {Illingworth}, G.~D., {et~al.} 2013, \apj,
  773, 75, \dodoi{10.1088/0004-637X/773/1/75}

\bibitem[{{Oesch} {et~al.}(2016){Oesch}, {Brammer}, {van Dokkum},
  {Illingworth}, {Bouwens}, {Labb{\'e}}, {Franx}, {Momcheva}, {Ashby}, {Fazio},
  {Gonzalez}, {Holden}, {Magee}, {Skelton}, {Smit}, {Spitler}, {Trenti}, \&
  {Willner}}]{Oesch2016}
{Oesch}, P.~A., {Brammer}, G., {van Dokkum}, P.~G., {et~al.} 2016, \apj, 819,
  129, \dodoi{10.3847/0004-637X/819/2/129}

\bibitem[{{Oke}(1974)}]{Oke74}
{Oke}, J.~B. 1974, \apjs, 27, 21, \dodoi{10.1086/190287}

\bibitem[{{Oke} \& {Gunn}(1983)}]{Oke83}
{Oke}, J.~B., \& {Gunn}, J.~E. 1983, \apj, 266, 713, \dodoi{10.1086/160817}

\bibitem[{{Pei}(1992)}]{Pei1992}
{Pei}, Y.~C. 1992, \apj, 395, 130, \dodoi{10.1086/171637}

\bibitem[{{P{\'e}rez-Gonz{\'a}lez} {et~al.}(2023){P{\'e}rez-Gonz{\'a}lez},
  {Costantin}, {Langeroodi}, {Rinaldi}, {Annunziatella}, {Ilbert}, {Colina},
  {N{\o}rgaard-Nielsen}, {Greve}, {{\"O}stlin}, {Wright}, {Alonso-Herrero},
  {{\'A}lvarez-M{\'a}rquez}, {Caputi}, {Eckart}, {Le F{\`e}vre}, {Labiano},
  {Garc{\'\i}a-Mar{\'\i}n}, {Hjorth}, {Kendrew}, {Pye}, {Tikkanen}, {van der
  Werf}, {Walter}, {Ward}, {Bik}, {Boogaard}, {Bosman}, {G{\'o}mez}, {Gillman},
  {Iani}, {Jermann}, {Melinder}, {Meyer}, {Moutard}, {van Dishoek}, {Henning},
  {Lagage}, {Guedel}, {Peissker}, {Ray}, {Vandenbussche},
  {Garc{\'\i}a-Argum{\'a}nez}, \& {Mar{\'\i}a M{\'e}rida}}]{PerezGonzalez2023}
{P{\'e}rez-Gonz{\'a}lez}, P.~G., {Costantin}, L., {Langeroodi}, D., {et~al.}
  2023, \apjl, 951, L1, \dodoi{10.3847/2041-8213/acd9d0}

\bibitem[{{Perrin} {et~al.}(2014){Perrin}, {Sivaramakrishnan}, {Lajoie},
  {Elliott}, {Pueyo}, {Ravindranath}, \& {Albert}}]{Perrin2014-WebbPSF}
{Perrin}, M.~D., {Sivaramakrishnan}, A., {Lajoie}, C.-P., {et~al.} 2014, in
  Society of Photo-Optical Instrumentation Engineers (SPIE) Conference Series,
  Vol. 9143, Space Telescopes and Instrumentation 2014: Optical, Infrared, and
  Millimeter Wave, ed. J.~{Oschmann}, Jacobus~M., M.~{Clampin}, G.~G. {Fazio},
  \& H.~A. {MacEwen}, 91433X, \dodoi{10.1117/12.2056689}

\bibitem[{{Perrin} {et~al.}(2012){Perrin}, {Soummer}, {Elliott}, {Lallo}, \&
  {Sivaramakrishnan}}]{Perrin2012-WebbPSF}
{Perrin}, M.~D., {Soummer}, R., {Elliott}, E.~M., {Lallo}, M.~D., \&
  {Sivaramakrishnan}, A. 2012, in Society of Photo-Optical Instrumentation
  Engineers (SPIE) Conference Series, Vol. 8442, Space Telescopes and
  Instrumentation 2012: Optical, Infrared, and Millimeter Wave, ed. M.~C.
  {Clampin}, G.~G. {Fazio}, H.~A. {MacEwen}, \& J.~{Oschmann}, Jacobus~M.,
  84423D, \dodoi{10.1117/12.925230}

\bibitem[{{Planck Collaboration} {et~al.}(2016){Planck Collaboration}, {Ade},
  {Aghanim}, {Arnaud}, {Ashdown}, {Aumont}, {Baccigalupi}, {Banday},
  {Barreiro}, {Bartlett}, {Bartolo}, {Battaner}, {Battye}, {Benabed},
  {Beno{\^\i}t}, {Benoit-L{\'e}vy}, {Bernard}, {Bersanelli}, {Bielewicz},
  {Bock}, {Bonaldi}, {Bonavera}, {Bond}, {Borrill}, {Bouchet}, {Boulanger},
  {Bucher}, {Burigana}, {Butler}, {Calabrese}, {Cardoso}, {Catalano},
  {Challinor}, {Chamballu}, {Chary}, {Chiang}, {Chluba}, {Christensen},
  {Church}, {Clements}, {Colombi}, {Colombo}, {Combet}, {Coulais}, {Crill},
  {Curto}, {Cuttaia}, {Danese}, {Davies}, {Davis}, {de Bernardis}, {de Rosa},
  {de Zotti}, {Delabrouille}, {D{\'e}sert}, {Di Valentino}, {Dickinson},
  {Diego}, {Dolag}, {Dole}, {Donzelli}, {Dor{\'e}}, {Douspis}, {Ducout},
  {Dunkley}, {Dupac}, {Efstathiou}, {Elsner}, {En{\ss}lin}, {Eriksen},
  {Farhang}, {Fergusson}, {Finelli}, {Forni}, {Frailis}, {Fraisse},
  {Franceschi}, {Frejsel}, {Galeotta}, {Galli}, {Ganga}, {Gauthier}, {Gerbino},
  {Ghosh}, {Giard}, {Giraud-H{\'e}raud}, {Giusarma}, {Gjerl{\o}w},
  {Gonz{\'a}lez-Nuevo}, {G{\'o}rski}, {Gratton}, {Gregorio}, {Gruppuso},
  {Gudmundsson}, {Hamann}, {Hansen}, {Hanson}, {Harrison}, {Helou},
  {Henrot-Versill{\'e}}, {Hern{\'a}ndez-Monteagudo}, {Herranz}, {Hildebrandt},
  {Hivon}, {Hobson}, {Holmes}, {Hornstrup}, {Hovest}, {Huang}, {Huffenberger},
  {Hurier}, {Jaffe}, {Jaffe}, {Jones}, {Juvela}, {Keih{\"a}nen}, {Keskitalo},
  {Kisner}, {Kneissl}, {Knoche}, {Knox}, {Kunz}, {Kurki-Suonio}, {Lagache},
  {L{\"a}hteenm{\"a}ki}, {Lamarre}, {Lasenby}, {Lattanzi}, {Lawrence}, {Leahy},
  {Leonardi}, {Lesgourgues}, {Levrier}, {Lewis}, {Liguori}, {Lilje},
  {Linden-V{\o}rnle}, {L{\'o}pez-Caniego}, {Lubin}, {Mac{\'\i}as-P{\'e}rez},
  {Maggio}, {Maino}, {Mandolesi}, {Mangilli}, {Marchini}, {Maris}, {Martin},
  {Martinelli}, {Mart{\'\i}nez-Gonz{\'a}lez}, {Masi}, {Matarrese}, {McGehee},
  {Meinhold}, {Melchiorri}, {Melin}, {Mendes}, {Mennella}, {Migliaccio},
  {Millea}, {Mitra}, {Miville-Desch{\^e}nes}, {Moneti}, {Montier}, {Morgante},
  {Mortlock}, {Moss}, {Munshi}, {Murphy}, {Naselsky}, {Nati}, {Natoli},
  {Netterfield}, {N{\o}rgaard-Nielsen}, {Noviello}, {Novikov}, {Novikov},
  {Oxborrow}, {Paci}, {Pagano}, {Pajot}, {Paladini}, {Paoletti}, {Partridge},
  {Pasian}, {Patanchon}, {Pearson}, {Perdereau}, {Perotto}, {Perrotta},
  {Pettorino}, {Piacentini}, {Piat}, {Pierpaoli}, {Pietrobon}, {Plaszczynski},
  {Pointecouteau}, {Polenta}, {Popa}, {Pratt}, {Pr{\'e}zeau}, {Prunet},
  {Puget}, {Rachen}, {Reach}, {Rebolo}, {Reinecke}, {Remazeilles}, {Renault},
  {Renzi}, {Ristorcelli}, {Rocha}, {Rosset}, {Rossetti}, {Roudier},
  {Rouill{\'e} d'Orfeuil}, {Rowan-Robinson}, {Rubi{\~n}o-Mart{\'\i}n},
  {Rusholme}, {Said}, {Salvatelli}, {Salvati}, {Sandri}, {Santos},
  {Savelainen}, {Savini}, {Scott}, {Seiffert}, {Serra}, {Shellard}, {Spencer},
  {Spinelli}, {Stolyarov}, {Stompor}, {Sudiwala}, {Sunyaev}, {Sutton},
  {Suur-Uski}, {Sygnet}, {Tauber}, {Terenzi}, {Toffolatti}, {Tomasi},
  {Tristram}, {Trombetti}, {Tucci}, {Tuovinen}, {T{\"u}rler}, {Umana},
  {Valenziano}, {Valiviita}, {Van Tent}, {Vielva}, {Villa}, {Wade}, {Wandelt},
  {Wehus}, {White}, {White}, {Wilkinson}, {Yvon}, {Zacchei}, \&
  {Zonca}}]{Planck2016}
{Planck Collaboration}, {Ade}, P.~A.~R., {Aghanim}, N., {et~al.} 2016, \aap,
  594, A13, \dodoi{10.1051/0004-6361/201525830}

\bibitem[{{Planck Collaboration} {et~al.}(2020){Planck Collaboration},
  {Aghanim}, {Akrami}, {Ashdown}, {Aumont}, {Baccigalupi}, {Ballardini},
  {Banday}, {Barreiro}, {Bartolo}, {Basak}, {Battye}, {Benabed}, {Bernard},
  {Bersanelli}, {Bielewicz}, {Bock}, {Bond}, {Borrill}, {Bouchet}, {Boulanger},
  {Bucher}, {Burigana}, {Butler}, {Calabrese}, {Cardoso}, {Carron},
  {Challinor}, {Chiang}, {Chluba}, {Colombo}, {Combet}, {Contreras}, {Crill},
  {Cuttaia}, {de Bernardis}, {de Zotti}, {Delabrouille}, {Delouis}, {Di
  Valentino}, {Diego}, {Dor{\'e}}, {Douspis}, {Ducout}, {Dupac}, {Dusini},
  {Efstathiou}, {Elsner}, {En{\ss}lin}, {Eriksen}, {Fantaye}, {Farhang},
  {Fergusson}, {Fernandez-Cobos}, {Finelli}, {Forastieri}, {Frailis},
  {Fraisse}, {Franceschi}, {Frolov}, {Galeotta}, {Galli}, {Ganga},
  {G{\'e}nova-Santos}, {Gerbino}, {Ghosh}, {Gonz{\'a}lez-Nuevo}, {G{\'o}rski},
  {Gratton}, {Gruppuso}, {Gudmundsson}, {Hamann}, {Handley}, {Hansen},
  {Herranz}, {Hildebrandt}, {Hivon}, {Huang}, {Jaffe}, {Jones}, {Karakci},
  {Keih{\"a}nen}, {Keskitalo}, {Kiiveri}, {Kim}, {Kisner}, {Knox},
  {Krachmalnicoff}, {Kunz}, {Kurki-Suonio}, {Lagache}, {Lamarre}, {Lasenby},
  {Lattanzi}, {Lawrence}, {Le Jeune}, {Lemos}, {Lesgourgues}, {Levrier},
  {Lewis}, {Liguori}, {Lilje}, {Lilley}, {Lindholm}, {L{\'o}pez-Caniego},
  {Lubin}, {Ma}, {Mac{\'\i}as-P{\'e}rez}, {Maggio}, {Maino}, {Mandolesi},
  {Mangilli}, {Marcos-Caballero}, {Maris}, {Martin}, {Martinelli},
  {Mart{\'\i}nez-Gonz{\'a}lez}, {Matarrese}, {Mauri}, {McEwen}, {Meinhold},
  {Melchiorri}, {Mennella}, {Migliaccio}, {Millea}, {Mitra},
  {Miville-Desch{\^e}nes}, {Molinari}, {Montier}, {Morgante}, {Moss}, {Natoli},
  {N{\o}rgaard-Nielsen}, {Pagano}, {Paoletti}, {Partridge}, {Patanchon},
  {Peiris}, {Perrotta}, {Pettorino}, {Piacentini}, {Polastri}, {Polenta},
  {Puget}, {Rachen}, {Reinecke}, {Remazeilles}, {Renzi}, {Rocha}, {Rosset},
  {Roudier}, {Rubi{\~n}o-Mart{\'\i}n}, {Ruiz-Granados}, {Salvati}, {Sandri},
  {Savelainen}, {Scott}, {Shellard}, {Sirignano}, {Sirri}, {Spencer},
  {Sunyaev}, {Suur-Uski}, {Tauber}, {Tavagnacco}, {Tenti}, {Toffolatti},
  {Tomasi}, {Trombetti}, {Valenziano}, {Valiviita}, {Van Tent}, {Vibert},
  {Vielva}, {Villa}, {Vittorio}, {Wandelt}, {Wehus}, {White}, {White},
  {Zacchei}, \& {Zonca}}]{Planck2020}
{Planck Collaboration}, {Aghanim}, N., {Akrami}, Y., {et~al.} 2020, \aap, 641,
  A6, \dodoi{10.1051/0004-6361/201833910}

\bibitem[{{Raiter} {et~al.}(2010{\natexlab{a}}){Raiter}, {Schaerer}, \&
  {Fosbury}}]{Raiter2010}
{Raiter}, A., {Schaerer}, D., \& {Fosbury}, R.~A.~E. 2010{\natexlab{a}}, \aap,
  523, A64, \dodoi{10.1051/0004-6361/201015236}

\bibitem[{{Raiter} {et~al.}(2010{\natexlab{b}}){Raiter}, {Schaerer}, \&
  {Fosbury}}]{Raiter2010b}
---. 2010{\natexlab{b}}, \aap, 523, A64, \dodoi{10.1051/0004-6361/201015236}

\bibitem[{{Rasmussen Cueto} {et~al.}(2023){Rasmussen Cueto}, {Hutter}, {Dayal},
  {Gottl{\"o}ber}, {Heintz}, {Mason}, {Trebitsch}, \& {Yepes}}]{Rasmussen2023}
{Rasmussen Cueto}, E., {Hutter}, A., {Dayal}, P., {et~al.} 2023, arXiv
  e-prints, arXiv:2312.12109, \dodoi{10.48550/arXiv.2312.12109}

\bibitem[{{Rawle} {et~al.}(2022){Rawle}, {Giardino}, {Franz}, {Rapp}, {te
  Plate}, {Zincke}, {Abul-Huda}, {Alves de Oliveira}, {Bechtold}, {Beck},
  {Birkmann}, {B{\"o}ker}, {Ehrenwinkler}, {Ferruit}, {Garland}, {Jakobsen},
  {Karakla}, {Karl}, {Keyes}, {Koehler}, {Nimisha}, {L{\"u}tzgendorf},
  {Manjavacas}, {Marston}, {Moseley}, {Mosner}, {Muzerolle}, {Ogle},
  {Proffitt}, {Sabbi}, {Sirianni}, {Wahlgren}, {Wislowski}, {Wright}, {Wu}, \&
  {Zeidler}}]{Rawle2022}
{Rawle}, T.~D., {Giardino}, G., {Franz}, D.~E., {et~al.} 2022, in Society of
  Photo-Optical Instrumentation Engineers (SPIE) Conference Series, Vol. 12180,
  Space Telescopes and Instrumentation 2022: Optical, Infrared, and Millimeter
  Wave, ed. L.~E. {Coyle}, S.~{Matsuura}, \& M.~D. {Perrin}, 121803R,
  \dodoi{10.1117/12.2629231}

\bibitem[{{Rieke} {et~al.}(2005){Rieke}, {Kelly}, \& {Horner}}]{Rieke2005}
{Rieke}, M.~J., {Kelly}, D., \& {Horner}, S. 2005, in Society of Photo-Optical
  Instrumentation Engineers (SPIE) Conference Series, Vol. 5904, Cryogenic
  Optical Systems and Instruments XI, ed. J.~B. {Heaney} \& L.~G. {Burriesci},
  1--8, \dodoi{10.1117/12.615554}

\bibitem[{{Rieke} {et~al.}(2023){Rieke}, {Kelly}, {Misselt}, {Stansberry},
  {Boyer}, {Beatty}, {Egami}, {Florian}, {Greene}, {Hainline}, {Leisenring},
  {Roellig}, {Schlawin}, {Sun}, {Tinnin}, {Williams}, {Willmer}, {Wilson},
  {Clark}, {Rohrbach}, {Brooks}, {Canipe}, {Correnti}, {DiFelice}, {Gennaro},
  {Girard}, {Hartig}, {Hilbert}, {Koekemoer}, {Nikolov}, {Pirzkal}, {Rest},
  {Robberto}, {Sunnquist}, {Telfer}, {Wu}, {Ferry}, {Lewis}, {Baum},
  {Beichman}, {Doyon}, {Dressler}, {Eisenstein}, {Ferrarese}, {Hodapp},
  {Horner}, {Jaffe}, {Johnstone}, {Krist}, {Martin}, {McCarthy}, {Meyer},
  {Rieke}, {Trauger}, \& {Young}}]{Rieke2023}
{Rieke}, M.~J., {Kelly}, D.~M., {Misselt}, K., {et~al.} 2023, \pasp, 135,
  028001, \dodoi{10.1088/1538-3873/acac53}

\bibitem[{{Rigby} {et~al.}(2015){Rigby}, {Bayliss}, {Gladders}, {Sharon},
  {Wuyts}, {Dahle}, {Johnson}, \& {Pe{\~n}a-Guerrero}}]{Rigby2015}
{Rigby}, J.~R., {Bayliss}, M.~B., {Gladders}, M.~D., {et~al.} 2015, \apjl, 814,
  L6, \dodoi{10.1088/2041-8205/814/1/L6}

\bibitem[{{Roberts-Borsani} {et~al.}(2024){Roberts-Borsani}, {Treu}, {Shapley},
  {Fontana}, {Pentericci}, {Castellano}, {Morishita}, {Bergamini}, \&
  {Rosati}}]{Roberts-Borsani2024}
{Roberts-Borsani}, G., {Treu}, T., {Shapley}, A., {et~al.} 2024, arXiv
  e-prints, arXiv:2403.07103, \dodoi{10.48550/arXiv.2403.07103}

\bibitem[{{Rodrigo} \& {Solano}(2020)}]{Rodrigo2020-SVO}
{Rodrigo}, C., \& {Solano}, E. 2020, in XIV.0 Scientific Meeting (virtual) of
  the Spanish Astronomical Society, 182

\bibitem[{{Rogers} {et~al.}(2013){Rogers}, {McLure}, \& {Dunlop}}]{Rogers2013}
{Rogers}, A.~B., {McLure}, R.~J., \& {Dunlop}, J.~S. 2013, \mnras, 429, 2456,
  \dodoi{10.1093/mnras/sts515}

\bibitem[{{Rogers} {et~al.}(2014){Rogers}, {McLure}, {Dunlop}, {Bowler},
  {Curtis-Lake}, {Dayal}, {Faber}, {Ferguson}, {Finkelstein}, {Grogin},
  {Hathi}, {Kocevski}, {Koekemoer}, \& {Kurczynski}}]{Rogers2014}
{Rogers}, A.~B., {McLure}, R.~J., {Dunlop}, J.~S., {et~al.} 2014, \mnras, 440,
  3714, \dodoi{10.1093/mnras/stu558}

\bibitem[{{Salim} \& {Narayanan}(2020)}]{Salim2020}
{Salim}, S., \& {Narayanan}, D. 2020, \araa, 58, 529,
  \dodoi{10.1146/annurev-astro-032620-021933}

\bibitem[{{Salpeter}(1955)}]{Salpeter1955}
{Salpeter}, E.~E. 1955, \apj, 121, 161, \dodoi{10.1086/145971}

\bibitem[{{Sanders} {et~al.}(2024){Sanders}, {Shapley}, {Topping}, {Reddy}, \&
  {Brammer}}]{Sanders2024}
{Sanders}, R.~L., {Shapley}, A.~E., {Topping}, M.~W., {Reddy}, N.~A., \&
  {Brammer}, G.~B. 2024, \apj, 962, 24, \dodoi{10.3847/1538-4357/ad15fc}

\bibitem[{{Sarangi} {et~al.}(2019){Sarangi}, {Dwek}, \&
  {Kazanas}}]{Sarangi2019}
{Sarangi}, A., {Dwek}, E., \& {Kazanas}, D. 2019, \apj, 885, 126,
  \dodoi{10.3847/1538-4357/ab46a9}

\bibitem[{{Sarmento} {et~al.}(2018){Sarmento}, {Scannapieco}, \&
  {Cohen}}]{Sarmento2018}
{Sarmento}, R., {Scannapieco}, E., \& {Cohen}, S. 2018, \apj, 854, 75,
  \dodoi{10.3847/1538-4357/aa989a}

\bibitem[{{Sarmento} {et~al.}(2019){Sarmento}, {Scannapieco}, \&
  {C{\^o}t{\'e}}}]{Sarmento2019}
{Sarmento}, R., {Scannapieco}, E., \& {C{\^o}t{\'e}}, B. 2019, \apj, 871, 206,
  \dodoi{10.3847/1538-4357/aafa1a}

\bibitem[{{Saxena} {et~al.}(2020){Saxena}, {Pentericci}, {Schaerer},
  {Schneider}, {Amorin}, {Bongiorno}, {Calabr{\`o}}, {Castellano}, {Cimatti},
  {Cullen}, {Fontana}, {Fynbo}, {Hathi}, {McLeod}, {Talia}, \&
  {Zamorani}}]{Saxena2020}
{Saxena}, A., {Pentericci}, L., {Schaerer}, D., {et~al.} 2020, \mnras, 496,
  3796, \dodoi{10.1093/mnras/staa1805}

\bibitem[{{Saxena} {et~al.}(2023{\natexlab{a}}){Saxena}, {Bunker}, {Jones},
  {Stark}, {Cameron}, {Witstok}, {Arribas}, {Baker}, {Baum}, {Bhatawdekar},
  {Bowler}, {Boyett}, {Carniani}, {Charlot}, {Chevallard}, {Curti},
  {Curtis-Lake}, {Eisenstein}, {Endsley}, {Hainline}, {Helton}, {Johnson},
  {Kumari}, {Looser}, {Maiolino}, {Rieke}, {Rix}, {Robertson}, {Sandles},
  {Simmonds}, {Smit}, {Tacchella}, {Williams}, {Willmer}, \&
  {Willott}}]{Saxena2023a}
{Saxena}, A., {Bunker}, A.~J., {Jones}, G.~C., {et~al.} 2023{\natexlab{a}},
  arXiv e-prints, arXiv:2306.04536, \dodoi{10.48550/arXiv.2306.04536}

\bibitem[{{Saxena} {et~al.}(2023{\natexlab{b}}){Saxena}, {Robertson}, {Bunker},
  {Endsley}, {Cameron}, {Charlot}, {Simmonds}, {Tacchella}, {Witstok},
  {Willott}, {Carniani}, {Curtis-Lake}, {Ferruit}, {Jakobsen}, {Arribas},
  {Chevallard}, {Curti}, {D'Eugenio}, {De Graaff}, {Jones}, {Looser}, {Maseda},
  {Rawle}, {Rix}, {Del Pino}, {Smit}, {{\"U}bler}, {Eisenstein}, {Hainline},
  {Hausen}, {Johnson}, {Rieke}, {Williams}, {Willmer}, {Baker}, {Bhatawdekar},
  {Bowler}, {Boyett}, {Chen}, {Egami}, {Ji}, {Kumari}, {Nelson}, {Perna},
  {Sandles}, {Scholtz}, \& {Shivaei}}]{Saxena2023b}
{Saxena}, A., {Robertson}, B.~E., {Bunker}, A.~J., {et~al.} 2023{\natexlab{b}},
  \aap, 678, A68, \dodoi{10.1051/0004-6361/202346245}

\bibitem[{{Schaerer}(2002)}]{Schaerer2002}
{Schaerer}, D. 2002, \aap, 382, 28, \dodoi{10.1051/0004-6361:20011619}

\bibitem[{{Schaerer}(2003)}]{Schaerer2003}
---. 2003, \aap, 397, 527, \dodoi{10.1051/0004-6361:20021525}

\bibitem[{{Schmidt} {et~al.}(2017){Schmidt}, {Huang}, {Treu}, {Hoag},
  {Brada{\v{c}}}, {Henry}, {Jones}, {Mason}, {Malkan}, {Morishita},
  {Pentericci}, {Trenti}, {Vulcani}, \& {Wang}}]{Schmidt2017}
{Schmidt}, K.~B., {Huang}, K.~H., {Treu}, T., {et~al.} 2017, \apj, 839, 17,
  \dodoi{10.3847/1538-4357/aa68a3}

\bibitem[{{Schneider} \& {Maiolino}(2023)}]{Schneider2023}
{Schneider}, R., \& {Maiolino}, R. 2023, arXiv e-prints, arXiv:2310.00053,
  \dodoi{10.48550/arXiv.2310.00053}

\bibitem[{{Senchyna} {et~al.}(2023){Senchyna}, {Plat}, {Stark}, \&
  {Rudie}}]{Senchyna2023}
{Senchyna}, P., {Plat}, A., {Stark}, D.~P., \& {Rudie}, G.~C. 2023, arXiv
  e-prints, arXiv:2303.04179, \dodoi{10.48550/arXiv.2303.04179}

\bibitem[{{Skelton} {et~al.}(2014){Skelton}, {Whitaker}, {Momcheva}, {Brammer},
  {van Dokkum}, {Labb{\'e}}, {Franx}, {van der Wel}, {Bezanson}, {Da Cunha},
  {Fumagalli}, {F{\"o}rster Schreiber}, {Kriek}, {Leja}, {Lundgren}, {Magee},
  {Marchesini}, {Maseda}, {Nelson}, {Oesch}, {Pacifici}, {Patel}, {Price},
  {Rix}, {Tal}, {Wake}, \& {Wuyts}}]{Skelton2014}
{Skelton}, R.~E., {Whitaker}, K.~E., {Momcheva}, I.~G., {et~al.} 2014, \apjs,
  214, 24, \dodoi{10.1088/0067-0049/214/2/24}

\bibitem[{{Stanway} \& {Eldridge}(2018)}]{Stanway2018-BPASS}
{Stanway}, E.~R., \& {Eldridge}, J.~J. 2018, \mnras, 479, 75,
  \dodoi{10.1093/mnras/sty1353}

\bibitem[{{Stark} {et~al.}(2017){Stark}, {Ellis}, {Charlot}, {Chevallard},
  {Tang}, {Belli}, {Zitrin}, {Mainali}, {Gutkin}, {Vidal-Garc{\'\i}a},
  {Bouwens}, \& {Oesch}}]{Stark2017}
{Stark}, D.~P., {Ellis}, R.~S., {Charlot}, S., {et~al.} 2017, \mnras, 464, 469,
  \dodoi{10.1093/mnras/stw2233}

\bibitem[{{Stecher} \& {Donn}(1965)}]{Stecher1965}
{Stecher}, T.~P., \& {Donn}, B. 1965, \apj, 142, 1681, \dodoi{10.1086/148461}

\bibitem[{{Stefanon} {et~al.}(2017){Stefanon}, {Labb{\'e}}, {Bouwens},
  {Brammer}, {Oesch}, {Franx}, {Fynbo}, {Milvang-Jensen}, {Muzzin},
  {Illingworth}, {Le F{\`e}vre}, {Caputi}, {Holwerda}, {McCracken}, {Smit}, \&
  {Magee}}]{Stefanon2017}
{Stefanon}, M., {Labb{\'e}}, I., {Bouwens}, R.~J., {et~al.} 2017, \apj, 851,
  43, \dodoi{10.3847/1538-4357/aa9a40}

\bibitem[{{Steinhardt} {et~al.}(2023){Steinhardt}, {Kokorev}, {Rusakov},
  {Garcia}, \& {Sneppen}}]{Steinhardt2023-HOT-IMFs}
{Steinhardt}, C.~L., {Kokorev}, V., {Rusakov}, V., {Garcia}, E., \& {Sneppen},
  A. 2023, \apjl, 951, L40, \dodoi{10.3847/2041-8213/acdef6}

\bibitem[{{Tacchella} {et~al.}(2022){Tacchella}, {Finkelstein}, {Bagley},
  {Dickinson}, {Ferguson}, {Giavalisco}, {Graziani}, {Grogin}, {Hathi},
  {Hutchison}, {Jung}, {Koekemoer}, {Larson}, {Papovich}, {Pirzkal},
  {Rojas-Ruiz}, {Song}, {Schneider}, {Somerville}, {Wilkins}, \&
  {Yung}}]{Tacchella2022}
{Tacchella}, S., {Finkelstein}, S.~L., {Bagley}, M., {et~al.} 2022, \apj, 927,
  170, \dodoi{10.3847/1538-4357/ac4cad}

\bibitem[{{Tang} {et~al.}(2023){Tang}, {Stark}, {Chen}, {Mason}, {Topping},
  {Endsley}, {Senchyna}, {Plat}, {Lu}, {Whitler}, {Robertson}, \&
  {Charlot}}]{Tang2023}
{Tang}, M., {Stark}, D.~P., {Chen}, Z., {et~al.} 2023, \mnras, 526, 1657,
  \dodoi{10.1093/mnras/stad2763}

\bibitem[{{Tang} {et~al.}(2024){Tang}, {Stark}, {Ellis}, {Sun}, {Topping},
  {Robertson}, {Tacchella}, {Arribas}, {Baker}, {Bhatawdekar}, {Boyett},
  {Bunker}, {Charlot}, {Chen}, {Chevallard}, {Jones}, {Kumari}, {Lyu},
  {Maiolino}, {Maseda}, {Saxena}, {Whitler}, {Williams}, {Willott}, \&
  {Witstok}}]{Tang2024}
{Tang}, M., {Stark}, D.~P., {Ellis}, R.~S., {et~al.} 2024, arXiv e-prints,
  arXiv:2402.06070, \dodoi{10.48550/arXiv.2402.06070}

\bibitem[{{Tepper-Garc{\'\i}a}(2006)}]{Tepper-Garcia2006}
{Tepper-Garc{\'\i}a}, T. 2006, \mnras, 369, 2025,
  \dodoi{10.1111/j.1365-2966.2006.10450.x}

\bibitem[{{Todini} \& {Ferrara}(2001)}]{Todini2001}
{Todini}, P., \& {Ferrara}, A. 2001, \mnras, 325, 726,
  \dodoi{10.1046/j.1365-8711.2001.04486.x}

\bibitem[{{Tomczak} {et~al.}(2014){Tomczak}, {Quadri}, {Tran}, {Labb{\'e}},
  {Straatman}, {Papovich}, {Glazebrook}, {Allen}, {Brammer}, {Kacprzak},
  {Kawinwanichakij}, {Kelson}, {McCarthy}, {Mehrtens}, {Monson}, {Persson},
  {Spitler}, {Tilvi}, \& {van Dokkum}}]{Tomczak2014}
{Tomczak}, A.~R., {Quadri}, R.~F., {Tran}, K.-V.~H., {et~al.} 2014, \apj, 783,
  85, \dodoi{10.1088/0004-637X/783/2/85}

\bibitem[{{Topping} {et~al.}(2022){Topping}, {Stark}, {Endsley}, {Plat},
  {Whitler}, {Chen}, \& {Charlot}}]{Topping2022}
{Topping}, M.~W., {Stark}, D.~P., {Endsley}, R., {et~al.} 2022, arXiv e-prints,
  arXiv:2208.01610.
\newblock \doarXiv{2208.01610}

\bibitem[{{Topping} {et~al.}(2023){Topping}, {Stark}, {Endsley}, {Whitler},
  {Hainline}, {Johnson}, {Robertson}, {Tacchella}, {Chen}, {Alberts}, {Baker},
  {Bunker}, {Carniani}, {Charlot}, {Chevallard}, {Curtis-Lake}, {DeCoursey},
  {Egami}, {Eisenstein}, {Ji}, {Maiolino}, {Williams}, {Willmer}, {Willott}, \&
  {Witstok}}]{Topping2023}
---. 2023, arXiv e-prints, arXiv:2307.08835, \dodoi{10.48550/arXiv.2307.08835}

\bibitem[{{Topping} {et~al.}(2024){Topping}, {Stark}, {Senchyna}, {Plat},
  {Zitrin}, {Endsley}, {Charlot}, {Furtak}, {Maseda}, {Smit}, {Mainali},
  {Chevallard}, {Molyneux}, \& {Rigby}}]{Topping2024}
{Topping}, M.~W., {Stark}, D.~P., {Senchyna}, P., {et~al.} 2024, arXiv
  e-prints, arXiv:2401.08764, \dodoi{10.48550/arXiv.2401.08764}

\bibitem[{{Totani} {et~al.}(2006){Totani}, {Kawai}, {Kosugi}, {Aoki}, {Yamada},
  {Iye}, {Ohta}, \& {Hattori}}]{Totani2006}
{Totani}, T., {Kawai}, N., {Kosugi}, G., {et~al.} 2006, \pasj, 58, 485,
  \dodoi{10.1093/pasj/58.3.485}

\bibitem[{{Treu} {et~al.}(2022){Treu}, {Roberts-Borsani}, {Bradac}, {Brammer},
  {Fontana}, {Henry}, {Mason}, {Morishita}, {Pentericci}, {Wang}, {Acebron},
  {Bagley}, {Bergamini}, {Belfiori}, {Bonchi}, {Boyett}, {Boutsia},
  {Calabr{\'o}}, {Caminha}, {Castellano}, {Dressler}, {Glazebrook}, {Grillo},
  {Jacobs}, {Jones}, {Kelly}, {Leethochawalit}, {Malkan}, {Marchesini},
  {Mascia}, {Mercurio}, {Merlin}, {Nanayakkara}, {Nonino}, {Paris},
  {Poggianti}, {Rosati}, {Santini}, {Scarlata}, {Shipley}, {Strait}, {Trenti},
  {Tubthong}, {Vanzella}, {Vulcani}, \& {Yang}}]{Treu2022-GLASS}
{Treu}, T., {Roberts-Borsani}, G., {Bradac}, M., {et~al.} 2022, \apj, 935, 110,
  \dodoi{10.3847/1538-4357/ac8158}

\bibitem[{{Umeda} {et~al.}(2023){Umeda}, {Ouchi}, {Nakajima}, {Harikane},
  {Ono}, {Xu}, {Isobe}, \& {Zhang}}]{Umeda2023}
{Umeda}, H., {Ouchi}, M., {Nakajima}, K., {et~al.} 2023, arXiv e-prints,
  arXiv:2306.00487, \dodoi{10.48550/arXiv.2306.00487}

\bibitem[{{Valiante} {et~al.}(2014){Valiante}, {Schneider}, {Salvadori}, \&
  {Gallerani}}]{Valiante2014}
{Valiante}, R., {Schneider}, R., {Salvadori}, S., \& {Gallerani}, S. 2014,
  \mnras, 444, 2442, \dodoi{10.1093/mnras/stu1613}

\bibitem[{Vallenari {et~al.}(2022)Vallenari, Brown, \& Prusti}]{GAIA-DR3}
Vallenari, A., Brown, A., \& Prusti, T. 2022, Astronomy \& Astrophysics

\bibitem[{{V{\'a}zquez} \& {Leitherer}(2005)}]{Vazquez2005}
{V{\'a}zquez}, G.~A., \& {Leitherer}, C. 2005, \apj, 621, 695,
  \dodoi{10.1086/427866}

\bibitem[{{Vijayan} {et~al.}(2019){Vijayan}, {Clay}, {Thomas}, {Yates},
  {Wilkins}, \& {Henriques}}]{Vijayan2019}
{Vijayan}, A.~P., {Clay}, S.~J., {Thomas}, P.~A., {et~al.} 2019, \mnras, 489,
  4072, \dodoi{10.1093/mnras/stz1948}

\bibitem[{{Vijayan} {et~al.}(2021){Vijayan}, {Lovell}, {Wilkins}, {Thomas},
  {Barnes}, {Irodotou}, {Kuusisto}, \& {Roper}}]{Vijayan2021_FLARES_II}
{Vijayan}, A.~P., {Lovell}, C.~C., {Wilkins}, S.~M., {et~al.} 2021, \mnras,
  501, 3289, \dodoi{10.1093/mnras/staa3715}

\bibitem[{Virtanen {et~al.}(2020)Virtanen, Gommers, Oliphant, Haberland, Reddy,
  Cournapeau, Burovski, Peterson, Weckesser, Bright, {van der Walt}, Brett,
  Wilson, Millman, Mayorov, Nelson, Jones, Kern, Larson, Carey, Polat, Feng,
  Moore, {VanderPlas}, Laxalde, Perktold, Cimrman, Henriksen, Quintero, Harris,
  Archibald, Ribeiro, Pedregosa, {van Mulbregt}, \& {SciPy 1.0
  Contributors}}]{2020SciPy-NMeth}
Virtanen, P., Gommers, R., Oliphant, T.~E., {et~al.} 2020, Nature Methods, 17,
  261, \dodoi{10.1038/s41592-019-0686-2}

\bibitem[{{Wang} {et~al.}(2023){Wang}, {Fujimoto}, {Labb{\'e}}, {Furtak},
  {Miller}, {Setton}, {Zitrin}, {Atek}, {Bezanson}, {Brammer}, {Leja}, {Oesch},
  {Price}, {Chemerynska}, {Cutler}, {Dayal}, {van Dokkum}, {Goulding},
  {Greene}, {Fudamoto}, {Khullar}, {Kokorev}, {Marchesini}, {Pan}, {Weaver},
  {Whitaker}, \& {Williams}}]{Wang2023}
{Wang}, B., {Fujimoto}, S., {Labb{\'e}}, I., {et~al.} 2023, \apjl, 957, L34,
  \dodoi{10.3847/2041-8213/acfe07}

\bibitem[{{Whitaker} {et~al.}(2019){Whitaker}, {Ashas}, {Illingworth}, {Magee},
  {Leja}, {Oesch}, {van Dokkum}, {Mowla}, {Bouwens}, {Franx}, {Holden},
  {Labb{\'e}}, {Rafelski}, {Teplitz}, \& {Gonzalez}}]{Whitaker2019-HLF}
{Whitaker}, K.~E., {Ashas}, M., {Illingworth}, G., {et~al.} 2019, \apjs, 244,
  16, \dodoi{10.3847/1538-4365/ab3853}

\bibitem[{{Whitler} {et~al.}(2023){Whitler}, {Stark}, {Endsley}, {Chen},
  {Mason}, {Topping}, \& {Charlot}}]{Whitler2023c}
{Whitler}, L., {Stark}, D.~P., {Endsley}, R., {et~al.} 2023, arXiv e-prints,
  arXiv:2305.16670, \dodoi{10.48550/arXiv.2305.16670}

\bibitem[{{Wilkins} {et~al.}(2016){Wilkins}, {Bouwens}, {Oesch}, {Labb{\'e}},
  {Sargent}, {Caruana}, {Wardlow}, \& {Clay}}]{Wilkins2016}
{Wilkins}, S.~M., {Bouwens}, R.~J., {Oesch}, P.~A., {et~al.} 2016, \mnras, 455,
  659, \dodoi{10.1093/mnras/stv2263}

\bibitem[{{Wilkins} {et~al.}(2013){Wilkins}, {Bunker}, {Coulton}, {Croft}, {di
  Matteo}, {Khandai}, \& {Feng}}]{Wilkins2013}
{Wilkins}, S.~M., {Bunker}, A., {Coulton}, W., {et~al.} 2013, \mnras, 430,
  2885, \dodoi{10.1093/mnras/stt096}

\bibitem[{{Wilkins} {et~al.}(2012){Wilkins}, {Gonzalez-Perez}, {Lacey}, \&
  {Baugh}}]{Wilkins2012}
{Wilkins}, S.~M., {Gonzalez-Perez}, V., {Lacey}, C.~G., \& {Baugh}, C.~M. 2012,
  \mnras, 424, 1522, \dodoi{10.1111/j.1365-2966.2012.21344.x}

\bibitem[{{Wilkins} {et~al.}(2023{\natexlab{a}}){Wilkins}, {Vijayan}, {Lovell},
  {Roper}, {Irodotou}, {Caruana}, {Seeyave}, {Kuusisto}, {Thomas}, \&
  {Parris}}]{Wilkins2023_FLARES_V}
{Wilkins}, S.~M., {Vijayan}, A.~P., {Lovell}, C.~C., {et~al.}
  2023{\natexlab{a}}, \mnras, 519, 3118, \dodoi{10.1093/mnras/stac3280}

\bibitem[{{Wilkins} {et~al.}(2023{\natexlab{b}}){Wilkins}, {Turner}, {Bagley},
  {Finkelstein}, {Amor{\'\i}n}, {Hautefort}, {Behroozi}, {Bhatawdekar},
  {Dekel}, {Donnellan}, {Drakos}, {Fortuni}, {Hathi}, {Hirschmann}, {Holwerda},
  {Irodotou}, {Koekemoer}, {Lovell}, {Merlin}, {Roper}, {Seeyave}, {Vijayan},
  \& {Yung}}]{Wilkins2023-colors}
{Wilkins}, S.~M., {Turner}, J.~C., {Bagley}, M.~B., {et~al.}
  2023{\natexlab{b}}, arXiv e-prints, arXiv:2311.08065,
  \dodoi{10.48550/arXiv.2311.08065}

\bibitem[{{Williams} {et~al.}(2018){Williams}, {Curtis-Lake}, {Hainline},
  {Chevallard}, {Robertson}, {Charlot}, {Endsley}, {Stark}, {Willmer},
  {Alberts}, {Amorin}, {Arribas}, {Baum}, {Bunker}, {Carniani}, {Crandall},
  {Egami}, {Eisenstein}, {Ferruit}, {Husemann}, {Maseda}, {Maiolino}, {Rawle},
  {Rieke}, {Smit}, {Tacchella}, \& {Willott}}]{Williams2018}
{Williams}, C.~C., {Curtis-Lake}, E., {Hainline}, K.~N., {et~al.} 2018, \apjs,
  236, 33, \dodoi{10.3847/1538-4365/aabcbb}

\bibitem[{{Windhorst} {et~al.}(2023){Windhorst}, {Cohen}, {Jansen}, {Summers},
  {Tompkins}, {Conselice}, {Driver}, {Yan}, {Coe}, {Frye}, {Grogin},
  {Koekemoer}, {Marshall}, {O'Brien}, {Pirzkal}, {Robotham}, {Ryan}, {Willmer},
  {Carleton}, {Diego}, {Keel}, {Porto}, {Redshaw}, {Scheller}, {Wilkins},
  {Willner}, {Zitrin}, {Adams}, {Austin}, {Arendt}, {Beacom}, {Bhatawdekar},
  {Bradley}, {Broadhurst}, {Cheng}, {Civano}, {Dai}, {Dole}, {D'Silva},
  {Duncan}, {Fazio}, {Ferrami}, {Ferreira}, {Finkelstein}, {Furtak}, {Gim},
  {Griffiths}, {Hammel}, {Harrington}, {Hathi}, {Holwerda}, {Honor}, {Huang},
  {Hyun}, {Im}, {Joshi}, {Kamieneski}, {Kelly}, {Larson}, {Li}, {Lim}, {Ma},
  {Maksym}, {Manzoni}, {Meena}, {Milam}, {Nonino}, {Pascale}, {Petric},
  {Pierel}, {del Carmen Polletta}, {R{\"o}ttgering}, {Rutkowski}, {Smail},
  {Straughn}, {Strolger}, {Swirbul}, {Trussler}, {Wang}, {Welch}, {B. Wyithe},
  {Yun}, {Zackrisson}, {Zhang}, \& {Zhao}}]{Windhorst2023}
{Windhorst}, R.~A., {Cohen}, S.~H., {Jansen}, R.~A., {et~al.} 2023, \aj, 165,
  13, \dodoi{10.3847/1538-3881/aca163}

\bibitem[{{Witstok} {et~al.}(2023{\natexlab{a}}){Witstok}, {Shivaei}, {Smit},
  {Maiolino}, {Carniani}, {Curtis-Lake}, {Ferruit}, {Arribas}, {Bunker},
  {Cameron}, {Charlot}, {Chevallard}, {Curti}, {de Graaff}, {D'Eugenio},
  {Giardino}, {Looser}, {Rawle}, {Rodr{\'\i}guez del Pino}, {Willott},
  {Alberts}, {Baker}, {Boyett}, {Egami}, {Eisenstein}, {Endsley}, {Hainline},
  {Ji}, {Johnson}, {Kumari}, {Lyu}, {Nelson}, {Perna}, {Rieke}, {Robertson},
  {Sandles}, {Saxena}, {Scholtz}, {Sun}, {Tacchella}, {Williams}, \&
  {Willmer}}]{Witstok2023}
{Witstok}, J., {Shivaei}, I., {Smit}, R., {et~al.} 2023{\natexlab{a}}, \nat,
  621, 267, \dodoi{10.1038/s41586-023-06413-w}

\bibitem[{{Witstok} {et~al.}(2023{\natexlab{b}}){Witstok}, {Smit}, {Saxena},
  {Jones}, {Helton}, {Sun}, {Maiolino}, {Kumari}, {Stark}, {Bunker}, {Arribas},
  {Baker}, {Bhatawdekar}, {Boyett}, {Cameron}, {Carniani}, {Charlot},
  {Chevallard}, {Curti}, {Curtis-Lake}, {Eisenstein}, {Endsley}, {Hainline},
  {Ji}, {Johnson}, {Looser}, {Nelson}, {Perna}, {Rix}, {Robertson}, {Sandles},
  {Scholtz}, {Simmonds}, {Tacchella}, {{\"U}bler}, {Williams}, {Willmer}, \&
  {Willott}}]{Witstok2023-LAEs}
{Witstok}, J., {Smit}, R., {Saxena}, A., {et~al.} 2023{\natexlab{b}}, arXiv
  e-prints, arXiv:2306.04627, \dodoi{10.48550/arXiv.2306.04627}

\bibitem[{{Wu} {et~al.}(2020){Wu}, {Dav{\'e}}, {Tacchella}, \&
  {Lotz}}]{Wu2020-SIMBA-EoR}
{Wu}, X., {Dav{\'e}}, R., {Tacchella}, S., \& {Lotz}, J. 2020, \mnras, 494,
  5636, \dodoi{10.1093/mnras/staa1044}

\bibitem[{{Xiao} {et~al.}(2018){Xiao}, {Stanway}, \& {Eldridge}}]{Xiao2018}
{Xiao}, L., {Stanway}, E.~R., \& {Eldridge}, J.~J. 2018, \mnras, 477, 904,
  \dodoi{10.1093/mnras/sty646}

\bibitem[{{Yan} {et~al.}(2022){Yan}, {Ma}, {Ling}, {Cheng}, {Huang}, \&
  {Zitrin}}]{Yan2022}
{Yan}, H., {Ma}, Z., {Ling}, C., {et~al.} 2022, arXiv e-prints,
  arXiv:2207.11558.
\newblock \doarXiv{2207.11558}

\bibitem[{{Yung} {et~al.}(2024){Yung}, {Somerville}, {Finkelstein}, {Wilkins},
  \& {Gardner}}]{Yung2024-SC-SAM-GUREFT}
{Yung}, L.~Y.~A., {Somerville}, R.~S., {Finkelstein}, S.~L., {Wilkins}, S.~M.,
  \& {Gardner}, J.~P. 2024, \mnras, 527, 5929, \dodoi{10.1093/mnras/stad3484}

\bibitem[{{Yung} {et~al.}(2023){Yung}, {Somerville}, {Nguyen}, {Behroozi},
  {Modi}, \& {Gardner}}]{Yung2023-SC-SAM-GUREFT}
{Yung}, L.~Y.~A., {Somerville}, R.~S., {Nguyen}, T., {et~al.} 2023, arXiv
  e-prints, arXiv:2309.14408, \dodoi{10.48550/arXiv.2309.14408}

\bibitem[{{Zackrisson} {et~al.}(2013){Zackrisson}, {Inoue}, \&
  {Jensen}}]{Zackrisson2013}
{Zackrisson}, E., {Inoue}, A.~K., \& {Jensen}, H. 2013, \apj, 777, 39,
  \dodoi{10.1088/0004-637X/777/1/39}

\bibitem[{{Zackrisson} {et~al.}(2011){Zackrisson}, {Rydberg}, {Schaerer},
  {{\"O}stlin}, \& {Tuli}}]{Zackrisson2011}
{Zackrisson}, E., {Rydberg}, C.-E., {Schaerer}, D., {{\"O}stlin}, G., \&
  {Tuli}, M. 2011, \apj, 740, 13, \dodoi{10.1088/0004-637X/740/1/13}

\bibitem[{{Zhukovska} {et~al.}(2008){Zhukovska}, {Gail}, \&
  {Trieloff}}]{Zhukovska2008}
{Zhukovska}, S., {Gail}, H.~P., \& {Trieloff}, M. 2008, \aap, 479, 453,
  \dodoi{10.1051/0004-6361:20077789}

\bibitem[{{Ziparo} {et~al.}(2023){Ziparo}, {Ferrara}, {Sommovigo}, \&
  {Kohandel}}]{Ziparo2023}
{Ziparo}, F., {Ferrara}, A., {Sommovigo}, L., \& {Kohandel}, M. 2023, \mnras,
  520, 2445, \dodoi{10.1093/mnras/stad125}

\end{thebibliography}
\bibliographystyle{aasjournal}



\end{document}